\documentclass[aps,prd,twocolumn,preprintnumbers,showpacs,nofootinbib,amssymb]{revtex4}
\usepackage{graphicx}
\usepackage{amsmath}
\usepackage{amssymb}
\usepackage{amsfonts}
\usepackage{bm}
\usepackage{color}

\def\be{\begin{equation}}
\def\ee{\end{equation}}
\def\bea{\begin{eqnarray}}
\def\eea{\end{eqnarray}}

\begin{document}

\title{A predictive model of BEC dark matter halos with a solitonic core \\
and an isothermal atmosphere}
\author{Pierre-Henri Chavanis}
\email{chavanis@irsamc.ups-tlse.fr}
\affiliation{Laboratoire de Physique Th\'eorique, Universit\'e de Toulouse,
CNRS, UPS, France}

\begin{abstract}

We develop a model of Bose-Einstein condensate dark matter halos with a
solitonic core 
and an isothermal atmosphere based on a generalized Gross-Pitaevskii-Poisson 
equation [P.H. Chavanis, Eur. Phys. J. Plus {\bf 132}, 248 (2017)].  This
equation provides a heuristic  
coarse-grained parametrization of the ordinary Gross-Pitaevskii-Poisson equation
accounting for violent relaxation and gravitational cooling. It involves
a cubic nonlinearity taking into account the self-interaction of the bosons, a
logarithmic nonlinearity associated with an effective temperature, and a source
of dissipation. It leads to superfluid dark matter halos with a
core-halo structure. The quantum potential or the 
self-interaction of the bosons generates a
solitonic core that solves the cusp
problem of the cold dark matter model. The logarithmic nonlinearity generates an
 isothermal atmosphere accounting for the  flat rotation curves of the galaxies.
The
dissipation ensures that the system relaxes towards an equilibrium
configuration. In the Thomas-Fermi approximation, the dark matter halo is
equivalent to a barotropic gas  with an equation of state $P=2\pi
a_s\hbar^2\rho^2/m^3+\rho k_B T/m$, where $a_s$ is the scattering length of the
bosons and $m$ is their individual mass. We
numerically solve the equation of hydrostatic equilibrium and determine the
corresponding density profiles and rotation curves. We impose that the surface
density of the halos has the universal value $\Sigma_0=\rho_0r_h=141\,
M_{\odot}/{\rm
pc}^2$ obtained from the observations. For a boson with ratio
$a_s/m^3=3.28\times 10^3 \, {\rm fm}/({\rm eV/c^2})^3$, we find a  minimum
halo mass  $(M_h)_{\rm min}=1.86\times 10^8\,
M_{\odot}$ and a minimum halo radius $(r_h)_{\rm min}=788\, {\rm
pc}$. This ultracompact halo
corresponds to a pure soliton which is the ground state of the
Gross-Pitaevskii-Poisson equation. For  $(M_h)_{\rm min}< M_h
<(M_h)_*=3.30\times 10^{9}\, M_{\odot}$ the soliton is surrounded by a
tenuous
isothermal atmosphere without plateau. For $M_h > (M_h)_*$ we find two branches
of
solutions corresponding to (i) pure isothermal halos without soliton and (ii)
isothermal
halos harboring a central soliton and presenting a plateau. The purely
isothermal halos (gaseous phase)
are stable.  For $M_h>(M_h)_c=6.86\times 10^{10}\, M_{\odot}$, they are
indistinguishable from the observational Burkert
profile. For $(M_h)_*<M_h<(M_h)_c$, the deviation from the isothermal law (most
probable state) may be explained by incomplete violent relaxation, tidal
effects, or stochastic forcing. The isothermal halos harboring a central soliton
(core-halo phase) are canonically unstable (having a negative
specific heat) but they are microcanonically stable so they are
long-lived. By
extremizing the free energy (or entropy) with respect to the
core mass, we find that the core mass scales as $M_c/(M_h)_{\rm min}=0.626 \,
(M_h/(M_h)_{\rm min})^{1/2}\ln(M_h/(M_h)_{\rm min})$. 
For a halo of mass
$M_h=10^{12}\, M_{\odot}$, similar to the mass of the halo that surrounds our
Galaxy, the solitonic core has a mass $M_c=6.39\times 10^{10}\, M_{\odot}$
 and a
radius $R_c=1\, {\rm kpc}$. The
solitonic core cannot mimic by itself a supermassive black hole at the center of
the Galaxy but it may represent a large bulge which is
either present now or may have, in the past, triggered
the
collapse of the surrounding gas, leading to a supermassive black hole
and a quasar. On the other hand, we argue that large halos
with a mass $M_h>10^{12}\, M_{\odot}$ may undergo a gravothermal catastrophe
leading ultimately to the formation of a supermassive black hole (for smaller
halos, the gravothermal catastrophe is inhibited by quantum effects). We
relate the bifurcation point and the point above which supermassive black holes
may form to the canonical
and microcanonical critical points $(M_h)_{\rm CCP}=3.27\times
10^9\, M_{\odot}$ and $(M_h)_{\rm MCP}\sim 2\times
10^{12}\, M_{\odot}$ of the thermal self-gravitating bosonic gas.
Our model has no free parameter so
it is completely predictive. Extension of this model to noninteracting bosons
and fermions will be presented in forthcoming papers.
Preliminary calculations show that our results
are in agreement with the results of Schive {\it et al.} [Phys. Rev. Lett {\bf
113}, 261302 (2014)] for noninteracting bosons and to the results of Ruffini
{\it et al.} [Mon. Not. R.
Astron. Soc. {\bf 451}, 622 (2015)] for fermions and that they provide
a thermodynamical justification to their core mass - halo mass
relations.

\end{abstract}


\maketitle

\section{Introduction}

The nature of dark matter (DM) is still unknown and remains one of the greatest
mysteries of modern cosmology. In the standard
cold dark matter (CDM) model, DM is assumed to be made of weakly
interacting massive particles (WIMPs) with a mass in the
GeV-TeV range. They may correspond to
supersymmetric (SUSY) particles \cite{susy}. These particles freeze out from
thermal
equilibrium in the
early universe and, as a consequence of this decoupling,  cool off rapidly as
the universe expands. As a result, DM can be represented by a
pressureless
gas at zero thermodynamical temperature ($T_{\rm th}=0$) described by the
Euler-Poisson equation or as a collisionless system of particles described by
the Vlasov-Poisson equation \cite{peeblesbook}. The CDM model
works remarkably well at large (cosmological) scales and is consistent with ever
improving measurements of the cosmic microwave background (CMB) from WMAP and
Planck missions \cite{planck2013,planck2016}. However, it encounters serious
problems at small (galactic) scales. In particular, it predicts that DM halos
should be cuspy \cite{nfw}, with a density diverging as $r^{-1}$ for
$r\rightarrow 0$,  while observations reveal that they have a flat 
core density \cite{observations}. On the other hand, the CDM model predicts
an over-abundance of small-scale structures (subhalos/satellites), much more
than what is observed around the Milky Way \cite{satellites}. These problems are
referred to as the ``cusp problem'' and  ``missing satellite problem''. The
expression ``small-scale crisis of CDM'' has been coined.

The small-scale problems of the CDM model are somehow related to the
assumption that DM
is pressureless. In order to remedy this
difficulty,\footnote{Other
possibilities to solve the CDM crisis invoke (i) self-interacting CDM with a
large scattering cross section but negligible annihilation or dissipation
\cite{spergelsteinhardt}, (ii) warm dark matter (WDM) where the dispersion of
the particles is responsible for a pressure force that can halt gravitational
collapse and prevent the formation of cusps \cite{wdm}, (iii) the feedback of
baryons that can transform cusps into cores \cite{romano}.} some authors have
proposed to take into account the quantum nature
of the DM particle.

If the DM particle is a  fermion, like a  massive
neutrino, as originaly suggested in \cite{markov,cmc1,cmc2}, gravitational
collapse is prevented by
the Pauli exclusion principle.
Fermionic DM halos are described by the Fermi-Dirac distribution
\cite{gao,stella,zcls,cls,cl,ir,gmr,merafina,imrs,vtt,bvn,bmv,csmnras,bvr,pt,
dark,bmtv,btv,rieutord,ijmpb,dvs1,dvs2,vss,urbano,rar,clm1,clm2,vs1,krut,vs2,
rsu}.
The fermionic DM halos
generically have a
core-halo structure
consisting in a completely degenerate core (fermion ball) with a polytropic
equation of state
$P=({1}/{20})({3}/{\pi})^{2/3}{h^2}\rho^{5/3}/{m^{8/3}}$  and an isothermal
atmosphere with an equation of state $P=\rho k_B T/m$. The core is
stabilized by
quantum mechanics and solves the cusp problem of the CDM model.\footnote{In the
case of
large DM halos, quantum mechanics may be negligible and the core may be
stabilized by
thermal effects (see Appendix \ref{sec_eff}).} On the other
hand, the
density decreases as $r^{-2}$ in the
isothermal halo, yielding flat rotation curves in agreement with the
observations \cite{flat}. This core-halo structure has been studied in detail in
\cite{gao,stella,zcls,cls,cl,ir,gmr,merafina,imrs,vtt,bvn,bmv,csmnras,bvr,pt,
dark,bmtv,btv,rieutord,ijmpb,dvs1,dvs2,vss,urbano,rar,clm1,clm2,vs1,krut,vs2,
rsu}. The mass of the fermions must
be of the
order of 
$m=170\, {\rm eV/c^2}$ (see Appendix D of \cite{suarezchavanis3}) to account
for the
size of ultracompact DM halos like Fornax  ($R\sim 1\, {\rm kpc}$ and $M\sim
10^{8}\,
M_\odot$) interpreted as the ground state ($T=0$) of the self-gravitating Fermi
gas.

In this paper, we shall assume that the DM particle  is a boson, like an
ultralight axion (ULA) \cite{marsh}. At very low temperatures, bosons form
self-gravitating Bose-Einstein condensates (BECs). In that case, DM halos can 
be viewed as gigantic bosonic atoms  at $T_{\rm th}=0$ where the bosonic
particles are
condensed in a single
macroscopic quantum state. They are described by a scalar field (SF)   that can be interpreted as the wavefunction $\psi({\bf r},t)$
of the condensate. The bosons may be noninteracting or self-interacting. The wave properties of the
SF are negligible at large (cosmological)
scales where the SF behaves as CDM, but they become relevant at small (galactic)
scales and can prevent gravitational collapse. However, for quantum mechanics to
manifest itself at the scale of DM halos, the mass of the DM particle must be
extremely small, of the order of $m=
2.92\times 10^{-22}\, {\rm eV}/c^2$  (see below).  These ultralight particles
are not excluded by
particle physics.  This model is referred
to as wave DM, fuzzy DM (FDM), BECDM, $\psi$DM,
SFDM \cite{baldeschi,khlopov,membrado,sin,jisin,leekoh,schunckpreprint,
matosguzman,sahni,
guzmanmatos,hu,peebles,goodman,mu,arbey1,silverman1,matosall,silverman,
lesgourgues,arbey,fm1,bohmer,fm2,bmn,fm3,sikivie,mvm,lee09,ch1,lee,prd1,prd2,
prd3,briscese,
harkocosmo,harko,abrilMNRAS,aacosmo,velten,pires,park,rmbec,rindler,lora2,
abrilJCAP,mhh,lensing,glgr1,ch2,ch3,shapiro,bettoni,lora,mlbec,madarassy,
abrilph,playa,stiff,guth,souza,freitas,alexandre,schroven,pop,eby,cembranos,
braaten,davidson,schwabe,fan,calabrese,bectcoll,cotner,chavmatos,hui,zhang,
tkachevprl,suarezchavanis3,shapironew,phi6,abriljeans,desjacques} (see the
introduction of
\cite{prd1} for a short historic of this model). In this model,  gravitational
collapse is prevented by the quantum pressure arising from the Heisenberg
uncertainty principle or from the scattering of the bosons.  This leads to DM halos presenting a central
core instead of a cusp. Since the Jeans scale is finite, this suppresses
the formation of small-scale structures even at $T_{\rm th}=0$. Therefore,
quantum
mechanics may be a
way to solve the small-scale problems of the CDM model such 
as the cusp problem and the missing satellite problem. The viability of this
model has been recently  demonstrated by the high
resolution simulations of Schive
{\it et al.}
\cite{ch2,ch3} and the comprehensive paper of Hui {\it
et al.}
\cite{hui}.

At the scale of  DM halos, Newtonian gravity
can be
used so the
evolution of the wave function of the condensate is governed by the
Gross-Pitaevskii-Poisson (GPP) equations \cite{prd1}:
\begin{eqnarray}
\label{intro1}
i\hbar \frac{\partial\psi}{\partial
t}=-\frac{\hbar^2}{2m}\Delta\psi+m\Phi\psi+
\frac{4\pi a_s\hbar^2}{m^2}|\psi|^{2}\psi,
\end{eqnarray}
\begin{equation}
\label{intro2}
\Delta\Phi=4\pi G |\psi|^2,
\end{equation}
where $\Phi$ is the gravitational potential, $m$ is the mass of the bosons,
and
$a_s$ is their scattering
length. The interaction between the bosons is repulsive when $a_s>0$ and
attractive when $a_s<0$. The mass density of the BECDM halo is $\rho=|\psi|^2$.
Its total mass is
$M=\int\rho\, d{\bf r}$.

A serious DM particle candidate  is the QCD axion \cite{kc} which has been
proposed as a solution
of the charge parity (CP) problem of quantum chromodynamics (QCD) \cite{pq}.
The QCD axion  is a spin-$0$ boson with a mass $m= 10^{-4}\, {\rm eV}/c^2$
and an attractive self-interaction $a_s= -5.8\times 10^{-53}\, {\rm m}$.
Since the self-interaction is attractive, self-gravitating axions can form
stable clusters only below a maximum mass $M_{\rm
max}=1.012{\hbar}/{\sqrt{Gm|a_s|}}$ and above a minimum radius
$R_{99}\ge 5.5\left ({|a_s|\hbar^2}/{Gm^3}\right )^{1/2}$
\cite{prd1,prd2}.
The equilibrium states result from the balance between the repulsive pressure
arising from the
Heisenberg uncertainty principle, the attractive self-interaction of the bosons,
and the  gravitational attraction. For QCD axions this maximum mass is
very small, of the order of $M_{\rm
max}=6.5\times
10^{-14}\, M_{\odot}$ (corresponding to a radius $R_{99}=3.3\times 10^{-4}\,
R_{\odot}=230\, {\rm km}$) \cite{bectcoll}. Obviously, QCD
axions cannot form DM halos. They may form  mini axion stars of the asteroid
size. These
mini axion stars could be the constituents  of DM halos in the form of
mini-MACHOS. However, since  they behave essentially as CDM, they cannot solve
the CDM small-scale crisis.

Other kinds of axions may exist with a much smaller mass. 
These ULAs could form DM halos similar to gigantic boson stars
(see Appendix D of \cite{suarezchavanis3} for numerical applications). If
they have a mass $m=2.19\times
10^{-22}\, {\rm eV/c^2}$ and an attractive self-interaction
$a_s=-1.11\times 10^{-62}\, {\rm fm}$, the maximum mass $M_{\rm max}$ and the
minimum radius $R_{99}$ become comparable to the size
of ultracompact DM halos like Fornax  ($R\sim 1\, {\rm kpc}$ and $M\sim 10^{8}\,
M_\odot$). If
they are noninteracting they must have a mass of the order of  $m=
2.92\times 10^{-22}\, {\rm eV}/c^2$ to account for the size of ultracompact DM
halos.  In that case, the equilibrium state results from
the balance between the repulsive pressure arising from the Heisenberg principle
and the gravitational attraction.  Finally, if they have a
repulsive self-interaction $a_s>0$, they can account for the size of
ultracompact DM halos with a larger mass $m$ because, in the
Thomas-Fermi (TF) limit, only the ratio
$a_s/m^3=3.28\times 10^3 \, {\rm fm}/({\rm eV/c^2})^3$ is constrained. In that
case, the equilibrium
state  results from the balance between the repulsive pressure arising from the
self-interaction of the bosons and the gravitational attraction. Cosmological
considerations suggest that the bosonic DM particle has a repulsive
self-interaction \cite{shapiro,suarezchavanis3}. A repulsive self-interaction
may also solve some tensions encountered in the non-interacting model (see the
Remark at the end of Appendix D.4 of \cite{suarezchavanis3}).

Although the GPP equations are simple to write down, they actually have a very
complicated dynamics. 
A self-gravitating BEC at $T_{\rm th}=0$ that is not initially in a steady state
undergoes gravitational collapse (Jeans instability), displays damped
oscillations, and finally settles down on a quasi stationary state (virialization) by radiating part
of the scalar field \cite{seidel94,gul0,gul}. This is the process of
gravitational cooling initially introduced by Seidel and Suen \cite{seidel94} in
the context of boson stars. As a result of gravitational cooling, the system 
reaches an equilibrium configuration with a core-halo structure. The condensed 
core (soliton/BEC) is stabilized by quantum mechanics and has a smooth density
profile. This is a stable stationary solution of the GPP equations at
$T_{\rm th}=0$ (ground state). Gravitational collapse is
prevented by the quantum potential arising from the Heisenberg principle or by
the pressure $P=2\pi a_s\hbar^2\rho^{2}/m^3$ arising  from the self-interaction
of the bosons. This
solitonic  core (ground state) is 
 surrounded by a halo of scalar radiation corresponding
to the quantum
interferences of excited states. As shown by Schive {\it et al.}
\cite{ch2,ch3}, these interferences produce time-dependent small-scale
density granules (of the size of the solitonic core) 
that counter self-gravity and create an effective thermal pressure. These
noninteracting excited states are analogous to collisionless particles in
classical mechanics. As a result, the halo behaves essentially as CDM and  is
approximately isothermal with an equation of state $P=\rho k_B T/m$
involving an effective temperature $T$ (not to be confused with
the thermodynamic temperature $T_{\rm th}$ which is equal to zero). The
solitonic core
solves the cusp
problem of the CDM model (see footnote 2)
and the 
isothermal halo where the density decreases as $r^{-2}$ yields flat rotation
curves  in agreement with the observations.\footnote{The halo cannot
be exactly
isothermal otherwise it would have an infinite
mass \cite{bt}. In reality, the density in the
halo decreases as $r^{-3}$, similarly to the Navarro-Frenk-White (NFW)
\cite{nfw} and Burkert
\cite{observations} profiles, instead of $r^{-2}$ corresponding to the
isothermal
sphere \cite{bt}. This extra-confinement may be due to incomplete
relaxation, tidal
effects, and stochastic
perturbations (see Sec. \ref{sec_comp} and Appendix \ref{sec_diff} for a more
detailed discussion).} This core-halo
structure (and the presence of granules) has been clearly
evidenced in the numerical
simulations of  Schive {\it et al.} \cite{ch2,ch3}.

Gravitational cooling  is a dissipationless relaxation mechanism similar in some
respect to the concept of violent
relaxation introduced by  Lynden-Bell \cite{lb}  in the context of
collisionless 
self-gravitating systems described by the Vlasov-Poisson
equation.
A collisionless self-gravitating
system  that
is not initially in a dynamically stable steady 
state undergoes gravitational collapse (Jeans instability), displays damped
oscillations, and finally settles down on a quasi stationary 
state (virialization) by sending some of the particles at large distances. This
process is related to phase mixing and
nonlinear Landau damping. Lynden-Bell \cite{lb} developed a statistical
mechanics of this
process and obtained, at the coarse-grained scale, an equilibrium distribution
similar to the Fermi-Dirac
distribution.\footnote{The theory of Lynden-Bell \cite{lb} applies to
collisionless classical particles like stars as well as to collisionless
quantum particles like fermions or bosons. Actually, in the fermionic DM
model mentioned at the begining of this introduction, the Fermi-Dirac
distribution is justified by the theory of violent relaxation (see
the
discussion in \cite{csmnras,clm1,clm2}), not by standard quantum mechanics.
Indeed, the relaxation
time towards the true Fermi-Dirac distribution with a temperature $T_{\rm
th}\neq 0$ is larger than the age of the
Universe by many orders of magnitude. Therefore, the DM halos cannot thermalize
by a ``collisional'' process and one must invoke a
process of violent collisionless relaxation \cite{lb}. As a result, the
temperature $T$ appearing in the Fermi-Dirac distribution is an effective
temperature (the true thermodynamic temperature  $T_{\rm th}$ is very small and
can be taken
equal to zero). It can be shown that the maximum
value of the distribution function $\eta_0$ appearing in the Lynden-Bell
distribution is of the same order as the
bound $m^4/h^3$ set by the Pauli exclusion principle (see
footnote 34 of \cite{clm2}). This  makes the analogy between the Lynden-Bell
distribution and the Fermi-Dirac distribution even closer.}
The Lynden-Bell distribution function takes into account a sort of
exclusion principle implied
by the Vlasov equation, similar to the Pauli exclusion principle for fermions,
but of a nonquantum origin. In Lynden-Bell's theory, the QSS has a core-halo
structure with a  completely degenerate core (effective fermion ball) and an
isothermal atmosphere  with an effective temperature, like in the fermionic
model. The equation of state in the core
is $P=(1/5)[3/(4\pi\eta_0)]^{2/3}\rho^{5/3}$ and the equation of state in the
halo is $P=\rho T_{\rm LB}/\eta_0$. In the analogy between the gravitational
cooling of
self-gravitating BECs and the violent
relaxation of collisionless self-gravitating  systems, the bosonic core
(BEC/soliton) corresponds to the
fermion ball and the halo made of scalar radiation corresponds to the isothermal
halo predicted by Lynden-Bell. Actually, since a collisionless system of bosons
is described by the Vlasov-Poisson equations at large scales (where quantum
effects become negligible), it is very likely that both processes (gravitational
cooling and violent relaxation) are at work in self-gravitating BECs and may
even correspond to the same
phenomenon.  As a
result, self-gravitating
BECs should have a core that is partly bosonic (soliton) and partly fermionic
(in the sense of Lynden-Bell), surrounded by an effective isothermal halo. In
conclusion, gravitational cooling and violent relaxation explain
how collisionless self-gravitating systems can rapidly thermalize and acquire a
large
effective temperature $T$ even if $T_{\rm th}=0$ fundamentally. 
 Gravitational cooling and violent relaxation may be at work during hierarchical
clustering, a process 
 by which small DM halos merge and form larger halos in a bottom-up
structure
formation scenario.
It is believed that DM halos acquire an approximately isothermal
profile, or more realistically a NFW or Burkert profile (see footnote 3),
as a result of successive mergings.

In view of these remarks, it is important to obtain a parametrization of the
process of violent
relaxation and gravitational cooling on a coarse-grained scale.

A heuristic
parametrization of violent relaxation for classical collisionless
self-gravitating
systems described by the Vlasov-Poisson equation has been proposed in
\cite{csr,mnras,dubrovnik}. It has the form of a fermionic Fokker-Planck (or
Landau) equation for the coarse-grained distribution
function $\overline{f}({\bf r},{\bf v},t)$ involving a diffusion term and a
friction term. This equation respects the Lynden-Bell exclusion principle.
The diffusion term accounts for effective thermal effects (fluctuations) and the
friction
term accounts for collisionless dissipation (nonlinear Landau damping). The
competition between these two terms establishes, at statistical equilibrium, the
Lynden-Bell
distribution\footnote{When coupled to the Poisson
equation, the Lynden-Bell (or Fermi-Dirac) distribution generates a halo with an
infinite mass like the classical isothermal sphere \cite{lb}. This is
because the Lynden-Bell
distribution does not take into account the escape of high energy particles. 
However, it is possible to
derive from the kinetic theory a truncated Lynden-Bell distribution taking
into account the escape of high energy particles \cite{mnras}. This
model, which can be viewed as a sort of fermionic King model \cite{clm1,clm2},
has a
finite mass.} in a process reminiscent of the
fluctuation-dissipation theorem.

In previous papers \cite{bdo,nottalechaos}, we have introduced a
heuristic
parametrization of gravitational
cooling and violent relaxation for self-gravitating BECs described by the GPP
equation. We proposed to model
these complicated processes on a
coarse-grained scale by the generalized GPP equations
\cite{bdo,nottalechaos}:\footnote{A detailed derivation
of these equations will be given in a forthcoming paper \cite{forthcoming}.}
\begin{eqnarray}
\label{intro3}
i\hbar \frac{\partial\psi}{\partial
t}=&-&\frac{\hbar^2}{2m}\Delta\psi+m(\Phi+\Phi_{\rm
ext})\psi+\frac{K\gamma
m}{\gamma-1}|\psi|^{2(\gamma-1)}\psi\nonumber\\
&&+\frac{m}{2}\left (\frac{3}{4\pi\eta_0}\right
)^{2/3}|\psi|^{4/3}\psi
+2k_B
T\ln|\psi|\psi\nonumber\\
&-&i\frac{\hbar}{2}\xi\left\lbrack \ln\left (\frac{\psi}{\psi^*}\right
)-\left\langle \ln\left (\frac{\psi}{\psi^*}\right
)\right\rangle\right\rbrack\psi,
\end{eqnarray}
\begin{equation}
\label{intro4}
\Delta\Phi=4\pi G |\psi|^2,
\end{equation}
where $\langle X\rangle=\frac{1}{M}\int \rho X\, d{\bf r}$ denotes a spatial
average over the halo. The terms on the first line of Eq. (\ref{intro3})
correspond
to the ordinary GP equation (\ref{intro1}). For the
sake of generality, we have introduced an
external potential $\Phi_{\rm ext}$ that could take into account
the presence of a central black hole\footnote{The case of an external potential
$\Phi_{\rm BH}=-{GM_{\rm BH}}/{r}$ created by a central black hole is
treated specifically in Ref. \cite{forthcoming}.} or model other effects of
astrophysical
interest. In the following, for illustration, we shall consider the harmonic
potential
\begin{equation}
\label{harmonic}
\Phi_{\rm ext}=\frac{1}{2}\omega_0^2 r^2.
\end{equation}
When $\omega_0^2>0$, it can mimic the tidal interactions arising from
neighboring
galaxies. When $\omega_0^2<0$, it can mimic a solid-body rotation of the system
or the effect of dark energy (cosmological constant). The last term on
the first line of Eq. (\ref{intro3}) takes into account the self-interaction
of the bosons. For
the sake of generality, we have considered an arbitrary power-law nonlinearity
instead of the cubic nonlinearity present in the ordinary GP
equation \cite{gross1,gross2,gross3,pitaevskii2}. In the theoretical part of
this paper, we shall give results valid for arbitrary values of $\gamma$ and
$K$. They can be useful in more general situations. However, in the
applications, we shall 
specifically consider the standard BEC model corresponding to 
\begin{equation}
\label{standard}
K=\frac{2\pi a_s\hbar^2}{m^3}\quad {\rm and}\quad \gamma=2. 
\end{equation}
The terms on the second and third lines of Eq. (\ref{intro3}) correspond to our
heuristic parametrization
of gravitational cooling and violent relaxation. The first term on the second
line of Eq. (\ref{intro3}) accounts for an effective
fermionic core and the 
second term on the second line of Eq. (\ref{intro3}) accounts for an
isothermal halo, with an effective temperature $T$, surrounding the
core. This fermionic
core and this isothermal halo are justified by
Lynden-Bell's theory of violent relaxation (the isothermal halo is also expected
from the process of gravitational cooling).\footnote{The effective temperature
appearing in Eq. (\ref{intro3}) is related to the Lynden-Bell temperature by
$k_B T/m=T_{\rm LB}/\eta_0$. Since the mass $m$ of the particles does not
matter in collisionless systems, only the ratio $k_B T/m$ makes sense. In other
words, the temperature is proportional to mass \cite{lb}.}
These
two terms could be combined
into a single nonlinearity expressed as an enthalpic function $h_{\rm
LB}(|\psi|^2)$ associated with the equation of state arising from the
Lynden-Bell distribution or from the fermionic King model (see
\cite{bdo,nottalechaos} for a
general formalism). However, in the
present paper, we shall assume that the system is nondegenerate in the sense of
Lynden-Bell and we shall accordingly neglect the contribution of the fermionic
core.\footnote{This term will be considered in a forthcoming paper
\cite{forthcoming}.} As
a result, we just consider the contribution of the isothermal halo and formally
take $\eta_0\rightarrow +\infty$.
Finally, the term on the third line
of Eq. (\ref{intro3})  is a damping term that ensures that the system relaxes
towards an
equilibrium state. This is guaranteed by an $H$-theorem for a generalized free
energy functional \cite{bdo,nottalechaos}. It is natural to have a friction
term
and a temperature term
in
the phenomenology of violent relaxation and gravitational cooling. This
manifests a sort of
fluctuation-dissipation theorem.\footnote{We show in Appendix \ref{sec_nott}
that these two
terms
emerge from a unified framework related to Nottale's theory of scale
relativity \cite{nottale}.} It can be shown \cite{nottalechaos,forthcoming} that
the hydrodynamic representation of
the generalized GPP equations (\ref{intro3}) and (\ref{intro4}) is consistent
with the hydrodynamic moments of the  fermionic Fokker-Planck equation
introduced in \cite{csr} to parameterize the classical process
of violent relaxation. In the case of BECs, quantum mechanics introduces
additional terms which are the quantum potential and the pressure associated
with the self-interaction of the bosons. When these terms become negligible at
large scales one recovers the hydrodynamic equations of
\cite{csr}.

This paper is organized as follows. In Sec. \ref{sec_prop}  we
review the main properties of the generalized GPP equations (\ref{intro3}) and
(\ref{intro4})
introduced in \cite{bdo}. In Secs.
\ref{sec_ch} and \ref{sec_complete} we show that the equilibrium states of
these equations determine a DM
halo with a core-halo structure made of a solitonic core and an isothermal
atmosphere. In Sec. \ref{sec_desch} we provide a semi-analytical
description
of this core-halo structure. We mention the analogy with the core-halo structure
of fermionic DM
halos studied in the past. In Sec. \ref{sec_mod1}, we introduce a first model of
BECDM
halos (model I) in which the core-halo structure does not present a plateau.
This model
describes ultracompact DM halos that are purely solitonic (quantum ground
state), small DM halos with a solitonic core and a tenuous isothermal atmosphere
(quantum phase), and large DM halos which are purely isothermal without a
solitonic core (gaseous phase). In Sec. \ref{sec_another}, we introduce a second
model of BECDM
halos (model II) in which the core-halo structure may present a plateau. This
model
describes large halos that are purely isothermal
without a solitonic core (gaseous phase) as in Model I and large halos with a
solitonic core and a massive isothernal atmosphere (core-halo phase). We argue 
that the solitonic core may represent a bulge but that it cannot mimic a
supermassive black hole. In Sec. \ref{sec_ptb}, we show that the previous
solutions
(quantum, gaseous, core-halo) can be recovered by studying the phase
transitions of a  ``thermal'' self-gravitating  boson gas in a box. We discuss
the
stability
of these solutions and determine the solitonic core mass as a function of the
halo mass from thermodynamical considerations. In Sec. \ref{sec_astapp}, we
consider
astrophysical applications of our model. In particular, we connect the
bifurcation between the gaseous phase and the core-halo phase to the canonical
critical point  $(M_h)_{\rm CCP}=3.27\times
10^9\, M_{\odot}$ and  we connect the possible formation of supermassive
black
holes at the centers of galaxies to the
microcanonical critical point  $(M_h)_{\rm MCP}\sim 2\times
10^{12}\, M_{\odot}$.

\section{Properties of the generalized GPP equations}
\label{sec_prop}

We propose to heuristically model the process of
gravitational cooling and violent relaxation of self-gravitating BECs at
$T_{\rm th}=0$
by the generalized GPP
equations (\ref{intro3}) and (\ref{intro4}) which include a logarithmic
nonlinearity
and a source of dissipation (damping). These equations provide an effective
description of the system's dynamics on a coarse-grained
scale. In other words, they provide a coarse-grained parametrization of the
fined-grained GPP equations (\ref{intro1}) and (\ref{intro2}). As mentioned
before, in this paper we assume that the system is nondegenerate (in the sense
of Lynden-Bell) and ignore the contribution of the effective fermionic core
($\eta_0\rightarrow +\infty$).

\subsection{Madelung transformation}
\label{sec_mad}

In order to enlighten the physical meaning of the generalized GPP equations
(\ref{intro3}) and (\ref{intro4}), we can write them in the form of hydrodynamic
equations by using the Madelung \cite{madelung} transformation. We write the
wavefunction as
\begin{equation}
\label{mad1}
\psi({\bf r},t)=\sqrt{{\rho({\bf r},t)}} e^{iS({\bf r},t)/\hbar},
\end{equation}
where  $\rho=|\psi|^2$ is the density and  $S({\bf
r},t)=-i(\hbar/2)\ln(\psi/\psi^*)$ is the real action. We note that the
effective
temperature term in the generalized GP equation (\ref{intro3}) can be written as
$k_B T\ln\rho\, \psi$
and the dissipative term  as $\xi
(S-\langle S\rangle)\psi$. Following Madelung, we introduce the
velocity field ${\bf u}={\nabla S}/{m}$. Since the velocity is potential, the
flow is irrotational: $\nabla\times {\bf u}={\bf 0}$. Substituting Eq.
(\ref{mad1}) into Eq. (\ref{intro3}) and separating the
real and imaginary parts, we find that the generalized GPP equations
(\ref{intro3}) and (\ref{intro4}) are
equivalent
to the hydrodynamic equations
\begin{equation}
\label{mad2}
\frac{\partial\rho}{\partial t}+\nabla\cdot (\rho {\bf u})=0,
\end{equation}
\begin{equation}
\label{mad3}
\frac{\partial {\bf u}}{\partial t}+({\bf u}\cdot \nabla){\bf
u}=-\frac{1}{\rho}\nabla P-\nabla\Phi-\nabla \Phi_{\rm ext}-\frac{1}{m}\nabla
Q-\xi{\bf u},
\end{equation}
\begin{equation}
\label{mad4}
\Delta\Phi=4\pi G\rho,
\end{equation}
where
\begin{equation}
\label{mad5}
Q=-\frac{\hbar^2}{2m}\frac{\Delta
\sqrt{\rho}}{\sqrt{\rho}}
\end{equation}
is the quantum potential and 
\begin{equation}
\label{mad6}
P=K\rho^{\gamma}+\rho \frac{k_B T}{m}\qquad (\gamma=1+1/n)
\end{equation}
is the total pressure. It is the sum of a polytropic equation of
state due to the
power-law nonlinearity  in the generalized GP equation
(\ref{intro3}) and an isothermal (linear) equation of
state due to the logarithmic term in the generalized GP equation
(\ref{intro3}). For the standard BEC model of Eq. (\ref{standard}), the 
 polytropic equation of state writes
\begin{equation}
\label{mad6b}
P=\frac{2\pi a_s\hbar^2}{m^3}\rho^2,
\end{equation}
corresponding to a polytropic index $\gamma=2$ ($n=1$). It takes into account
the self-interaction of the bosons. In that case, the total equation of state
(\ref{mad6}) becomes
\begin{equation}
\label{mad6bq}
P=\frac{2\pi a_s\hbar^2}{m^3}\rho^2+\rho \frac{k_B T}{m}.
\end{equation}
Equation (\ref{mad2}) is the continuity
equation, Eq.
(\ref{mad3}) is the  momentum equation, and Eq. (\ref{mad4}) is the
Poisson
equation. We note that the momentum equation involves a damping term,
proportional and opposite to the velocity, corresponding  to the last term in
the generalized GPP equation (\ref{intro3}). Using the
continuity
equation (\ref{mad2}), the momentum equation (\ref{mad3}) can can be rewritten
as
\begin{equation}
\label{mad7}
\frac{\partial}{\partial t}(\rho {\bf u})+\nabla(\rho {\bf u}\otimes {\bf u})
=-\nabla P-\rho\nabla\Phi-\rho\nabla\Phi_{\rm ext}-\frac{\rho}{m}\nabla
Q-\xi\rho {\bf u}.
\end{equation}
Equations (\ref{mad2})-(\ref{mad7}) form the quantum damped barotropic Euler
equations. When the quantum potential is neglected (TF approximation), we
recover the classical damped barotropic Euler equations. These equations
do not involve viscous terms since they are equivalent to the
generalized GPP equations (\ref{intro3}) and (\ref{intro4}). As a result, they
describe a superfluid. For dissipationless systems ($\xi=0$), we recover the 
quantum and classical  barotropic Euler
equations.

\subsection{Connection with the equations of Brownian theory}
\label{sec_brown}

In the overdamped limit $\xi\rightarrow +\infty$, we can
formally neglect the inertial term in Eq. (\ref{mad3}) so that
\begin{equation}
\label{brown1}
\xi{\bf u}\simeq -\frac{1}{\rho}\nabla P-\nabla\Phi-\nabla\Phi_{\rm ext}
-\frac{1}{m}\nabla Q.
\end{equation}
Substituting this relation into the continuity equation (\ref{mad2}), we 
obtain the quantum barotropic Smoluchowski  equation \cite{pre11}:
\begin{equation}
\label{brown3}
\xi\frac{\partial\rho}{\partial t}=\nabla\cdot\left (\nabla
P+\rho\nabla\Phi+\rho\nabla \Phi_{\rm ext}+\frac{\rho}{m}\nabla Q\right ).
\end{equation}
When the quantum potential is neglected (TF approximation), we obtain the
classical barotropic Smoluchowski equation which arises in the context of
nonlinear Fokker-Planck equations and generalized thermodynamics
\cite{nfp,entropy}. The
isothermal equation of state $P=\rho k_B T/m$ yields an ordinary diffusion term
with a diffusion
coefficient
given by the Einstein formula $D=k_B T/\xi m$. The polytropic equation
of state  $P=K\rho^{\gamma}$ leads to anomalous diffusion. If
we neglect the advection term $\nabla(\rho {\bf u}\otimes {\bf u})$ in Eq.
(\ref{mad7}), but retain the term $\partial (\rho{\bf u})/\partial t$, and
combine the
resulting
equation with the continuity equation (\ref{mad2}), we obtain the quantum
telegraph equation
\begin{equation}
\label{brown4}
\frac{\partial^2\rho}{\partial t^2}+\xi\frac{\partial\rho}{\partial
t}=\nabla\cdot\left (\nabla P+\rho\nabla\Phi+\rho\nabla\Phi_{\rm ext}
+\frac{\rho}{m}\nabla Q\right ),
\end{equation}
which can be seen as a generalization of the quantum Smoluchowski equation
(\ref{brown3}) taking inertial (or memory) effects into account.
When the quantum potential is neglected, we recover the
classical telegraph equation.

It is interesting to recover the equations of Brownian theory from the 
generalized GP equation (\ref{intro3}) in the strong friction limit
$\xi\rightarrow +\infty$. In this sense, the generalized GP equation
(\ref{intro3}) makes the connection between quantum mechanics and
Brownian theory. However, the analogy with Brownian theory is essentially
effective as discussed in more detail in \cite{bdo,nottalechaos}. The 
Smoluchowski-Poisson
equations describing self-gravitating Brownian particles in the strong friction
limit have been studied in \cite{sp} and subsequent papers. If the
strong friction limit is relevant,\footnote{Some arguments in favor of the
strong friction limit are given in \cite{nottalechaos}.} these equations may
find a new
application (with a different interpretation) in the context of DM halos.

\subsection{Generalization of the CDM model}
\label{sec_cdm}

The hydrodynamic equations associated with the CDM model correspond to Eqs.
(\ref{mad2})-(\ref{mad4}) with $P=Q=\xi=0$. Therefore,
the fluid equations (\ref{mad2})-(\ref{mad4})
associated with the GPP equations (\ref{intro3}) and (\ref{intro4})
generalize the hydrodynamic equations of the CDM model in different respects.
First, the Euler equation (\ref{mad3}) includes a quantum force ${\bf
F}_Q=-(1/m)\nabla Q$
that takes into account the Heisenberg uncertainty principle. This force is
equivalent to an anisotropic pressure force of the form
$(F_Q)_i=-(1/\rho)\partial_jP_{ij}$,
where $P_{ij}=-(\hbar^2/4m^2)\rho\partial_i\partial_j\ln\rho$ is the quantum
pressure tensor. The Euler
equation (\ref{mad3}) also
includes an isotropic pressure force with a polytropic equation of state
$P=K\rho^{\gamma}$ due to the power-law nonlinearity in the generalized GP
equation (\ref{intro3}). For the standard BEC model (\ref{standard}), this
pressure force takes
into account the self-interaction of the bosons.  These two terms
(quantum force and self-interaction) are already present in the hydrodynamic
equations associated with the standard
GPP equations  (\ref{intro1}) and (\ref{intro2}) \cite{prd1}. The  hydrodynamic
equations (\ref{mad2})-(\ref{mad4}) associated with the
generalized GPP equations  (\ref{intro3}) and (\ref{intro4}) involve in
addition a pressure force with
an isothermal equation of state $P=\rho k_B T/m$ due to the logarithmic
nonlinearity present in the generalized GP equation (\ref{intro3}). As a result,
the complete equation of state is given by Eq. (\ref{mad6}). 
Finally, the Euler equation (\ref{mad3}) includes a damping term $-\xi {\bf u}$.
The damping term ensures that the system relaxes towards an equilibrium state.
This result is guaranteed by the existence of an $H$-theorem as discussed in the
next section. 

\subsection{$H$-theorem}
\label{sec_ht}

The free energy
associated with the generalized GPP equations can
be written in the usual form $F=E_0-TS_B$, where $E_0$ is the total energy, $T$
the effective temperature, and 
$S_B=-k_B \int (\rho/m)(\ln\rho-1)\, d{\bf r}$ the Boltzmann entropy.
The total energy $E_0=\Theta_c+\Theta_Q+W+W_{\rm ext}+U$  
is the sum of the classical kinetic energy $\Theta_c=(1/2)\int\rho {\bf
u}^2\, d{\bf r}$, the quantum kinetic energy $\Theta_Q=(1/m)\int \rho
{Q}\, d{\bf
r}$, the gravitational
potential energy $W=({1}/{2})\int\rho\Phi\, d{\bf r}$, the external
potential energy $W_{\rm ext}=\int\rho\Phi_{\rm ext}\, d{\bf
r}=(1/2)\omega_0^2I$ (where $I=\int\rho r^2\, d{\bf r}$ is the moment of
inertia) and the internal energy $U=[K/(\gamma-1)]\int
\rho^{\gamma}\, d{\bf r}$ arising from the self-interaction of the
bosons.\footnote{The free energy can also be
written as $F=E_*-TS_B-KS_\gamma$, where $E_*=\Theta_c+\Theta_Q+W+W_{\rm
ext}$ is the ideal energy (without the self-interaction term),
$S_{\gamma}=-[1/(\gamma-1)]\int (\rho^{\gamma}-\rho)\, d{\bf r}$
is the Tsallis entropy of index $\gamma$ (the standard BEC
model corresponds to a quadratic entropy with $\gamma=2$), and
$K$ is the polytropic temperature. We can introduce a mixed
entropy combining the Boltzmann and Tsallis entropies as discussed
in \cite{bdo}.}
The generalized GPP equations satisfy an $H$-theorem \cite{bdo}:
\begin{eqnarray}
\label{ht1}
\dot F=-\xi\int \rho {\bf u}^2\, d{\bf r}\le 0.
\end{eqnarray}
When $\xi=0$, the generalized GPP equations (\ref{intro3}) and (\ref{intro4})
conserve the free energy ($\dot F=0$). When $\xi>0$, the free energy decreases
monotonically ($\dot F\le 0$). We note that $\dot F=0$ if, and
only if, ${\bf u}={\bf 0}$. Therefore, the dissipative term ensures that the
system relaxes
towards an equilibrium state  for $t\rightarrow +\infty$.\footnote{This result
assumes that $F$ is bounded from below. For isothermal self-gravitating
systems this is not the case. There is no minimum of free energy at fixed mass
because the system can always loose free energy by evaporating. However,
evaporation is
a slow process. In practice, the system relaxes towards a quasiequilibrium
state which occupies a finite region of space. When necessary,  we shall
artificially confine the system within a box, where the size of the box
represents the typical size of the system \cite{paddy,ijmpb}.} In this
sense, the generalized GPP equations can account, at least heuristically, for
the
complicated processes of violent
relaxation and gravitational cooling.

{\it Remark:} the generalized GPP equations (\ref{intro3}) and
(\ref{intro4}) are associated with the canonical ensemble (fixed temperature
$T$). However, they can be extended to the microcanonical ensemble (fixed energy
$E$) as discussed
in Appendix I of \cite{bdo}.

\subsection{Equilibrium state}
\label{sec_es}

From Lynapunov's direct method based on Eq. (\ref{ht1}), we know that a stable
equilibrium state
is a minimum of free energy $F$ at fixed mass $M$. Therefore, it
satisfies the
variational principle $\delta F-(\mu/m)\delta M=0$, where $\mu$ is a
Lagrange multiplier (chemical potential) taking into account the conservation of
mass. This gives the relation \cite{bdo}:
\begin{equation}
\label{es1}
Q+m\Phi+\frac{1}{2}m\omega_0^2 r^2+k_B T\ln\rho+\frac{K\gamma
m}{\gamma-1}\rho^{\gamma-1}=\mu
\end{equation}
that coincides with a static state $\psi({\bf r},t)=\phi({\bf
r})e^{-iEt/\hbar}$ of the generalized GPP equations
(\ref{intro3}) and (\ref{intro4}) provided
that we make the identification between the eigenenergy
and the chemical potential: $E=\mu$  \cite{bdo}.  Equation (\ref{es1}) can
be
rewritten as
\begin{equation}
\label{es2}
\rho=\left\lbrace \frac{k_B T}{|K|\gamma m}W\left\lbrack \frac{|K|\gamma m}{k_B
T}
e^{-\frac{\gamma-1}{k_B T}(m\Phi+Q+\frac{1}{2}m\omega_0^2r^2-\mu)}
\right\rbrack\right\rbrace^{\frac{1}{\gamma-1}},
\end{equation}
where $W(z)$ is a (generalized) Lambert function defined implicitly by the
equation
$We^W=z$ when $K>0$ and $We^{-W}=z$ when $K<0$.
This equation determines the
relation between the density $\rho$ and the gravitational potential
$\Phi$ at equilibrium.\footnote{We note that Eq. (\ref{es2}) is a complicated
differential equation because of the presence of the quantum potential that
involves derivatives of $\rho$. It is only in the TF approximation ($Q=
0$) that the relationship between $\rho$ and $\Phi$  is
explicit.}
When $K=0$ ($W=z$), we obtain the quantum Boltzmann distribution
\begin{equation}
\label{es3}
\rho=e^{-\frac{1}{k_B T}(m\Phi+Q+\frac{1}{2}m\omega_0^2r^2-\mu)}
\end{equation}
associated with the isothermal equation of state. When $T=0$ ($W=\ln z$), we
obtain the quantum Tsallis distribution
\begin{equation}
\label{es4}
\rho=\left\lbrack -\frac{\gamma-1}{K\gamma
m}\left (m\Phi+Q+\frac{1}{2}m\omega_0^2 r^2-\mu\right
)\right\rbrack^{\frac{1}{\gamma-1}}
\end{equation}
associated with the polytropic equation of state. For the standard BEC model,
corresponding to $\gamma=2$, the relationship between
$\rho$ and $\Phi$ is linear.
Combining Eq. (\ref{es2}) with the Poisson equation (\ref{mad4}), we
obtain a differential equation that determines $\Phi$ and $\rho$. When $K=0$,
it reduces to the quantum Boltzmann-Poisson equation and when $T=0$ it reduces
to the quantum Tsallis-Poisson equation.

\subsection{Virial theorem}
\label{sec_vt}

In order to understand qualitatively how the system  relaxes towards
equilibrium, it may be useful to consider the virial theorem. From the damped
quantum barotropic Euler-Poisson equations (\ref{mad2})-(\ref{mad7}), we can
derive the time-dependent scalar virial theorem \cite{bdo}:
\begin{equation}
\label{vt1}
\frac{1}{2}\ddot I+\frac{1}{2}\xi\dot I+\omega_0^2 I=2(\Theta_c+\Theta_Q)+3\int
P\, d{\bf r}+W.
\end{equation}
This equation, together with the $H$-theorem (\ref{ht1}), shows that the system
generically converges towards an equilibrium state (or a quasiequilibrium
state) by exhibiting damped oscillations.  These damped oscillations are
consistent with the phenomenology of gravitational
cooling \cite{seidel94,gul0,gul} and violent relaxation \cite{lb}.

\subsection{Gaussian ansatz}
\label{sec_ga}

\begin{figure}
\begin{center}
\includegraphics[clip,scale=0.3]{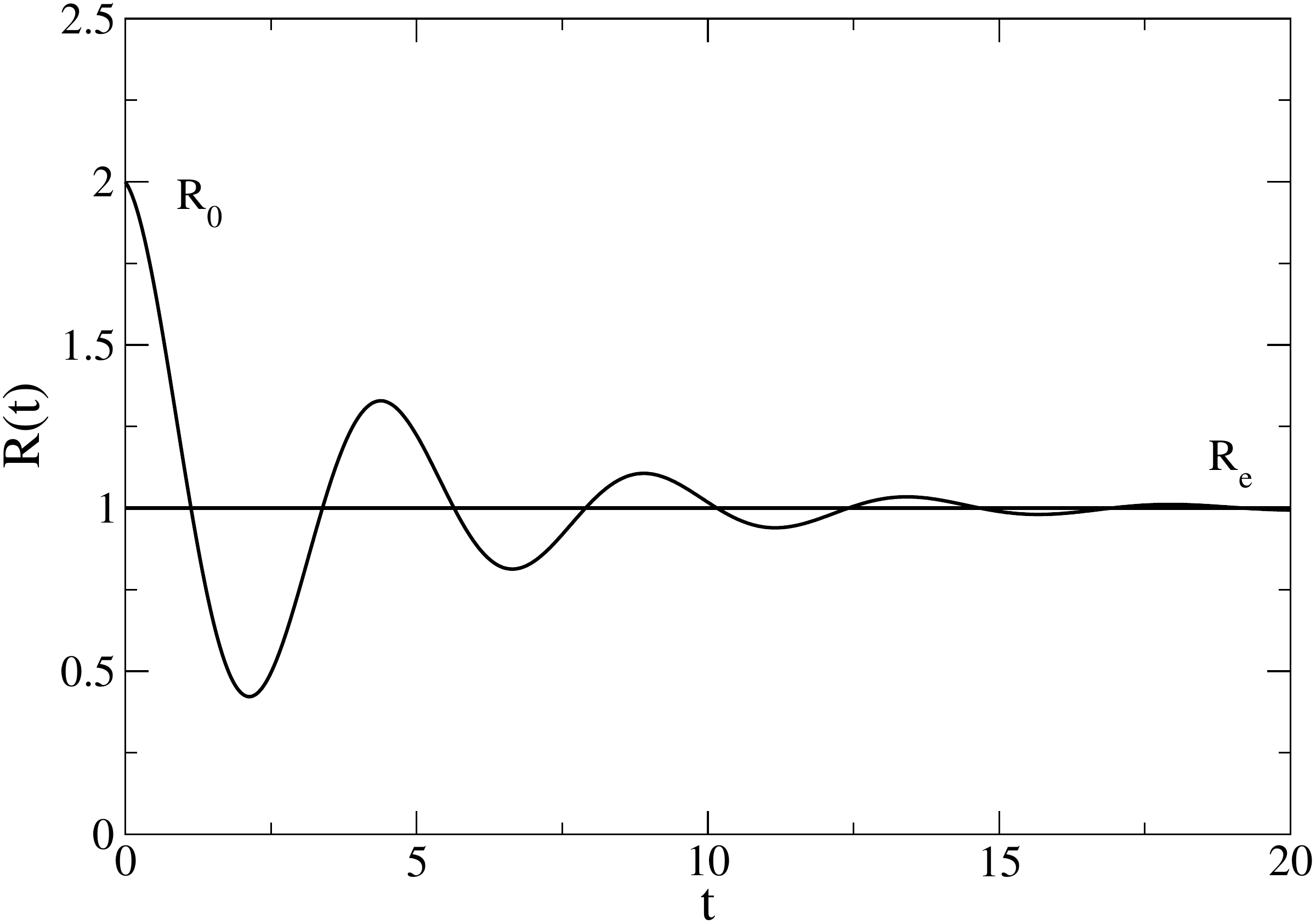}
\caption{Damped oscillations of the radius of the
self-gravitating BEC during its relaxation towards
equilibrium (schematic evolution based on the Gaussian ansatz).
}
\label{oscill}
\end{center}
\end{figure}

In order to determine accurately the dynamical evolution of a
self-gravitating BEC, one must solve the (generalized) GPP equations
(\ref{intro3}) and (\ref{intro4})
numerically. However, an approximate analytical solution  can be obtained by
making a Gaussian ansatz for the wave function. From the virial theorem, we can
obtain a simple differential equation governing the temporal evolution of the
typical radius
$R(t)$ of the BEC. It is given
by \cite{bdo}:
\begin{eqnarray}
\label{ga1}
\alpha M\frac{d^2R}{dt^2}+\xi\alpha M\frac{d R}{dt}+\alpha\omega_0^2
MR=2\sigma\frac{\hbar^2 M}{m^2 R^3}+3\frac{M k_B T}{m R}\nonumber\\
+3\zeta
\frac{K M^{\gamma}}{R^{3(\gamma-1)+1}}-\nu
\frac{G M^2}{R^{2}}.\qquad
\end{eqnarray}
The coefficients
are $\alpha={3}/{2}$,
$\sigma={3}/{4}$, $\zeta=\pi^{-3(\gamma-1)/2}\gamma^{-3/2}$ and
$\nu={1}/{\sqrt{2\pi}}$. 
At equilibrium ($\dot R=\ddot R=0$), this equation determines an approximate
expression of the
mass-radius
relation of the self-gravitating BEC. In many cases, this approximate
mass-radius relation gives a good agreement
with the exact
mass-radius relation obtained by solving the GPP equations numerically
\cite{prd1,prd2}.
On the other hand, as illustrated in Fig. \ref{oscill}, 
the dynamical equation (\ref{ga1})  confirms that the system
generically relaxes
towards the equilibrium state by undergoing damped oscillations.\footnote{We
note that if we make the Gaussian
ansatz on the usual GPP equations (\ref{intro1}) and (\ref{intro2}), we miss the
important processes of 
gravitational cooling and violent relaxation because the resulting differential
equation for $R(t)$
does not exhibit damped oscillations \cite{prd1} while the GPP equations
(\ref{intro1}) and (\ref{intro2}) do \cite{seidel94,gul0,gul}. This is an
interest of our heuristic parametrization relying on the 
generalized  GPP equations  (\ref{intro3}) and (\ref{intro4}).}

\subsection{Complex pulsation}
\label{sec_ledoux}

Close to equilibrium, the complex pulsation can be determined approximately
from the Gaussian ansatz. It is given by \cite{bdo}:
\begin{equation}
\label{ledoux1}
\omega^2=\frac{6\Theta_Q+3(\gamma-1)(3\gamma-2)  U+2
W+\omega_0^2 I+3Nk_B T}{I}.
\end{equation}
Alternative expressions  of
the pulsation can be obtained by using the equilibrium form of the free energy
and of the virial theorem ($\ddot I=\dot I=\Theta_c=0$):
\begin{eqnarray}
\label{ledoux2}
F=\Theta_Q+U+W+\frac{1}{2}\omega_0^2 I-TS_B,
\end{eqnarray}
\begin{equation}
\label{ledoux3}
2\Theta_Q+3\int P\, d{\bf r}+W-\omega_0^2I=0.
\end{equation}
Particular cases are
considered specifically in \cite{bdo}. In the TF
approximation, they agree with the approximate expression of the pulsation given
by the Ledoux formula \cite{ledoux}.

\section{Core-halo structure of the equilibrium states}
\label{sec_ch}

\subsection{Fundamental equation of quantum hydrostatic equilibrium}
\label{sec_chg}

In order to determine the equilibrium states of a self-gravitating BECDM halo,
instead of solving the coupled equations (\ref{mad4}) and (\ref{es2}), it is
more convenient to
proceed as follows. We first take the gradient of Eq. (\ref{es1}) and obtain
the condition of quantum hydrostatic equilibrium \cite{bdo}: 
\begin{eqnarray}
\label{ch1}
\frac{\rho}{m}\nabla Q+\nabla P+\rho\nabla\Phi
+\rho\nabla\Phi_{\rm ext}={\bf 0}.
\end{eqnarray}
This equation also corresponds to a static state ($\partial_t\rho=0$, ${\bf
u}={\bf 0}$) of the damped quantum Euler equations (\ref{mad2}) and
(\ref{mad3}). It
describes the balance
between the quantum potential arising from the Heisenberg uncertainty
principle,  the pressure due to short-range interactions
(scattering), the pressure due to the effective temperature, the
gravitational attraction, and the
external potential. Combining Eq.
(\ref{ch1}) with the Poisson equation (\ref{mad4}), we obtain the fundamental
differential equation of quantum hydrostatic equilibrium \cite{bdo}:
\begin{equation}
\label{ch2}
\frac{\hbar^2}{2m^2}\Delta
\left (\frac{\Delta\sqrt{\rho}}{\sqrt{\rho}}\right )-\nabla\cdot \left
(\frac{\nabla P}{\rho}\right )=4\pi G\rho+3\omega_0^2.
\end{equation}
For the
equation of state (\ref{mad6}),  it takes the form
 \begin{eqnarray}
\label{ch3}
\frac{\hbar^2}{
2m^2}\Delta
\left (\frac{\Delta\sqrt{\rho}}{\sqrt{\rho}}\right
)-\frac{K\gamma}{\gamma-1}\Delta\rho^{\gamma-1}-\frac{k_B
T}{m}\Delta\ln\rho\nonumber\\
=4\pi
G\rho+3\omega_0^2.
\end{eqnarray}
This differential equation determines the general equilibrium density profile
$\rho({\bf r})$ of a BECDM halo in our model. This profile generically has a
core-halo
structure with a solitonic core and an isothermal halo (we assume that
$\gamma>1$). In the following, for simplicity, we take
$\omega_0=0$ and consider spherically symmetic distributions.

\subsection{Solitonic core}
\label{sec_core}

In the core, where the density is high, the equation of state (\ref{mad6})
is dominated by the polytropic (self-interaction) term and the thermal
term can be neglected ($T=0$). The
gravitational attraction is counterbalanced by the quantum
potential and by the self-interaction of the bosons.  In
that
case, Eq. (\ref{ch3})  reduces to
\begin{equation}
\label{ch4}
\frac{\hbar^2}{
2m^2}\Delta
\left (\frac{\Delta\sqrt{\rho}}{\sqrt{\rho}}\right
)-\frac{K\gamma}{\gamma-1}\Delta\rho^{\gamma-1}=4\pi
G\rho.
\end{equation}
The solution of Eq.
(\ref{ch4}) is called a soliton\footnote{In the TF approximation (see below),
we will still call the solution of Eq. (\ref{ch4}) a soliton although this
terminology may be abusive since the effect of the quantum potential which
usually gives rise to the soliton in the absence of self-interaction is
neglected.} 
because it corresponds to the static state of the ordinary GPP equations
(\ref{intro1})
and (\ref{intro2}). The ground state corresponds to the density profile that has
no node.

\subsubsection{Noninteracting limit}

\begin{figure}[!h]
\begin{center}
\includegraphics[clip,scale=0.3]{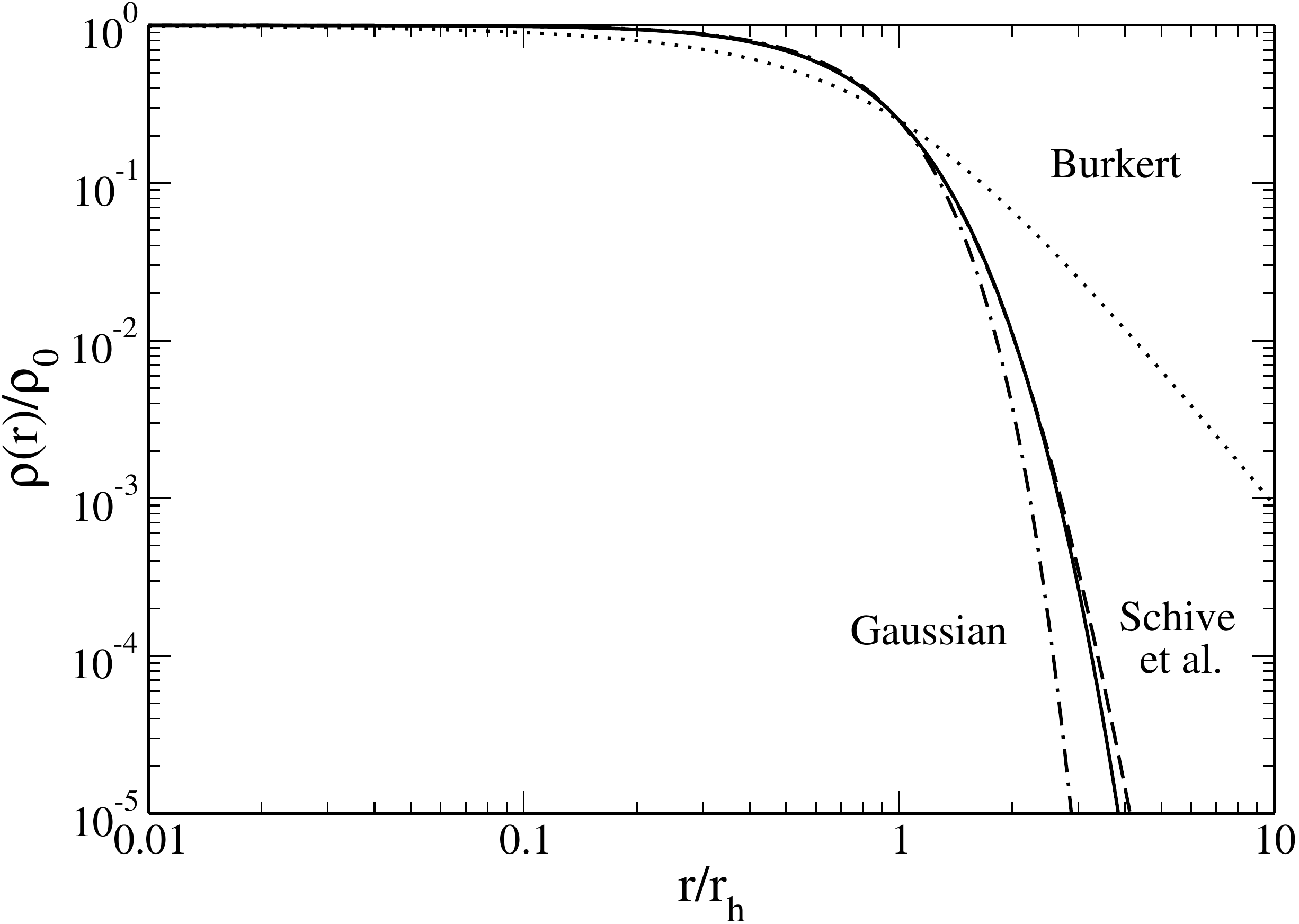}
\caption{Normalized density profile of the soliton. It is compared with the
Gaussian profile (\ref{ch8}) considered in \cite{prd1}, the profile (\ref{ch9})
proposed by Schive {\it et al.}  \cite{ch2,ch3}, and the Burkert
 profile \cite{observations}.}
\label{fitASnull}
\end{center}
\end{figure}

In the noninteracting limit ($K=0$), the gravitational attraction is
counterbalanced by the quantum
potential. In that case, Eq. (\ref{ch4})  reduces to
\begin{equation}
\label{ch5}
\frac{\hbar^2}{
2m^2}\Delta
\left (\frac{\Delta\sqrt{\rho}}{\sqrt{\rho}}\right )=4\pi
G\rho.
\end{equation}
 This equation has been 
numerically solved in \cite{rb,membrado,gul0,gul,prd2,ch2,ch3,pop} and the exact
density profile has been obtained in these papers. The result of \cite{prd2}
is reproduced in Fig. \ref{fitASnull} where the density is normalized by the
central
density $\rho_0$ and the radial distance is normalized by the halo radius
$r_h$ defined by Eq. (\ref{hr1}).
This profile  has not a compact support, i.e., it extends to infinity. The
exact mass-radius relation is given by 
\cite{rb,membrado,prd2}:
\begin{equation}
\label{ch7}
R_{99}=9.946 \frac{\hbar^2}{GMm^2},
\end{equation}
where $R_{99}$ is the radius of the configuration containing $99\%$
of the mass. 

In \cite{prd1} we have approximated the solitonic density profile by a
Gaussian:
\begin{equation}
\label{ch8}
\rho=\rho_0e^{-r^2/R^2}.
\end{equation}
From this distribution, we can obtain the following results. The total mass is
given
by $M=5.57 \rho_0R^3$. The radius containing $99\%$ of the mass is
$R_{99}=2.38R$. The halo radius where the central density is divided by $4$ is
$r_h=1.18 R$ and the core 
radius where the central density is divided by $2$ is $r_c=0.833 R$. We also
find that $M_h/(\rho_0 r_h^3)=1.95$, where $M_h$ is the halo mass defined by Eq.
(\ref{hr3}).  Using Eq. (\ref{ch7}), we
obtain
$r_h=2.82\hbar^2/(Gm^2M_h)$ and $\rho_0=1.45 \, \hbar^2/(Gm^2 r_h^4)$.

On the other hand, Schive {\it et al.}  \cite{ch2,ch3} have introduced
a fit of a form
\begin{equation}
\label{ch9}
\rho=\frac{\rho_0}{\lbrack 1+(r/R)^2\rbrack^8}.
\end{equation}
From this distribution, we can obtain the following results. The total mass is
given by $M=0.318 \rho_0R^3$. The radius containing $99\%$ of the mass is
$R_{99}=1.151R$. The halo radius
where the central density is divided by $4$ is $r_h=0.435 R$ and the core 
radius where the central density is divided by $2$ is $r_c=0.301 R$. We also
find that $M_h/(\rho_0 r_h^3)=1.91$. Using Eq. (\ref{ch7}), we obtain
$r_h=1.85\hbar^2/(Gm^2M_h)$ and $\rho_0=0.969 \, \hbar^2/(Gm^2 r_h^4)$.

These fits are compared with the
exact density profile from
\cite{prd2} in Fig. \ref{fitASnull}. The Gaussian
profile \cite{prd1} works very well up to the halo radius $r_h$. The
fit of  Schive {\it et al.}  \cite{ch2,ch3} is
valid on a slightly longer distance $\sim 2.5 r_h$.  For comparison,
we have  plotted the Burkert profile (see
Appendix \ref{sec_burkert}).

\subsubsection{TF approximation}
\label{sec_tfw}

In this section, we assume that the bosons have a repulsive self-interaction
($K>0$). In the TF
approximation ($Q=0$), Eq. (\ref{ch4})  reduces to
\begin{equation}
\label{ch10}
-\frac{K\gamma}{\gamma-1}\Delta\rho^{\gamma-1}=4\pi
G\rho.
\end{equation}
This equation is equivalent to the Tsallis-Poisson
equation obtained by combining Eqs. (\ref{mad4}) and (\ref{es4}). It is also
equivalent to the Lane-Emden equation (\ref{p5}) \cite{chandra}. It
describes the balance between the gravitational attraction and
the repulsion due to the short-range interactions.

For the standard BEC model (\ref{standard}), the system is equivalent to a
polytrope of index
$n=1$. In that case, Eq. (\ref{ch10}) becomes
\begin{equation}
\label{ch11}
\Delta\rho+\frac{Gm^3}{a_s\hbar^2}\rho=0.
\end{equation}
This equation has a simple analytical solution given by \cite{chandra}:
\begin{equation}
\rho(r)=\rho_0\frac{\sin(\pi r/R)}{\pi r/R},
\label{ch12}
\end{equation}
where $\rho_0$ is the central density  and  
\begin{equation}
\label{ch13}
R=\pi\left (\frac{a_s\hbar^2}{Gm^3}\right )^{1/2}
\end{equation}
is the radius of the configuration at which the density vanishes (the
density profile has a compact
support). The radius of a polytrope $n=1$ is independent of its mass
\cite{chandra}. The central
density is related to the mass by
\begin{equation}
\label{ch14}
\rho_0=\frac{\pi M}{4R^3}=\frac{M}{4\pi^2}\left (\frac{Gm^3}{a_s\hbar^2}\right
)^{3/2}.
\end{equation}
These results have been derived by several authors in the
context of self-gravitating BECs and SFs
\cite{tkachev,maps,leekoh,goodman,arbey,bohmer,prd1}.

Using Eq. (\ref{ch12}), we find that the accumulated mass and circular velocity
profiles defined by Eqs. (\ref{hr2}) and (\ref{hr4}) are given by 
\begin{equation}
M(r)=\frac{4\rho_0 R^3}{\pi^2}\left \lbrack \sin\left
(\frac{\pi r}{R}\right )-\frac{\pi r}{R}\cos\left (\frac{\pi r}{R}\right )\right
\rbrack,
\label{ch15}
\end{equation}
\begin{equation}
\label{ch16}
v^2(r)=\frac{4G\rho_0 R^2}{\pi}\left \lbrack \frac{R}{\pi r}\sin\left
(\frac{\pi r}{R}\right )-\cos\left (\frac{\pi r}{R}\right )\right \rbrack.
\end{equation}
For $r\rightarrow 0$, the velocity increases linearly with $r$ as for a uniform
sphere with density $\rho_0$: $v(r)\sim (4\pi\rho_0 G/3)^{1/2} r$. For
$r\ge
R$, we recover the Keplerian law $v(r)=({GM}/{r})^{1/2}$. 

The halo radius, the halo mass and the circular velocity at the halo radius  are
given by (see Appendix \ref{sec_p}):
\begin{equation}
\label{ch17}
\frac{r_h}{R}=0.788,\qquad \frac{M_h}{\rho_0 r_h^3}=2.12,
\end{equation}
\begin{equation}
\frac{v_h^2}{4\pi G \rho_0 r_h^2}=0.169.
\label{ch18}
\end{equation}

The density and  circular velocity
profiles are plotted in Figs. \ref{becdens} and \ref{becvit}. For comparison,
we have plotted the Burkert profile (see Appendix \ref{sec_burkert}).
We recall that
the Burkert profile is empirical. In particular, the Burkert density profile 
behaves like
$r$ instead of $r^2$ as $r\rightarrow 0$, which is not physical for
spherically symmetic systems. This explains the disagreement between the two
density profiles for $r\le r_h$. In spite of this difference, the circular
velocity profiles are relatively close to
each other up to the halo radius $r_h$. As
discussed in Sec. \ref{sec_comp}, the solitonic solution of the BECDM model is
expected to
provide a better description of ultracompact DM
halos than the Burkert profile that is
more adapted to describe larger DM halos.

\begin{figure}[!h]
\begin{center}
\includegraphics[clip,scale=0.3]{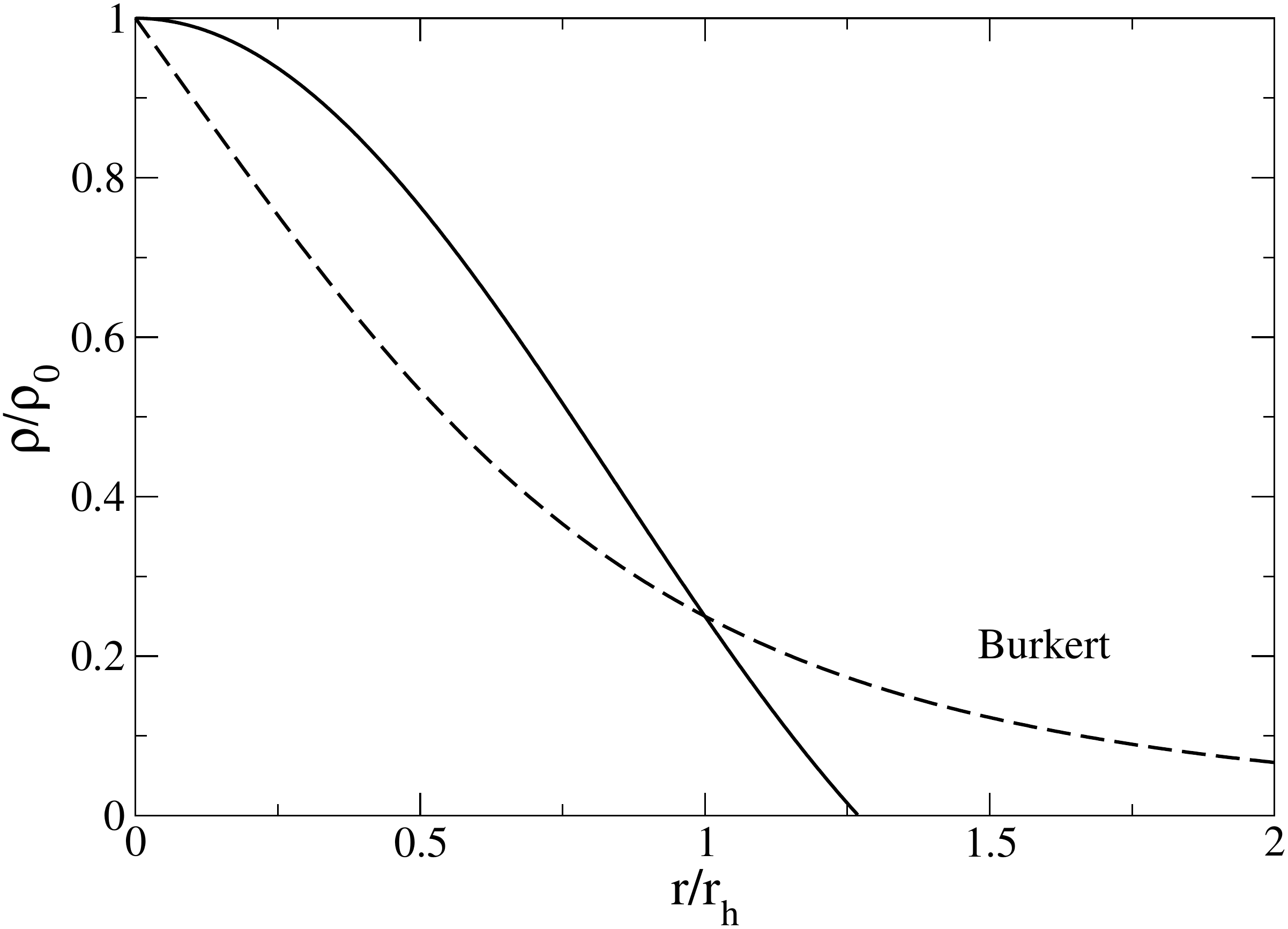}
\caption{Density profile of a self-gravitating BEC with a repulsive
self-interaction in the TF limit (polytrope $n=1$). It is compared to the
Burkert profile (dashed line). }
\label{becdens}
\end{center}
\end{figure}

\begin{figure}[!h]
\begin{center}
\includegraphics[clip,scale=0.3]{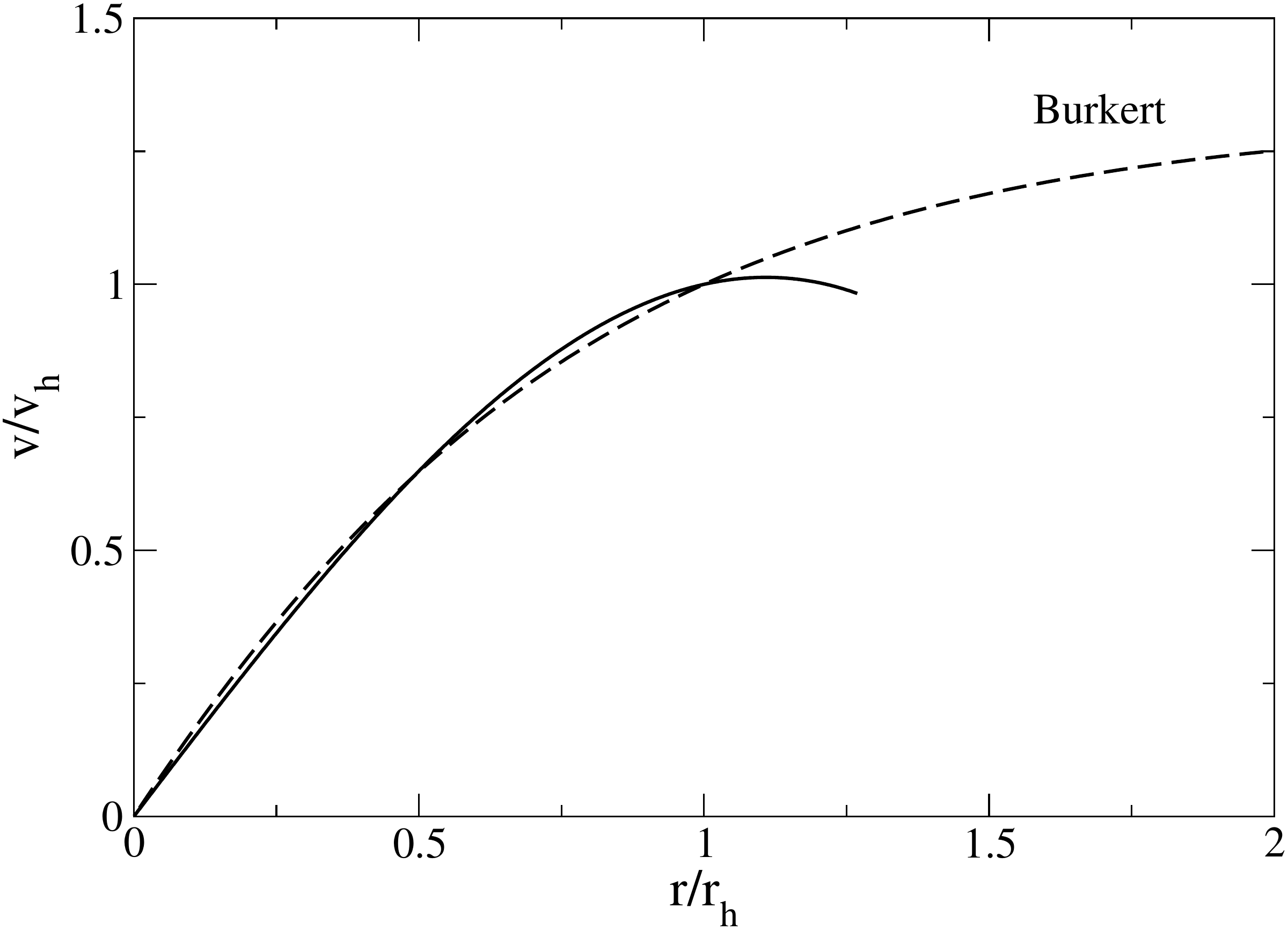}
\caption{Same as
Fig. \ref{becdens} for the circular velocity profile. }
\label{becvit}
\end{center}
\end{figure}

\subsubsection{General case}

In the general case, Eq. (\ref{ch4}) has been solved numerically in
\cite{gul,prd2} for the
standard BEC model ($\gamma=2$). It leads to a general soliton profile taking
the quantum potential
and the self-interaction into account. This profile is relatively
well-approximated by a Gaussian distribution, especially when the
self-interaction is weak. The solitonic density profile does not
diverge at the origin. This may solve the cusp problem of the CDM model. 
The mass-radius relation of self-gravitating BECs (representing the solitonic
core of DM halos) has been obtained analytically (approximately) and
numerically (exactly) in \cite{prd1,prd2} for the standard BEC model
($\gamma=2$). The case of a self-gravitating BEC with an attractive
self-interaction ($a_s<0$) has also  been considered in these papers. It
is found that stable
configurations only exist below a maximum mass and above a
minimum radius given by \cite{prd1,prd2}:
\begin{eqnarray}
\label{ch19}
M_{\rm max}=1.012\frac{\hbar}{\sqrt{Gm|a_s|}},
\end{eqnarray}
\begin{eqnarray}
\label{ch19b}
R_{99}= 5.5\left (\frac{|a_s|\hbar^2}{Gm^3}\right )^{1/2}.
\end{eqnarray}

\subsection{Isothermal halo}
\label{sec_halo}

In the halo, where the density is low, the equation of  state (\ref{mad6}) is
dominated by the linear (thermal) term  and the self-interaction of the
bosons can
be neglected ($K=0$). The quantum potential can also be neglected ($Q=0$). In
that case, Eq.
(\ref{ch3}) reduces to
\begin{equation}
\label{ch20}
-\frac{k_B
T}{m}\Delta\ln\rho=4\pi
G\rho.
\end{equation}
This equation is equivalent to the Boltzmann-Poisson equation obtained by
combining Eqs. (\ref{mad4}) and (\ref{es3}). It is also equivalent to the Emden
equation (\ref{i5}) \cite{chandra}.\footnote{The
Boltzmann-Poisson equation and the Emden equation describe a
classical self-gravitating gas with an isothermal
equation of state \cite{chandra}. They also arise in
the statistical mechanics of stellar systems \cite{paddy,ijmpb}. In
these two cases, they correspond to a collisional relaxation due to
strong collisions (gas) or to weak two-body gravitational encounters (stellar
systems). In the
present case, the effective isothermal halo is justified by Lynden-Bell's
statistical
mechanics of collisionless violent relaxation \cite{lb}.} It
describes the
balance between the gravitational attraction and the thermal pressure.
This
equation has no simple analytical solution and must be solved
numerically. However, its asymptotic behavior is known
analytically \cite{chandra}. The
density of a self-gravitating isothermal
halo decreases as $\rho(r)\sim k_B T/(2\pi G m r^{2})$ for $r\rightarrow
+\infty$, corresponding to an accumulated mass $M(r)\sim 2k_B Tr/(Gm)$
increasing
linearly with $r$. This leads to flat rotation curves
$v^2(r)=GM(r)/r\rightarrow v_{\infty}^2=2k_B T/m$ in agreement with the
observations \cite{bt}.

We note that the isothermal profile has not a compact support so that it
extends up to infinity. Furthermore its total mass is infinite \cite{bt}. This
is why self-gravitating systems have no statistical equilibrium state in an
unbounded domain (see \cite{paddy,ijmpb} and footnote 13). In practice, the
isothermal equation of state is not valid
at arbitrarily large distances and the halo is confined by other effects (see
Appendix  \ref{sec_diff}). From Eq. (\ref{ch20}) we can define a
characteristic
radius
\begin{equation}
\label{chp3}
r_0=\left (\frac{k_B T}{4\pi G\rho_0 m}\right )^{1/2}
\end{equation}
that we shall call the thermal core radius. It
represents the typical core radius of an isothermal halo of
central density $\rho_0$.

The halo mass, the temperature and the circular velocity at the halo radius
are given by (see Appendix \ref{sec_i}):
\begin{eqnarray}
\label{ch21}
\frac{r_h}{r_0}=3.63,\qquad \frac{M_h}{\rho_0 r_h^3}=1.76,
\end{eqnarray}
\begin{eqnarray}
\label{ch22}
 \frac{k_B T}{Gm\rho_0
r_h^2}=0.954,\qquad \frac{v_h^2}{4\pi G\rho_0r_h^2}=0.140.
\end{eqnarray}

The density and  circular velocity
profiles of a purely isothermal halo are plotted in Figs.
\ref{isodensnormaliseELARGI} and \ref{isovitnormalise} (see Appendix
\ref{sec_i}). For comparison, we have
plotted the empirical (observational) Burkert profile
\cite{observations} (see Appendix \ref{sec_burkert}). The isothermal profile is
close 
to the  Burkert profile up to $r/r_h=6$. We have also plotted some
analytical
profiles that have been introduced in the literature to fit the isothermal
profile. The pseudo
isothermal profile  (see Appendix
\ref{sec_pseudo}) provides a good
fit of the isothermal profile
up to
$r/r_h=1$. The modified
Hubble profile \cite{bt} (see Appendix
\ref{sec_hubble}) provides a
good fit of the isothermal profile up to $r/r_h=3$. The Natarajan and
Lynden-Bell  profile \cite{nlb} (see Appendix
\ref{sec_nlb}) provides a good fit
of the isothermal profile for all radii.

\begin{figure}[!h]
\begin{center}
\includegraphics[clip,scale=0.3]{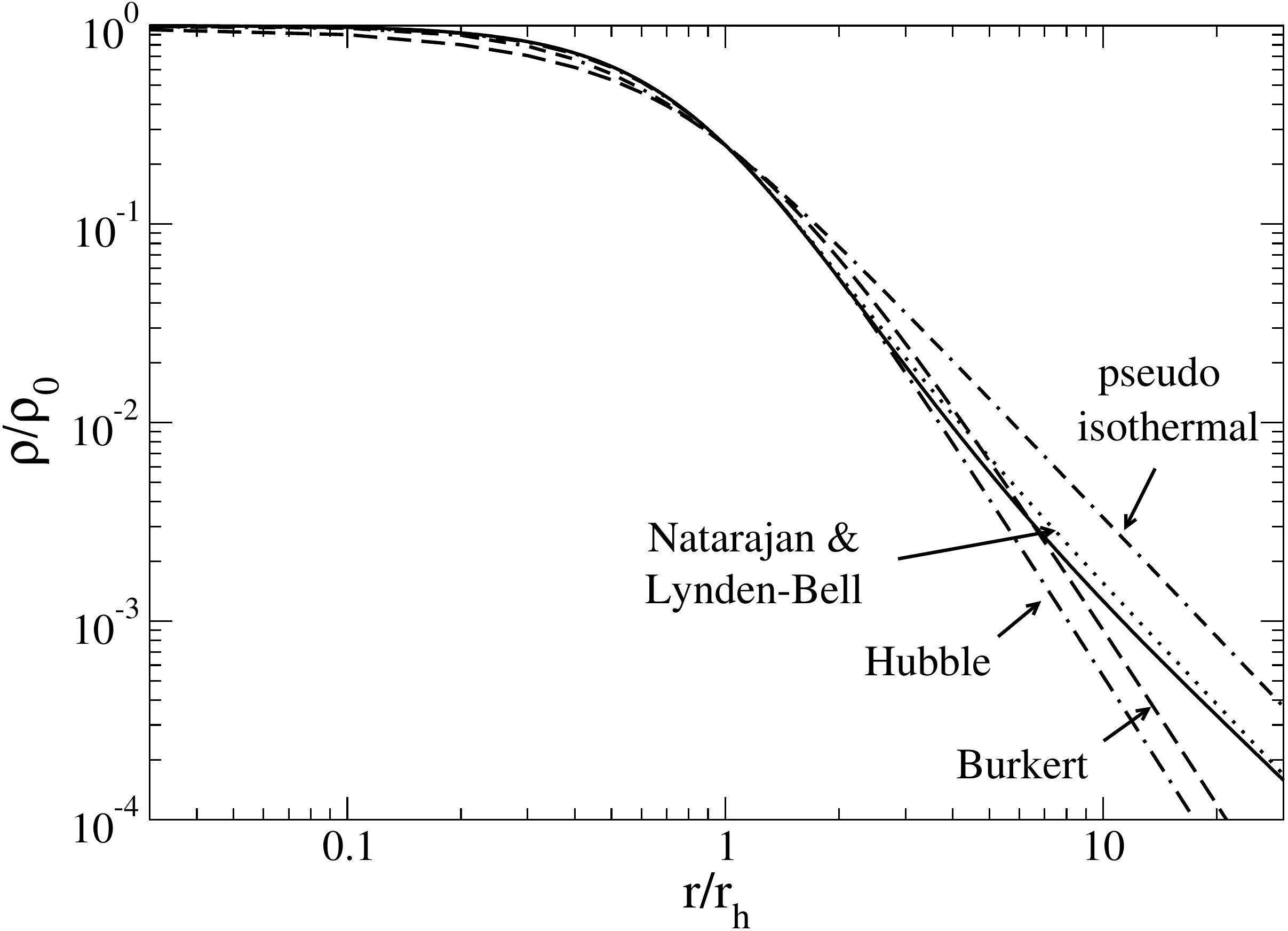}
\caption{Normalized isothermal density profile up to $30\, r_h$. It is
compared to the Burkert profile and to other profiles introduced in the
literature
(see Appendix \ref{sec_famous}).}
\label{isodensnormaliseELARGI}
\end{center}
\end{figure}

\begin{figure}[!h]
\begin{center}
\includegraphics[clip,scale=0.3]{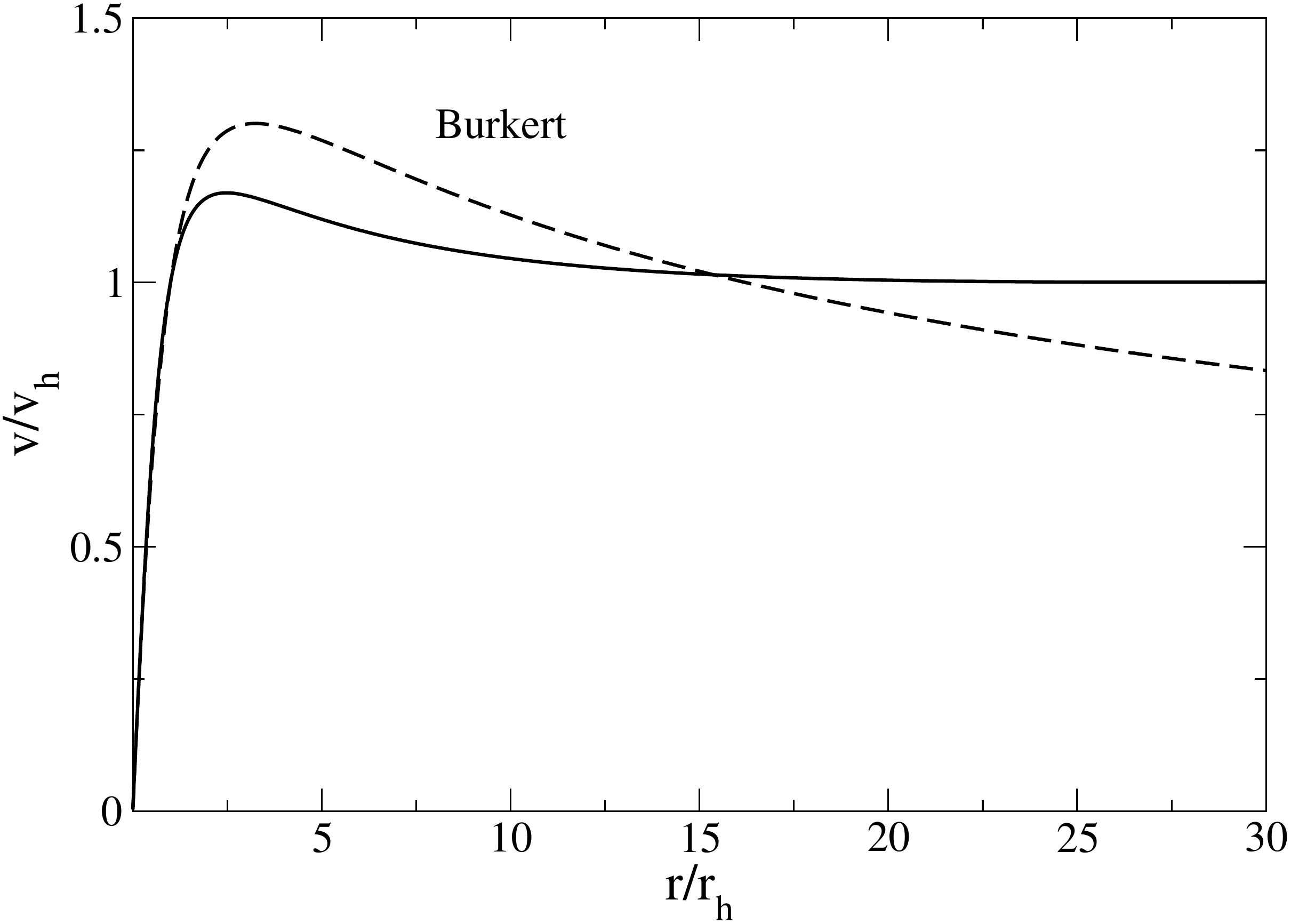}
\caption{Normalized circular velocity profile corresponding to the isothermal
sphere (the circular velocity reaches its maximum value $v_{\rm max}/v_h=1.17$
at $r_*/r_h=2.48$). It is compared to the Burkert profile ($v_{\rm
max}/v_h=1.30$ at $r_*/r_h=3.24$).}
\label{isovitnormalise}
\end{center}
\end{figure}

{\it Remark:} The modified Hubble profile provides a good fit to the isothermal
profile
for $r/r_h\le 3$. In particular, it provides a better fit than the pseudo
isothermal profile even though the pseudo isothermal profile
decreases asymptotically as  $r^{-2}$, like the isothermal profile,  while
the  
modified Hubble profile
decreases  asymptotically  as  $r^{-3}$. The reason
is that, for  $r/r_h\le 3$, we are not in the asymptotic limit where the
 isothermal density profile displays a logarithmic slope $-2$. This remark may
explain why,
in certain circumstances, we observe a density profile with an effective
logarithmic slope $-3$ (like for the observational Burkert profile  and for the
numerical NFW profile) instead of $-2$ (corresponding to  the isothermal profile
predicted by statistical mechanics). Indeed, at intermediate distances
$r/r_h\le 3$,  the isothermal profile
presents an effective logarithmic slope
$-3$ (see Fig. \ref{isodensnormaliseELARGI}). In this sense, the Burkert and NFW
profiles are not in contradiction with
the isothermal profile although their asymptotic slopes  (for $r\rightarrow
+\infty$) are different. This remark is important since the Burkert and NFW
profiles are purely empirical while the isothermal profile is justified by
statistical mechanics (in the sense of Lynden-Bell). This argument may provide
a physical justification of the   Burkert and NFW
profiles. A detailed comparison between the isothermal and Burkert profiles is
made is Sec. \ref{sec_comp}.

\section{Complete core-halo solution}
\label{sec_complete}

When studying BECDM halos, many authors \cite{ch2,ch3,marsh,hui} assume that
the bosons are noninteracting ($a_s=0$). However, cosmological constraints 
impose that the bosons should have a repulsive
self-interaction \cite{shapiro,suarezchavanis3}. A repulsive
self-interaction may also solve some tensions encountered in the noninteracting
model (see the Remark at the end of Appendix D.4 of \cite{suarezchavanis3}).
Therefore, a self-interacting SF may be more relevant than a
noninteracting SF. In this paper, we consider BECDM halos with a repulsive
self-interaction. We assume that the self-interaction is
sufficiently strong so that the TF approximation, which amounts to neglecting
the quantum potential, is applicable.\footnote{The considerations developed in
Appendix D.4 of \cite{suarezchavanis3} indicate that DM halos with an
attractive self-interaction may be just at the transition between the
noninteracting regime and the TF regime. Therefore, the TF
approximation may be just marginally applicable. In principle, we should take
into account both the quantum
potential and the self-interaction of the bosons as done in \cite{prd1,prd2}
for the self-gravitating BEC model at $T=0$. In this paper, for simplicity, we
ignore the contribution of the quantum potential.} We shall
obtain the complete core-halo profile of BECDM halos in the TF approximation.
Other situations (noninteracting bosons, bosons with an attractive
self-interaction, fermions...) will be considered in forthcoming papers
\cite{forthcoming}.

\subsection{Generalized Emden equation}
\label{sec_geq}

\begin{figure}
\begin{center}
\includegraphics[clip,scale=0.3]{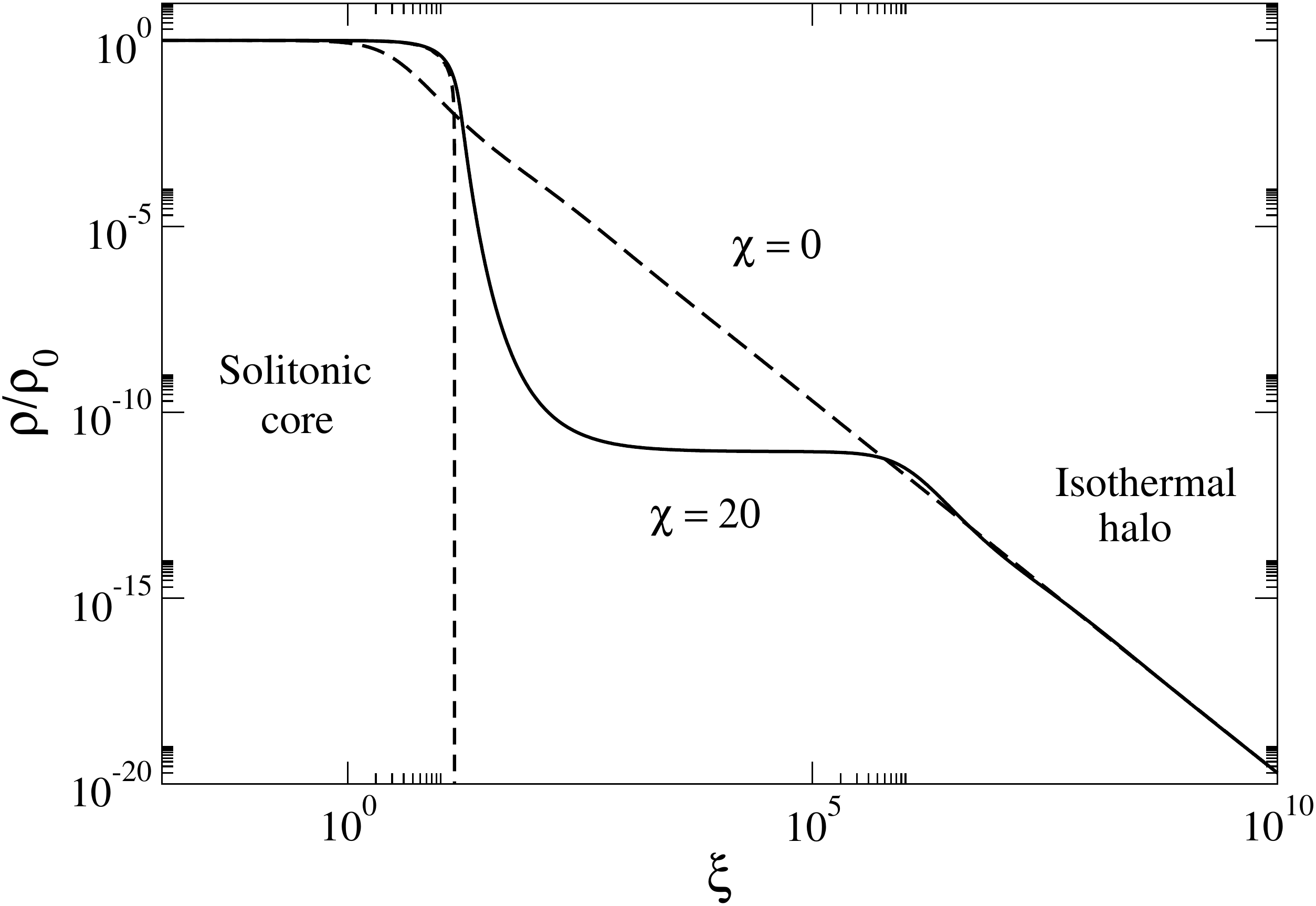}
\caption{Normalized density  profile corresponding to  $\chi=20$. It presents a
core-halo structure with a solitonic core and an isothermal halo (see Sec.
\ref{sec_desch}). The
long-dashed line corresponds to a pure isothermal halo ($\chi=0$).
The short-dashed line corresponds  to a pure solitonic profile whose analytical
expression is  $\rho/\rho_0=(\sqrt{\chi}/\xi)\sin(\xi/\sqrt{\chi})$.}
\label{densite}
\end{center}
\end{figure}

We start from the general equation (\ref{ch3}) determining the complete
core-halo structure of the system. We consider the standard BEC model
(\ref{standard}). We assume that the bosons have a repulsive self-interaction
($a_s>0$) and we
make the TF approximation ($Q=0$).  We also ignore the harmonic potential
($\omega_0=0$)
and restrict ourselves to spherically symmetric configurations. In
that case, Eq. (\ref{ch3}) reduces to
\begin{eqnarray}
\label{ch3sw}
-\frac{4\pi a_s\hbar^2}{m^3}\Delta\rho-\frac{k_B
T}{m}\Delta\ln\rho=4\pi
G\rho.
\end{eqnarray}
This equation is equivalent
to the equation obtained by combining Eq. (\ref{es2}) with the Poisson equation
(\ref{mad4}). We write
\begin{equation}
\label{ch23sw}
\rho=\rho_0
e^{-\psi}\quad {\rm and}\quad \xi=\frac{r}{r_0},
\end{equation}
where $\rho_0$ is the central density and $r_0$ is the thermal
core radius defined by Eq.
(\ref{chp3}).
We also introduce the dimensionless parameter
\begin{equation}
\label{ch27}
\chi=\frac{4\pi a_s\hbar^2\rho_0}{m^2 k_B T},
\end{equation}
which is a measure of the central density $\rho_0$ for a given value of the
temperature $T$. We call it the concentration parameter. Equation (\ref{ch3sw})
then takes the form of a
generalized Emden equation
\begin{equation}
\label{ch25sw}
\Delta\psi+\chi\nabla\cdot\left ( e^{-\psi}\nabla\psi\right
)=e^{-\psi}.
\end{equation}
The ordinary Emden equation (\ref{i5}) is recovered for $\chi=0$.
Another transformation in which   Eq. (\ref{ch3sw}) takes the form of a
generalized Lane-Emden equation is proposed in Appendix \ref{sec_gde}. For a
spherically
symmetric configuration, the generalized
Emden equation (\ref{ch25sw}) takes the form
\begin{equation}
\label{ch26}
\frac{1}{\xi^2}\frac{d}{d\xi}\left
(\xi^2\frac{d\psi}{d\xi}\right )+\frac{\chi}{\xi^2}\frac{d}{d\xi}\left
(\xi^2 e^{-\psi}\frac{d\psi}{d\xi}\right )=e^{-\psi},
\end{equation}
or, equivalently,
\begin{equation}
\label{ch28}
\frac{d^2\psi}{d\xi^2}+\frac{2}{\xi}\frac{d\psi}{d\xi}=\frac{\chi \left
(\frac{d\psi}{d\xi}\right )^2+1}{\chi+e^{\psi}}.
\end{equation}
For a given value of $\chi$, this equation can be solved numerically with
the boundary conditions $\psi(0)=\psi'(0)=0$. We note that
$\psi''(0)=1/[3(1+\chi)]$. The density profile is plotted in Fig.
\ref{densite} for $\chi=20$.

\subsection{Mass and circular velocity profiles}
\label{sec_mcvp}

\begin{figure}
\begin{center}
\includegraphics[clip,scale=0.3]{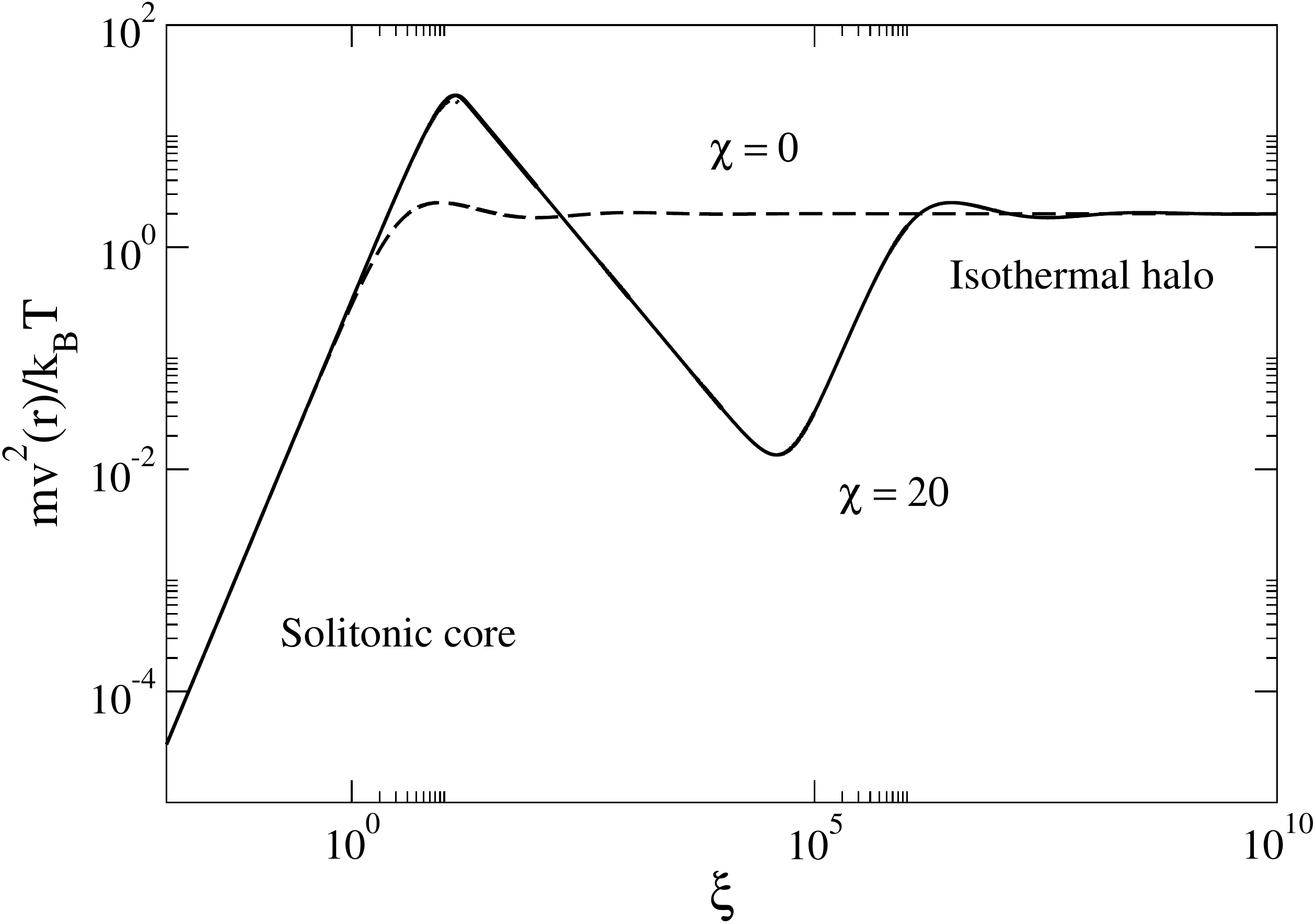}
\caption{Normalized circular velocity profile corresponding to $\chi=20$. It
displays a
dip due to the presence of the solitonic
core (see Sec. \ref{sec_desch}). The long-dashed
line corresponds to a pure isothermal profile ($\chi=0$). The
short-dashed line corresponds to
a pure solitonic profile  whose analytical
expression is $mv_c^2/k_B T=\chi\lbrack
(\sqrt{\chi}/\xi)\sin(\xi/\sqrt{\chi})-\cos(\xi/\sqrt{\chi})\rbrack$ (it can
hardly be distinguished from the solid line up to the border of the soliton).}
\label{vitesse}
\end{center}
\end{figure}

According to Eqs. (\ref{ch23sw}) and (\ref{hr2}),  the mass contained within a
sphere of
radius $r$ is
\begin{eqnarray}
\label{ch29}
M(r)=\rho_0 r_0^3 \int_0^{\xi}e^{-\psi}4\pi{\xi'}^2\, d\xi'.
\end{eqnarray} 
Using the generalized Emden equation (\ref{ch26}), we get
\begin{equation}
\label{ch30}
\frac{M(r)}{4\pi\rho_0 r_0^3}=\xi^2\psi'(\xi)\left \lbrack1+\chi
e^{-\psi(\xi)}\right
\rbrack.
\end{equation}
The circular velocity defined by Eq. (\ref{hr4}) is given by
\begin{equation}
\label{ch31}
\frac{v^2(r)}{4\pi G\rho_0r_0^2}=\xi \psi'(\xi)\left \lbrack
1+\chi e^{-\psi(\xi)}\right
\rbrack.
\end{equation}
Using Eq. (\ref{chp3}), we find that the temperature satisfies the
relation 
\begin{eqnarray}
\label{ch32}
\frac{k_B T}{m}=4\pi G\rho_0 r_0^2.
\end{eqnarray} 
Therefore, we can rewrite  Eq. (\ref{ch31}) as
\begin{equation}
\label{ch32b}
\frac{m v^2(r)}{k_B T}=\xi \psi'(\xi)\left \lbrack 1+\chi e^{-\psi(\xi)}\right
\rbrack.
\end{equation}
The circular velocity 
profile is plotted in Fig.
\ref{vitesse} for $\chi=20$.

\subsection{Normalized halo parameters}
\label{sec_hp}

The halo radius defined by Eq.
(\ref{hr1}) is given by
\begin{eqnarray}
\label{ch33}
r_h=\xi_h r_0,
\end{eqnarray} 
where the function $\xi_h(\chi)$ is determined by the equation
\begin{eqnarray}
\label{ch34}
e^{-\psi(\xi_h)}=\frac{1}{4}.
\end{eqnarray}
Using Eqs. (\ref{ch30}) and (\ref{ch33}) the halo mass defined by Eq.
(\ref{hr3}) is
given by
\begin{eqnarray}
\label{ch35}
\frac{M_h}{\rho_0 r_h^3}=4\pi\frac{\psi'(\xi_h)}{{\xi_h}}\left \lbrack 1+\chi
e^{-\psi(\xi_h)}\right
\rbrack.
\end{eqnarray} 
Using Eqs. (\ref{ch31}) and (\ref{ch33}) the circular velocity at the halo
radius defined by Eq. (\ref{hr5}) is given by
\begin{eqnarray}
\label{ch36}
\frac{v_h^2}{4\pi G\rho_0r_h^2}=\frac{\psi'(\xi_h)}{\xi_h}\left \lbrack 1+\chi
e^{-\psi(\xi_h)}\right
\rbrack.
\end{eqnarray}
Finally, using Eqs. (\ref{ch32}) and (\ref{ch33}), the normalized temperature
satisfies the relation
\begin{eqnarray}
\label{ch37}
\frac{k_B T}{Gm\rho_0 r_h^2}=\frac{4\pi}{\xi_h^2}.
\end{eqnarray}

\begin{figure}
\begin{center}
\includegraphics[clip,scale=0.3]{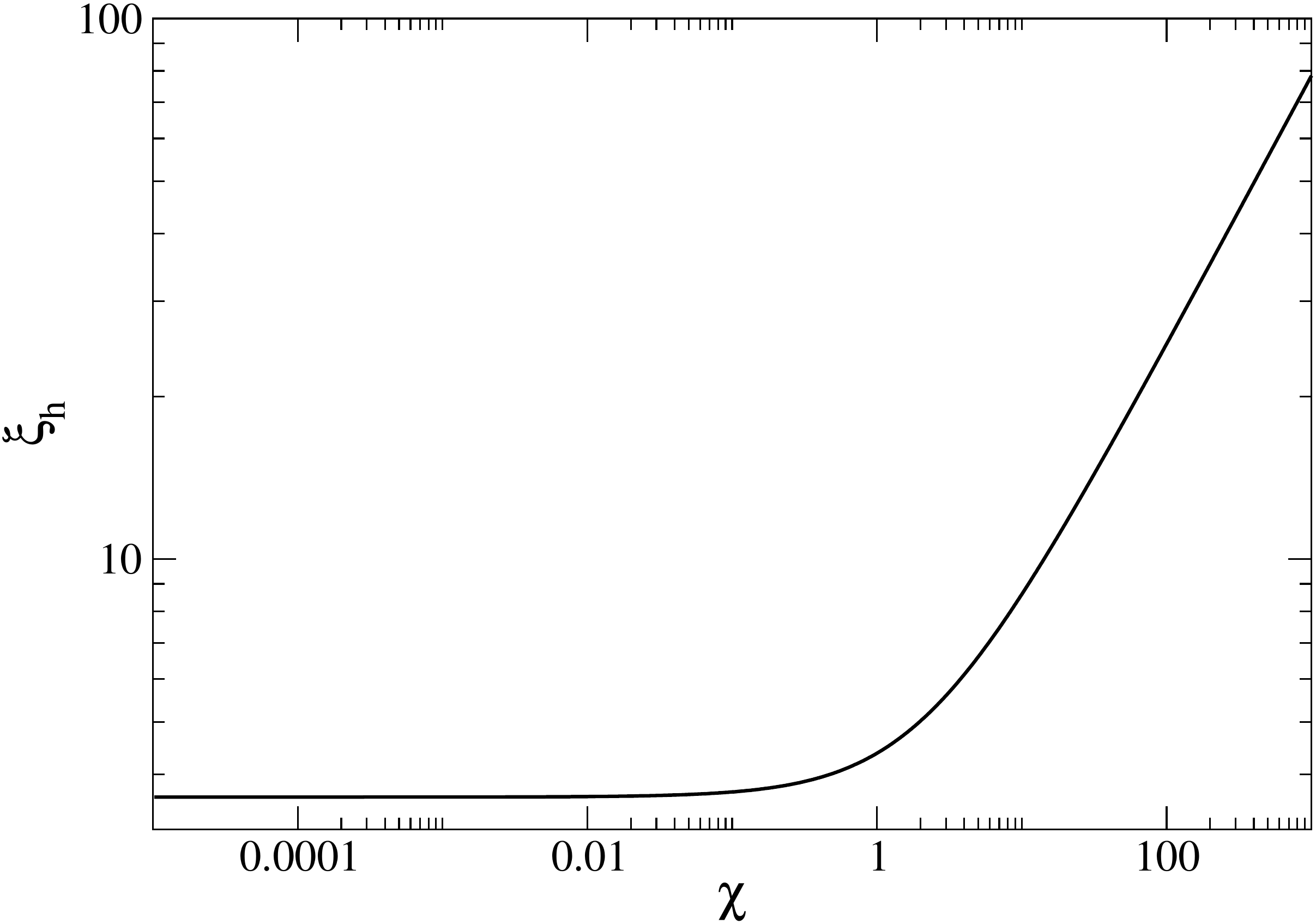}
\caption{Normalized halo radius $\xi_h$ vs $\chi$. For $\chi\rightarrow 0$,
$\xi_h\rightarrow 3.63$.  For $\chi\rightarrow +\infty$, $\xi_h\sim
2.47\sqrt{\chi}$.}
\label{xih}
\end{center}
\end{figure}

\begin{figure}
\begin{center}
\includegraphics[clip,scale=0.3]{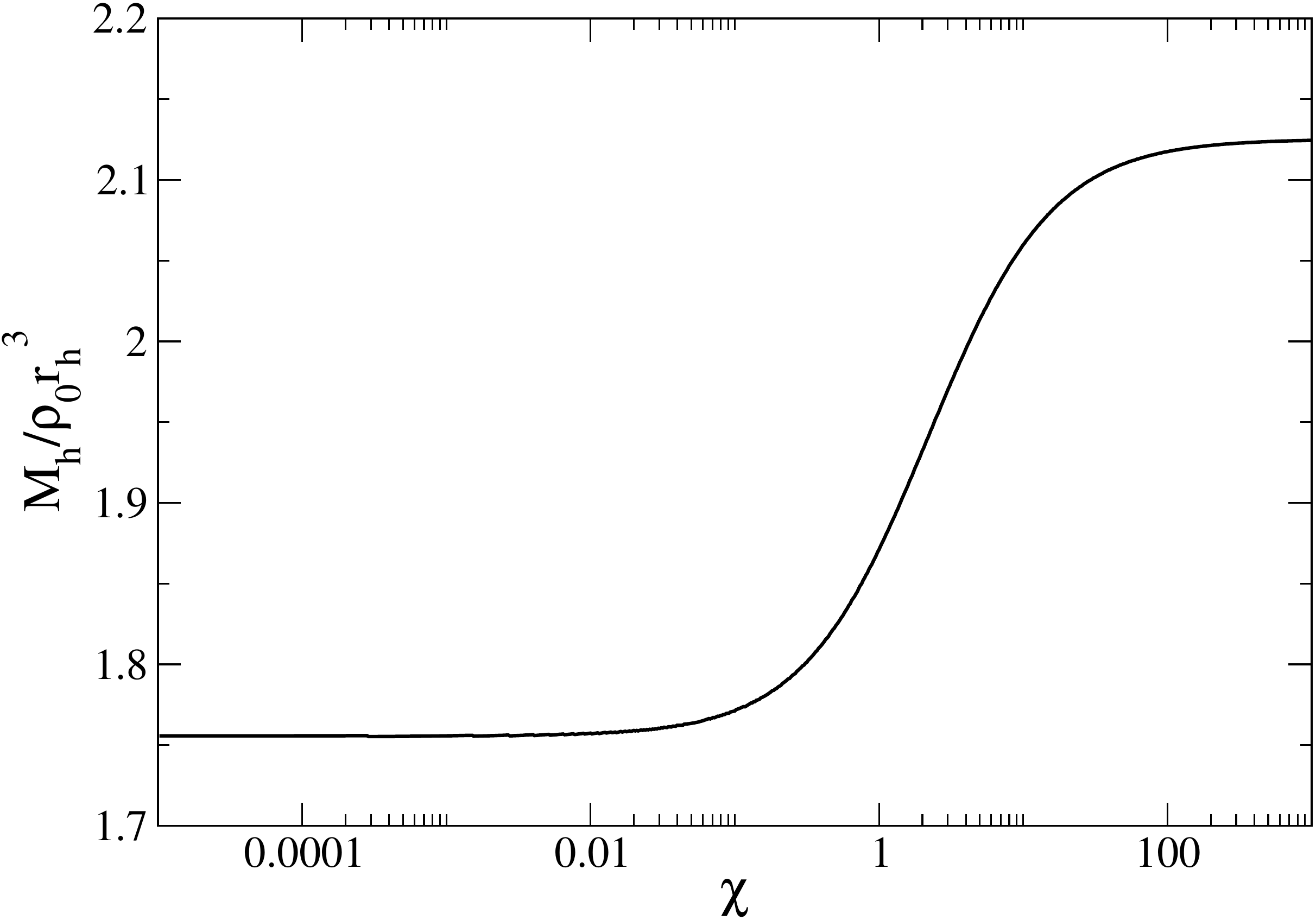}
\caption{Normalized halo mass $M_h/\rho_0 r_h^3$ vs $\chi$. It slightly changes 
from the value $1.76$ corresponding to the isothermal profile ($\chi\rightarrow
0$)
to the value $2.125$  corresponding to the solitonic profile ($\chi\rightarrow
+\infty$).
}
\label{massh}
\end{center}
\end{figure}

The normalized halo radius $\xi_h=r_h/r_0$ and the normalized halo mass
$M_h/\rho_0r_h^3$ are plotted as a function of $\chi$ in Figs.
\ref{xih} and \ref{massh}. The evolution with $\chi$ of the  normalized circular
velocity at the halo radius $v_h^2/4\pi G\rho_0 r_h^2$ and the evolution with
$\chi$ of the normalized temperature $k_B T/Gm\rho_0r_h^2$ can be easily deduced
from these curves.

\begin{figure}
\begin{center}
\includegraphics[clip,scale=0.3]{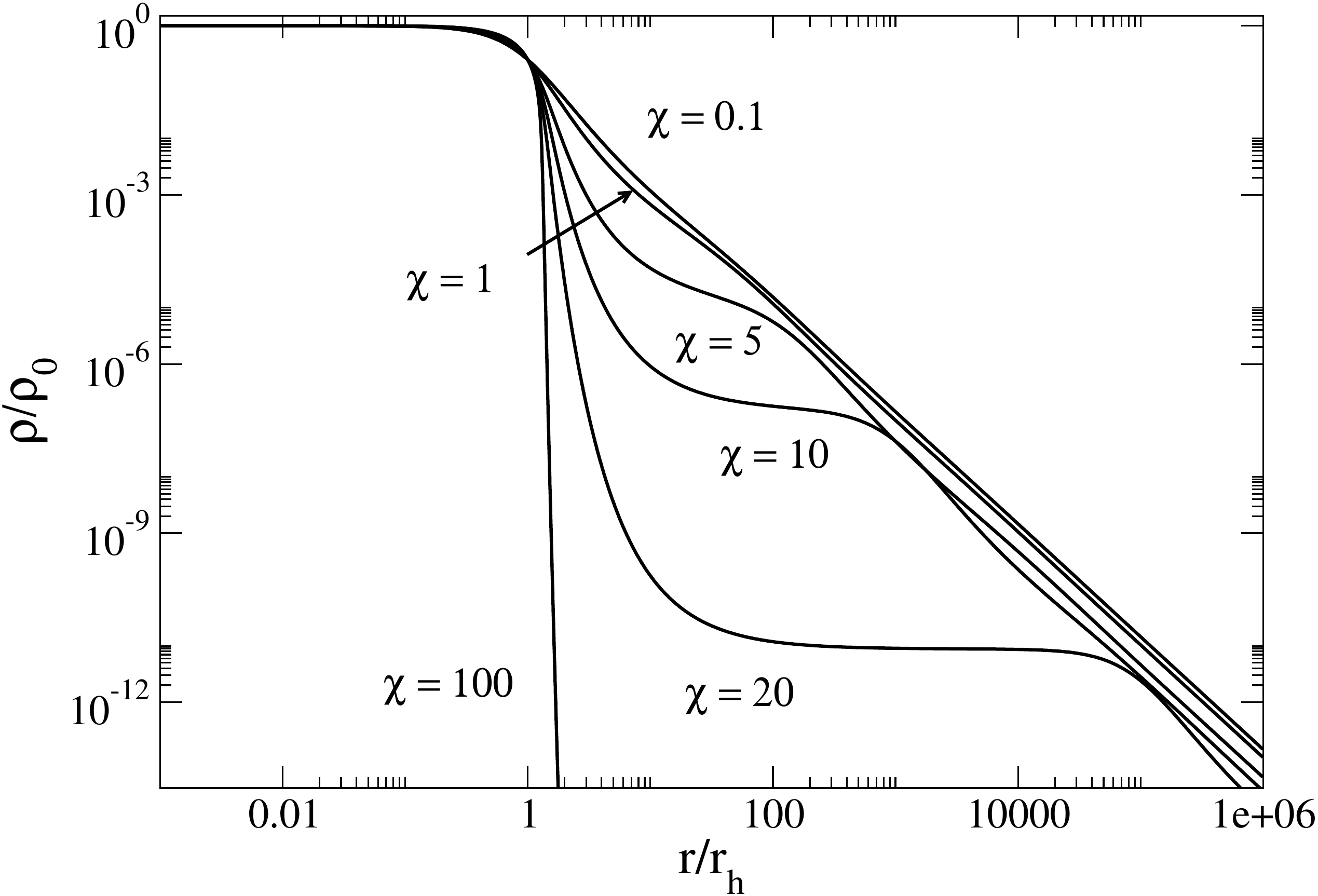}
\caption{Normalized density profiles for different values of $\chi$. For
$\chi\rightarrow 0$ we recover the purely isothermal profile of Fig.
\ref{isodensnormaliseELARGI} (the density profiles with $\chi\le \chi_*=0.1$ are
indistinguishable from the isothermal profile).  For
$\chi\rightarrow +\infty$ we recover the purely solitonic profile of Fig.
\ref{becdens}. 
For intermediate values of $\chi$, the density profiles have a core-halo
structure (see Sec. \ref{sec_desch}).  
}
\label{intrinsicprofiles}
\end{center}
\end{figure}

\begin{figure}
\begin{center}
\includegraphics[clip,scale=0.3]{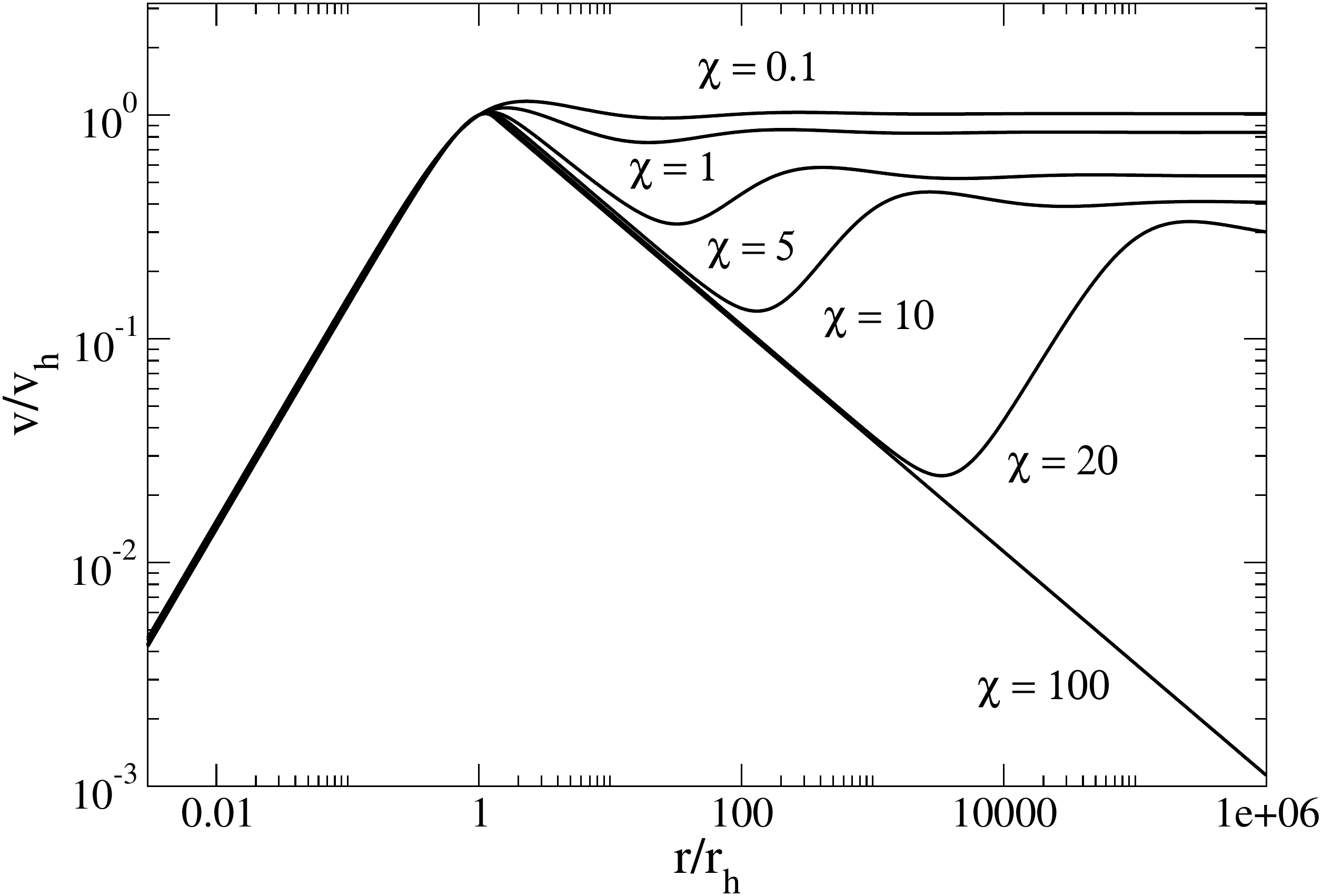}
\caption{Normalized circular velocity profiles for different values of $\chi$.
For
$\chi\rightarrow 0$ we recover the purely isothermal profile of Fig.
\ref{isovitnormalise}. For
$\chi\rightarrow +\infty$ we recover the purely solitonic profile of Fig.
\ref{becvit}. 
For intermediate values of $\chi$, the circular velocity profiles present a 
dip due to the presence of the solitonic core (see Sec.
\ref{sec_desch}).}
\label{intrinsicvitesse}
\end{center}
\end{figure}

The normalized density profile $\rho(r)/\rho_0$ and the normalized circular
velocity profile $v(r)/v_h$ are plotted as a function of the normalized
radial distance 
\begin{eqnarray}
\label{fou1}
\frac{r}{r_h}=\frac{\xi}{\xi_h}
\end{eqnarray} 
in Figs. \ref{intrinsicprofiles} and
\ref{intrinsicvitesse}.

The purely isothermal halo ($a_s=0$) corresponds to $\chi\rightarrow 0$. In that
limit, we recover the results of
Sec. \ref{sec_halo}. We also have (see Appendix \ref{sec_i}):
\begin{equation}
\label{ch38q}
\xi_h\rightarrow 3.63 \qquad
(\chi\rightarrow 0).
\end{equation}

The pure
soliton
(polytrope of index $n=1$ with $T=0$) corresponds to $\chi\rightarrow +\infty$.
In that limit, we recover the results of Sec. \ref{sec_tfw}. 
We must be
careful that the definition of $\xi$ adopted in the present section
differs from the
one used in Appendix \ref{sec_p}. It is easy to
see that they are related to each other by $\xi=\tilde\xi \sqrt{\chi}$,
where $\tilde\xi$ refers to the $\xi$ used in Appendix \ref{sec_p}. Since 
$\tilde\xi_h=2.4746$ (see Appendix \ref{sec_p}), we get
\begin{equation}
\label{ch38}
\xi_h\sim 2.4746\sqrt{\chi} \qquad (\chi\rightarrow +\infty).
\end{equation}

\section{Semi-analytical description of the core-halo structure}
\label{sec_desch}

In the previous section, the differential equation (\ref{ch3}) has been 
solved numerically in the TF approximation. The corresponding density and
circular velocity profiles are plotted in Figs. \ref{intrinsicprofiles} and
\ref{intrinsicvitesse}. They present a striking
core-halo structure with a solitonic core and an isothermal halo. The presence
of the core solves the cusp problem of the CDM model. The
presence of the isothermal atmosphere leads to flat rotation curves in
agreement with the observations. The circular velocity profile shows a dip
due to the presence of the
solitonic core.

\subsection{A short historic}

Historically, this core-halo structure was first obtained in models
where DM is made of fermions. In that case, the core corresponds
to a fermion ball which is a
completely degenerate nucleus at $T=0$  in which the gravitational
attraction is balanced by the quantum pressure arising from the Pauli
exclusion principle. In the nonrelativistic limit, the fermion ball is
equivalent to a polytrope of index $n=3/2$ and its mass-radius relation is
given by $R= 0.114\, {h^2}/{(Gm^{8/3}M^{1/3})}$ \cite{chandra}. The fermion ball
is surrounded
by an isothermal halo in which the gravitational
attraction is balanced by thermal pressure. An isothermal halo, with a true
thermodynamic temperature $T_{\rm th}$, corresponds to the statistical
equilibrium state of a gas of self-gravitating fermions resulting from a
collisional relaxation. However, the collisional relaxation time is generally
much larger than the age of the Universe. This is a problem to justify the
Fermi-Dirac distribution in a cosmological context. However, a quasistationary
state having  a core-halo structure made of an effective fermion ball surrounded
by an isothermal halo with an effective temperature $T$ may also be justified by
the statistical mechanics of violent relaxation \cite{lb} which takes place on 
a much
shorter timescale. This may be the correct justification of the Fermi-Dirac
distribution in a cosmological context as proposed in \cite{csmnras,clm1,clm2}
(see footnote 4).

The core-halo structure of the self-gravitating Fermi gas at nonzero
temperature, in the nonrelativistic and relativistic regimes,
was found by numerous authors.\footnote{We focus our review on papers that
explicitly display density profiles similar to those reported in Fig.
\ref{intrinsicprofiles}. For a more general bibliography, see Refs.
\cite{gao,stella,zcls,cls,cl,ir,gmr,merafina,imrs,vtt,bvn,bmv,csmnras,bvr,pt,
dark,bmtv,btv,rieutord,ijmpb,dvs1,dvs2,vss,urbano,rar,clm1,clm2,vs1,krut,vs2,
rsu}.}
The density profiles of a
partially degenerate gas of self-gravitating fermions at nonzero temperature
(like electrons in white
dwarfs, neutrons in neutron stars or massive neutrinos in DM halos) were first
computed by Wares 
\cite{wares}, Margrave \cite{margrave},
Hertel and Thirring \cite{ht}, Bludman and Van Riper 
\cite{bvriper}, Edwards and Merilan \cite{em} and
Edwards \cite{edwards} in
the context of stellar structure; by Ruffini and Stella  \cite{stella},
Chau {\it et al.}  \cite{cls},
Ingrosso and
Ruffini \cite{ir}, Gao {\it et
al.} \cite{gmr}, Merafina \cite{merafina} and Ingrosso {\it et
al.} \cite{imrs} in the context of DM made of massive neutrinos; and by
Chavanis and Sommeria 
\cite{csmnras} in
the context of
Lynden-Bell's theory of violent relaxation. Bilic
{\it et al.} \cite{bmtv,btv} considered a general relativistic
Fermi gas at
nonzero temperature, obtained a core-halo profile, and proposed
that the fermion ball may mimic a supermassive black hole at the center of
the galaxies. This idea was taken up again recently by Ruffini {\it et al.}
\cite{rar} and further developed by Arg\"uelles {\it et al.} \cite{krut}.
Chavanis {\it et
al.} \cite{clm2} studied phase
transitions in the fermionic King model and obtained a core-halo profile similar
to that of Fig. \ref{intrinsicprofiles} with a confinement due to tidal
effects.

A similar core-halo profile was obtained by Slepian and Goodman 
\cite{sg} in a model where DM is made of bosons with a repulsive
self-interaction (see also the figures in \cite{bdo,nottalechaos}). In that
case, the core corresponds to a soliton which is the
ground state of the self-gravitating BEC at $T=0$. In the TF limit, it is
equivalent to a polytrope of index $n=1$ with a radius $R=\pi
({a_s\hbar^2}/{Gm^3})^{1/2}$ independent of the mass. The solitonic core is
surrounded by an isothermal halo.\footnote{In the study of Slepian and Goodman
\cite{sg} the isothermal halo is justified by ordinary thermodynamics. This,
however, poses a timescale problem, related to the establishment of a
statistical equilibrium state by collisional relaxation on a relevant timescale
(shorter than the Hubble time) as discussed above in the case of
fermions. This also implies that the temperature in their model is
the true
thermodynamic temperature $T_{\rm th}$. As
a result, in order to derive the
equation of state of the boson gas, Slepian and Goodman \cite{sg} must consider
the case where the
bosons are condensed ($\rho>\rho_c$) or not ($\rho<\rho_c$). The
resulting equation of state (see their Fig. 1) presents a plateau after $\rho_c$
for weakly self-interacting bosons ($\theta \ll 1$) leading to an impossibility
to construct BECDM halos that match the observations (see Appendix
\ref{sec_sg}). In our approach, the isothermal halo has a
different origin. We assume since the start that the true thermodynamic
temperature is rigorously equal to
zero, or $T_{\rm th}\ll T_c$ (see Appendix
\ref{sec_cond}),
so the bosons
are always condensed and the fundamental equation is the GPP equation at
$T_{\rm th}=0$. The core-halo structure of the system (with a solitonic core
and a effective isothermal halo) is then an out-of-equilibrium
virialized structure justified by the process of
gravitational cooling and violent relaxation as explained in the Introduction.
As a result, the problems raised by  Slepian and Goodman \cite{sg} do not arise
in our model since the two models are phyically different.}
This core-halo structure was also observed in the numerical simulations of
Schive {\it et al.} \cite{ch2,ch3} for noninteracting bosons. In that
case, the halo is fitted by a NFW profile. In the very recent works of Lin {\it
et al.} \cite{lin} and Mocz {\it et al.} \cite{moczchavanis}, it
is shown
that the halo is relatively close to an isothermal distribution or to a
fermionic King distribution \cite{mnras,clm2} in which the
degeneracy is due to Lynden-Bell's type
of exclusion principle, as suggested in \cite{bdo,nottalechaos}.

In the following subsections, we provide a semi-analytic description of the
core-halo structure of self-gravitating bosons at nonzero
temperature by analogy with our previous work on fermions \cite{csmnras}. We
assume that $\chi\gg 1$ so that a clear separation exists between the core and
the halo marked by the presence of a plateau (see Sec. \ref{sec_another}).

\subsection{Properties of the density profile}
\label{sec_pchs}

We first consider the density profile. In the core, thermal effects are
negligible as compared to quantum
effects (self-interaction) so the system is equivalent to a pure soliton with
central density $\rho_0$, radius $R_c$ and mass $M_c$ (see Sec. \ref{sec_tfw}).
The soliton radius is
given by 
\begin{equation}
\label{rc}
R_c=\pi\left (\frac{a_s\hbar^2}{Gm^3}\right )^{1/2}
\end{equation}
and the soliton mass is given by 
\begin{equation}
\label{pchs1}
M_c=\frac{4}{\pi}\rho_0R_c^3.
\end{equation}

At larger distances, quantum effects are negligible as compared to thermal
effects so the system presents an isothermal halo with a density profile [see
Eq.
(\ref{es3})]:
\begin{equation}
\label{pchs2}
\rho=A\, e^{-\beta m \Phi}.
\end{equation}
Close to the core, the gravitational potential $\Phi$ is dominated by the
contribution of the central body (soliton) so that 
\begin{equation}
\label{pchs3}
\Phi\sim -\frac{GM_c}{r}.
\end{equation}
Therefore, the density profile can be approximated by
\begin{equation}
\label{pchs4}
\rho=B\rho_0 e^{\beta G M_c m \left (\frac{1}{r}-\frac{1}{R_c}\right )},
\end{equation}
where $B$ is a dimensionless prefactor of order unity which can be obtained
numerically (see below). 

When $r\rightarrow R_c$, the foregoing equation reduces to
\begin{equation}
\label{pchs5}
\rho=B\rho_0 e^{\beta G M_c m (R_c-r)/R_c^2}.
\end{equation}
Therefore, at the contact with the solitonic core, the density decreases
exponentially rapidly, and the system develops a {\it spike} (see
Fig. \ref{densite}) on a
typical lengthscale
\begin{equation}
\label{pchs6}
l=\frac{R_c^2}{\beta GM_c m}.
\end{equation}
The spike extends up to $R_s=R_c+l$. From Eqs. (\ref{ch27}), (\ref{rc}),
(\ref{pchs1}) and
(\ref{pchs6}), we get
\begin{equation}
\label{pchs7}
\frac{l}{R_c}=\frac{1}{\chi}.
\end{equation}
We note that $l\rightarrow 0$ when $\chi\rightarrow +\infty$ so that $R_s\simeq
R_c$ in that limit.

When $r\rightarrow +\infty$, the density profile given by Eq. (\ref{pchs4})
tends towards a constant
\begin{equation}
\label{pchs8}
\rho_c=B\rho_0 e^{-\beta G M_c m / R_c}.
\end{equation}
Therefore, when $r>R_s$, the density profile forms a {\it plateau} (see
Fig. \ref{densite}) with a
constant density $\rho_c$.\footnote{We emphasize that $\rho_c$ is different from
the
transition density 
\begin{equation}
\label{pchs9}
\rho_i=\frac{k_B Tm^2}{2\pi a_s\hbar^2}
\end{equation}
obtained
by equating the pressure $P=2\pi a_s\hbar^2\rho^2/m^3$ in the solitonic
core and the pressure $P=\rho
k_B T/m$ in the isothermal halo. When $\rho\gg \rho_i$ the
thermal pressure can be neglected and when $\rho\ll \rho_i$ the quantum pressure
can be neglected. This is similar to the Sommerfeld criterion for
fermions (the transition temperature for a given density is $k_BT_i=2\pi
a_s\hbar^2\rho/m^2$). We note that
\begin{equation}
\label{pchs10}
\frac{\rho_i}{\rho_0}=\frac{2}{\chi}.
\end{equation}
We also note that $\rho_i=(\pi/4)v_{\infty}^2/GR_c^2$, 
where $v_{\infty}^2=2k_B T/m$ is the constant circular velocity in an isothermal
halo.} Using Eq. (\ref{pchs6}), this
density may be rewritten as
\begin{equation}
\label{pchs11}
\rho_c=B\rho_0 e^{-R_c/l}.
\end{equation}
From Eqs. (\ref{pchs7}) and (\ref{pchs11}) we get
\begin{equation}
\label{pchs12}
\frac{\rho_c}{\rho_0}=B e^{-\chi}.
\end{equation}
We have numerically computed the ratio $\rho_c/\rho_0$ as a function of $\chi$
and found an excellent agreement with the theoretical prediction from Eq.
(\ref{pchs12}). This numerical study (not reported) allows us to obtain the
value of the
prefactor:
\begin{equation}
\label{pchs13}
B=4.24\times 10^{-3}.
\end{equation} 
Equation (\ref{pchs12})  is valid for $\chi\gg 1$. Actually, our numerical study
shows that Eq.
(\ref{pchs12}) with $B$ given by Eq. (\ref{pchs13}) is valid in good
approximation for $\chi \gtrsim 1$. For $\chi\lesssim 1$ the separation between
a core and a halo is not clear cut (the plateau disappears) so Eq.
(\ref{pchs12}) does not really make sense. In order to connect the result
$\rho_c=\rho_0$ when $\chi=0$ (no solitonic core) to the result from Eq.
(\ref{pchs12})
 when $\chi\gg  1$, we introduce a convenient
interpolation formula of
the form
\begin{equation}
\label{pchs14a}
\frac{\rho_c}{\rho_0}=B(\chi) e^{-\chi}
\end{equation} 
with
\begin{equation}
\label{pchs14b}
B(\chi)=1+(B-1)\tanh(\chi).
\end{equation} 
This interpolation formula is of course purely {\it ad hoc} but it has the
correct limiting behaviors and it will
facilitate the analysis of
Sec. \ref{sec_another}.

The plateau extends from $R_s$ up to a distance  $R_p$ after which it is not
possible to neglect the self-gravity of the halo as compared to the attraction
of the core. Therefore, $R_p$ is determined by the condition
\begin{equation}
\label{pchs15}
\frac{4}{3}\pi\rho_c(R_p^3-R_s^3)\simeq M_c.
\end{equation}
Making the approximation $R_p\gg R_s$ valid for $\chi\gg 1$, we get 
\begin{equation}
\label{pchs16}
R_p=\left (\frac{3M_c}{4\pi\rho_c}\right )^{1/3}.
\end{equation}
From Eqs. (\ref{pchs1}), (\ref{pchs12}) and (\ref{pchs16}), we obtain
\begin{equation}
\label{pchs17}
\frac{R_p}{R_c}=\left \lbrack \frac{3}{B(\chi)\pi^2}\right
\rbrack^{1/3}e^{\chi/3}.
\end{equation}

When $r\gg R_p$, we can neglect the gravitational attraction of the solitonic
core. In that case, the system is asymptotically equivalent to a purely
isothermal halo with a density profile decreasing as $\rho\sim k_B T/(2\pi Gm
r^{2})$ with damped oscillations superimposed \cite{chandra}.

In conclusion, the density profile represented in Fig. 
\ref{densite}  can be
divided in four regions:

(i) a purely solitonic core of almost constant density,

(ii) a spike,

(iii) a plateau of constant density,

(iv) a purely isothermal halo where the density decreases as
$r^{-2}$ with some oscillations.

This core-halo structure is similar to the one discussed in the case of
self-gravitating
fermions \cite{csmnras} with the difference that the solitonic
core
replaces the
degenerate fermion ball.

\subsection{Properties of the circular velocity profile}
\label{sec_pchv}

We now consider the circular velocity profile. In the solitonic core, the
density is approximately constant with value
$\rho_0$. Therefore, the mass contained within the sphere of radius $r$ is
$M(r)\simeq (4/3)\pi\rho_0 r^3$. This leads to a circular
velocity profile of the form
\begin{equation}
\label{pchv1}
v^2(r)\sim \frac{4}{3}\pi G\rho_0 r^2.
\end{equation}
The circular velocity increases linearly with the distance (see
Fig. \ref{vitesse}).

In the spike and at the beginning of the plateau, the mass almost does not
change so that $M(r)\simeq M_c$. This leads to a circular
velocity profile of the form
\begin{equation}
\label{pchv2}
v^2(r)\sim \frac{GM_c}{r}.
\end{equation}
The circular velocity has a Keplerian decay $\propto r^{-1/2}$ (see
Fig. \ref{vitesse}).

On the plateau, the density is constant with value
$\rho_c$. Therefore, at the end of the plateau where we can ignore the
contribution of the central mass $M_c$, the mass contained within the sphere of
radius $r$ is $M(r)\simeq (4/3)\pi\rho_c r^3$. This leads to a circular
velocity profile of the form
\begin{equation}
\label{pchv3}
v^2(r)\sim \frac{4}{3}\pi G\rho_c r^2.
\end{equation}
The circular velocity increases linearly with the distance as in the core (see
Fig. \ref{vitesse}).

At large distances, the system is purely isothermal and the mass
increases as $M(r)\sim 2k_B Tr/Gm$ leading to flat rotation curves:
\begin{equation}
\label{pchv4}
v^2(r)\rightarrow  \frac{2k_B T}{m}.
\end{equation}

In conclusion, the circular velocity profile represented in Fig. 
\ref{vitesse} can be divided in four regions:

(i) a  region where $v\propto r$ associated with the purely solitonic core,

(ii) a region where $v\propto r^{-1/2}$ associated with the spike,

(iii) a region where $v\propto r$ associated with the plateau, 

(iv)  a region where the velocity tends to a constant after some
oscillations associated with the purely isothermal halo.

This profile reflects the core-halo structure of the system. In particular, it
presents a dip due to the presence of the solitonic core.

\section{Mass-radius relation of DM halos and their physical
characteristics (Model I)}
\label{sec_mod1}

In this section, we express the previous results in terms of physical variables
appropriate to make a detailed comparison with observations. We determine the
mass-radius relation of DM halos and discuss their physical
characteristics.

\subsection{The constant surface density}
\label{sec_cst}

It is an
observational evidence that the surface density of DM halos is independent of
their mass and size and has a  universal value \cite{kormendy,spano,donato}:
\begin{equation}
\label{observation}
\Sigma_0=\rho_0r_h=141\, M_{\odot}/{\rm
pc}^2.
\end{equation}
This result is valid for all the galaxies even if their sizes and masses vary by
several orders of magnitude (up to $14$ orders of magnitude in luminosity). The
reason for this universality is not known but it is crucial to take this result
into account in any modeling of DM halos. Therefore, we shall assume this
relation as an empirical fact.

\subsection{Ultracompact halos: solitonic profile \\
(ground state)}
\label{sec_gs}

Ultrasmall DM halos such as dwarf spheroidal
galaxies (dSph) like Fornax ($R\sim 1\, {\rm kpc}$ and $M\sim 10^8\,
M_\odot$) are very compact and do not have an atmosphere, or just a tiny one.
The BECDM model predicts that there exists a minimum halo radius and a minimum
halo mass corresponding to a pure soliton without atmosphere ($T=0$). This is 
the
ground state of the GPP equations (\ref{intro1}) and (\ref{intro2}).
In the TF approximation, the soliton
radius $R_c$ where the density vanishes is given by Eq. (\ref{rc}) and the
soliton mass $M_c$ is given by Eq. (\ref{pchs1}). The halo
radius $r_h$ and the halo mass $M_h$ are given by Eq. 
(\ref{ch17}) where $R=R_c$. The halo radius is entirely determined by the
physical
properties of the bosons through the ratio $a_s/m^3$. The halo mass depends on
the central density $\rho_0$. However, since the central density is determined
by the halo radius according to Eq. (\ref{observation}), we find that the
halo mass is determined by the ratio $a_s/m^3$ and by the universal
value of $\Sigma_0$. Therefore, in the BECDM model, the   minimum halo radius
and the minimum halo
mass are given by
\begin{equation}
\label{pv1}
(r_h)_{\rm min}=0.788\, R_c,\quad (M_h)_{\rm min}= 1.32 \, \Sigma_0 R_c^2.
\end{equation}
Using Eqs. (\ref{ch18}) and (\ref{observation}), we find that
the maximum central
density and the minimum halo circular velocity are
\begin{equation}
\label{pv3b}
(\rho_0)_{\rm max}=1.27 \,
\frac{\Sigma_0}{R_c},\qquad (v_h^2)_{\rm min}= 1.67\, G\Sigma_0R_c.
\end{equation}
They can be explicitly rewritten  as
\begin{equation}
\label{pv2}
(r_{h})_{\rm min}=2.47\left (\frac{a_s\hbar^2}{Gm^3}\right )^{1/2},
\end{equation}
\begin{equation}
\label{pv3}
(M_{h})_{\rm min}=13.0
\frac{a_s\hbar^2\Sigma_0}{Gm^3},
\end{equation}
\begin{equation}
\label{pv3c}
(\rho_0)_{\rm max}=0.404\, \left
(\frac{Gm^3\Sigma_0^2}{a_s\hbar^2}\right )^{1/2},
\end{equation}
\begin{equation}
\label{pv3d}
(v_h^2)_{\rm min}= 5.25\,  \left (\frac{a_s\hbar^2G\Sigma_0^2}{m^3}\right
)^{1/2}.
\end{equation}

If we know the parameters $m$ and $a_s$ of the DM particle, we can
determine the  minimum halo radius $(r_h)_{\rm min}$ and the minimum halo mass
$(M_h)_{\rm min}$ from the foregoing expressions. 
However, we shall proceed the other way round. We assume that the
smallest
halo that we know (say Fornax to fix the ideas)  represents the ground state of
the GPP
equations (pure soliton without atmosphere) and we determine the characteristics
of the DM particle from the observed parameters of this halo. For
convenience, we adopt the
following value for its radius\footnote{It is not clear whether Fornax really
corresponds to the ground state of the GPP equations (there can be a little
atmosphere due to quantum interferences of excited states).
Furthermore, its radius $R_c$ is not
exactly
given by Eq. (\ref{pv4}). Therefore, the values of the minimum halo and of
$a_s/m^3$ that we obtain below are approximate. However, a more accurate
characterization of the ground state and of the value of $R_c$ will not
crucially affect our results.}
\begin{equation}
\label{pv4}
R_c=1\, {\rm kpc}.
\end{equation}
Using Eqs.  (\ref{observation}), (\ref{pv1}), (\ref{pv3b})
and (\ref{pv4}), we get 
\begin{equation}
\label{pv5}
(r_{h})_{\rm min}=788\, {\rm pc},\qquad (M_{h})_{\rm min}=1.86\times 10^8\,
M_{\odot},
\end{equation}
\begin{equation}
\label{pv6}
(\rho_0)_{\rm max}=0.179\,
M_{\odot}/{\rm pc}^3,\qquad (v_{h})_{\rm min}=31.9\, {\rm km/s}.
\end{equation}
The ratio $a_s/m^3$ characterizing the DM particle can then be
obtained from Eq. (\ref{rc}) yielding 
\begin{eqnarray}
\frac{a_s}{\rm fm}\left (\frac{{\rm eV}/c^2}{m}\right )^3=3.28\times 10^3.
\label{pv7}
\end{eqnarray}
Inversely, if we assume that DM halos are made of bosons with a ratio $a_s/m^3$
given by Eq. (\ref{pv7}), then we find that the minimum halo radius and the
minimum halo mass (ground state) are given by Eq. (\ref{pv5}). These values are
remarkably consistent with the mass and size of dSphs like Fornax.\footnote{In
the BECDM model, the order of magnitude of the minimum halo mass $(M_h)_{\rm
min}$ and minimum halo radius $(r_{h})_{\rm min}$ can be determined by a
Jeans stability analysis as detailed in Ref. \cite{abriljeans}. On the other
hand, the maximum
mass $(M_h)_{\rm max}\sim 10^{13}\, M_{\odot}$ of the DM halos (superclusters)
may be connected to the maximum Jeans mass at the transition between the
ultrarelativistic regime and the nonrelativistic regime (see Appendix F.7 of
\cite{abriljeans}).}

{\it Remark:} If we do the same reasoning for noninteracting
bosons and for fermions we get $m=2.92\times 10^{-22}\, {\rm eV}/c^2$ and $m=170
\, {\rm eV}/c^2$ respectively (see Appendix D of \cite{suarezchavanis3}).
The order of magnitude of these values is consistent with the values obtained by
other authors using more precise methods. We stress that, in this paper, we are
more interested in developing a general theory of DM halos (and presenting
basic ideas) rather than determining the characteristics of the DM particle
accurately. Therefore, orders of magnitudes are sufficient for our purposes.

\subsection{Large halos: isothermal profile}
\label{sec_lh}

For large halos like the Medium Spiral ($R\sim
10\, {\rm kpc}$ and $M\sim 10^{11}\, M_\odot$), the
mass of the solitonic core is negligible (see below) and it is a good
approximation
to assume that the halo is purely isothermal. In that case, using Eqs.
(\ref{ch21}), (\ref{ch22}) and  (\ref{observation}), we get 
\begin{equation}
\label{pv8}
M_h=1.76 \Sigma_0 r_h^2,\qquad 
\frac{k_B T}{m}=0.954 G\Sigma_0r_h,
\end{equation}
\begin{equation}
\label{pv9}
v_h^2=1.76 G\Sigma_0r_h,\qquad \rho_0=\frac{\Sigma_0}{r_h}.
\end{equation}
We can rewrite these equations as
\begin{equation}
\label{pv8q}
\frac{M_h}{\Sigma_0R_c^2}=1.76 \left (\frac{r_h}{R_c}\right )^2,\qquad 
\frac{k_B T}{mG\Sigma_0R_c}=0.954 \frac{r_h}{R_c},
\end{equation}
\begin{equation}
\label{pv9q}
\frac{v_h^2}{G\Sigma_0R_c}=1.76 \frac{r_h}{R_c},\qquad
\frac{\rho_0}{\Sigma_0/R_c}=\frac{R_c}{r_h}.
\end{equation}

The halo mass scales with the size as
$M_h\propto r_h^2$ and the temperature as $k_B T/m\propto r_h$. For
a halo of mass $M_h=10^{11}\, M_\odot$, we find
$r_h=20.1\, {\rm kpc}$, $\rho_0=7.02\times 10^{-3}M_{\odot}/{\rm
pc}^3$, $(k_B T/m)^{1/2}=108\, {\rm km/s}$, and 
$v_h=(GM_h/r_h)^{1/2}=146\,  {\rm km/s}$ (we also have $v_\infty=153\, {\rm
km/s}$).  We stress that these results
are independent of the characteristics of the DM particle.

{\it Remark:} If we consider an ultralight boson of mass $m\sim
10^{-22}\, {\rm
eV}/c^2$, we find that the temperature of
large halos such as the Medium Spiral is $T\sim 10^{-25}\, {\rm K}$.
Such a small temperature may not be
physical. This strongly
suggests that $T$ is not the true thermodynamic temperature. It may
rather
represent an effective temperature as we have argued in the Introduction.

\subsection{Small halos: core-halo profile}
\label{sec_chp}

We now consider the general case where the DM halos have a core-halo profile
with a solitonic core (polytrope $n=1$) and an isothermal halo. We shall
determine $r_h$, $\rho_0$, $M_h$, $v_h$, $(k_B
T/m)^{1/2}$ as a function of $\chi$ (see Sec. \ref{sec_complete}). We shall
express these quantities in units
of $R_c$, $\Sigma_0/R_c$, $\Sigma_0
R_c^2$ and $(G\Sigma_0 R_c)^{1/2}$ in order to be general (recall that $R_c$
can itself be expressed in terms of $a_s/m^3$ according to Eq. (\ref{rc})).
However, for numerical
applications we will use the values of $\Sigma_0$ and $R_c$ given by Eqs.
(\ref{observation}) and (\ref{pv4}) yielding 
\begin{equation}
\label{chp1}
R_c=1\, {\rm kpc},\qquad  \Sigma_0/R_c= 0.141\,
M_{\odot}/{\rm pc}^3,
\end{equation}
\begin{equation}
\label{chp2}
\Sigma_0 R_c^2=1.41\times
10^8\, M_{\odot},\qquad (G\Sigma_0 R_c)^{1/2}=24.6\, {\rm km/s}.
\end{equation}

\subsubsection{The thermal core radius}
\label{sec_chi}

Using Eqs. (\ref{chp3}), (\ref{ch27}) and (\ref{rc}), the concentration
parameter $\chi$ can be rewritten as
\begin{equation}
\label{chp4}
\chi=\frac{1}{\pi^2}\left (\frac{R_c}{r_0}\right
)^2.
\end{equation}
It can be seen as the ratio between the soliton radius $R_c$ and the thermal
core radius $r_0$. Therefore,
the thermal core radius is given in terms of $\chi$ by
\begin{equation}
\label{chp5}
\frac{r_0}{R_c}=\frac{1}{\pi\sqrt{\chi}}.
\end{equation}
Since $\xi=r/r_0$, the normalized radial distance can be expressed as
\begin{equation}
\label{chp6}
\frac{r}{R_c}=\frac{\xi}{\pi\sqrt{\chi}}.
\end{equation}

\subsubsection{The halo radius}
\label{sec_phr}

\begin{figure}
\begin{center}
\includegraphics[clip,scale=0.3]{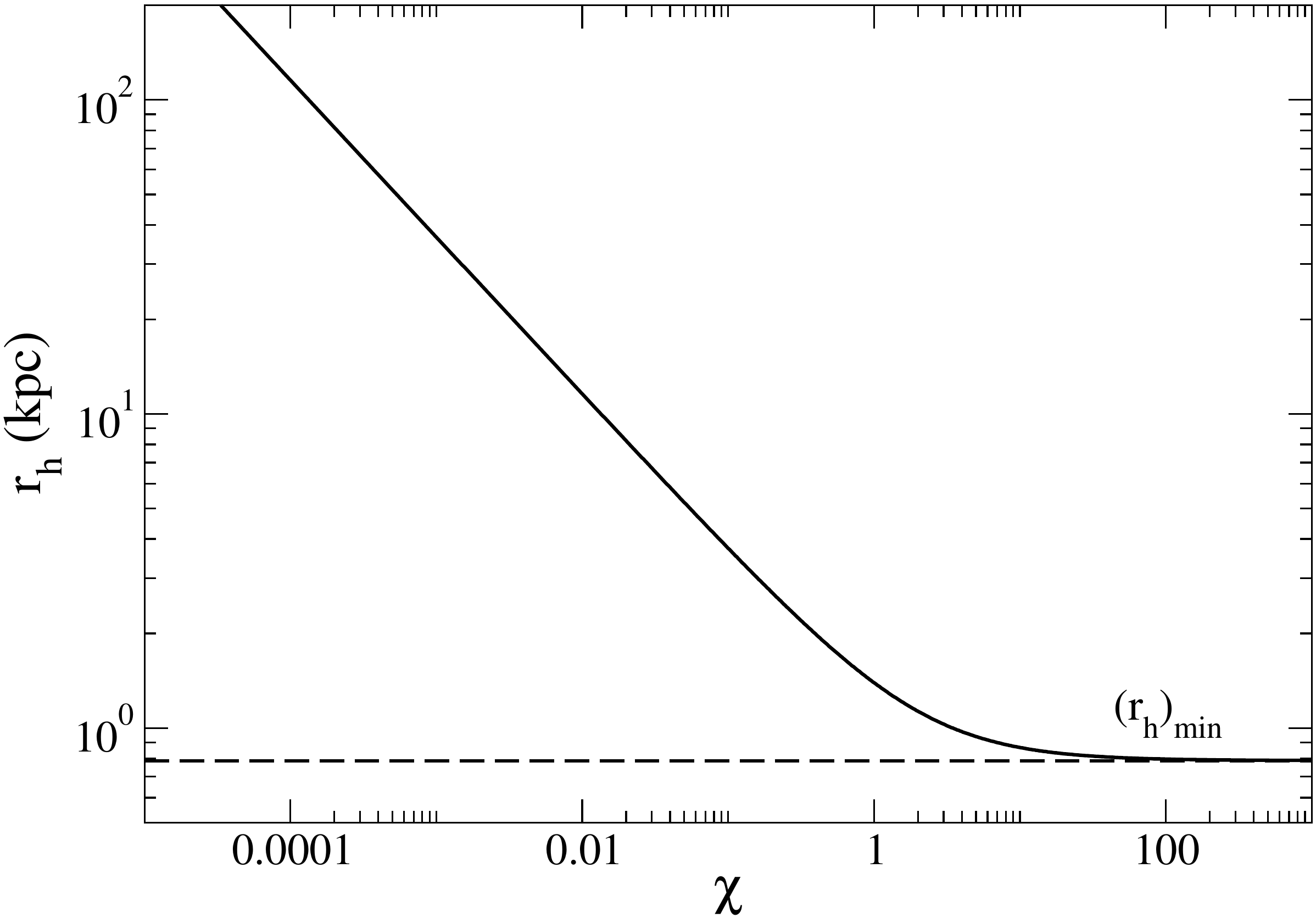}
\caption{Halo radius as a function of $\chi$. We note, parenthetically, that
$r_h=R_c$ for
$\chi=3.44$.}
\label{rayon}
\end{center}
\end{figure}

Using Eq. (\ref{chp6}), we find that the halo radius is given by
\begin{equation}
\label{phr1}
\frac{r_h}{R_c}=\frac{\xi_h}{\pi\sqrt{\chi}},
\end{equation}
where $\xi_h$ is defined in Sec. \ref{sec_complete}.
For $\chi\rightarrow
0$:
\begin{equation}
\label{phr2}
\frac{r_h}{R_c}\sim \frac{1.16}{\sqrt{\chi}}.
\end{equation}
For $\chi\rightarrow +\infty$:
\begin{equation}
\label{phr3}
\frac{r_h}{R_c}\rightarrow 0.788.
\end{equation}
The halo radius is
represented as a function of $\chi$ in Fig. \ref{rayon}. We see that large
halos correspond to small values of $\chi$ (i.e. they are isothermal)  and small
halos correspond to large values of $\chi$ (i.e. they are solitonic).

\subsubsection{The central density}
\label{sec_pcd}

\begin{figure}
\begin{center}
\includegraphics[clip,scale=0.3]{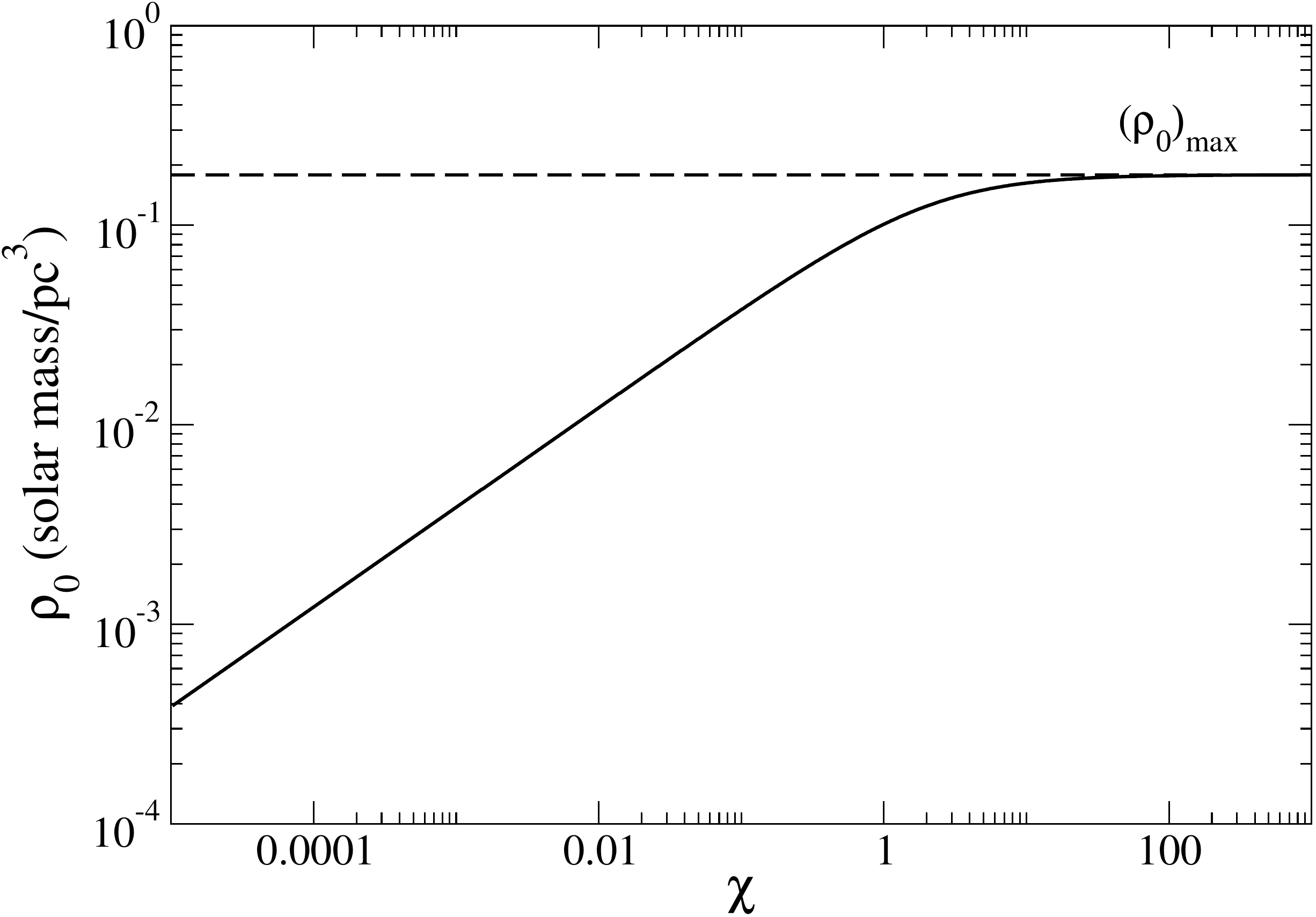}
\caption{Central density  as a function of $\chi$. }
\label{den}
\end{center}
\end{figure}

The central density can be obtained from  Eqs. (\ref{observation}) and
(\ref{phr1}) giving
\begin{equation}
\label{pcd1}
\frac{\rho_0R_c}{\Sigma_0}=\frac{\pi\sqrt{\chi}}{\xi_h}.
\end{equation}
For $\chi\rightarrow 0$:
\begin{equation}
\label{pcd2}
\frac{\rho_0R_c}{\Sigma_0}=0.865\sqrt{\chi}.
\end{equation}
For $\chi\rightarrow +\infty$:
\begin{equation}
\label{pcd3}
\frac{\rho_0R_c}{\Sigma_0}\rightarrow 1.27.
\end{equation}
The central density is
represented as a function of $\chi$ in Fig. \ref{den}. Large halos (small
$\chi$) have lower central
densities than small halos (large $\chi$).

\subsubsection{The halo mass}
\label{sec_phm}

\begin{figure}
\begin{center}
\includegraphics[clip,scale=0.3]{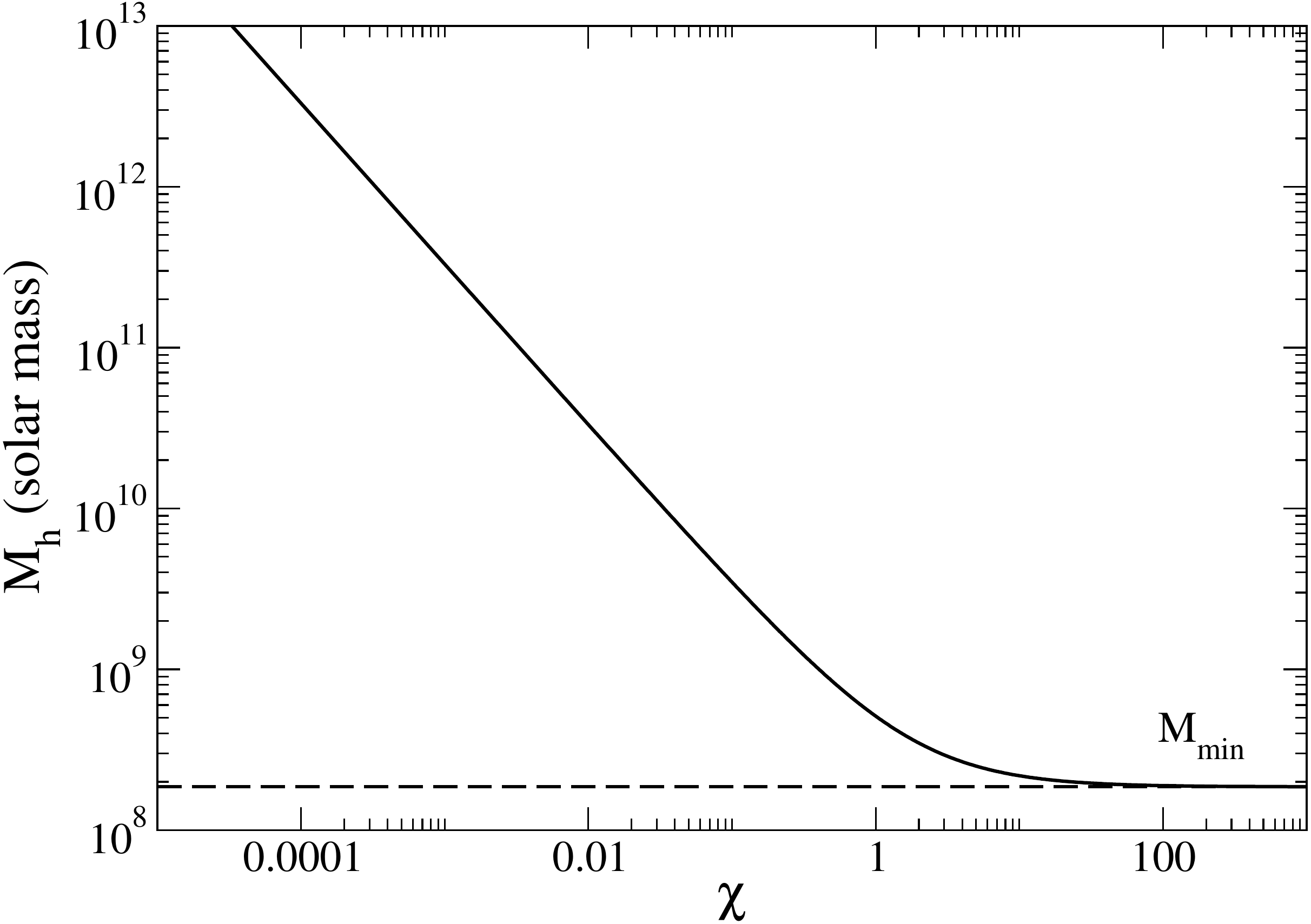}
\caption{Halo mass as a function of $\chi$.}
\label{masse}
\end{center}
\end{figure}

Using Eqs. (\ref{ch35}), (\ref{observation}) and (\ref{phr1}), the halo mass is
given by
\begin{equation}
\label{phm1}
\frac{M_h}{\Sigma_0R_c^2}=\frac{4\psi'(\xi_h)\xi_h}{\pi\chi}\left \lbrack 1+\chi
e^{-\psi(\xi_h)}\right
\rbrack.
\end{equation}
For $\chi\rightarrow 0$:
\begin{equation}
\label{phm2}
\frac{M_h}{\Sigma_0R_c^2}\sim \frac{2.34}{\chi}.
\end{equation}
For $\chi\rightarrow +\infty$:
\begin{equation}
\label{phm3}
\frac{M_h}{\Sigma_0R_c^2}\rightarrow 1.32.
\end{equation}
The halo mass is represented as a function of $\chi$ in Fig. \ref{masse}. Large
halos (small
$\chi$) have a larger mass than small halos (large $\chi$).

\begin{figure}
\begin{center}
\includegraphics[clip,scale=0.3]{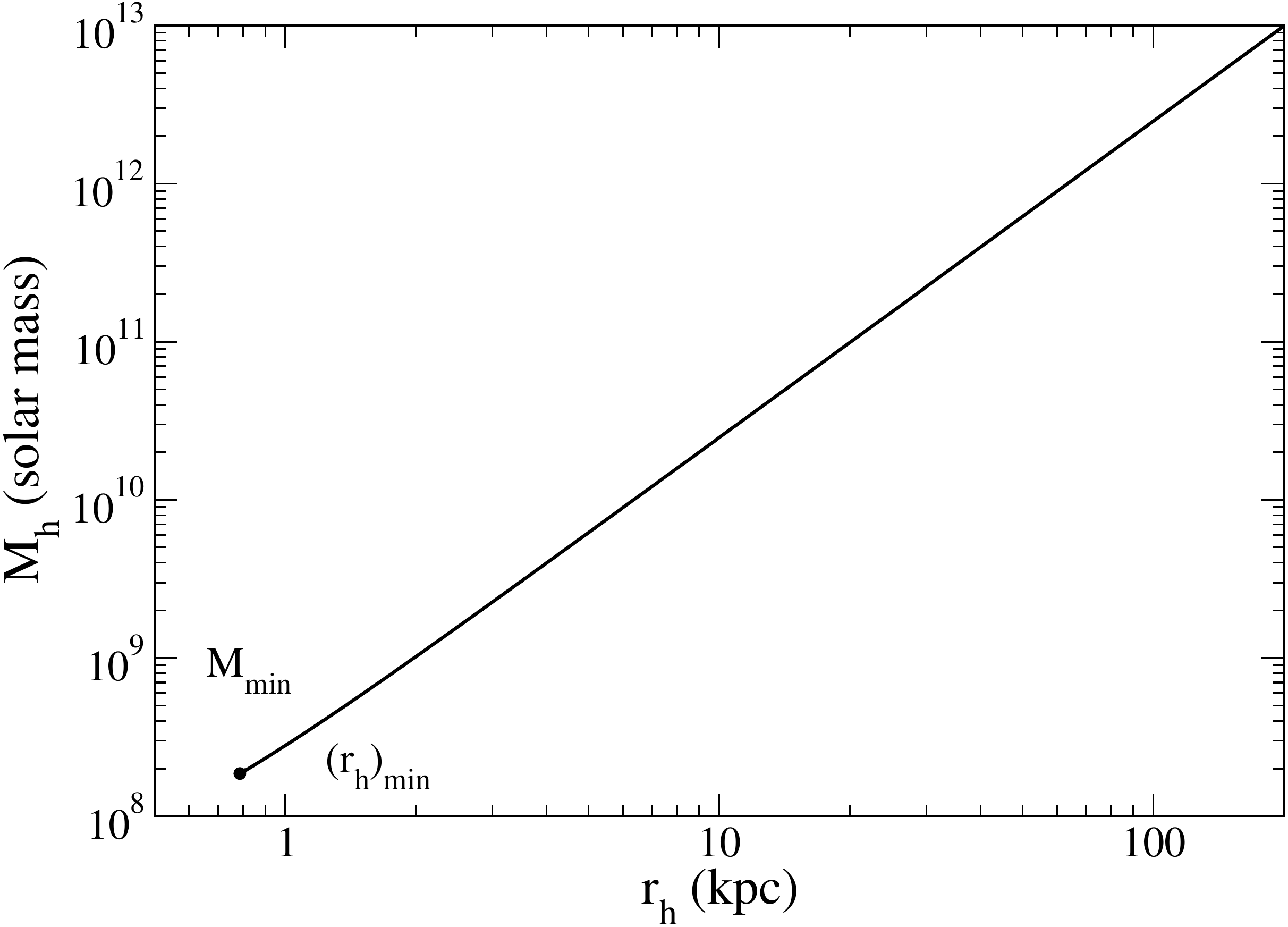}
\caption{Mass-radius relation of BECDM halos. Quantum mechanics is important
only
close to the ground state (bullet) where the halos have a solitonic core. Larger
halos
are purely isothermal without a solitonic core.}
\label{massradius}
\end{center}
\end{figure}

From Eqs. (\ref{phr1}) and (\ref{phm1}) we can obtain the mass-radius
relation $M_h(r_h)$ of the DM halos in parametric form (with parameter $\chi$).
It is
represented in Fig. \ref{massradius}.
It starts from 
$(r_h)_{\rm min}=788\, {\rm pc}$ and  $(M_{h})_{\rm min}=1.86\times 10^8\,
M_{\odot}$ (ground state) and behaves as $M_h\sim 1.76 \, \Sigma_0 r_h^2$ for
$r_h\rightarrow +\infty$ (large halos). Taking
$r_h=(r_h)_{\rm min}=788\, {\rm pc}$ and using the isothermal
mass-radius relation $M_h\sim 1.76 \, \Sigma_0 r_h^2$ [see Eq.
(\ref{pv8})], we find
$M_h=1.54\times 10^{8}\,
M_{\odot}$ which is very close to the value of the ground state $(M_{h})_{\rm
min}=1.86\times 10^8\,
M_{\odot}$ [see Eq. (\ref{pv5})]. Therefore, the difference between the exact
mass-radius relation
and its asymptotic behavior given by Eq. (\ref{pv8}), corresponding to purely
isothermal halos, is imperceptible in
a log-log
plot. It is only
close to the ground state ($(r_h)_{\rm min},(M_{h})_{\rm min}$) that quantum
effects (producing a solitonic core) are appreciable.
Actually, the main effect of quantum mechanics is to provide  an
origin (see the bullet in Fig. \ref{massradius}) to the mass-radius relation,
corresponding to a ``minimum halo'' (ground state). There is no minimum
halo in the $\Lambda$CDM model.

\subsubsection{The circular velocity}
\label{sec_phv}

\begin{figure}
\begin{center}
\includegraphics[clip,scale=0.3]{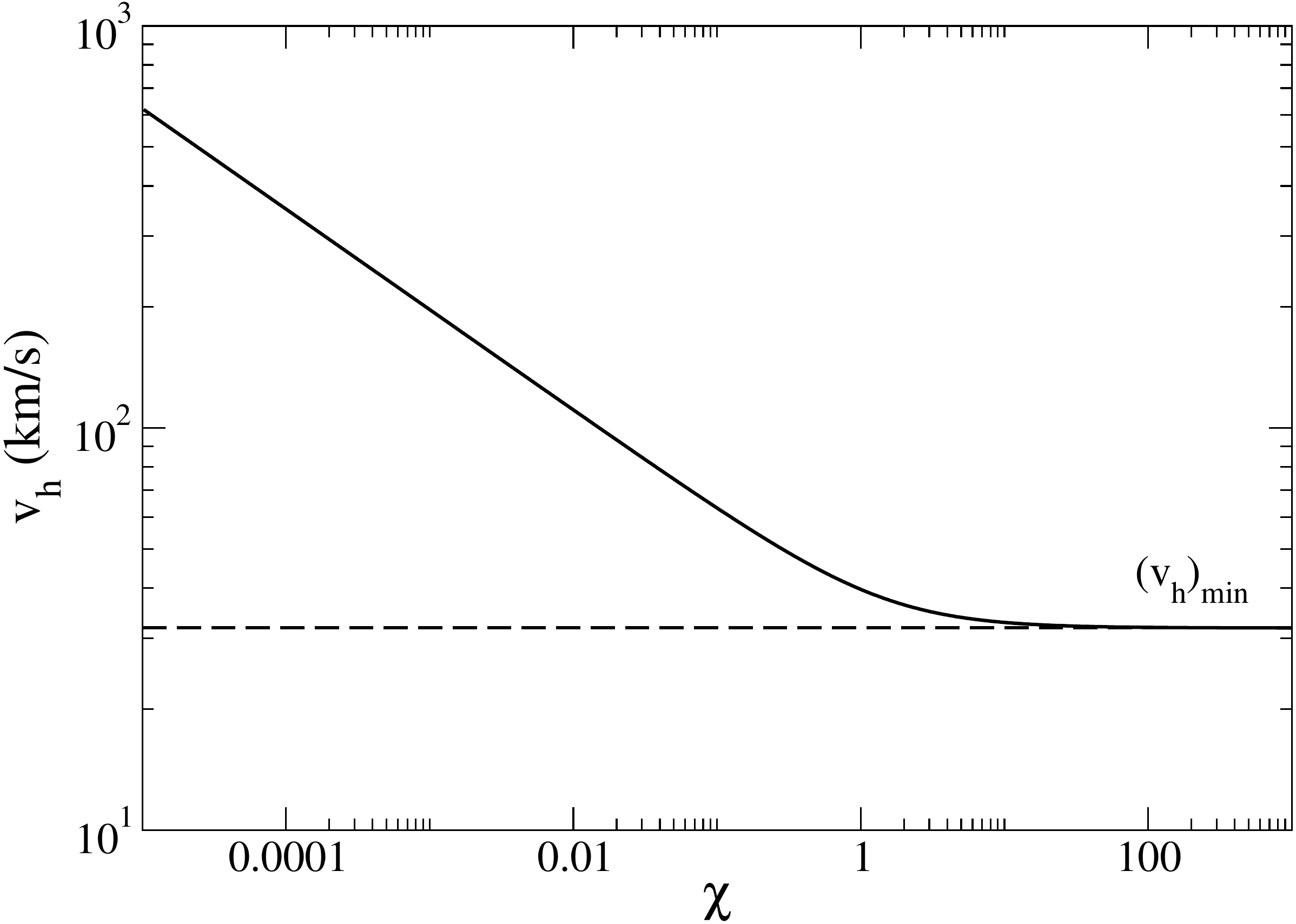}
\caption{Circular velocity as a function
of $\chi$. }
\label{vitesse2}
\end{center}
\end{figure}

Using Eqs. (\ref{ch36}), (\ref{observation}) and (\ref{phr1}),  the
halo velocity is given by
\begin{equation}
\label{phv1}
\frac{v_h^2}{G\Sigma_0R_c}=\frac{4\psi'(\xi_h)}{\sqrt{\chi}}\left \lbrack
1+\chi
e^{-\psi(\xi_h)}\right
\rbrack.
\end{equation}
For $\chi\rightarrow 0$:
\begin{equation}
\label{phv2}
\frac{v_h^2}{G\Sigma_0R_c}\sim \frac{2.03}{\sqrt{\chi}}.
\end{equation}
For $\chi\rightarrow +\infty$:
\begin{equation}
\label{phv3}
\frac{v_h^2}{G\Sigma_0R_c}\rightarrow 1.67.
\end{equation}
The halo velocity is represented as a function of $\chi$ in Fig. \ref{vitesse2}.
Large halos (small $\chi$) have a larger velocity than small halos
(large $\chi$).

\subsubsection{The effective temperature}
\label{sec_pet}

\begin{figure}
\begin{center}
\includegraphics[clip,scale=0.3]{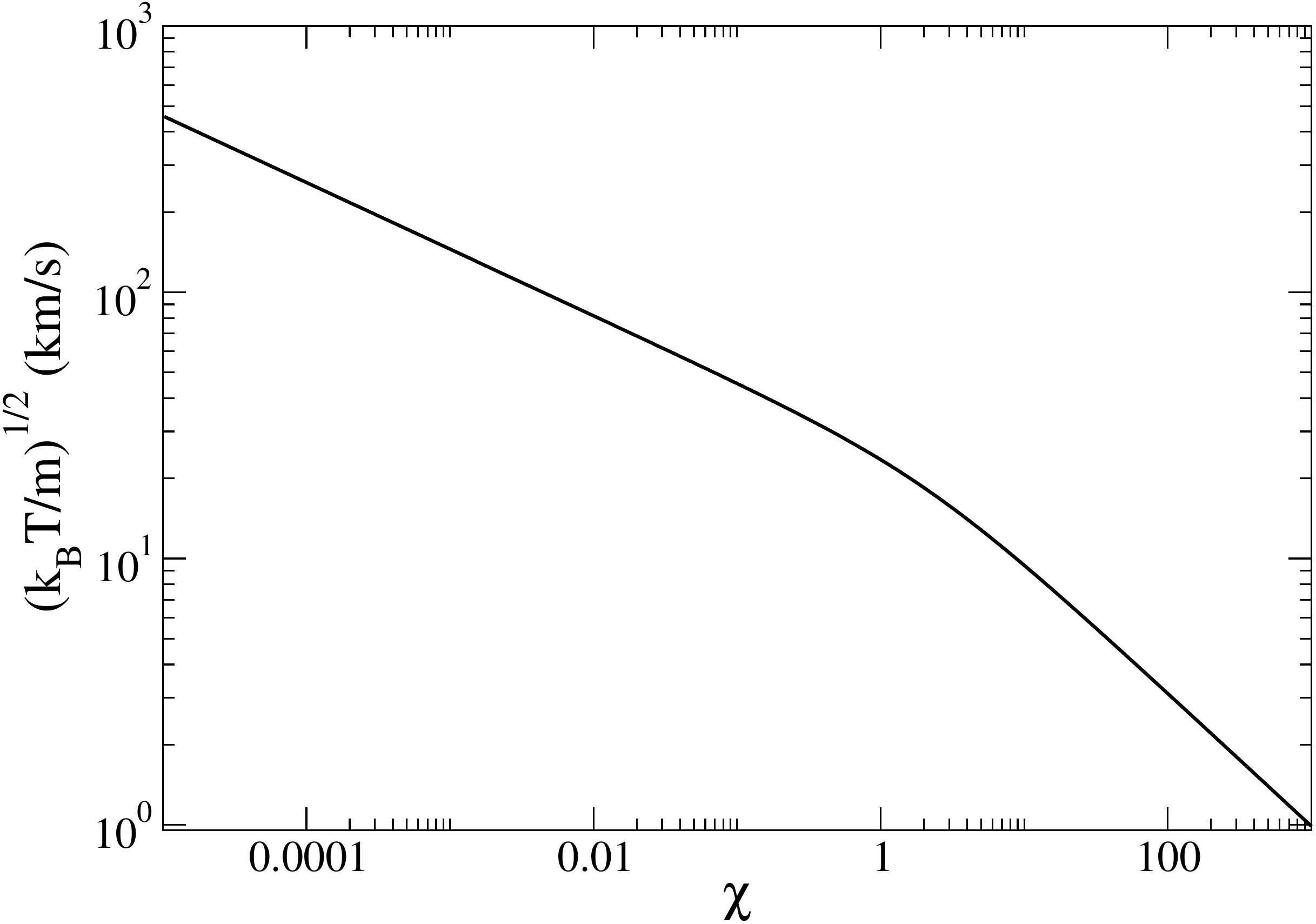}
\caption{Effective temperature as a function of $\chi$.}
\label{temperature}
\end{center}
\end{figure}

Using Eqs. (\ref{ch37}), (\ref{observation}) and (\ref{phr1}),
the effective temperature of the halos is given by
\begin{equation}
\label{pet1}
\frac{k_B T}{Gm\Sigma_0R_c}=\frac{4}{\xi_h\sqrt{\chi}}.
\end{equation}
For $\chi\rightarrow 0$:
\begin{equation}
\label{pet2}
\frac{k_B T}{Gm\Sigma_0R_c}\sim \frac{1.10}{\sqrt{\chi}}.
\end{equation}
For $\chi\rightarrow +\infty$:
\begin{equation}
\label{pet3}
\frac{k_B T}{Gm\Sigma_0R_c}\sim \frac{1.62}{\chi}.
\end{equation}
The effective temperature of the halo is represented as a function of $\chi$ in
Fig. \ref{temperature}. Large halos (small $\chi$)  have a larger effective
temperature than
small halos (large $\chi$). Actually, the temperature tends to zero when
$\chi\rightarrow
+\infty$ (ground state).

\begin{figure}
\begin{center}
\includegraphics[clip,scale=0.3]{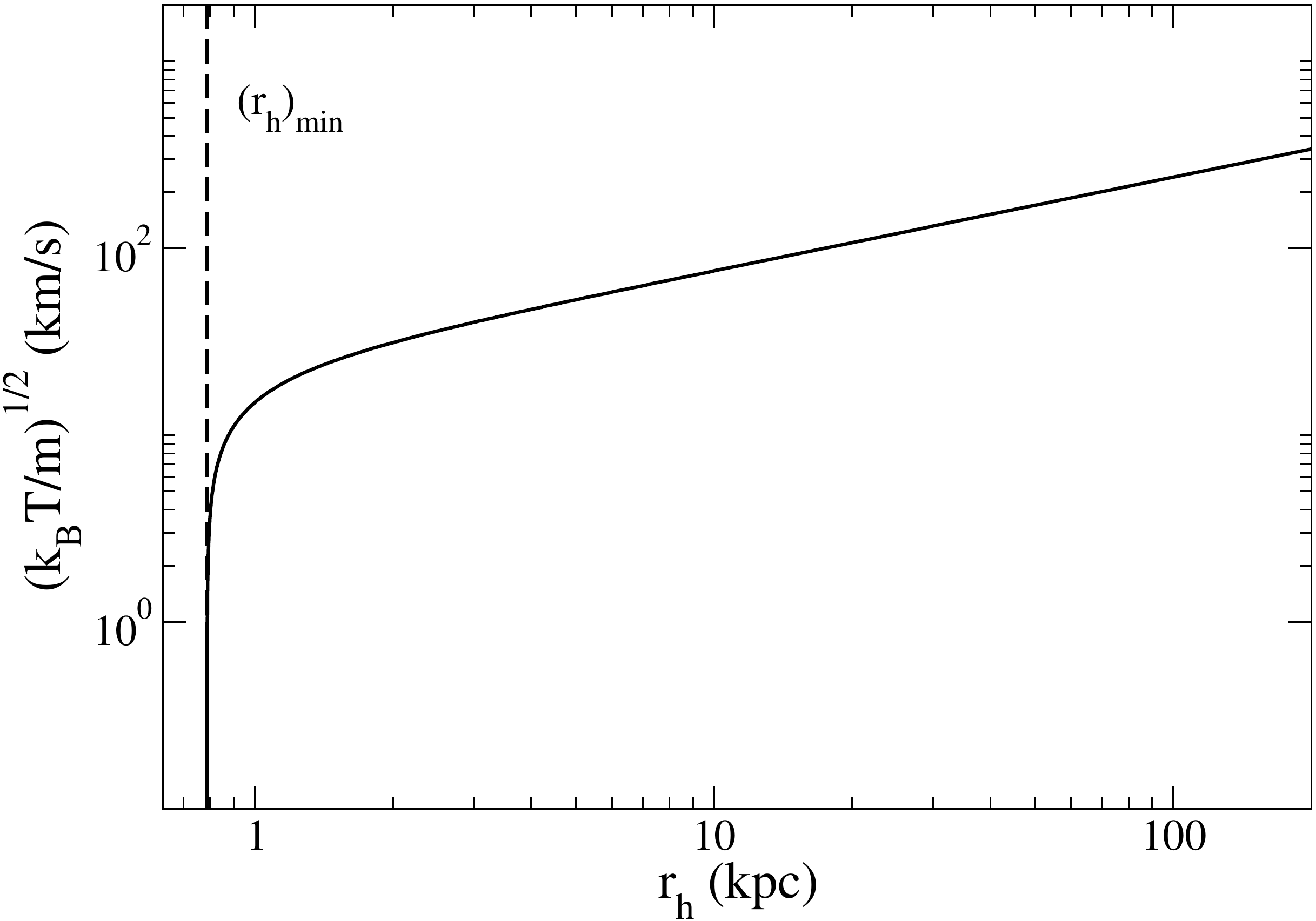}
\caption{Temperature-radius relation. A fit
for $r_h\rightarrow
(r_h)_{\rm min}$ gives $k_B T/Gm\Sigma_0R_c\sim 1.42
\lbrack (r_h-(r_h)_{\rm min})/R_c\rbrack^{0.888}$.}
\label{rayontemp}
\end{center}
\end{figure}

From Eqs. (\ref{phr1}) and (\ref{pet1}) we can obtain the
temperature-radius relation $T(r_h)$ in
parametric form (with parameter $\chi$). It is represented in
Fig. \ref{rayontemp}. It starts from $(r_h)_{\rm min}=788\, {\rm pc}$ and 
$T_{\rm min}=0$ (ground state)  and behaves as ${k_B T}/{m}\sim 0.954\,
G\Sigma_0r_h$ for $r_h\rightarrow +\infty$ (large
halos).

\subsubsection{The soliton radius and the soliton mass}
\label{sec_srsm}

The size of the solitonic core $R_c$ is given by Eq. (\ref{rc}). We note that
the
solitonic core always has the same radius, whatever the halo mass $M_h$, since
it only depends on the ratio  $a_s/m^3$ which is a propery of
the DM particle. The soliton mass $M_c$ is
given by Eq. (\ref{pchs1}). The soliton mass depends on the halo mass $M_h$
since it is proportional to the
central density $\rho_0$. From Eqs. (\ref{pchs1}) and (\ref{pcd1}), we find
that
the soliton mass is given by
\begin{equation}
\label{pcd1z}
\frac{M_c}{\Sigma_0R_c^2}=\frac{4\sqrt{\chi}}{\xi_h}.
\end{equation}
For $\chi\rightarrow 0$:
\begin{equation}
\label{pcd2z}
\frac{M_c}{\Sigma_0R_c^2}\sim 1.10\sqrt{\chi}.
\end{equation}
For $\chi\rightarrow +\infty$:
\begin{equation}
\label{pcd3z}
\frac{M_c}{\Sigma_0R_c^2}\rightarrow 1.62.
\end{equation}
The evolution of the soliton mass $M_c$ as a function of $\chi$ can be easily
deduced from Fig. 
\ref{den} since $M_c\propto \rho_0$. Large halos (small
$\chi$) have a less massive solitonic core than small halos (large $\chi$). For
large halos, using Eqs. (\ref{phm2}) and (\ref{pcd2z}) we find that
\begin{equation}
\label{pcd2zb}
\frac{M_c}{\Sigma_0R_c^2}\sim 1.68 \sqrt{\frac{\Sigma_0R_c^2}{M_h}}.
\end{equation}
According to this relation, the soliton mass decreases as
$M_c\propto
M_h^{-1/2}$ with the halo mass. Actually, for $\chi\lesssim 0.1$, there is no
well-defined solitonic core (see below) so that the relation from Eq.
(\ref{pcd2zb}) is meaningless. Large DM halos are purely
isothermal, without a solitonic core. In that case, Eq. (\ref{pcd2zb}) gives the
mass of the isothermal core within a sphere of radius $R_c$.

\subsubsection{The mass $M_{300}$}
\label{sec_mtc}

It is an observational evidence that all dwarf spheroidal galaxies (dSphs) of
the Milky Way have the same total DM mass contained within a radius
$r_u=300\, {\rm pc}$. From the observations, Strigari {\it et al.}
\cite{strigari} obtained $\log(M_{300}/M_{\odot})=7.0_{-0.4}^{+0.3}$. Let us see
how this result compares with our model.

Using Eqs. (\ref{ch30}), (\ref{observation}), (\ref{chp5}) and the relation
$\xi_h=r_h/r_0$ we obtain
\begin{equation}
\label{mtc1}
\frac{M_{300}}{\Sigma_0 R_c^2}=
\left(\frac{r_u}{R_c}\right)^2\frac{4\pi}{\xi_h}\psi'\left(\frac{r_u\pi\sqrt {
\chi}}{R_c} \right)\left
\lbrack 1+\chi
e^{-\psi(r_u\pi\sqrt{\chi}/R_c)}\right
\rbrack.
\end{equation}
For $\chi\rightarrow 0$:
\begin{equation}
\label{mtc2}
\frac{M_{300}}{\Sigma_0 R_c^2}\sim \frac{4\pi^2}{3}\left (\frac{r_u}{R_c}\right
)^3\frac{\sqrt{\chi}}{\xi_h}\sim 0.0979\sqrt{\chi}.
\end{equation}
For $\chi\rightarrow +\infty$:
\begin{equation}
\frac{M_{300}}{\Sigma_0 R_c^2}\rightarrow \frac{4R_c}{\pi^2 r_h}\left \lbrack
\sin\left
(\frac{\pi r_u}{R_c}\right )-\frac{\pi r_u}{R_c}\cos\left (\frac{\pi
r_u}{R_c}\right )\right
\rbrack=0.131.
\label{mtc3}
\end{equation}
To obtain Eq. (\ref{mtc2}) we have used $\psi\sim \xi^2/6$ when
$\xi\rightarrow 0$ \cite{chandra} and to obtain Eq. (\ref{mtc3}) we
have 
used Eqs. (\ref{ch15}) and (\ref{observation}).

\begin{figure}
\begin{center}
\includegraphics[clip,scale=0.3]{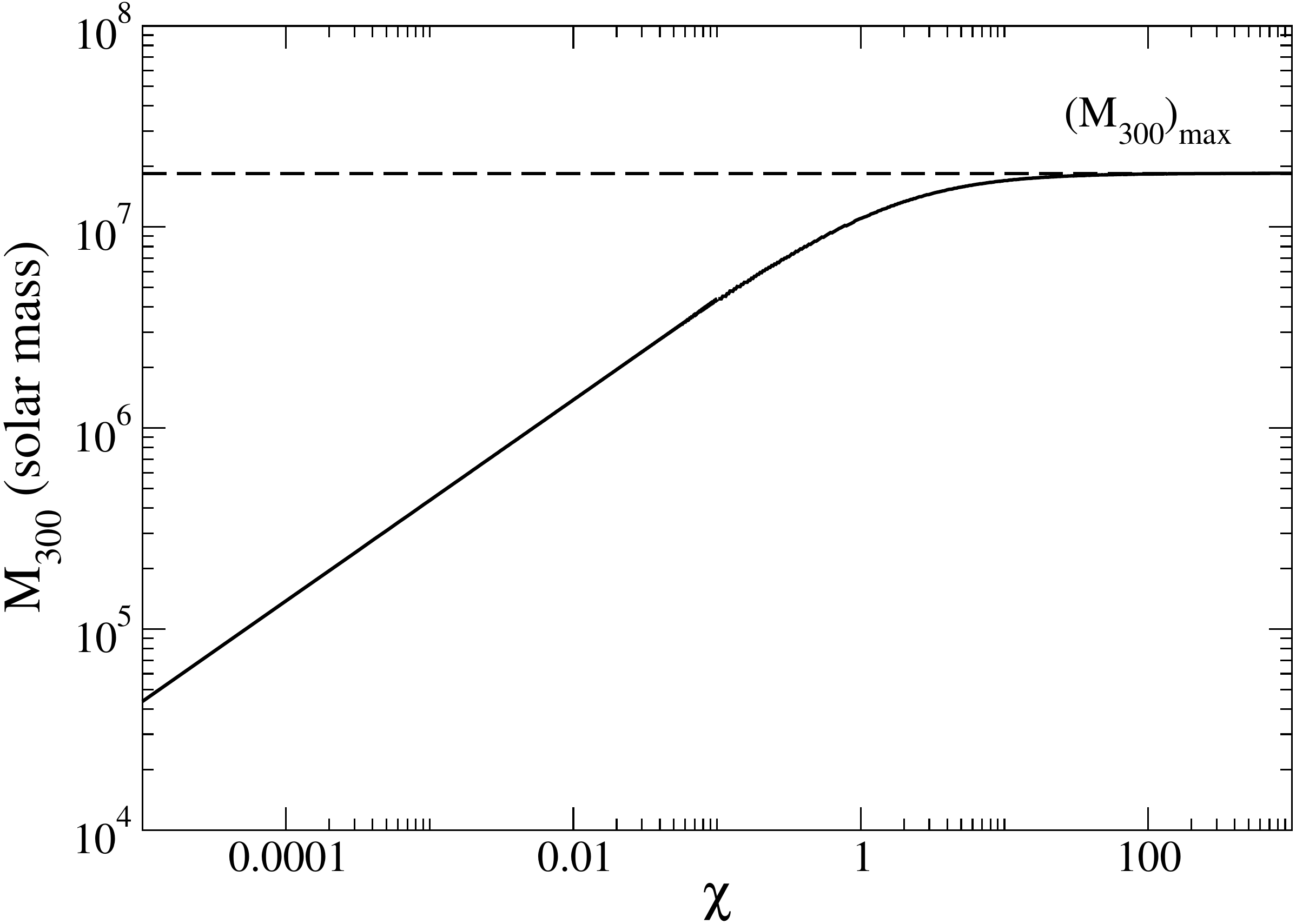}
\caption{Mass within a sphere of radius $300\, {\rm pc}$ as a function of
$\chi$.}
\label{masse300}
\end{center}
\end{figure}

The evolution of the mass $M_{300}$ as a function of $\chi$ is
represented in Fig. \ref{masse300}. We note that $M_{300}$ decreases as the
halo size increases ($\chi$ decreases). For small halos ($\chi\rightarrow
+\infty$), we get $(M_{300})_{\rm max}=0.131
\Sigma_0 R_c^2=0.0992 (M_h)_{\rm min}=1.85\times 10^7\,
M_{\odot}$. Therefore $\log((M_{300})_{\rm
max}/M_{\odot})=7.27$ in good agreement with the upper bound of the
observational result of 
Strigari {\it et al.} \cite{strigari} quoted above.\footnote{Our theoretical
result is valid up to $\chi\sim 1$ corresponding
to a halo mass $(M_h)_t=5.12\times
10^8\, M_{\odot}$ (see below).} For
large halos ($\chi\rightarrow 0$), using Eqs. (\ref{pv8}), (\ref{phm2}) and
(\ref{mtc2}) we find that
\begin{equation}
\label{mtc5}
\frac{M_{300}}{\Sigma_0R_c^2}\sim 0.150
\sqrt{\frac{\Sigma_0R_c^2}{M_h}}\sim 0.114 \frac{R_c}{r_h}.
\end{equation}
According to this relation, $M_{300}$ decreases as
$M_{300}\propto
M_h^{-1/2}\propto r_h^{-1}$ with the halo mass and halo radius. 

{\it Remark:} We note that, for large (isothermal) halos,
\begin{equation}
\label{mtc5b}
M_{300}\sim 5.54 \frac{\Sigma_0^{3/2}r_u^3}{M_h^{1/2}} \sim
\frac{4\pi}{3}\frac{\Sigma_0 r_u^3}{r_h}.
\end{equation}
The second equivalent can be obtained  directly from $M_{300}\sim (4\pi/3)\rho_0
r_u^3$ and Eq. (\ref{observation}).

\subsubsection{Transition between small and large halos}
\label{sec_t}

In our model, a DM halo is entirely characterized by the concentration parameter
$\chi$.
Indeed, for a given value of $\chi$, we can obtain all the characteristics of
the
halo such as $r_h$, $\rho_0$, $M_h$, $v_h$ and $T$.\footnote{Physically,
we can choose to characterize a halo by its mass $M_h$. The corresponding value
of
$\chi$ is then determined by Eq. (\ref{phm1}). From the knowledge of $\chi$
we can determine the other characteristics of the halo. Therefore, a DM halo
is entirely characterized by its mass $M_h$. In this sense, there is no free
parameter in our model except for the value of the ratio $a_s/m^3$ which
determines the
minimum halo radius from Eq. (\ref{pv2}).} Small halos (that contain a 
solitonic core) correspond to $\chi\gg 1$. Large halos (that are
essentially isothermal without a solitonic core) correspond to $\chi\ll 1$.
The transition between large and small halos  ($\chi_t=1$) corresponds to 
\begin{equation}
\label{t1}
(r_h)_t=1.39 R_c=1.39\, {\rm kpc}, 
\end{equation}
\begin{equation}
\label{t2}
(M_h)_t=3.63 \Sigma_0 R_c^2=5.12\times
10^8\, M_{\odot},
\end{equation}
\begin{equation}
\label{t3}
(\rho_0)_t=0.716  \Sigma_0/R_c=0.101\, M_{\odot}/{\rm pc}^3,
\end{equation}
\begin{equation}
\label{t4}
(v_h)_t=1.61 (G\Sigma_0 R_c)^{1/2}=39.7\, {\rm km/s},
\end{equation}
\begin{equation}
\label{t5}
(k_BT/m)_t^{1/2}=0.955 (G\Sigma_0 R_c)^{1/2}=23.5\, {\rm
km/s}. 
\end{equation}
We see that the transition is
very close to the ground state $(r_h)_{\rm min}=788\, {\rm pc}$ and 
$(M_{h})_{\rm min}=1.86\times 10^8\,
M_{\odot}$. This means that most of the DM halos are purely isothermal, except
the dwarf
halos that are very close to the ground state. This is in agreement with our
previous
observation (see Fig. \ref{massradius}).

\subsubsection{Physical density profiles}
\label{sec_pdp}

Using the preceding results, the physical density and circular
velocity profiles of a BECDM halo characterized by the concentration parameter
$\chi$ can be written as
\begin{equation}
\label{pdp1}
\frac{\rho(r)}{\Sigma_0/R_c}=\frac{\rho(\xi)}{\rho_0}\frac{\pi\sqrt{\chi}}{
\xi_h},
\end{equation}
\begin{equation}
\label{pdp2}
\frac{v^2(r)}{G\Sigma_0R_c}=\frac{v^2(\xi)}{4\pi
G\rho_0r_0^2}\frac{4}{\xi_h\sqrt{\chi}},
\end{equation}
\begin{equation}
\label{pdp3}
\frac{r}{R_c}=\frac{\xi}{\pi\sqrt{\chi}},
\end{equation}
where $\rho(\xi)/\rho_0$, $v^2(\xi)/4\pi
G\rho_0r_0^2$ and $\xi_h(\chi)$ are given by Eqs. (\ref{ch23sw}),
(\ref{ch31}) and (\ref{ch34}).

\begin{figure}
\begin{center}
\includegraphics[clip,scale=0.3]{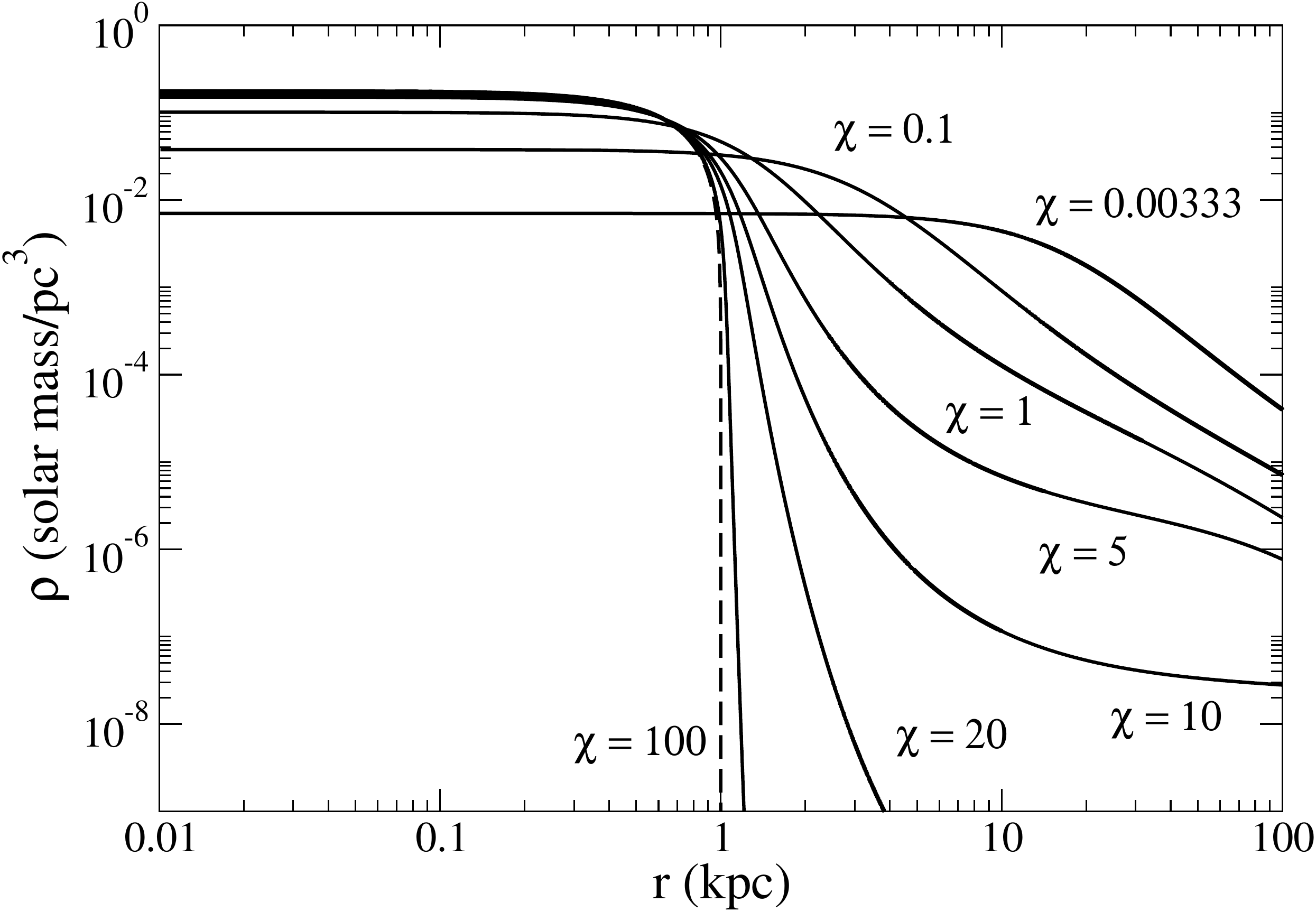}
\caption{Density profiles of different halos characterized by the
concentration parameter $\chi$. For $\chi=0.00333$
($M_h=10^{11}\, M_\odot$) and $\chi=0.1$
($M_h=3.30\times 10^9\, M_\odot$)  the halo is essentially
isothermal without a solitonic core (large halos). For $\chi=1$
($M_h=5.12\times
10^8\, M_{\odot}$) and $\chi=5$ ($M_h=2.50\times 10^8\,
M_\odot$)  the
density
profiles present a core-halo structure (small halos). For
$\chi=10$ ($M_h=2.18\times 10^8\, M_\odot$),
$\chi=20$ ($M_h=2.02\times 10^8\, M_\odot$) and $\chi=100$
($M_h=1.89\times 10^8\, M_\odot$) the halo is
dominated by the solitonic
core (ultracompact halo). The
dashed line corresponds to the pure soliton with $\chi\rightarrow
+\infty$ ($(M_{h})_{\rm min}=1.86\times 10^8\,
M_{\odot}$). }
\label{physicalprofiles}
\end{center}
\end{figure}

\begin{figure}
\begin{center}
\includegraphics[clip,scale=0.3]{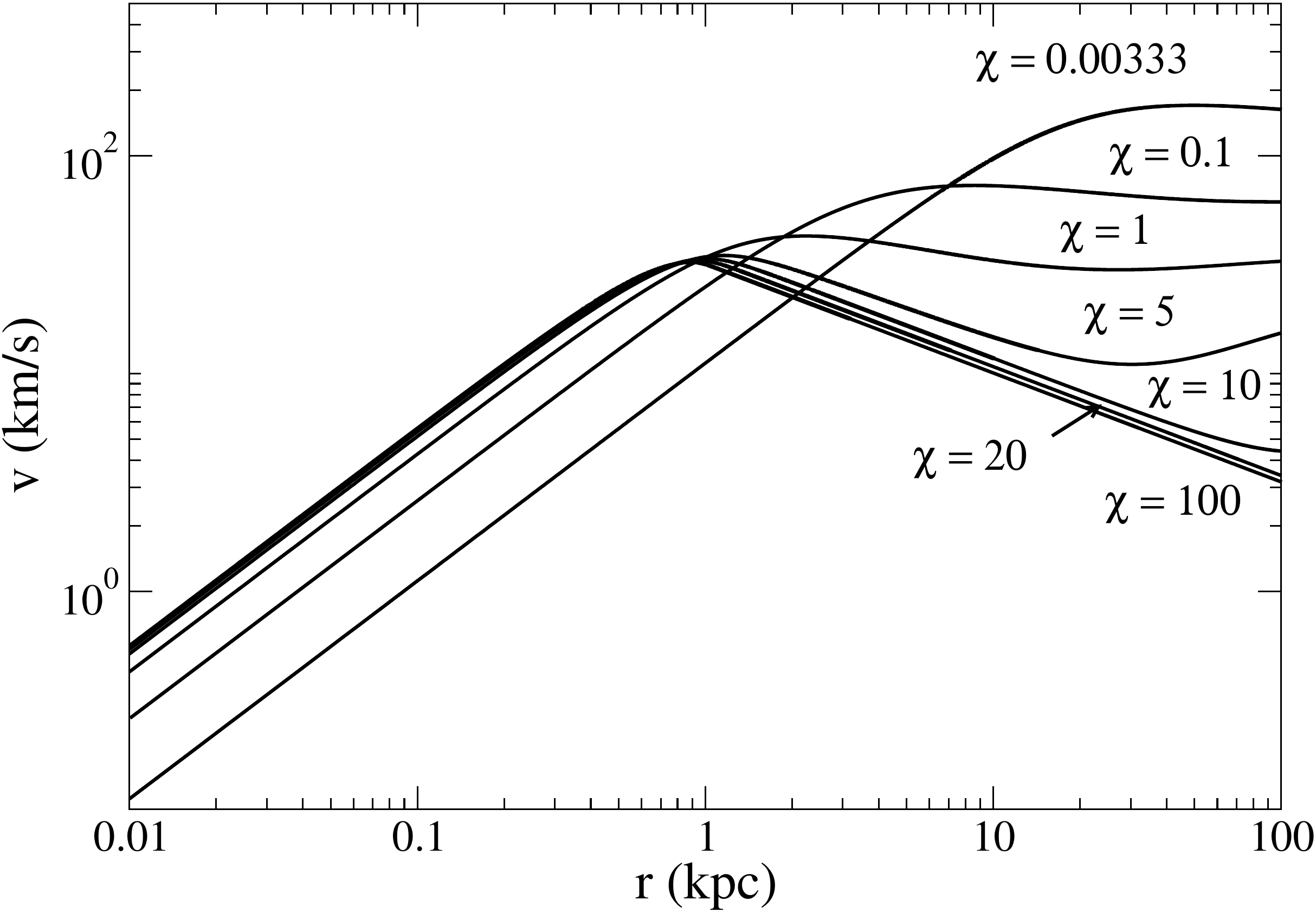}
\caption{Same as Fig. \ref{physicalprofiles} for the circular velocity
profiles.}
\label{physicalvitesse}
\end{center}
\end{figure}

These profiles are are represented in Figs. \ref{physicalprofiles} and
\ref{physicalvitesse} for different values of $\chi$. For
$\chi\rightarrow 0$, we obtain a purely isothermal halo without solitonic core.
For $\chi\rightarrow
+\infty$, we obtain a pure soliton without isothermal halo (ground state). For
intermediate values of
$\chi$, the
profile has a core-halo structure with a solitonic core and an isothermal halo.
At each value of $\chi$ corresponds a
halo whose characteristics ($M_h$, $r_h$...) can be determined from the
equations given in the
previous sections.  The transition between small and large halos
corresponds to $\chi_t=1$ (see Sec. \ref{sec_t}). On the other hand, as
indicated in the caption of Fig. \ref{intrinsicprofiles}, large halos with
$\chi\le \chi_*=0.1$ are almost
indistinguishable from a purely isothermal profile without a solitonic core.
This
corresponds to
\begin{equation}
\label{pdp4}
(r_h)_*=3.67 R_c=3.67\, {\rm kpc}, 
\end{equation}
\begin{equation}
\label{pdp5}
(M_h)_*=23.4 \Sigma_0 R_c^2=3.30\times
10^9\, M_{\odot}, 
\end{equation}
\begin{equation}
\label{pdp6}
(\rho_0)_*=0.2735  \Sigma_0/R_c=0.0386\, M_{\odot}/{\rm pc}^3,
\end{equation}
\begin{equation}
\label{pdp7}
(v_h)_*=2.53 (G\Sigma_0 R_c)^{1/2}=62.2\, {\rm km/s},
\end{equation}
\begin{equation}
\label{pdp8}
(k_BT/m)_*^{1/2}=1.865 (G\Sigma_0 R_c)^{1/2}=45.9\, {\rm
km/s}. 
\end{equation}

\subsection{The three types of DM halos in Model I}

\begin{figure}
\begin{center}
\includegraphics[clip,scale=0.3]{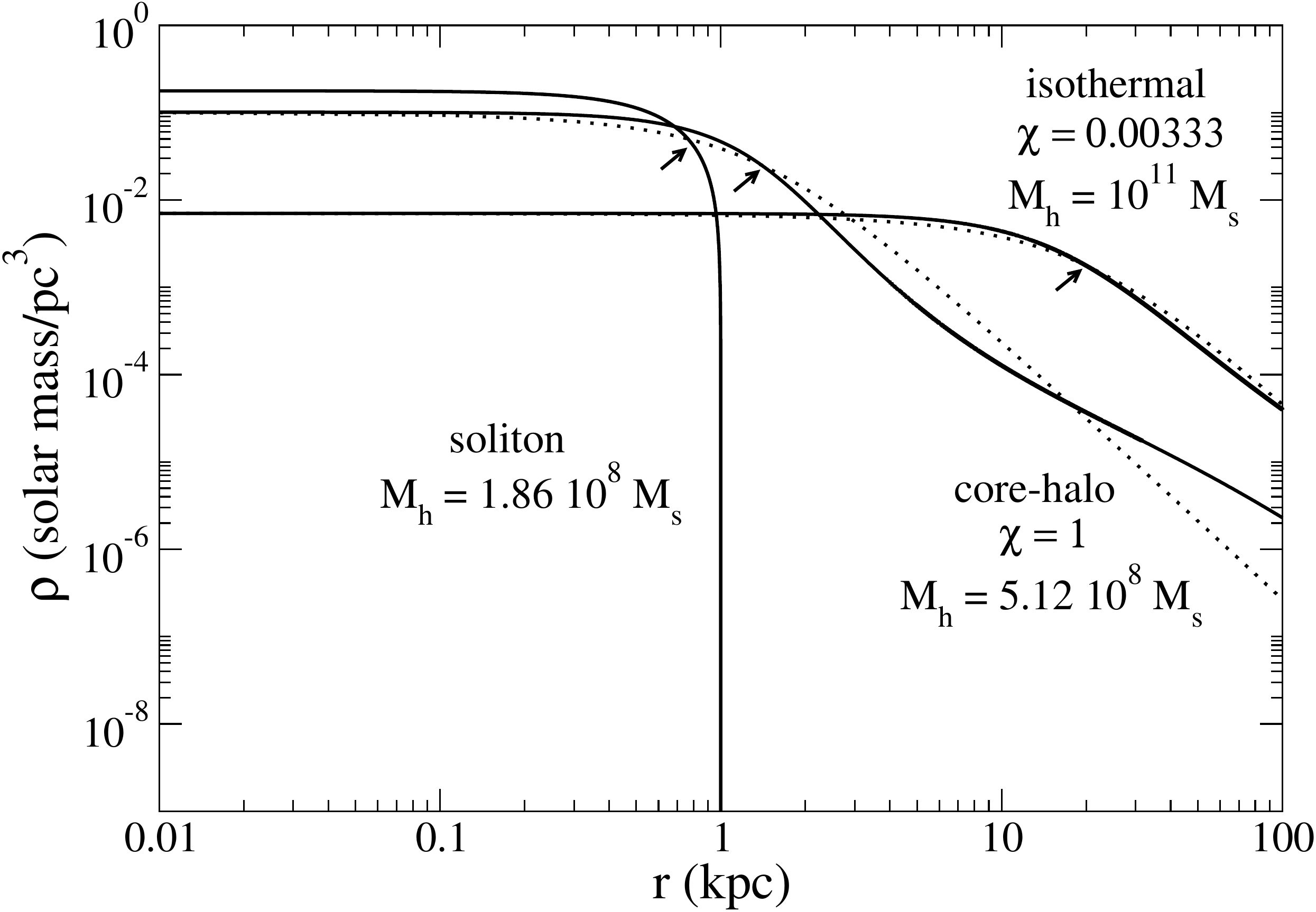}
\caption{Density profiles of DM halos of
different mass. We have represented the purely solitonic profile (ground
state of the BEC model) which is similar to an ultracompact halo like Fornax
($(M_{h})_{\rm min}=1.86\times 10^8\, M_{\odot}$), the
core-halo profile of a small halo ($M_{h}=5.12\times 10^8\,
M_{\odot}$), and the almost isothermal profile of a large halo
like the Medium Spiral ($M_{h}=10^{11}\, M_{\odot}$). In each
case, we have indicated the halo radius $r_h$ (where the central density is
divided by $4$) by an arrow. We have also plotted the Burkert profile for
comparison (dotted lines). }
\label{pd}
\end{center}
\end{figure}

\begin{figure}
\begin{center}
\includegraphics[clip,scale=0.3]{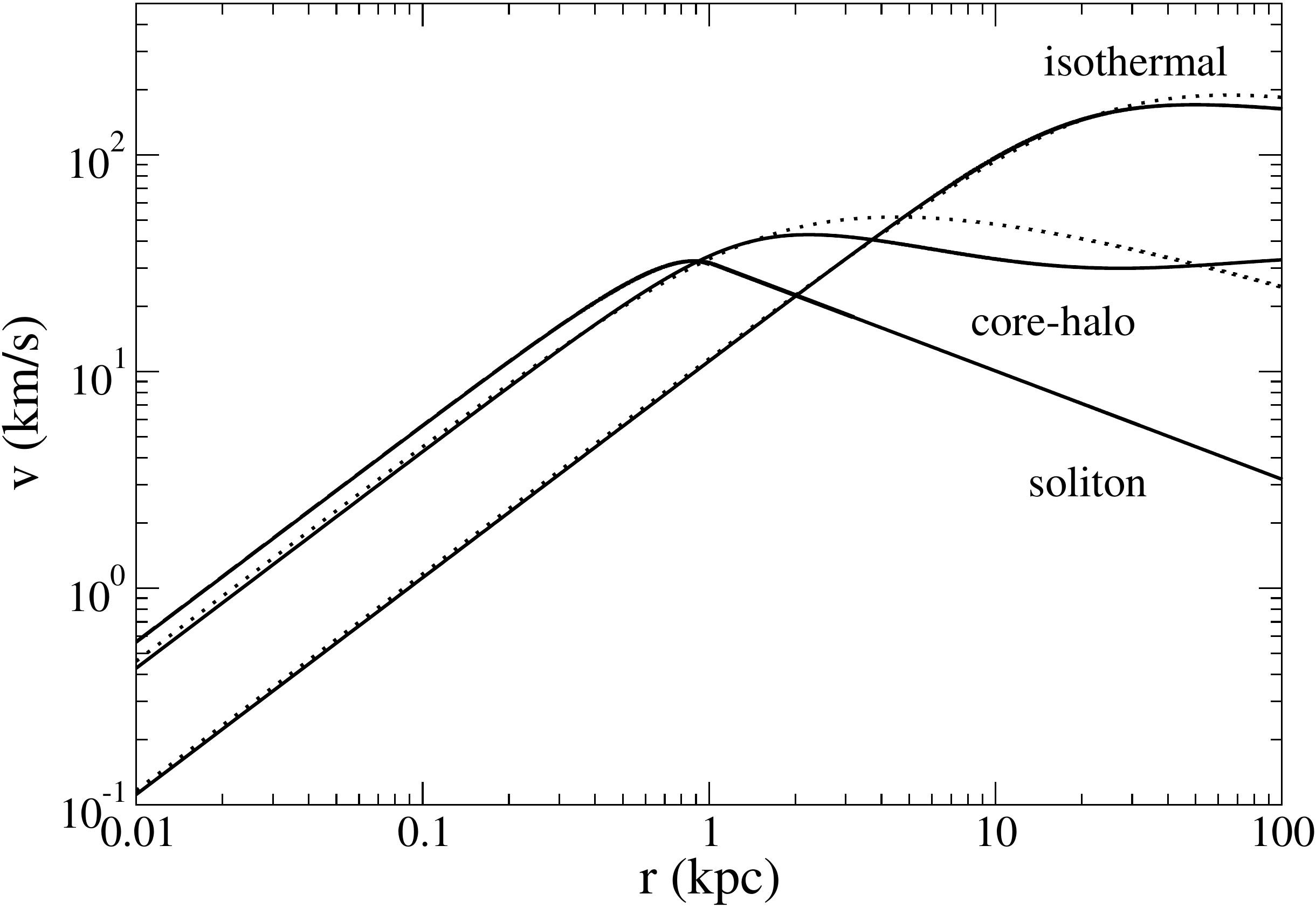}
\caption{Rotation curves of DM halos of
different mass (the conventions are the same as in Fig. \ref{pd}).
}
\label{pv}
\end{center}
\end{figure}

In this section, we illustrate the previous results by showing examples
of BECDM halos with a purely solitonic core (ultracompact halo), a core-halo
structure (small halo), and a purely isothermal halo (large halo). Their
density and circular velocity profiles are represented in Figs. \ref{pd} and
\ref{pv}.

Let us first consider a small DM halo with a concentration parameter
$\chi=1$. Its physical characteristics obtained from our model are $r_h=1.39\,
{\rm kpc}$, $M_h=5.12\times
10^8\, M_{\odot}$, $\rho_0=0.101\, M_{\odot}/{\rm pc}^3$, $v_h=39.7\, {\rm
km/s}$ and
$(k_BT/m)^{1/2}=23.5\, {\rm km/s}$. This DM halo has a core-halo structure with
a solitonic core and an isothermal atmosphere. The solitonic core has a radius
$R_c= 1\, {\rm kpc}$ and a mass
$M_c=1.29\times  10^8\, M_{\odot}$. We note that the velocity profile exhibits a
small
dip due to the
presence of the solitonic core. Nonmonotonic velocity profiles (oscillations)
are sometimes observed in real rotation curves of galaxies. We suggest that 
they could, in certain cases, be the
manifestation of a solitonic core.

For comparison, we have plotted the density and velocity profiles of the DM halo
with
the minimum mass corresponding to the ground state of the BECDM model. This
ultracompact DM halo ($\chi\rightarrow +\infty$) is
a pure
soliton, without
isothermal atmosphere ($T=0$). Its physical characteristics obtained from our
model
are $(r_{h})_{\rm min}=788\, {\rm pc}$, $(M_{h})_{\rm min}=1.86\times 10^8\,
M_{\odot}$, $(\rho_0)_{\rm
max}=0.179\, M_{\odot}/{\rm pc}^3$ and $(v_{h})_{\rm min}=31.9\, {\rm km/s}$
similar to the characteristics of dSphs like Fornax. We note that the density
profile of the soliton has a larger central density
than the profile of the DM halo with a core-halo structure.

Finally, we have plotted the density and velocity profiles corresponding to a
large DM halo with
a
concentration parameter $\chi=3.33\times 10^{-3}$. In that case, there is no
solitonic core and the
profiles coincide with those of the isothermal sphere. For large DM
halos, the flat
density core (for $r\rightarrow 0$) is due to the effective temperature, not to
the self-interaction of
the bosons (see Appendix \ref{sec_eff}). The physical
characteristics of
this halo 
obtained from our model
are $r_h=2.01\times
10^{4}\, {\rm pc}$, $M_h=10^{11}\, M_\odot$, $\rho_0=7.02\times
10^{-3}M_{\odot}/{\rm pc}^3$,
$v_h=(GM_h/r_h)^{1/2}=146\, {\rm km/s}$, and  $(k_B T/m)^{1/2}=108\, {\rm
km/s}$ similar to the Medium
Spiral. We note that the density profile of the purely isothermal halo has a
smaller central density than the profile of the DM halo with a
core-halo structure.

{\it Remark:} As soon as $M_h>(M_h)_{\rm min}$ (i.e. $T>0$) the
soliton is surrounded by an extended isothermal halo whose density
decreases slowly as $\rho\propto r^{-2}$. This yields an infinite mass if it
is extended to infinity. In practice, the halo is tidally truncated
(see Appendix \ref{sec_diff}).

\subsection{Comparison with observations}
\label{sec_comp}

In this section, we make a first comparison between the results of our model
and observations. A more detailed comparison will be made in future works.

\subsubsection{Preliminary remarks}

Classical numerical simulations of CDM lead to DM halos that are well-fitted by
the NFW profile \cite{nfw}:
\begin{equation}
\label{comp1}
\rho(r)\propto \frac{1}{\frac{r}{r_s}\left
(1+\frac{r}{r_s}\right )^2},
\end{equation}
where $r_s$ is a scale radius that varies from halo to halo.
The density decreases as $r^{-3}$ for $r\rightarrow +\infty$ and
diverges as $r^{-1}$ for $r\rightarrow 0$. This singular behavior is not
consistent with observations that reveal that DM halos possess a core, not
a cusp. Observed DM halos are better fitted by the Burkert profile
\cite{observations}:
\begin{equation}
\label{comp2}
\rho(r)=\frac{\rho_0}{\left (1+\frac{r}{r_h}\right )\left
(1+\frac{r^2}{r_h^2}\right )},
\end{equation}
where $\rho_0$ is the central density and $r_h$ is the halo radius defined by
Eq.
(\ref{hr1}). This density  profile decreases as $r^{-3}$ for $r\rightarrow
+\infty$,
like the NFW profile, but displays a flat core for $r\rightarrow
0$ instead of a cusp. In the following,  we shall
compare the theoretical
profiles of BECDM halos obtained from our model with the Burkert profile that
fits a lot of
observations.

Some preliminary remarks can be made:

(i) The Burkert profile is
empirical and does not rely on a theory. It is therefore important to
see if this profile is consistent with a profile obtained from a theoretical
model such as the one presented in this paper.

(ii) The profiles of DM halos are not expected to be universal.
Small halos, with a mass $\sim 10^8\, M_{\odot}$, are very compact. In the BECDM
model, they correspond to the solitonic solution of the GPP equations
(ground state). This solution is substantially different from the Burkert
profile (see below). However, the solitonic profile may be closer to
the observations of ultracompact DM halos than the Burkert profile. The
Burkert profile is expected to be valid only for relatively large halos. 

(iii) The DM halos of
our model behave as the isothermal sphere at large distances so their density
profiles decrease as $r^{-2}$ for $r\rightarrow +\infty$ while the Burkert
profile decreases as $r^{-3}$ for $r\rightarrow +\infty$. Therefore, if we
compare the halos of our model with the Burkert profile at arbitrarily large
distances, we will
clearly find  a difference of slope. However, in practice, the halos do not
extend to infinity so that both the isothermal profile and the Burkert profile
cease to be
valid above a certain distance. Furthermore, observational data are only
obtained within
a limited range of radial distances: $0\le r\le r_{\rm max}$. Therefore, we must
take this constraint into account when comparing the DM halos of our model with
the Burkert profile.

\subsubsection{Large halos: isothermal profile}

\begin{figure}
\begin{center}
\includegraphics[clip,scale=0.3]{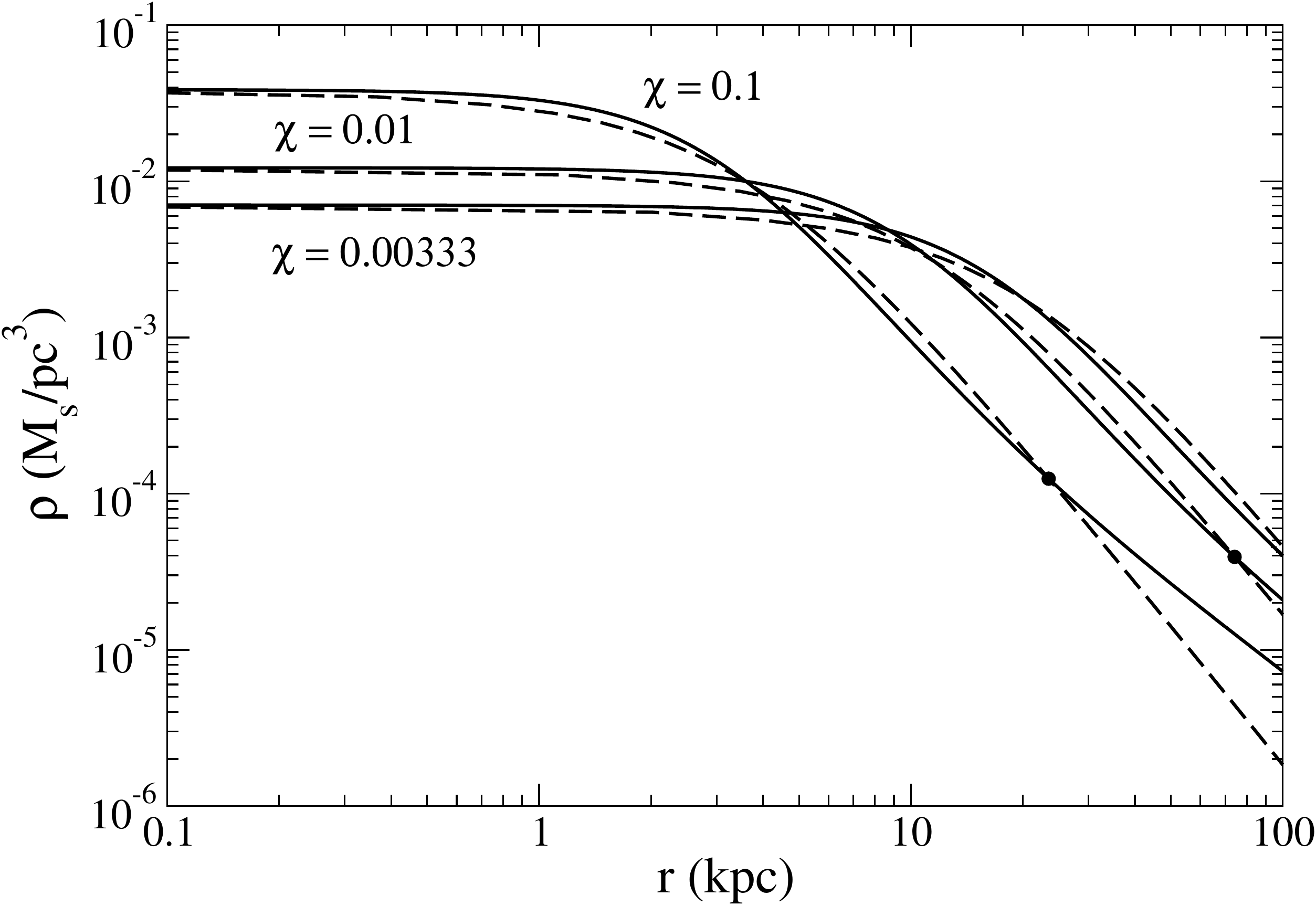}
\caption{Comparison between the purely isothermal profile (solid lines) valid
for $M_h>(M_h)_*=3.30\times 10^9\, M_{\odot}$ ($\chi\le
\chi_*=0.1$)  and the Burkert profile
(dashed lines) in the range $0\le r\le 100\, {\rm kpc}$
corresponding to the observations. For a large halo of mass $M_h=10^{11}
M_{\odot}$ ($\chi=0.00333<\chi_c=0.00481$) corresponding to the Medium Spiral,
the two profiles are almost indistinguishable. For a smaller
halo of mass
 $M_h=3.30\times 10^{10}\, M_{\odot}$ ($\chi=0.01\gtrsim
\chi_c=0.00481$), a deviation starts to appear close to the maximum
distance of observation (in the present case at $r=69.6\, {\rm kpc}$).
For a halo of mass $(M_h)_*=3.30\times
10^9\, M_{\odot}$ ($\chi_*=0.1>\chi_c=0.00481$), the
difference becomes very pronounced at
large distances where we clearly see the two slopes
$r^{-2}$
(isothermal) and
$r^{-3}$ (Burkert). The difference appears at $r=22.0\, {\rm kpc}$ (see the
bullet). For
smaller halos with a core-halo profile, the disagreement is even more
pronounced (see Fig. \ref{pd}). However, for small DM halos,
the BEC profile may be more relevant than the Burkert
profile.}
\label{compbukisodens}
\end{center}
\end{figure}

We first consider large DM halos that are essentially isothermal with a
negligible solitonic core. As we
have seen in Sec. \ref{sec_pdp}, these halos have a mass $M_h\ge
(M_h)_*=3.30\times 10^9\,
M_{\odot}$ corresponding to $\chi\le \chi_*=0.1$. The intrinsic isothermal
profile giving  the density normalized
by the central density,  $\rho/\rho_0$, as a function of the distance
normalized by
the halo radius,  $r/r_h$, is represented in Fig. \ref{isodensnormaliseELARGI}
together with the intrinsic Burkert profile (and other profiles that we do not
consider here). The rotation curves of DM halos are usually
measured up to a typical
distance $r_{\rm max}=100\, {\rm kpc}$. Therefore, for a large DM halo
characterized
by a concentration parameter $\chi\le \chi_*=0.1$, we have to make
the comparison between the intrinsic isothermal  profile and the intrinsic 
Burkert profile up to a maximum normalized distance $r_{\rm
max}/r_h=86.5\sqrt{\chi}$, where we have used Eq. (\ref{phr2}). The smallest
purely isothermal halo,
corresponding to
$\chi_*=0.1$, has a mass $(M_h)_*=3.30\times 10^9\, M_{\odot}$. For this halo,
we have to make the comparison with the Burkert profile up to $(r_{\rm
max}/r_h)_*=27.3$.  For larger
halos, we have to make the comparison up to a smaller normalized distance
$r_{\rm
max}/r_h=86.5\sqrt{\chi}\le
27.3$. This is why we have plotted the intrinsic density
profiles in Fig. \ref{isodensnormaliseELARGI}  up to $30\, r_h$ in order to
cover all possibilities. In Fig.
\ref{isodensnormaliseELARGI}, we see that the isothermal profile
is very close to the Burkert profile for $r/r_h\le 6$ and that it departs from
it for $r/r_h\ge 6$. Therefore, for large DM halos such that $\chi\le
\chi_c=(6/86.5)^2=0.00481$, corresponding to 
\begin{equation}
\label{pdp4n}
(r_h)_c=16.7 R_c=16.7\, {\rm kpc}, 
\end{equation}
\begin{equation}
\label{pdp5n}
(M_h)_c=486 \Sigma_0 R_c^2=6.86\times
10^{10}\, M_{\odot}, 
\end{equation}
\begin{equation}
\label{pdp6n}
(\rho_0)_c=0.0600 \Sigma_0/R_c=0.00846\, M_{\odot}/{\rm pc}^3,
\end{equation}
\begin{equation}
\label{pdp7n}
(v_h)_c=5.41 (G\Sigma_0 R_c)^{1/2}=133\, {\rm km/s},
\end{equation}
\begin{equation}
\label{pdp8n}
(k_BT/m)_c^{1/2}=3.98 (G\Sigma_0 R_c)^{1/2}=98.0\, {\rm
km/s},
\end{equation}
the isothermal profile is almost indistinguishable from the Burkert profile up
to the maximum distance of observation $r_{\rm
max}=100\, {\rm kpc}$. By contrast, for smaller isothermal halos with 
$\chi_c=0.00481\le \chi\le \chi_*=0.1$, corresponding to a
mass range $(M_h)_*=3.30\times 10^9\, M_{\odot}\le M_h\le (M_h)_c=6.86\times
10^{10}\,
M_{\odot}$, we can
see a difference between the isothermal profile and the Burkert profile at
large distances. This difference appears at $r/r_h>6$, corresponding to
$r>6.96\, \chi^{-1/2}\, {\rm kpc}$. Beyond this distance, the isothermal
profile decreases as $r^{-2}$ while the Burkert profile
 descreases as $r^{-3}$.\footnote{This difference of slope may
be explained by incomplete violent relaxation, tidal effects and stochastic
forcing, as discussed in Refs. \cite{clm1,clm2} in the context
of the King model (see also Appendix \ref{sec_diff}). In the
present model, the logarithmic 
slope of the density profile could be corrected heuristically by
tuning the external potential $\omega_0$ in the generalized GPP equations
(\ref{intro3}) and (\ref{intro4}). This may be a simple way to take into
account tidal effects and other nonideal effects.} These results are
illustrated in Fig. \ref{compbukisodens} where we have represented the
isothermal and Burkert
density profiles in physical scales for halos of different mass (see also Figs.
\ref{pd} and \ref{pv}).

\subsubsection{Small halos: core-halo profile}
\label{sec_qmg}

We now consider small halos with a core-halo profile. As we have seen in
Sec. \ref{sec_pdp}, they have a mass
$(M_h)_{\rm min}=1.86\times 10^8\, M_{\odot}< M_h\le
(M_h)_*=3.30\times
10^9\,
M_{\odot}$ corresponding to $\chi_*=0.1\le \chi<+\infty$. In that case, the
difference
between the core-halo profile and the Burkert profile is very significant
at large distances as shown in
Figs. \ref{pd} and \ref{pv}  for $\chi=1$ corresponding to $M_h=5.12\times
10^{8} M_{\odot}$. However, for small DM halos,
the BEC profile may be more relevant than the Burkert
profile.

\subsubsection{Ultracompact halos: ground state}

Ultracompact halos close to the ground state $(M_h)_{\rm
min}=1.86\times 10^8\, M_{\odot}$, corresponding to $\chi\rightarrow +\infty$, 
have no atmosphere (they are purely
solitonic) so it is not relevant to compare their density profile with the
Burkert profile at large distances ($r_{\rm max}=100\, {\rm kpc}$) since the
density drops to zero at $R_c=1\, {\rm
kpc}$. For these halos, the solitonic profile should
provide a better agreement with observations than the Burkert profile. We note,
however, that the two profiles (soliton and Burkert) are relatively close to
each other, especially for what concerns the rotation curve, up to the halo
radius (see Figs. \ref{becdens}
and \ref{becvit}).\footnote{This is also true for the larger halos considered
previously.} At larger distances, the density of the soliton drops to
zero and the circular velocity decreases according to the Kepler law, so it is
no more relevant to continue the
comparison.

\subsubsection{Conclusion}

In conclusion, our results are qualitatively consistent with the observations.
Quantum mechanics explains why there is a minimum halo mass $(M_{h})_{\rm
min}=1.86\times 10^8\,
M_{\odot}$ and a minimum halo
radius $(r_{h})_{\rm min}=788\, {\rm pc}$ (ground state). On the other hand, the
isothermal density profile of large DM halos ($M_h>(M_h)_c=6.86\times
10^{10}\,
M_\odot$) is indistinguishable from the empirical Burkert
\cite{observations} profile up to
$100\, {\rm kpc}$ (see Figs. \ref{pd}-\ref{compbukisodens}). Therefore, these
profiles
quantitatively agree with the structure of the observed
halos, at least in an average sense (recall that the Burkert profile is obtained
by
averaging over many rotation curves of galaxies). One interest of our model is
that there is
no
undetermined (free) parameter. For a given halo mass $M_h$, all the parameters
can be determined from our equations.
It would be therefore important to carry out a more detailed comparison of our
model
with observations. This will be considered in future works.

\section{Another family of solutions with a persistent solitonic core and a
plateau (Model II)}
\label{sec_another}

In the model developed in the previous section (Model I), we have seen that the
solitonic core disappears progressively as the halo mass increases so that large
DM halos are purely isothermal without a solitonic core. In the present section,
we develop another model (Model II) in which large DM halos present a 
persistent solitonic core and a plateau.

\subsection{A new definition of the halo radius}
\label{sec_nhr}

In Secs. \ref{sec_complete}-\ref{sec_mod1}, we have defined the halo radius
$r_h$ by Eq. (\ref{hr1}). However, in the case where there is a
strong separation between a
small solitonic core and a large isothermal halo ($\chi\gg 1$), as in Fig.
\ref{densite}, it
is more
relevant to define the halo radius  $r_h$ by
\begin{eqnarray}
\label{nhr1}
\frac{\rho(r_h)}{\rho_c}=\frac{1}{4},
\end{eqnarray} 
where $\rho_c$ is the density of the plateau given by Eq. (\ref{pchs14a}),
not the central
density $\rho_0$. Similarly, the (universal) surface density of DM halos should
be
defined 
by
\begin{equation}
\label{nhr2}
\Sigma_0=\rho_c r_h=141\, M_{\odot}/{\rm
pc}^2
\end{equation}
instead of Eq. (\ref{observation}). Indeed, when the size of the soliton is
small with respect
to
the size of the halo, the soliton may not be sufficiently well resolved in
observations and what we regard as being the ``central'' density is actually the
density of
the plateau $\rho_c$, not the density of the soliton $\rho_0$. As we shall see,
this change of definition leads to a new  family of solutions that appears
above a critical mass $(M_h)_b\sim 10^{9}\, M_{\odot}$. Contrary
to the family of solutions constructed in
Sec. \ref{sec_mod1}
this new family of solutions presents a persistent solitonic core as the halo
mass
increases. In a sense, it corresponds to a bifurcation from the branch of
solutions of Model I. In Sec. \ref{sec_ptb} we shall interprete this bifurcation
in relation to phase transitions in a ``thermal'' self-gravitating  boson gas in
a box.

\subsection{Physical parameters: exact expressions}
\label{sec_enr}

We can easily generalize the results of Sec. \ref{sec_mod1} with the new
definition of the
halo radius  from
Eq. (\ref{nhr1}). The halo radius $r_h$ is now given by
\begin{eqnarray}
\label{enr1}
\frac{r_h}{R_c}=\frac{\xi_h}{\pi\sqrt{\chi}},
\end{eqnarray}
where the
function $\xi_h(\chi)$ is determined by the equation
\begin{eqnarray}
\label{enr2}
e^{-\psi(\xi_h)}=\frac{B(\chi) e^{-\chi}}{4}.
\end{eqnarray}
The density of the plateau is given by
\begin{equation}
\label{enr3}
\frac{\rho_cR_c}{\Sigma_0}=\frac{\pi\sqrt{\chi}}{\xi_h}.
\end{equation}
The halo mass is given by
\begin{equation}
\label{enr4}
\frac{M_h}{\Sigma_0R_c^2}=\frac{4\psi'(\xi_h)\xi_h
e^{\chi}}{B(\chi)\pi\chi}\left
\lbrack 1+\chi
e^{-\psi(\xi_h)}\right
\rbrack.
\end{equation}
The
halo velocity is given by
\begin{equation}
\label{enr5}
\frac{v_h^2}{G\Sigma_0R_c}=\frac{4\psi'(\xi_h)e^{\chi}}{B(\chi)\sqrt{\chi}}\left
\lbrack
1+\chi
e^{-\psi(\xi_h)}\right
\rbrack.
\end{equation}
The effective temperature of the halo is given by
\begin{equation}
\label{enr6}
\frac{k_B T}{Gm\Sigma_0R_c}=\frac{4e^{\chi}}{B(\chi)\xi_h\sqrt{\chi}}.
\end{equation}
From Eqs. (\ref{pchs14a}) and (\ref{enr3}), we find that the
central density is
given by
\begin{equation}
\label{enr7}
\frac{\rho_0R_c}{\Sigma_0}=\frac{\pi\sqrt{\chi}}{B(\chi)\xi_h}e^{\chi}.
\end{equation}
From Eq. (\ref{pchs1}), we find that the soliton mass is given by
\begin{equation}
\label{enr8}
\frac{M_c}{\Sigma_0R_c^2}=\frac{4\sqrt{\chi}}{B(\chi)\xi_h}e^{\chi}.
\end{equation}
For a DM halo characterized by its concentration parameter $\chi$, all the halo
parameters $r_h$,
$\rho_c$, $M_h$, $v_h$, $T$, $\rho_0$ and $M_c$ are determined by
Eqs. (\ref{enr1})-(\ref{enr8}).

\subsection{Physical parameters: approximate analytical expressions}
\label{sec_aar}

\begin{figure}
\begin{center}
\includegraphics[clip,scale=0.3]{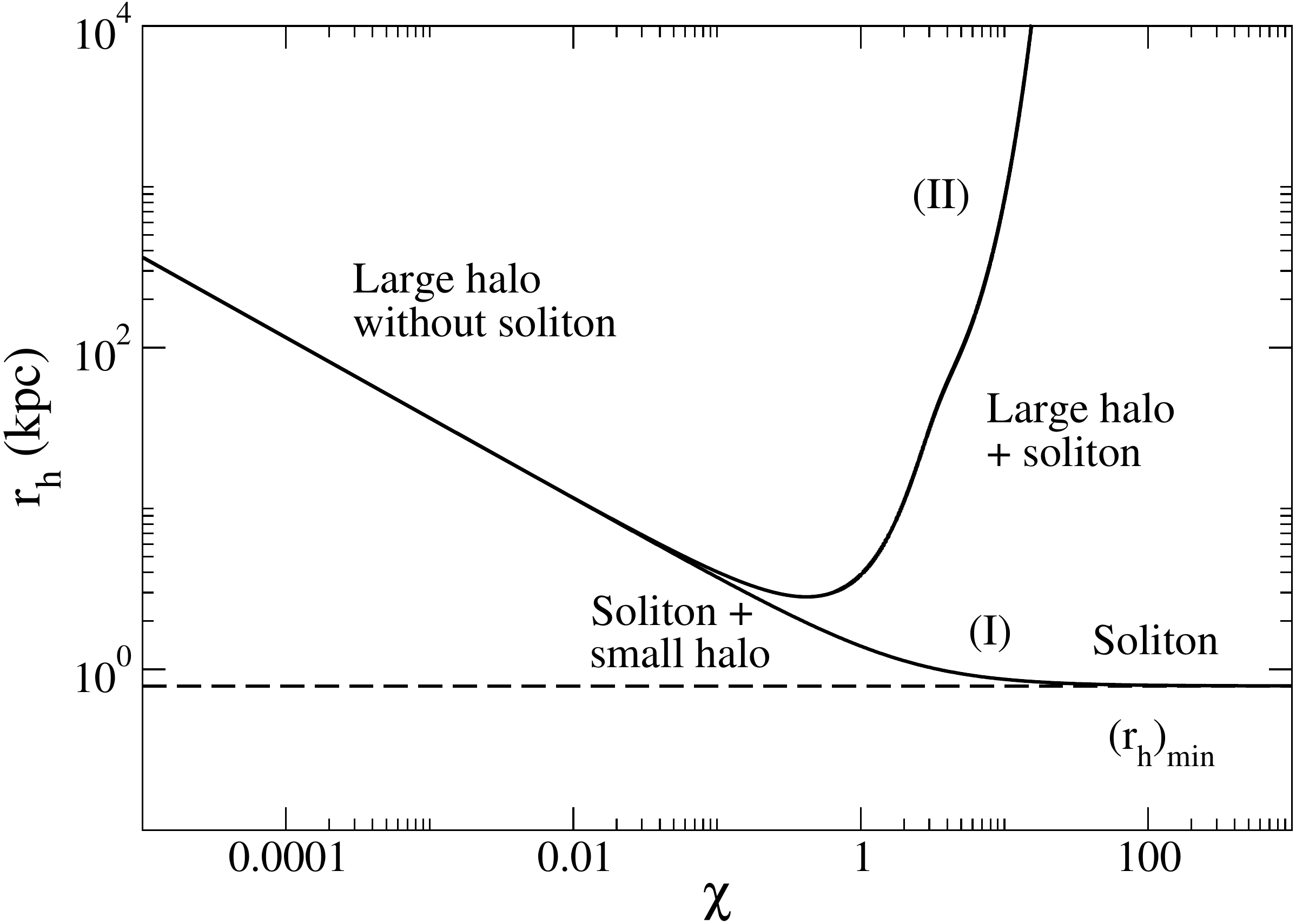}
\caption{Halo radius $r_h$ as a function of $\chi$ for model II.
We have also plotted the  curve corresponding to model I.}
\label{CHrayon}
\end{center}
\end{figure}

\begin{figure}
\begin{center}
\includegraphics[clip,scale=0.3]{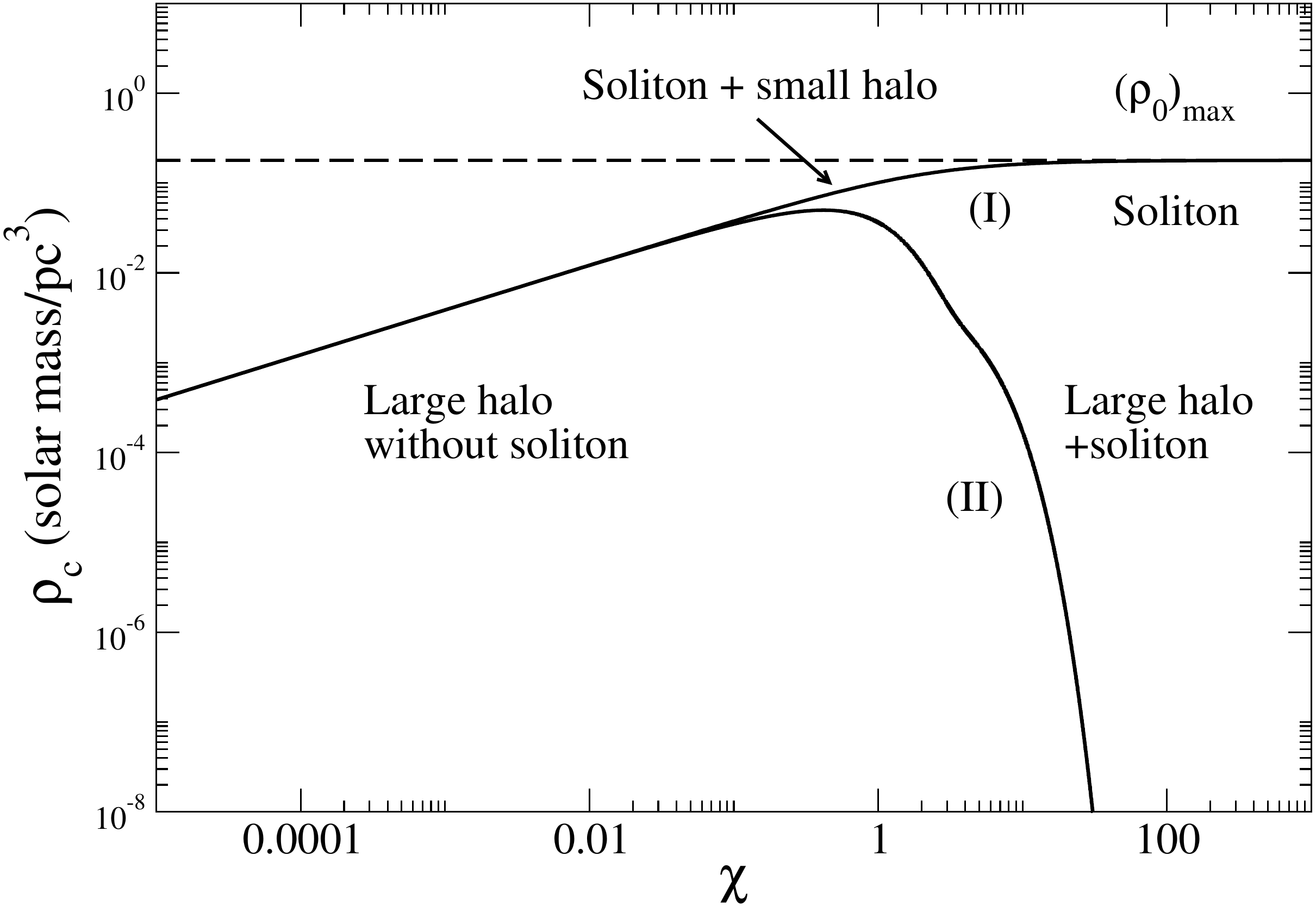}
\caption{Plateau density $\rho_c$ as a function of $\chi$ for models I and
II. When
there is no plateau, $\rho_c$ represents the central density $\rho_0$.}
\label{CHden}
\end{center}
\end{figure}

\begin{figure}
\begin{center}
\includegraphics[clip,scale=0.3]{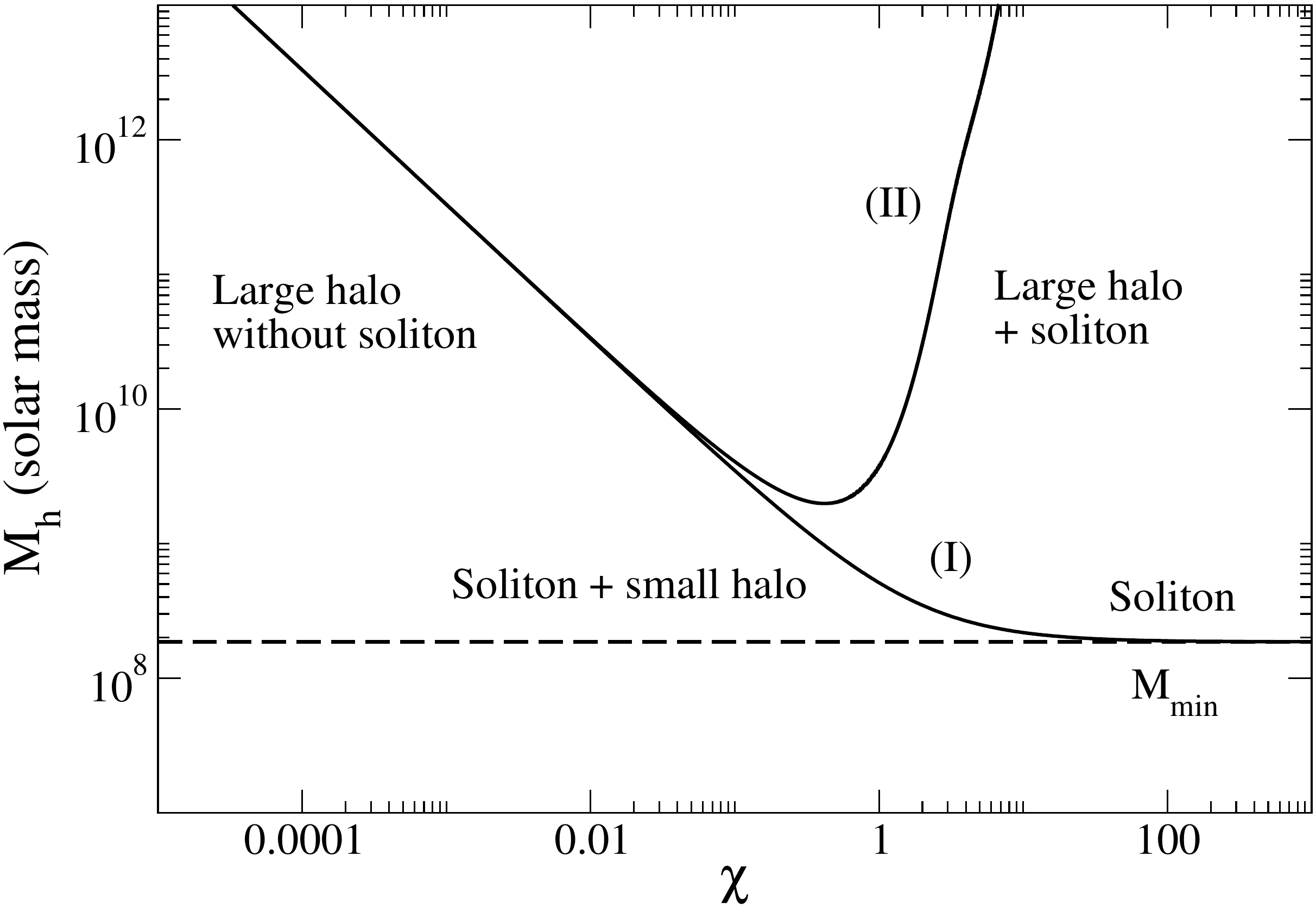}
\caption{Halo mass $M_h$ as a function of $\chi$ for models I and II.}
\label{CHmasse}
\end{center}
\end{figure}

\begin{figure}
\begin{center}
\includegraphics[clip,scale=0.3]{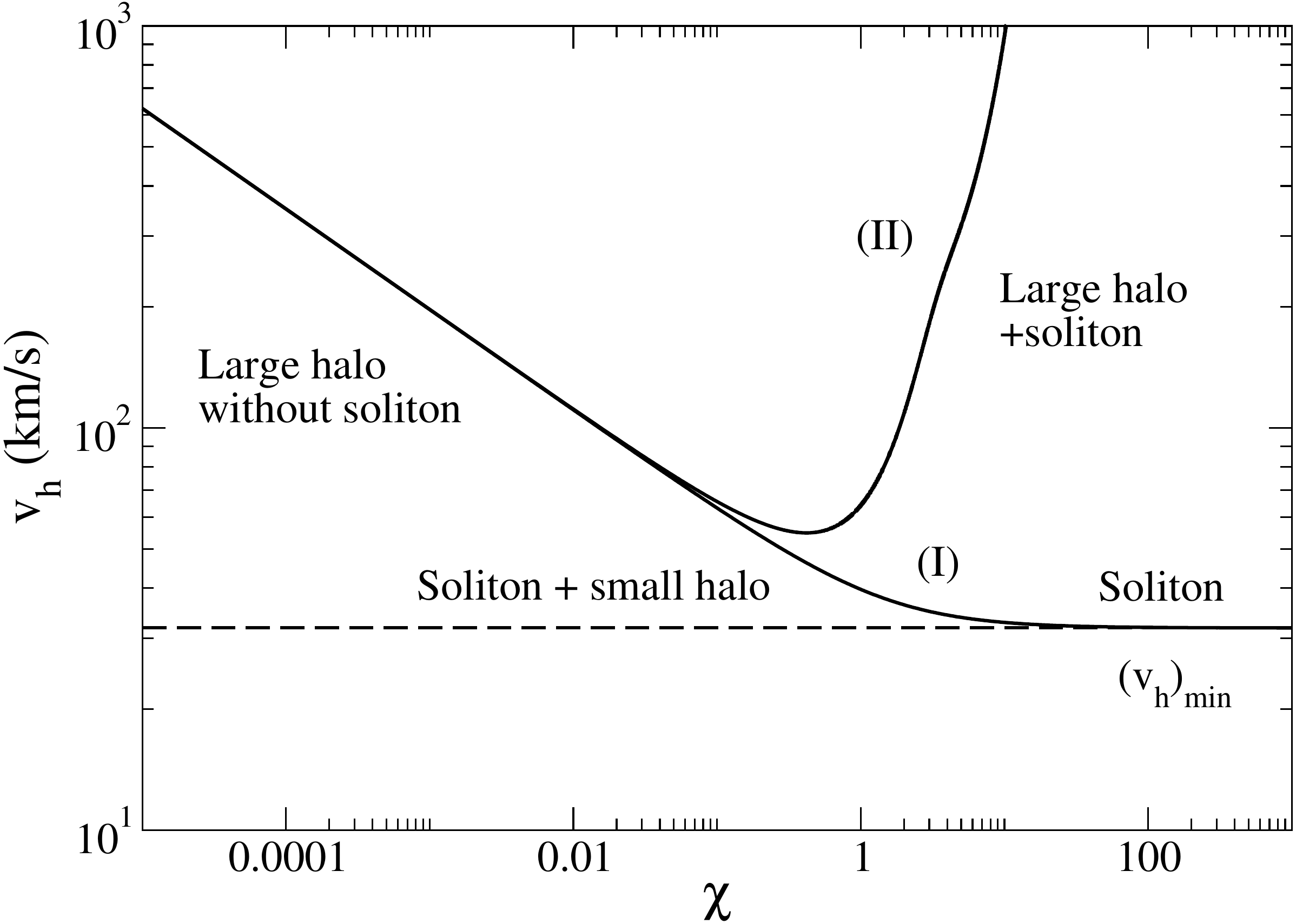}
\caption{Circular velocity $v_h$ as a function of $\chi$ for models I and II.}
\label{CHvitesse2}
\end{center}
\end{figure}

\begin{figure}
\begin{center}
\includegraphics[clip,scale=0.3]{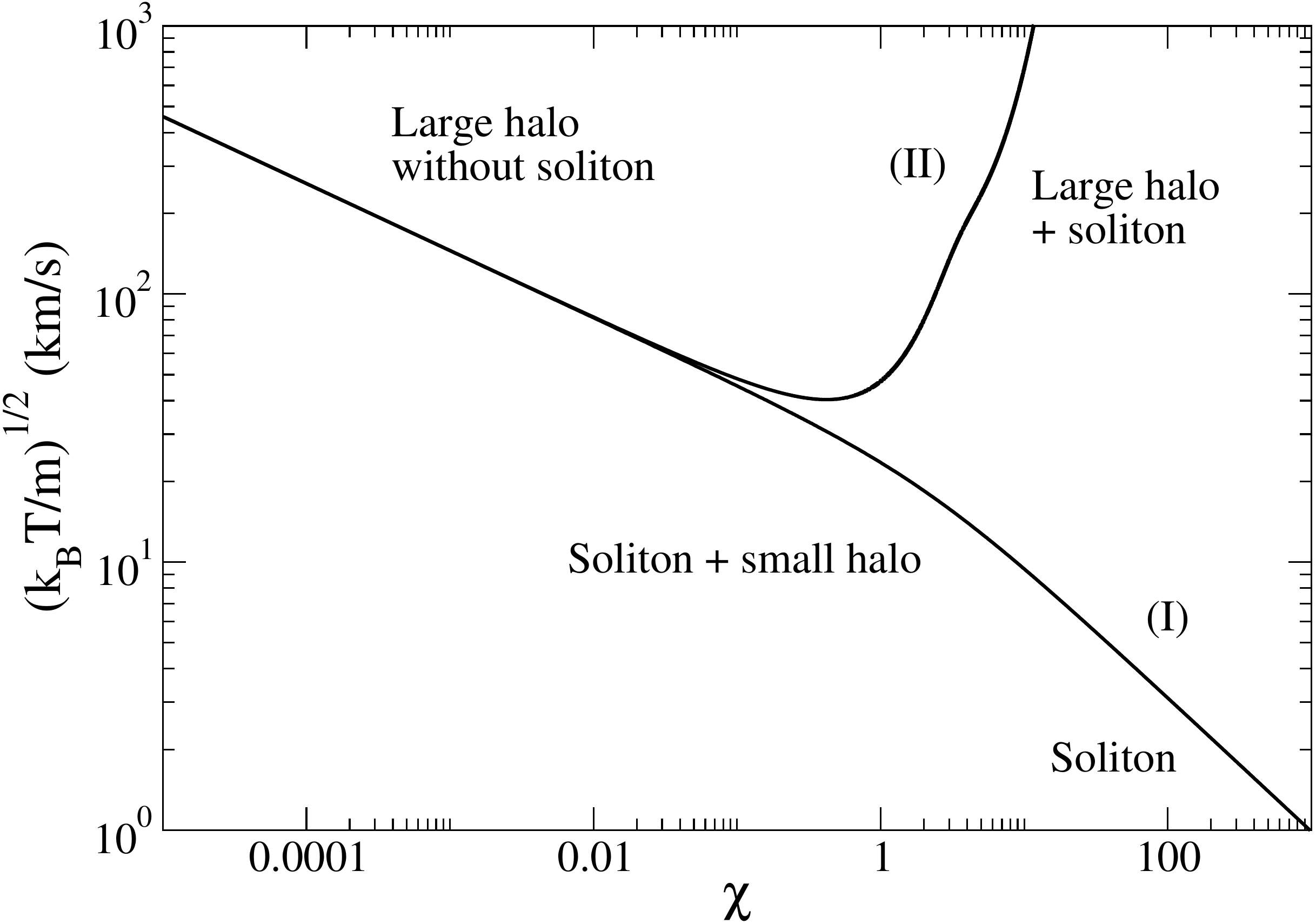}
\caption{Effective temperature $T$ as a function of $\chi$ for models I and 
II.}
\label{CHtemperature}
\end{center}
\end{figure}

\begin{figure}
\begin{center}
\includegraphics[clip,scale=0.3]{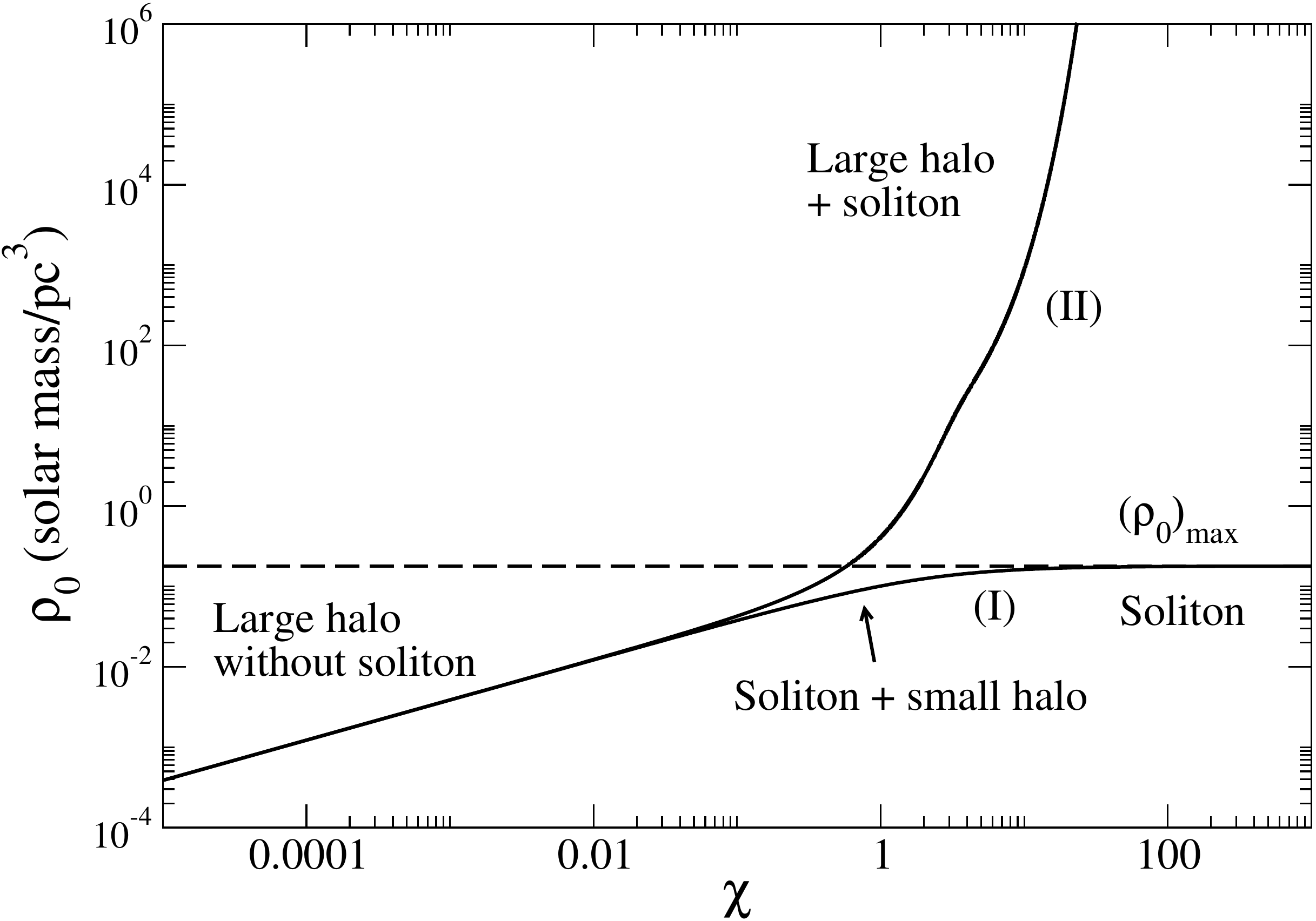}
\caption{Central density $\rho_0$ as a function of $\chi$ for models I and II.}
\label{CHSden}
\end{center}
\end{figure}

\begin{figure}
\begin{center}
\includegraphics[clip,scale=0.3]{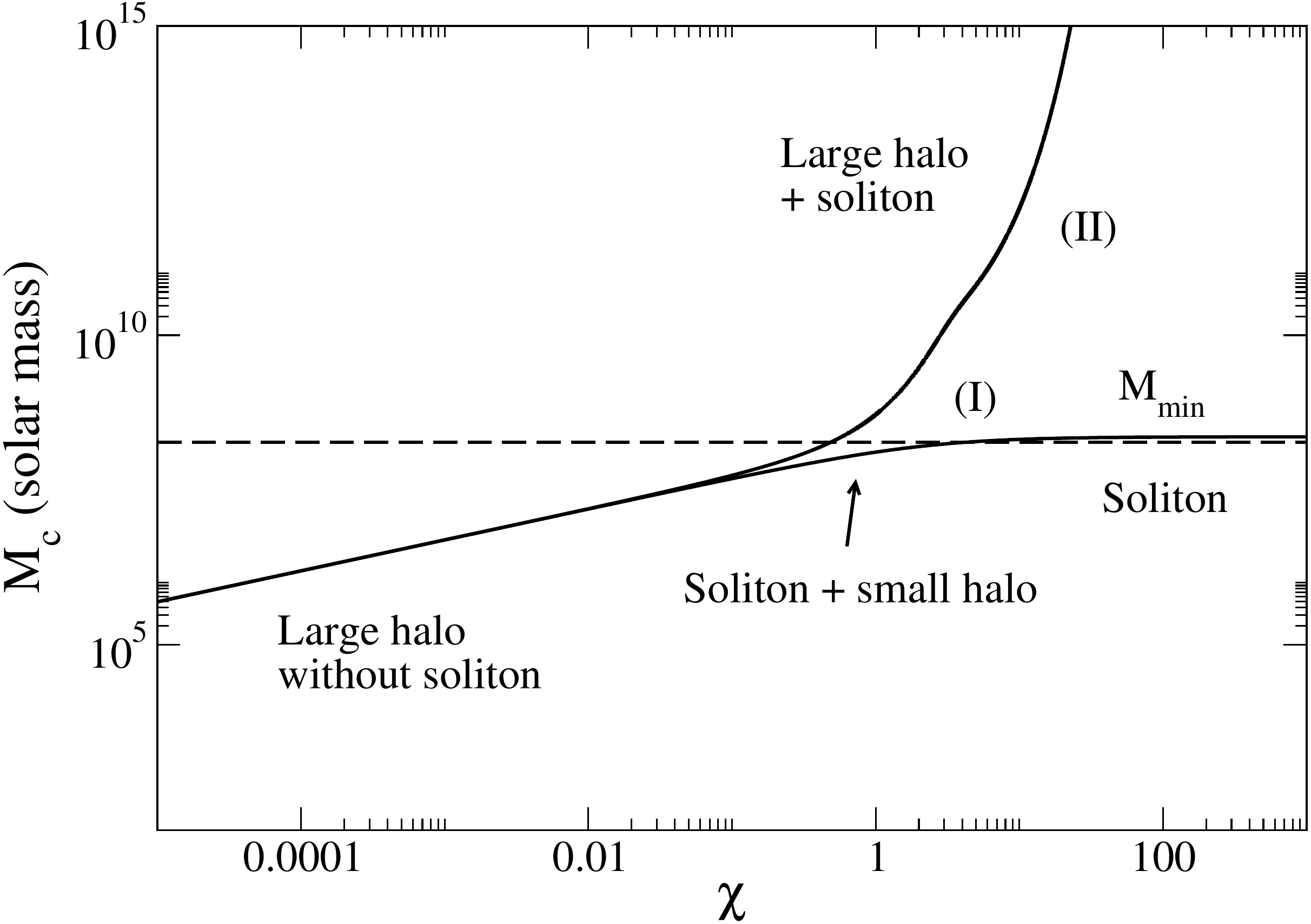}
\caption{Soliton mass  as a function of $\chi$ for models I and II.}
\label{CHSmasse}
\end{center}
\end{figure}

Actually, it is possible to obtain approximate analytical expressions of these
parameters. Indeed, for
large halos, the solitonic core is small compared to the halo radius so that,
from the ``outside'', everything happens as if the halo were purely isothermal
(i.e. we do not ``see'' the soliton). Said differently, the solitonic core is
not
expected to affect the properties of the halo at sufficiently large distances
$r\gg R_c$. Therefore, the ``external'' halo parameters 
$\rho_c$,
$M_h$, $v_h$ and $T$ should be given in good approximation by the purely
isothermal expressions
\begin{equation}
\label{aar1}
\frac{M_h}{\Sigma_0R_c^2}=1.76 \left (\frac{r_h}{R_c}\right )^2,\qquad 
\frac{k_B T}{mG\Sigma_0R_c}=0.954 \frac{r_h}{R_c},
\end{equation}
\begin{equation}
\label{aar2}
\frac{v_h^2}{G\Sigma_0R_c}=1.76 \frac{r_h}{R_c},\qquad
\frac{\rho_c}{\Sigma_0/R_c}=\frac{R_c}{r_h}.
\end{equation}
as in the absence of the soliton (see Sec. \ref{sec_lh} and note that $\rho_0$
has been replaced by $\rho_c$). In particular, the temperature
of the halo
should not depend whether there is a solitonic core or not. Identifying Eqs.
(\ref{enr6})
and (\ref{aar1}), and using  Eq. (\ref{enr1}), we obtain
\begin{equation}
\label{aar3}
\xi_h=\frac{3.63}{\sqrt{B(\chi)}} \, e^{\chi/2}.
\end{equation}
We then find that the halo radius is
given by
\begin{eqnarray}
\label{aar4}
\frac{r_h}{R_c}=\frac{1.155}{\sqrt{B(\chi)}}\, \frac{e^{\chi/2}}{\sqrt{\chi}}.
\end{eqnarray}
Using Eqs. (\ref{aar1}), (\ref{aar2}) and (\ref{aar4}), we get
\begin{equation}
\label{aar5}
\frac{M_h}{\Sigma_0R_c^2}=\frac{2.35}{B(\chi)}\, \frac{e^{\chi}}{\chi},
\end{equation}
\begin{equation}
\label{aar6}
\frac{v_h^2}{G\Sigma_0R_c}=\frac{2.03}{\sqrt{B(\chi)}} \,
\frac{e^{\chi/2}}{\sqrt{\chi}},
\end{equation}
\begin{equation}
\label{aar7}
\frac{k_B T}{Gm\Sigma_0R_c}=\frac{1.10}{\sqrt{B(\chi)}} \,
\frac{e^{\chi/2}}{\sqrt{\chi}},
\end{equation}
\begin{equation}
\label{aar8}
\frac{\rho_cR_c}{\Sigma_0}= 0.866\sqrt{B(\chi)} \sqrt{\chi}e^{-\chi/2}.
\end{equation}
Finally, from Eqs. (\ref{enr7}), (\ref{enr8}) and (\ref{aar3}), we find
that 
\begin{equation}
\label{aar9}
\frac{\rho_0R_c}{\Sigma_0}= \frac{0.866}{\sqrt{B(\chi)}} \sqrt{\chi}e^{\chi/2},
\end{equation}
\begin{equation}
\label{aar10}
\frac{M_c}{\Sigma_0R_c^2}=\frac{1.10}{\sqrt{B(\chi)}}\, \sqrt{\chi}e^{\chi/2}.
\end{equation}
These functions are plotted in Figs. \ref{CHrayon}-\ref{CHSmasse}. We see that
they define two branches of solutions. For $\chi<1$, the system is
equivalent to a purely isothermal halo without solitonic core.  For 
$\chi> 1$ the system has a core-halo  structure with a small solitonic
core and an extended isothermal atmosphere. They are separated by a plateau. If
we add the branch of solutions
obtained in Sec. \ref{sec_mod1} (Model I), we see that a sort of bifurcation
occurs at a
typical mass\footnote{The
precise value of  $(M_h)_b$ should not be given too much importance since it
partly relies on the (ad hoc) fitting procedure used in Sec. \ref{sec_pchs} to
estimate
$B(\chi)$ for
small $\chi$. It is sufficient to say that $(M_h)_b$ is larger than
$(M_h)_{\rm min}$ by about one order of magnitude,
i.e., $(M_h)_b/(\Sigma_0R_c^2)\sim 10$. It is convenient
to identify $(M_h)_b$ to $(M_h)_*$ so that the halo parameters at the
bifurcation are given by Eqs. (\ref{pdp4})-(\ref{pdp8}).} 
\begin{equation}
\label{aar10b}
(M_h)_b\sim 10^{9}\, M_{\odot}.
\end{equation}

When
$M_h=(M_h)_{\rm min}$ the DM halo is a pure
soliton (ground state). When $(M_h)_{\rm min}<M_h<(M_h)_b$ the DM halo has the
form
of a soliton with a tenuous isothermal halo. There is no plateau between
them. When $M_h>(M_h)_b$ two types of
solutions
are possible: a solution where the DM halo is purely isothermal without a
central
soliton (it corresponds to the branch studied in Sec. \ref{sec_mod1} and
recovered in the
present analysis for $\chi\lesssim 1$) and a solution where the DM halo has a
core-halo structure with a small solitonic core and an extended isothermal
atmosphere separated by a plateau.

The DM halo parameters corresponding to the purely isothermal branch (for
$M_h>(M_h)_b$) are given analytically in terms of $r_h$ by Eqs.  (\ref{aar1})
and (\ref{aar2}) with $\rho_c=\rho_0$, i.e., by Eqs. (\ref{pv8q}) and
(\ref{pv9q}).

The DM halo parameters corresponding to the core-halo branch (for
$M_h>(M_h)_b$) are given analytically in terms of $r_h$ by Eqs.  (\ref{aar1})
and (\ref{aar2}) for what concerns their ``external'' structure (isothermal
halo), and by
\begin{equation}
\label{aar11}
\frac{\rho_0R_c}{\Sigma_0}=1.50\, \frac{r_h}{R_c} \ln\left
(\frac{r_h}{R_c}\right ),
\end{equation}
\begin{equation}
\label{aar12}
\frac{M_c}{\Sigma_0R_c^2}=1.90\, \frac{r_h}{R_c} \ln\left
(\frac{r_h}{R_c}\right )
\end{equation}
for what concerns their ``internal'' structure (soliton). These latter
equations, which
determine the
soliton density and the soliton
mass have been obtained by eliminating $\chi$ from  Eqs.
(\ref{aar4}), (\ref{aar9}) and (\ref{aar10}).\footnote{We have neglected 
sublogarithmic corrections and taken $\chi\sim 2\ln(r_h/R_c)$
at leading order.} It is interesting to note that, apart from logarithmic
corrections, these
formulae are independent of $B$. Therefore, we can
consider that these results have been obtained in a purely
analytical manner since we only used the numerics in Sec. \ref{sec_pchs} to
determine 
the constant $B$ and we do not need its value here.

\subsection{The relation between the soliton  mass and the halo mass}
\label{sec_ap}

\begin{figure}
\begin{center}
\includegraphics[clip,scale=0.3]{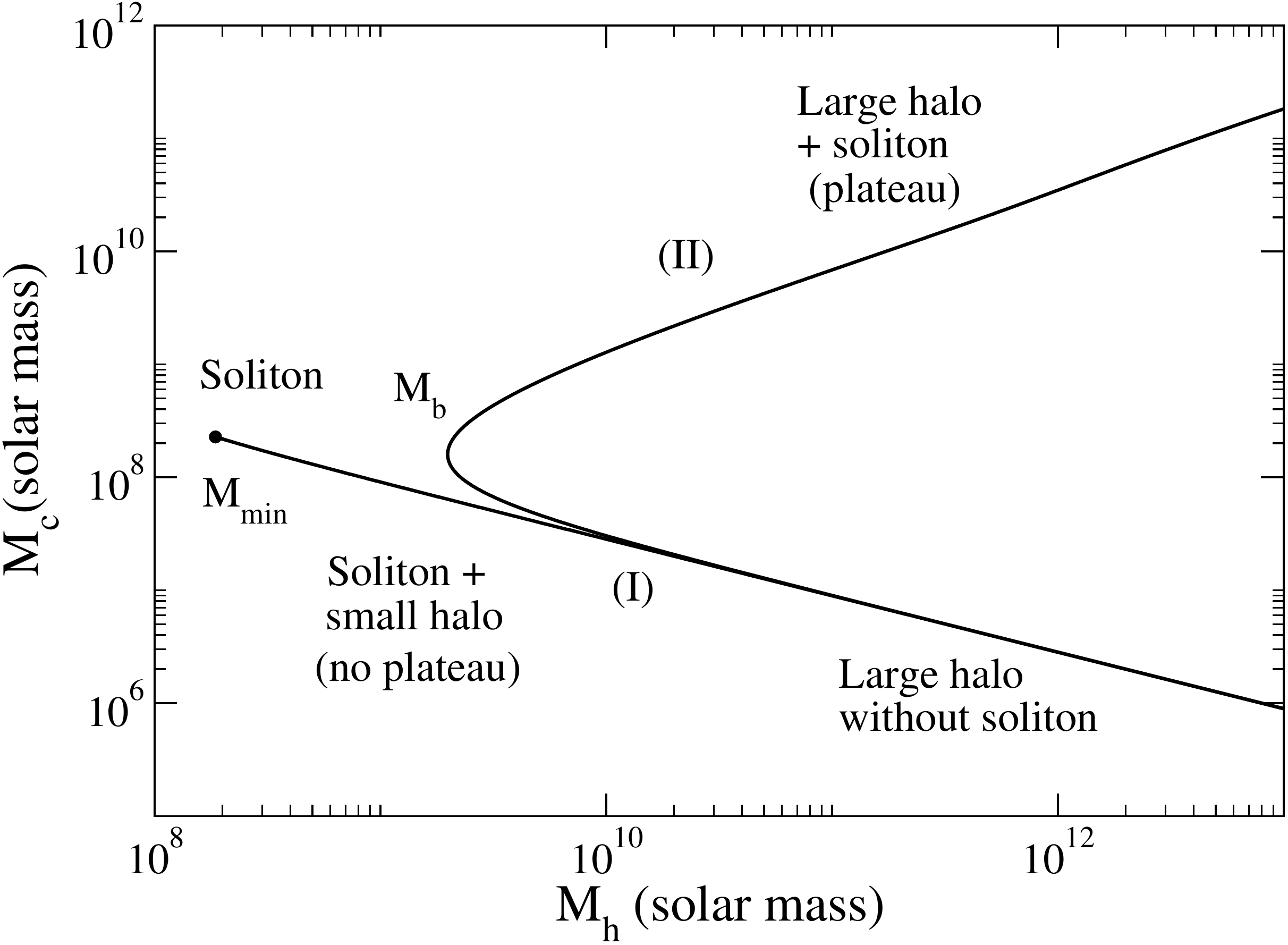}
\caption{Relation between the soliton mass $M_c$ and the halo
mass $M_h$ (according to Eq. (\ref{ap1}), we
approximately have $M_c/M_{\odot}\sim 8.54\times 10^{3}
(M_h/M_{\odot})^{1/2}$).}
\label{mhmc}
\end{center}
\end{figure}

From Eqs. (\ref{aar1}) and (\ref{aar12}), we can obtain the relation between the
soliton  mass and the
halo mass: 
\begin{equation}
\label{ap1}
\frac{M_c}{\Sigma_0R_c^2}=0.719\, \left
(\frac{M_h}{\Sigma_0R_c^2}\right )^{1/2} \ln \left
(\frac{M_h}{\Sigma_0R_c^2}\right ).
\end{equation}
It is plotted in Fig. \ref{mhmc}.
Introducing the minimum halo mass $(M_h)_{\rm
min}=1.32\Sigma_0R_c^2=1.86\times 10^8\, M_{\odot}$, we can rewrite the
foregoing equation as
\begin{equation}
\label{ap2}
\frac{M_c}{(M_h)_{\rm min}}=0.626\, \left
(\frac{M_h}{(M_h)_{\rm min}}\right )^{1/2} \ln \left
(\frac{M_h}{(M_h)_{\rm min}}\right ).
\end{equation}
We note that
\begin{equation}
\label{ap1b}
M_c=0.719\, R_c\sqrt{\Sigma_0 M_h} \ln \left
(\frac{M_h}{\Sigma_0R_c^2}\right ).
\end{equation}
These relations are valid  on the core-halo branch for $M_h>
(M_h)_b$ and sublogarithmic corrections have been neglected. At leading order,
the core mass increases as $M_c\propto M_h^{1/2}$. Logarithmic corrections
may slightly  change the apparent scaling exponent.

The bifurcation is clearly visible in Fig.  \ref{mhmc}. When $(M_h)_{\rm
min}<M_h<(M_h)_b$ there is only one solution which corresponds to a solitonic
core surrounded by a small isothermal halo without plateau. For $M_h>(M_h)_b$
there are two
solutions: a large isothermal halo without soliton and a large isothermal halo
with a solitonic core and a plateau.

{\it Remark:} in a recent work, Lin {\it
et al.} \cite{lin} found that ``small'' halos ($M_h<10^{10}\,
M_{\odot}$) have a core-halo
structure without a plateau while ``large halos'' ($M_h>10^{10}\,
M_{\odot}$) have a core-halo structure
exhibiting a plateau. This is qualitatively similar to the bifurcation
that we have independently obtained in our study.

\subsection{Astrophysical consequences: formation of a solitonic bulge}
\label{sec_ac}

\begin{figure}
\begin{center}
\includegraphics[clip,scale=0.3]{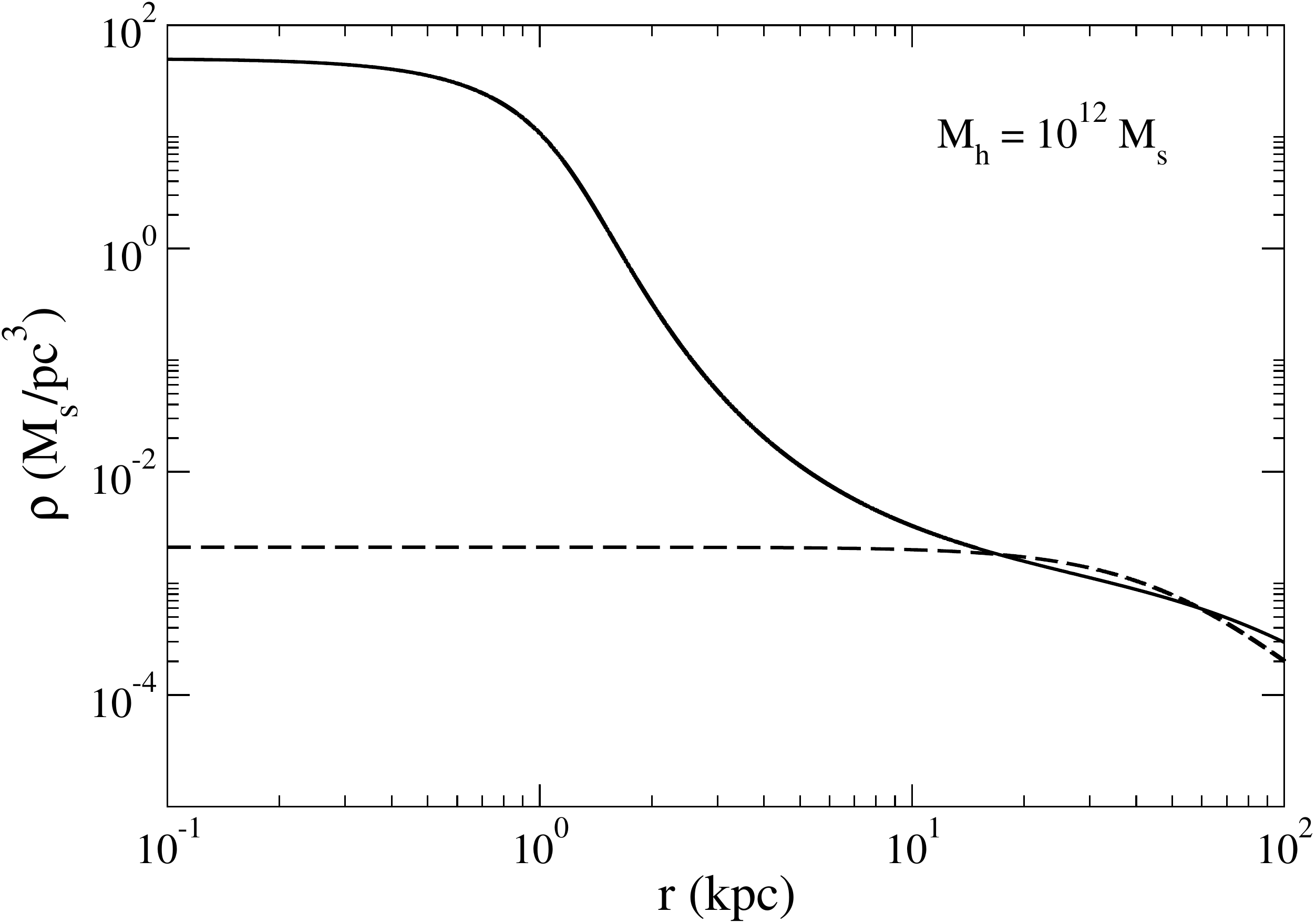}
\caption{Density profile of a DM halo of
mass $M_h=10^{12}\, M_{\odot}$. The full line corresponds to the
core-halo solution of Model II. The concentration parameter $\chi=4.54$ is
not very large. As a result, the plateau is not very clear cut and  the
definition of the ``plateau density'' $\rho_c$ is a bit ambiguous. It
approximately corresponds to the inflexion point of
the core-halo profile.  The dashed line corresponds to the purely isothermal
solution without a soliton. We see that the two profiles approximately match
each other for $r\gtrsim 10\, {\rm kpc}$.}
\label{MW}
\end{center}
\end{figure}

\begin{figure}
\begin{center}
\includegraphics[clip,scale=0.3]{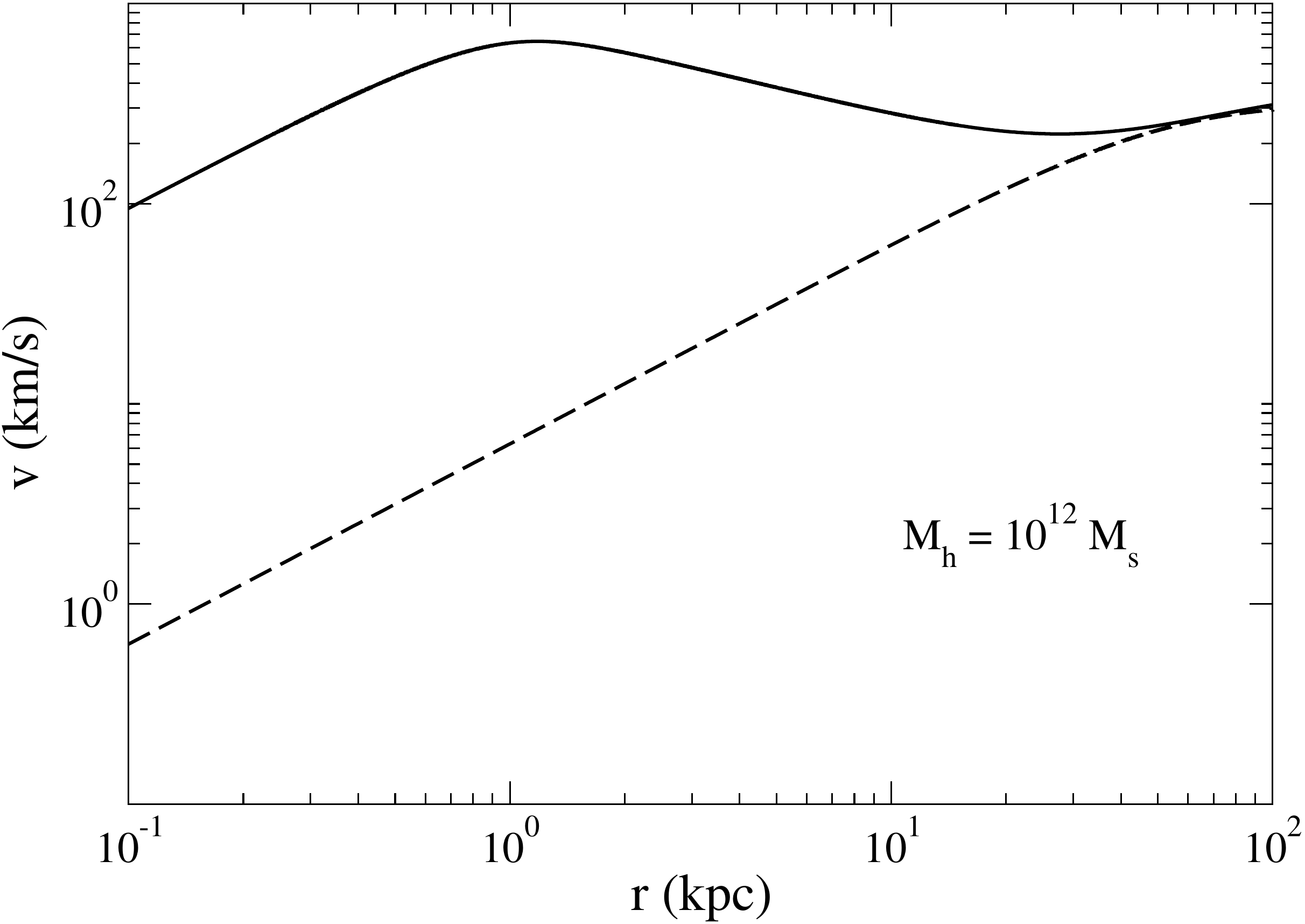}
\caption{Same as Fig. \ref{MW} for the circular velocity.}
\label{MWv}
\end{center}
\end{figure}

Let us study the astrophysical consequences of these results. For a large DM
halo of mass $M_h=10^{12}\,
M_{\odot}$,  considering the
core-halo solution, we find that $\chi=4.54$.
Then, we get $r_h=61.9\, {\rm kpc}$,
$v_h=263\, {\rm km/s}$, $\sqrt{k_B
T/m}=246\, {\rm
km/s}$, $\rho_c=2.27\times 10^{-3}\, M_{\odot}/{\rm pc}^3$,
$\rho_0=50.2\, M_{\odot}/{\rm pc}^3$, and  $M_c=6.39\times
10^{10}\, M_{\odot}$.\footnote{To determine these values, we have used the exact
expressions from Eqs. (\ref{enr1})-(\ref{enr8}), not the approximate analytical
expressions from Eqs. (\ref{aar4})-(\ref{aar10}) because $\chi$ is not large
enough to fully justify their validity.} Considering now a purely isothermal
DM halo ($\chi=0$)
with the same mass and using
Eqs. (\ref{pv8q}) and (\ref{pv9q}), we find that
$r_h=63.5\, {\rm kpc}$, $v_h=260\, {\rm km/s}$, $\sqrt{k_B T/m}=191\, {\rm
km/s}$, and $\rho_c=2.22\times 10^{-3}\, M_{\odot}/{\rm pc}^3$ (the
differences between the ``external'' parameters $r_h$, $v_h$, $T/m$ and
$\rho_c$ in the two cases are due to
the fact that
$\chi$ is relatively small).
The  density and velocity profiles are represented in Figs.
\ref{MW} and \ref{MWv}.

Our model II predicts that a large DM halo of mass $M_h=10^{12}\, M_{\odot}$,
such
as the one surrounding our Galaxy, should possess a solitonic core of radius
$R_c=1\, {\rm kpc}$, mass $M_c= 6.39\times 10^{10}\, M_{\odot}$, and central
density $\rho_0=50.2\, M_{\odot}/{\rm pc}^3$ (see Fig. \ref{MW}). In the
solitonic  core,
the circular velocity is much larger than for a purely isothermal
distribution  (see Fig. \ref{MWv}). Let us
consider different implications of this prediction.

In the most favorable scenario, the solitonic core is physical. This scenario
is supported by the numerical simulations of Schive {\it et al.}
\cite{ch2,ch3} that reveal the presence
of an extended solitonic core at the centres of DM halos.\footnote{These
simulations are made for a noninteracting BEC but similar
results should be obtained for a self-interacting BEC.} For real galaxies, this
solitonic core
may still exist now or may  have existed only in the past and has disappeared
since then (see below). Because of its deep gravitational potential, the
solitonic core could have acted as a ``seed'' for the formation of early
spheroids and quasars. It could have helped forming a stellar
bulge or a galactic nucleus.\footnote{In this respect, we note that the mass 
$M_c= 6.39\times 10^{10}\, M_{\odot}$ and the size $R_c=1\, {\rm kpc}$ of the
soliton are compatible with the mass and size of stellar bulges and  galactic
nuclei.} Indeed, the gravitational force created by the
soliton can quickly
attract a large amount of gas into a small central region, thereby
creating an
ultra-dense gas favorable for major starbursts and for the formation of
supermassive
black holes \cite{ch2,ch3}. We shall show in the following
section that this
core-halo solution is thermodynamically
unstable in the canonical ensemble.
Therefore, the solitonic core
may have formed only temporarily in the past, but long enough to constitute a
stellar bulge (possibly triggering the formation of a black hole and a quasar),
before
disappearing on a
longer timescale.\footnote{Note that the core-halo solution is dynamically
stable so that it is relatively persistent. It is also thermodynamically
stable in the microcanonical ensemble (see Secs. \ref{sec_ineq}
and \ref{sec_astapp}). This may increase its lifetime if the microcanonical
ensemble is the correct ensemble to consider in our problem. Therefore, it is
very likely that the core-halo solution is physically
relevant (see the Remark at the end of this section).}

In the most  defavorable scenario, the solitonic core is not physical. Indeed,
the solitonic core should produce a clear signature on
the velocity curves marked by the presence of a dip (see Fig. \ref{MWv}).
Apparently, this dip has not been clearly observed as mentioned by Slepian
and Goodman \cite{sg} (see, however,
the Remark at the end of this section). Assuming that this is not a
problem of measurement (or that the dip/soliton has not disappeared during the
evolution of the halo),
this raises the following possibilities:

(i) The first possibility is that the BECDM model with a repulsive
self-interaction is ruled out. This is essentially the conclusion of 
Slepian
and Goodman \cite{sg}.

(ii) There is, however, another possibility. The core-halo solution forms just
one possible
solution of the self-gravitating BEC model. Another solution
exists in which the DM halo is
purely isothermal without a solitonic core. It is
possible that this purely isothermal solution is selected instead
of the core-halo one.  We shall show in the following section that the purely
isothermal  solution is stable (minimum of free energy at fixed mass) while the
core-halo
solution is thermodynamically unstable  in the canonical ensemble (saddle point
of free
energy). From
these thermodynamical considerations, the purely isothermal
solution with no soliton is more probable than the core-halo
one (assuming that the canonical ensemble rather than the microcanonical
ensemble applies in our
problem).

Clearly, the confirmation of the presence or the absence of a ``solitonic''
bulge of mass $M_c= 6.39\times 10^{10}\, M_{\odot}$ and size $R_c=1\, {\rm
kpc}$ at the center of DM halos of mass  $M_h=10^{12}\, M_{\odot}$ and size
$r_h=61.9\, {\rm kpc}$ would be of considerable interest. This is a challenge
for astrophysical observations.

{\it Remark:} During the redaction of our manuscript, we
came accross the recent paper of De Martino {\it et al.} \cite{martino} who
show that the central motion of bulge stars in the Milky Way implies the
presence of a dark matter core  of mass $\simeq 10^{9}\, M_{\odot}$ and radius
$\simeq 100 \, {\rm pc}$ that they interprete as a soliton. Their result is
based on the measures of dispersion velocity by Zoccali {\it et al.}
\cite{zoccali} and Portail {\it et al.} \cite{portail} who construct a fully
dynamical model of the bulge and find the need for a compact mass of $\simeq
2\times 10^{9}\, M_{\odot}$.  These results are qualitatively consistent with
our model which predicts a large solitonic core of radius
$R_c=1\, {\rm kpc}$ and mass $M_c= 6.39\times 10^{10}\, M_{\odot}$ in a DM halo
of mass $M_h=10^{12}\, M_{\odot}$. The values of $M_c$ and $R_c$ are different
from De Martino {\it et al.} \cite{martino} because our model is different: we
are considering self-interacting bosons in the TF limit while De Martino {\it et
al.} \cite{martino} consider noninteracting bosons. Furthermore, our DM halo
of mass $M_h=10^{12}\, M_{\odot}$ is more massive than the DM halo of the  Milky
Way
implying a
larger soliton mass. Neverthless, this qualitative
agreement is encouraging and shows that a solitonic core can really be present,
even now, at the centers of the galaxies (from a thermodynamical point of view
this would imply that the microcanonical ensemble is more relevant than the
canonical one, see footnote 33). It will be important to extend our model to the
case of noninteracting bosons \cite{forthcoming} to see if the agreement with De
Martino {\it et
al.}
\cite{martino} improves. This may help discriminating between different types of
bosons, i.e., noninteracting bosons  versus self-interacting bosons.

\subsection{Can the soliton mimic a supermassive black hole?}
\label{sec_cheat}

There is strong observational evidence that very massive objects reside at
the centers of galaxies. These objects are usually considered to
be supermassive black
holes (SMBHs). For example, Sagittarius A* (Sgr A*), a bright and very compact
astronomical radio source that resides at the center of our Galaxy is thought to
be the location of a SMBH of mass $M=4.2\times 10^6\,
M_{\odot}$ and Schwarzschild radius $R_S=4.02\times 10^{-7}\, {\rm pc}$.
Whatever the object may be, it must be enclosed within a radius  $R_{\rm
P}=6\times 10^{-4} \,
{\rm pc}$ ($R_{\rm P}=1492\, R_S$), the S2 star pericenter
\cite{gillessen}.\footnote{The radius of the compact object  must satisfy
$R_{*}\le R_P$ from the observations.  This implies
$R/R_S\le 1492$. This
object is not necessarily  a black hole unless its radius
is much smaller than $R_P=6\times 10^{-4}\, {\rm pc}$, namely $R_*\sim
R_S=4.02\times 10^{-7}\, {\rm pc}$.} Similar
objects are expected to reside at the centers of most spiral and
elliptical galaxies, in active galactic nuclei (AGN). Although it is commonly
believed that these objects are SMBHs \cite{gillessen,nature,reid,genzel}, this
is
not yet established on a firm
observational basis in all cases. As an alternative to the black hole
hypothesis, it has been proposed that such objects could be fermion balls
\cite{bvn,bvr,bmtv,btv,rar,krut} or 
boson stars \cite{torres2000,guzmanbh} that could mimic a black hole.
Let us consider
this possibility in the framework of the present model. More precisely, let us
investigate if a solitonic core can mimic a supermassive
black hole at the
center of the Milky Way.

To be specific, let us consider a DM halo of mass $M_h=10^{11}\, M_{\odot}$
similar to the one that
surrounds the Wilky Way. Using the results of Sec. \ref{sec_aar}, we find that
this halo
should contain a solitonic core of mass  $M_c= 1.77\times 10^{10}\,
M_{\odot}$ and radius
$R_c=1\, {\rm kpc}$. Clearly, the soliton is too extended to
mimic a black hole. As discussed in Sec. \ref{sec_ac} the solitonic core is
more
likely to represent a bulge which is either present now or which, in the
past, may have triggered the formation of a black
hole.

\begin{figure}
\begin{center}
\includegraphics[clip,scale=0.3]{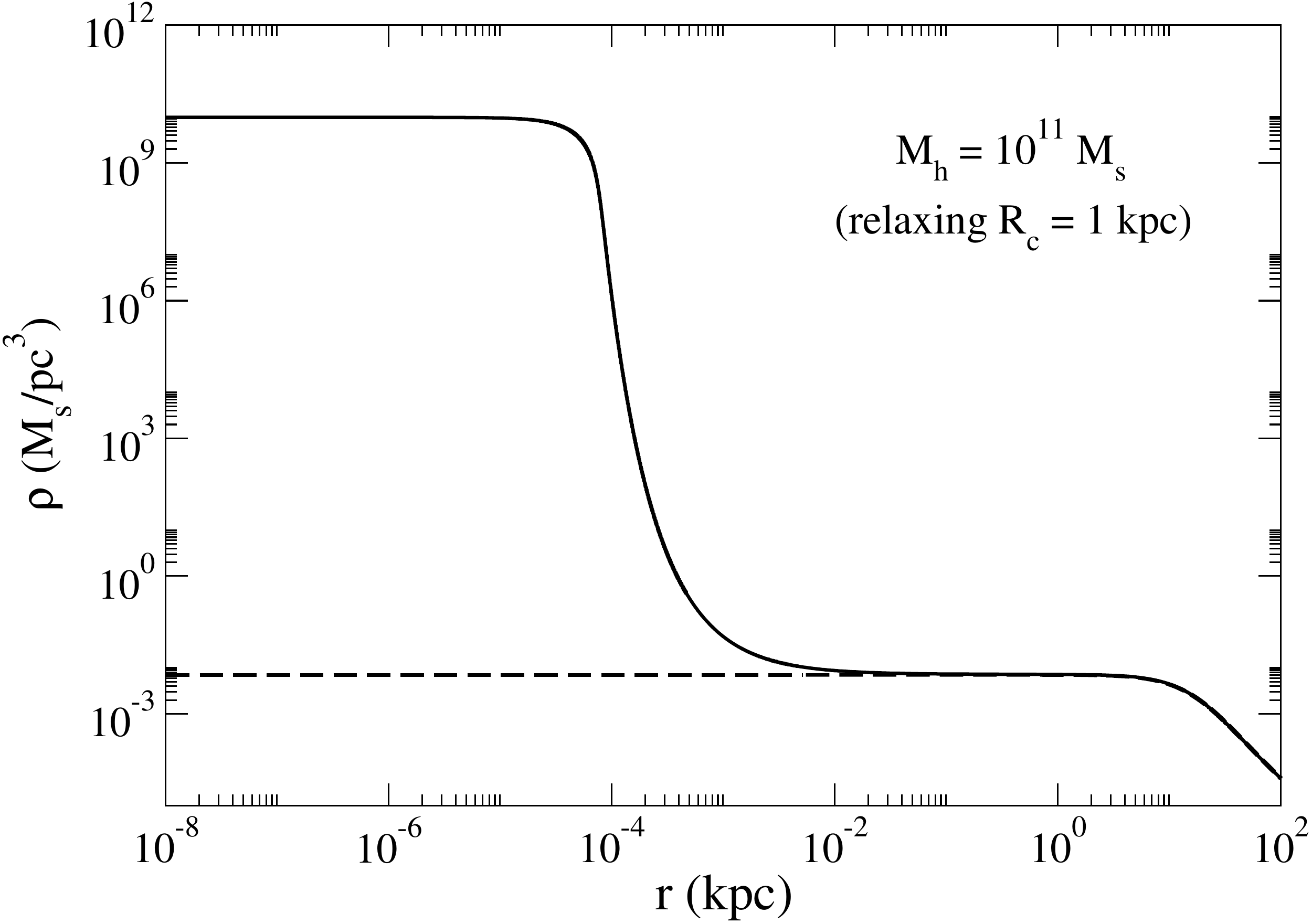}
\caption{Core-halo density profile  in the
framework of Model II corresponding to a DM halo of
mass $M_h=10^{11}\, M_{\odot}$ and
concentration parameter $\chi=22.5$ when the condition $R_c=1\, {\rm kpc}$ is
(arbitrarily) relaxed. It is compared with a purely isothermal
profile (dashed
line). The solitonic core cannot  model a SBH because it is nonrelativistic
and, above all, its radius is too large.}
\label{MWwrong}
\end{center}
\end{figure}

\begin{figure}
\begin{center}
\includegraphics[clip,scale=0.3]{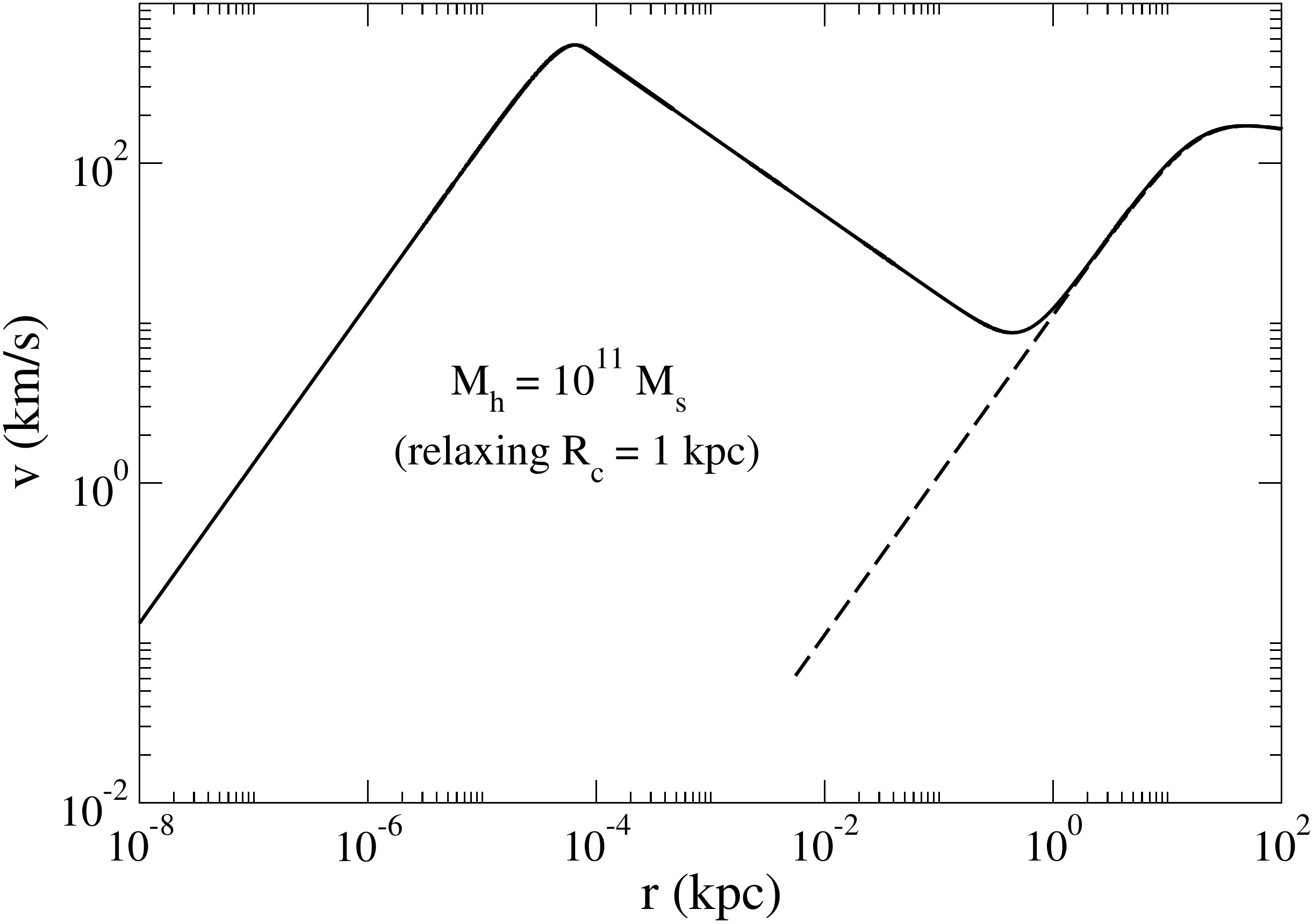}
\caption{Same as Fig. \ref{MWwrong} for the circular velocity.}
\label{MWvwrong}
\end{center}
\end{figure}

Nevertheless, let us try to push our  model to its limits by relaxing certain
assumptions. We relax the value of $R_c=1\, {\rm
kpc}$ that was fixed by the size of ultracompact DM halos (see Sec.
\ref{sec_gs}) and we impose that 
$M_h=10^{11}\, M_{\odot}$ and  
$M_c=4.2\times 10^6\, M_{\odot}$. Using the relation
$M_h/M_c=(2.14/\sqrt{B(\chi)})
e^{\chi/2}/\chi^{3/2}$ obtained from Eqs. (\ref{aar5}) and (\ref{aar10}) we
obtain $\chi=22.5$. Then, we find from Eq. (\ref{aar5}) that
$R_c=0.652\sqrt{B(\chi)\chi}e^{-\chi/2}(M_h/\Sigma_0)^{1/2}$, giving
$R_c=6.98\times 10^{-2} \, {\rm
pc}$.\footnote{This corresponds to a ratio $a_s/m^3=1.60\times
10^{-5}\, {\rm fm/(eV/c^2)^3}$ [see Eq. (\ref{rc})].}
Finally, Eqs. (\ref{aar4}), (\ref{aar6}), (\ref{aar7}), (\ref{aar8}) and
(\ref{aar9}) imply $r_h=2.00\times 10^{4}\,
{\rm pc}$, $\rho_c=7.03\times 10^{-3}M_{\odot}/{\rm
pc}^3$, $(k_B T/m)^{1/2}=108\, {\rm km/s}$, $v_h=(GM_h/r_h)^{1/2}=146\,  {\rm
km/s}$, and $\rho_0=9.80\times
10^{9}M_{\odot}/{\rm
pc}^3$. The values of the ``external'' parameters exactly match the values of a
purely isothermal halo (see Sec. \ref{sec_lh}). This is because $\chi\gg 1$
making our
approximate analytical expressions accurate. As a result, we obtain a
core-halo profile (see Figs. \ref{MWwrong} and \ref{MWvwrong}) that is
consistent with the observations from the outside, and that contains a solitonic
core of mass $M_c=4.2\times 10^6\, M_{\odot}$ similar to the mass of the compact
object at the center of the Galaxy. Unfortunately,  the radius of the
soliton, $R_c=6.98\times 10^{-2} \, {\rm
pc}$, is
$100$ times larger than the maximum size of this object, $R_P=6\times 10^{-4} \,
{\rm pc}$, deduced from the observations.\footnote{Instead of imposing
the core mass, we could impose the core radius $R_c=6\times 10^{-4} \, {\rm
pc}$. From Eq.
(\ref{aar5}) with  $M_h/(\Sigma_0R_c^2)=1.97\times
10^{15}$, we obtain  $\chi=32.4$. Then, Eq. (\ref{aar4}) and 
Eqs. (\ref{aar6})-(\ref{aar10}) give $r_h=2.02\times 10^{4}\,
{\rm pc}$, $\rho_c=6.95\times 10^{-3}M_{\odot}/{\rm
pc}^3$, $(k_B T/m)^{1/2}=108\, {\rm km/s}$,  
$v_h=(GM_h/r_h)^{1/2}=147\,  {\rm km/s}$, $\rho_0=1.93\times
10^{14}M_{\odot}/{\rm
pc}^3$, and $M_c=5.30\times 10^{4}\,
M_{\odot}$. This time, the core mass, $M_c=5.30\times 10^{4}\,
M_{\odot}$, is about $100$ times smaller than the
mass of the
central object, $M=4.2\times 10^6\,
M_{\odot}$, estimated from the observations.} More generally, we
show in Appendix \ref{sec_mimic} that the solitonic core is never relativistic
so it
can never
mimic a SMBH.

\begin{figure}
\begin{center}
\includegraphics[clip,scale=0.3]{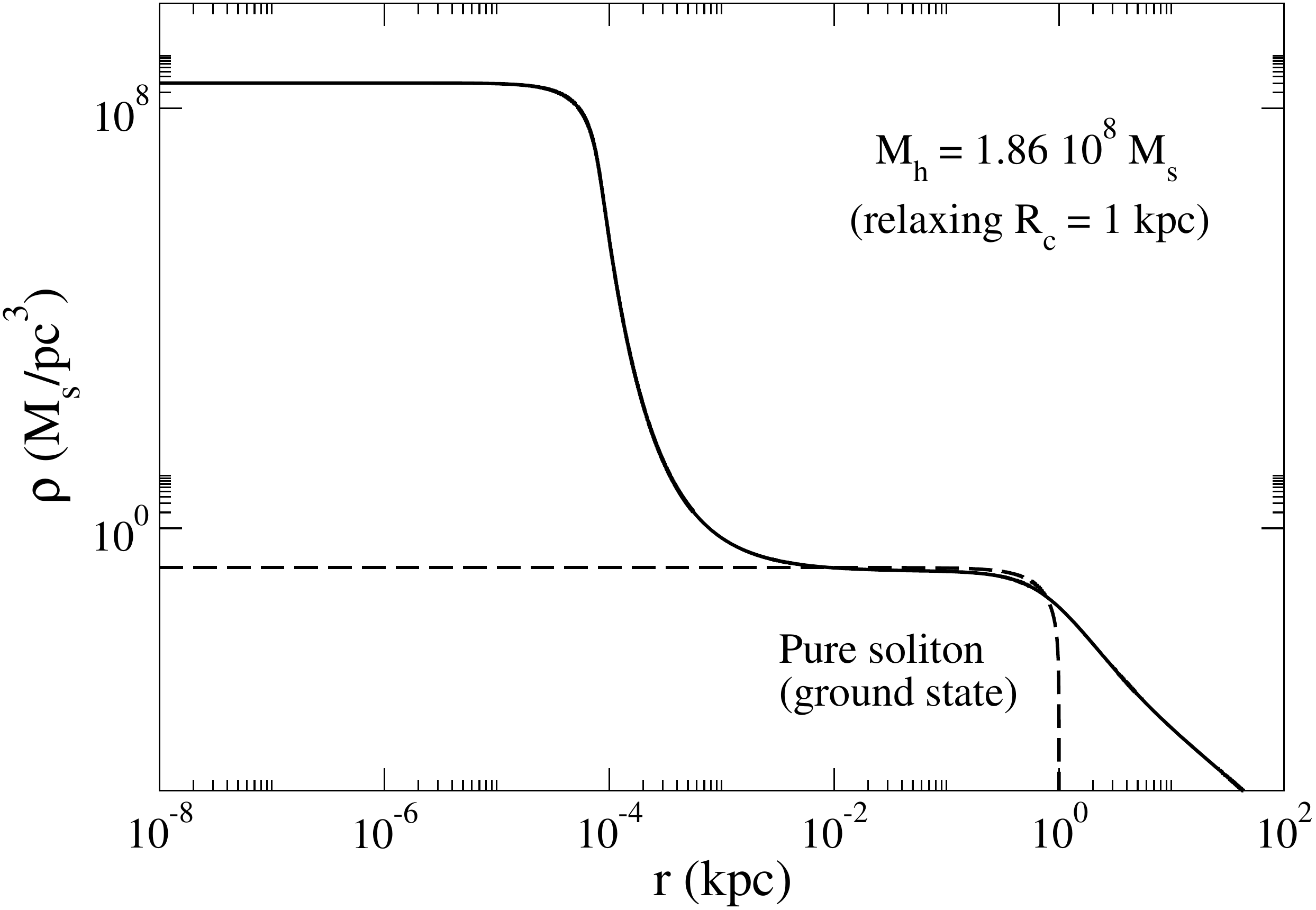}
\caption{Core-halo density profile in the
framework of Model II corresponding to a DM
halo of
mass $M_h=1.86\times 10^{8}\,
M_{\odot}$ (such as Fornax) and concentration parameter  $\chi=15.9$  when the
condition $R_c=1\, {\rm kpc}$ is
(arbitrarily) relaxed. As explained in the text, this
profile has to be rejected in favor of the pure
soliton  (dashed line) corresponding to the ground state of the BECDM model. }
\label{Fornaxwrong}
\end{center}
\end{figure}

\begin{figure}
\begin{center}
\includegraphics[clip,scale=0.3]{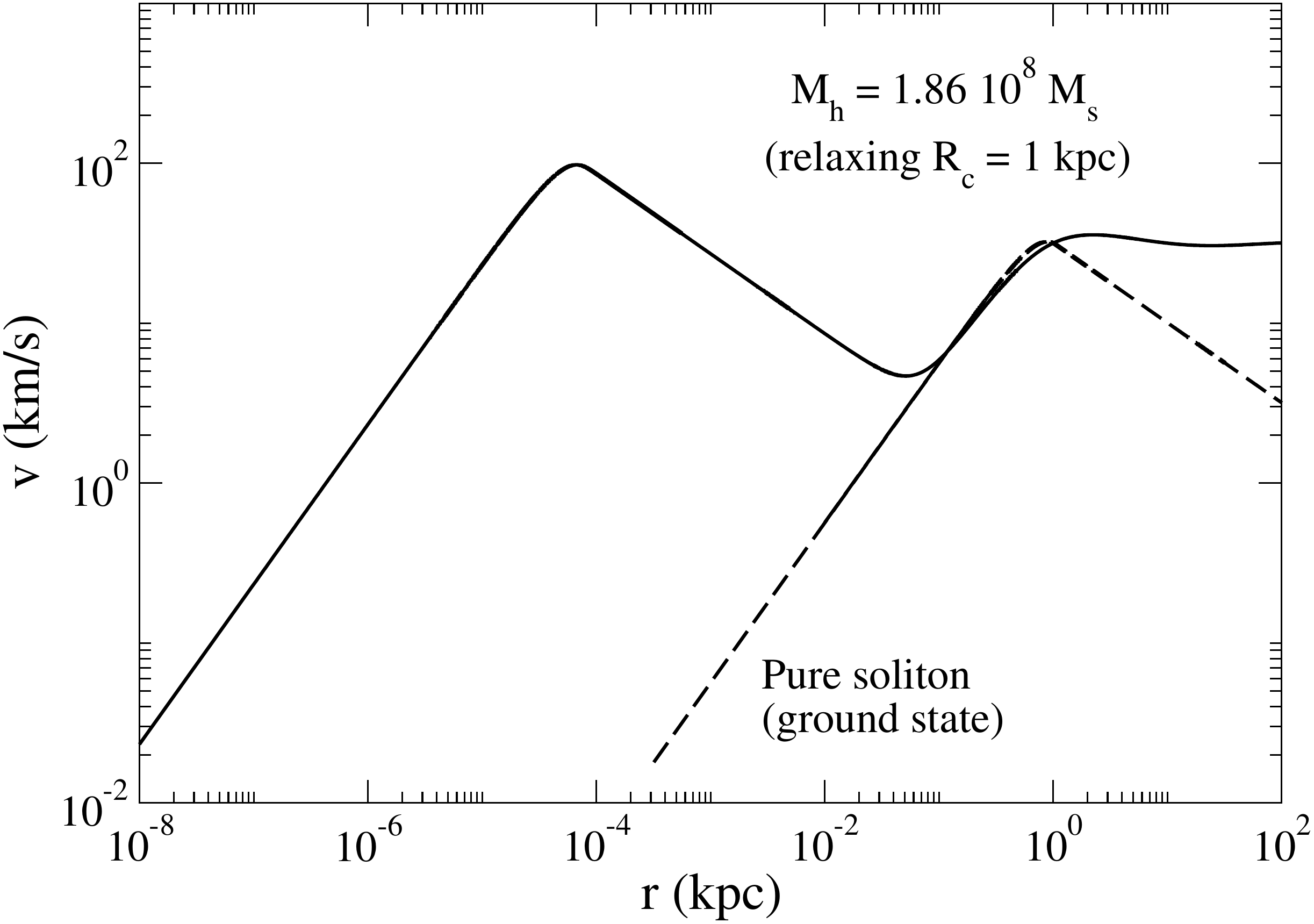}
\caption{Same as Fig. \ref{Fornaxwrong} for the circular
velocity. }
\label{Fornaxvwrong}
\end{center}
\end{figure}

In addition of being unable to mimic a central black hole, there is another
problem with the profile constructed previously. Since we have changed the
value of $R_c=6.98\times 10^{-2} \, {\rm
pc}$ with respect to its original value  $R_c=1\, {\rm
kpc}$ in order to impose a core mass $M_c=4.2\times 10^6\,
M_{\odot}$, the ground state of the self-gravitating BEC model has been changed
accordingly. For $R_c=6.98\times 10^{-2} \, {\rm
pc}$, the ground state
now corresponds to $(r_h)_{\rm min}=5.50\times 10^{-2}\, {\rm pc}$ and 
$(M_h)_{\rm min}=0.906\, M_{\odot}$ [see Eq. (\ref{pv1})]. The minimum halo
radius and
the minimum halo mass are much smaller than the radius and the mass of typical
dSphs like Fornax. Such small halos are not observed, suggesting that the ground
state of DM halos is at a much larger scale of the order of $1\, {\rm kpc}$
as we have initially assumed (see Sec. \ref{sec_gs}). If we
ignore this
difficulty and nevertheless apply the model with
$R_c=6.98\times
10^{-2} \, {\rm pc}$ to a halo of mass 
$M_h=1.86\times 10^{8}\, M_{\odot}$ (such as Fornax) we find from Eq.
(\ref{aar5}) that $\chi=15.9$. We then obtain from Eq. (\ref{aar4}) and 
Eqs. (\ref{aar6})-(\ref{aar10}) that $r_h=880\, {\rm pc}$, $\rho_c=0.160\,
M_{\odot}/{\rm
pc}^3$,  $(k_B T/m)^{1/2}=22.5\, {\rm km/s}$, $v_h=(GM_h/r_h)^{1/2}=30.6\,  {\rm
km/s}$, $\rho_0=3.04\times
10^{8}M_{\odot}/{\rm
pc}^3$ and $M_c=1.31\times 10^5\, M_{\odot}$. This DM halo has 
a core-halo structure with a very small nucleus of mass $M_c=1.31\times 10^5\,
M_{\odot}$ and radius $R_c=6.98\times
10^{-2} \, {\rm pc}$ and an extended isothermal halo (see
Figs. \ref{Fornaxwrong} and \ref{Fornaxvwrong}). This is very different from
the structure that we have considered in Sec.  \ref{sec_gs} consisting of a
pure soliton of mass $M_h=1.86\times 10^{8}\, M_{\odot}$ and radius $r_h=788\,
{\rm pc}$ (see the pure soliton in  Figs. \ref{pd} and \ref{pv}).

The previous arguments (and the results of Appendix \ref{sec_mimic}) lead to the
conclusion
that the solitonic core of Model II cannot mimic a SMBH.  The
solitonic core, is more likely to represent a
large central
bulge or a galactic nucleus (see Sec. \ref{sec_ac}). This conclusion is
important in view of the different
attempts that have
been made in the past to describe
the compact object that resides  at the center of our Galaxy, presumably a
SMBH \cite{nature,reid,gillessen,genzel}, by an object of another
nature like a fermion ball
\cite{bvn,bvr,bmtv,btv,rar,krut} or a boson
star \cite{torres2000,guzmanbh}.\footnote{Bilic {\it
et al.} \cite{btv} developed a model of fermionic
DM (with a fermion mass $m=15\, {\rm
keV/c^2}$) that describes both the center and the halo of the Galaxy.
They found a (nonrelativistic) fermion ball of mass $M=2.27\times 10^6\,
M_{\odot}$ and radius
$R=18\, {\rm mpc}$. Unfortunately, its radius is larger by a
factor $100$ than the bound
$R_P=6\times
10^{-4}\, {\rm pc}$ set by later observations \cite{gillessen}.
The same problem was encountered by Ruffini {\it
et al.} \cite{rar} who developed a similar model with a fermion mass $m\sim
10\, {\rm
keV/c^2}$. Very
recently, Arg\"uelles {\it et al.} \cite{krut} considered the fermionic
King model \cite{clm2} (accounting for a tidal confinement) with a fermion mass
$m=48\, {\rm
keV/c^2}$ and 
found a core-halo solution with a  (nonrelativistic) fermion ball of mass
$M=4.2\times 10^6\,
M_{\odot}$ and radius
$R=R_P=6\times 10^{-4}\, {\rm pc}$ consistent with the
observations. This core-halo state is Vlasov
dynamically stable. Therefore, if we ``prepare'' the system in this 
state, it will remain in this state for a very long time. However, we
have
argued in \cite{clm2,ac} that this core-halo
solution is thermodynamically unstable in all statistical ensembles.
As a result, it  is very unlikely to appear {\it spontaneously} in a
thermodynamical
sense. The fermion ball corresponds to a sort of  ``critical droplet'' in the
langage of phase transitions (saddle point of entropy) which has a very
low probability. Furthermore, the model of
Arg\"uelles {\it et al.} \cite{krut} faces the ground state problem reported
in the
fourth paragraph of this section. Assuming that Fornax is the ground state of
the self-gravitating Fermi gas imposes that $m=170 \, {\rm eV/c^2}$ (see the
Remark at the end of Sec. \ref{sec_gs}). This mass is much smaller than the mass
$m=48\, {\rm keV/c^2}$ taken in Ref. \cite{krut}. If we object that Fornax may
not be the ground state of the self-gravitating Fermi gas this would imply
that (i)
much smaller halos should exist, (ii) Fornax should have a core-halo
structure with a small central fermion ball. To our knowledge, these two
features  have not
been observed.} In
future works \cite{forthcoming}, we shall adapt our model to the case of
noninteracting bosons and self-gravitating fermions to see whether we reach the
same conclusion. We will then decide whether the boson star or fermion ball
scenario (as a SMBH mimicker) is ruled out or if we need to modify our model.
Any definite conclusion is premature for the moment.

\section{Phase transitions of a thermal self-gravitating boson gas in a box}
\label{sec_ptb}

In this section, we study the nature of phase transitions in a self-gravitating
BEC described by the equation of state (\ref{mad6bq}). This system generically
has a core-halo structure with a solitonic core and an isothermal halo.  Since
the halo is isothermal, the density decreases at large distances as $r^{-2}$.
This implies that the total mass of the system is infinite. A solution to avoid
the
infinite mass problem is to confine the system within a spherical box of radius
$R$. This box model  will allow us to recover the bifurcation between the purely
isothermal state and the core-halo state  obtained in the
preceding section and to interpret this bifurcation in terms
of a phase transition which is related to the existence of a canonical
critical point. It will also allow us to show that purely isothermal
configurations are thermodynamically stable (minima of
free energy) while 
core-halo configurations are thermodynamically unstable  in the canonical
ensemble (saddle points of
free
energy). We shall also discuss their microcanonical stability.

\subsection{Basic equations}
\label{sec_be}

The equilibrium states of the self-gravitating BEC are determined by the
generalized Emden equation (\ref{ch26}). Let us  denote by
$\alpha$ the value of $\xi$ at the box radius $R$.  According to Eqs.
(\ref{chp3}) and (\ref{ch23sw}), the normalized box
radius
$\alpha$ is given by
\begin{equation}
\label{be1}
\alpha=(4\pi G\rho_0\beta m)^{1/2}R.
\end{equation}
We then have $r=\xi R/\alpha$. The total mass enclosed within the box is
$M=M(R)$. From Eq. (\ref{ch30}) we find that the normalized inverse temperature 
\begin{equation}
\label{be2}
\eta=\frac{\beta G Mm}{R}
\end{equation}
is given by
\begin{equation}
\label{be3}
\eta=\alpha\left\lbrack 1+\chi
e^{-\psi(\alpha)}\right\rbrack\psi'(\alpha).
\end{equation}
We introduce the control parameter\footnote{This parameter is
the counterpart of the parameter $\mu=\eta_0\sqrt{512\pi^4G^3MR^3}$ with
$\eta_0=gm^4/h^3$ (where $g=2s+1$ is the spin multiplicity of the
quantum states) introduced in the study of self-gravitating fermions
\cite{ijmpb}.}
\begin{equation}
\label{be4}
\mu=\frac{Gm^3R^2}{a_s\hbar^2}.
\end{equation}
Using Eq. (\ref{rc}), it can be written as
\begin{equation}
\label{be5}
\mu=\pi^2\left (\frac{R}{R_{c}}\right )^2.
\end{equation}
Therefore, it measures the size of the system (represented by $R$) as compared
to the size $R_c$ of
the solitonic core. We note that the condition $R>R_c$ corresponds to
$\mu>\mu_{\rm min}=\pi^2=9.87$. From
Eqs. (\ref{ch27}) and (\ref{be1}), we find that
\begin{equation}
\label{be6}
\mu=\frac{\alpha^2}{\chi}.
\end{equation}
On the other hand, combining Eqs. (\ref{be1}), (\ref{be2}) and (\ref{be6}), we
find that the
normalized central density is given by
\begin{equation}
\label{be7}
\frac{4\pi \rho_0
R^3}{M}=\frac{\alpha^2}{\eta}=\frac{\mu\chi}{\eta}.
\end{equation}
According to Eqs. (\ref{pchs1}), (\ref{be5}) and (\ref{be7}), the mass of the
solitonic core
is 
\begin{equation}
\label{be8}
\frac{M_c}{M}=\frac{\pi\chi}{\eta\sqrt{\mu}}.
\end{equation}
Finally, the density profile is given by Eq. (\ref{ch23sw}). It can be written
as
\begin{equation}
\frac{4\pi R^3}{M}\rho=\frac{\alpha^2}{\eta}e^{-\psi(\alpha r/R)}.
\end{equation}

\subsection{Series of equilibria}
\label{sec_soe}

\begin{figure}
\begin{center}
\includegraphics[clip,scale=0.3]{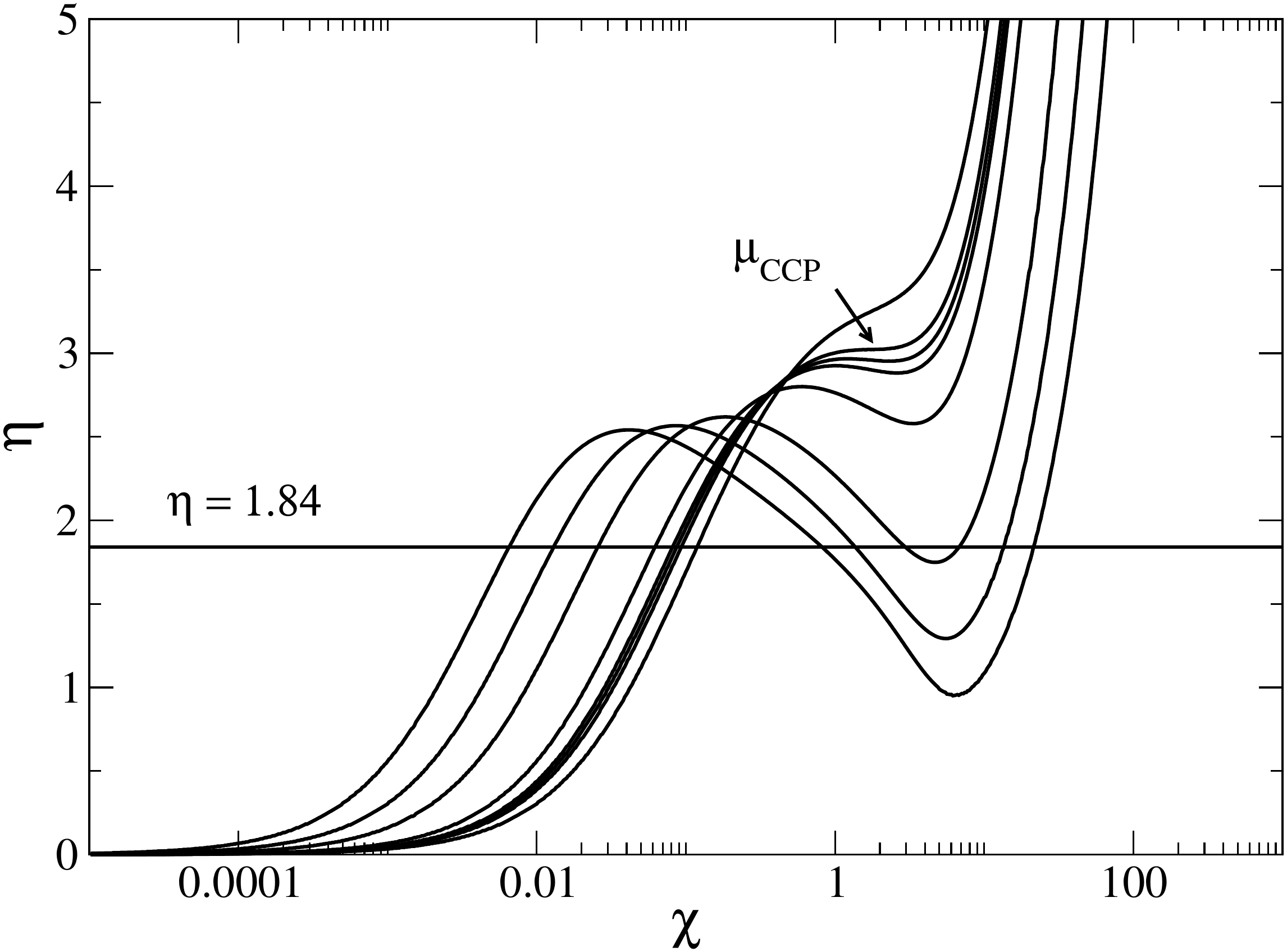}
\caption{Series of equilibria $\eta(\chi)$ for different values of $\mu$ (we
have taken $\mu=100,130,140,150,200,500,1000,2000$ - top to bottom - for
illustration). We find
the
existence of a canonical critical point $\mu_{\rm CCP}\simeq 130$ above which
bifurcations and phase transitions appear. They are associated with a
multiplicity of solutions
for the same value of the inverse temperature $\eta$.}
\label{chietaMUmulti}
\end{center}
\end{figure}

Let us prescribe a value of $\mu$. For a given value of $\chi$ we can solve the
differential equation (\ref{ch26}) up to the normalized box radius
$\alpha=\sqrt{\mu\chi}$ [see Eq. (\ref{be6})]. The corresponding normalized
inverse temperature $\eta$ is then given by Eq. (\ref{be3}). By varying the
value of
$\chi$
from $0$ to $+\infty$, we can obtain the series of equilibria $\eta(\chi)$ for a
fixed value of $\mu$. Examples of such curves are given in Fig.
\ref{chietaMUmulti}. For a given inverse temperature $\eta$ there may exist one
or several solutions with different values of $\chi$ (concentration). This
multiplicity of
solutions leads to bifurcations and phase transitions. Among all possible
solutions, we must
select the stable ones, i.e., those that correspond to (local) minima of free
energy.\footnote{In this section, we work in the canonical ensemble. The
microcanonical ensemble is considered in Sec. \ref{sec_ineq}.} Similarly to the
case of self-gravitating fermions \cite{ijmpb}, we find the existence of a
canonical critical point (see Fig.
\ref{chietaMUmulti}):
\begin{equation}
\label{be8b}
\mu_{\rm CCP}\simeq 130
\end{equation}
above which phase transitions
appear in the canonical ensemble. As
we shall see, this
canonical critical point  is connected to the bifurcation observed in Sec.
\ref{sec_another}. Below, we provide a preliminary study of phase transitions in
the
thermal
self-gravitating BEC gas, limiting ourselves to the canonical ensemble (fixed
$T$). A more
detailed study will be the subject of a future paper.

\subsubsection{$\mu<\mu_{\rm CCP}$}
\label{sec_aa}

\begin{figure}
\begin{center}
\includegraphics[clip,scale=0.3]{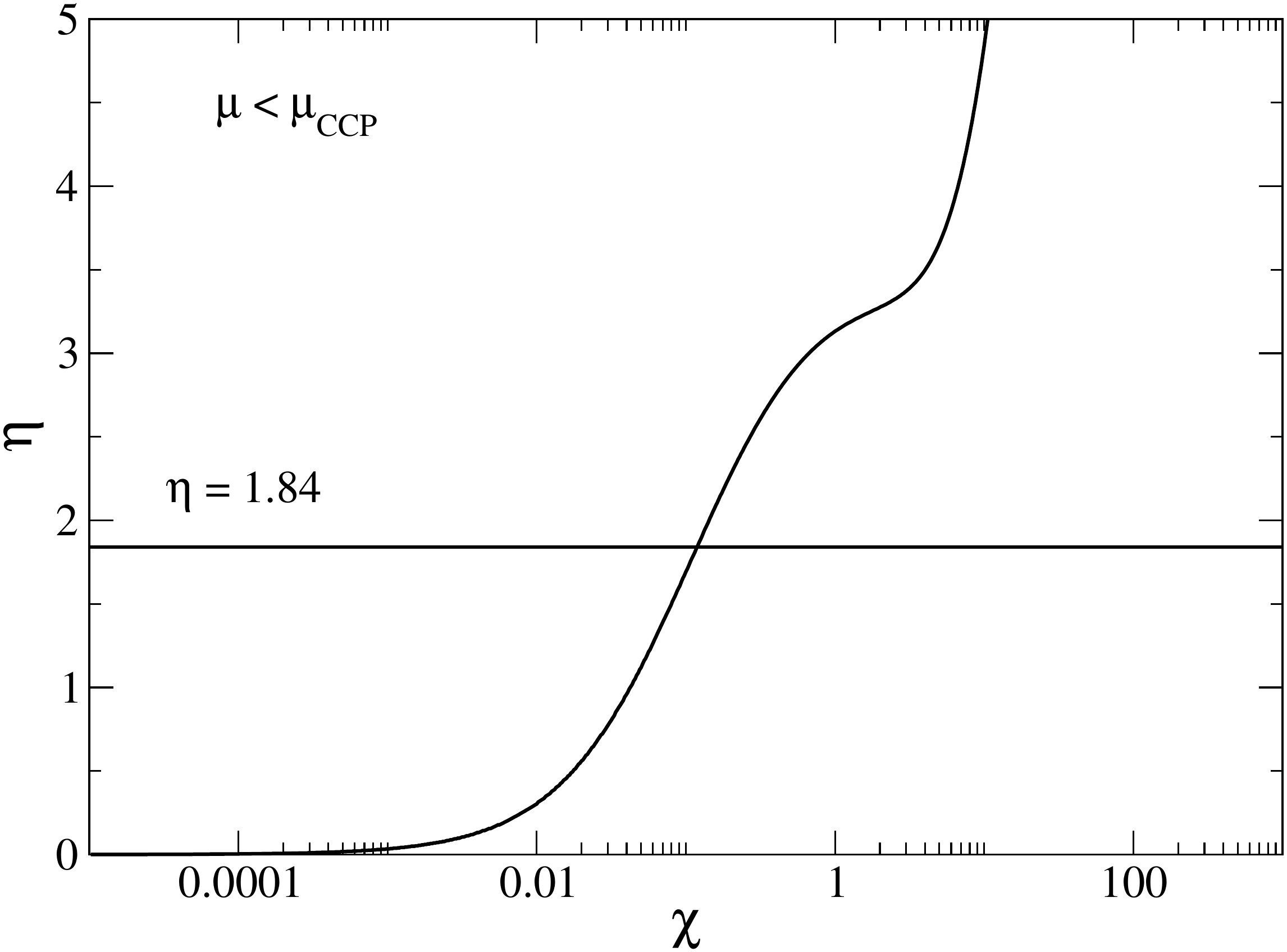}
\caption{Series of equilibria $\eta(\chi)$ for $\mu<\mu_{\rm CCP}\simeq 130$
(here $\mu=100$). There is only one
equilibrium state for each temperature. }
\label{chietaMU100}
\end{center}
\end{figure}

When $\mu<\mu_{\rm CCP}\simeq 130$ there is only one equilibrium state for each
temperature
(see Fig. \ref{chietaMU100}).  It corresponds to
a pure soliton surrounded by a tiny isothermal atmosphere. This structure is
thermodynamically stable. There
is no phase transition.

\subsubsection{$\mu>\mu_{\rm CCP}$}
\label{sec_bb}

\begin{figure}
\begin{center}
\includegraphics[clip,scale=0.3]{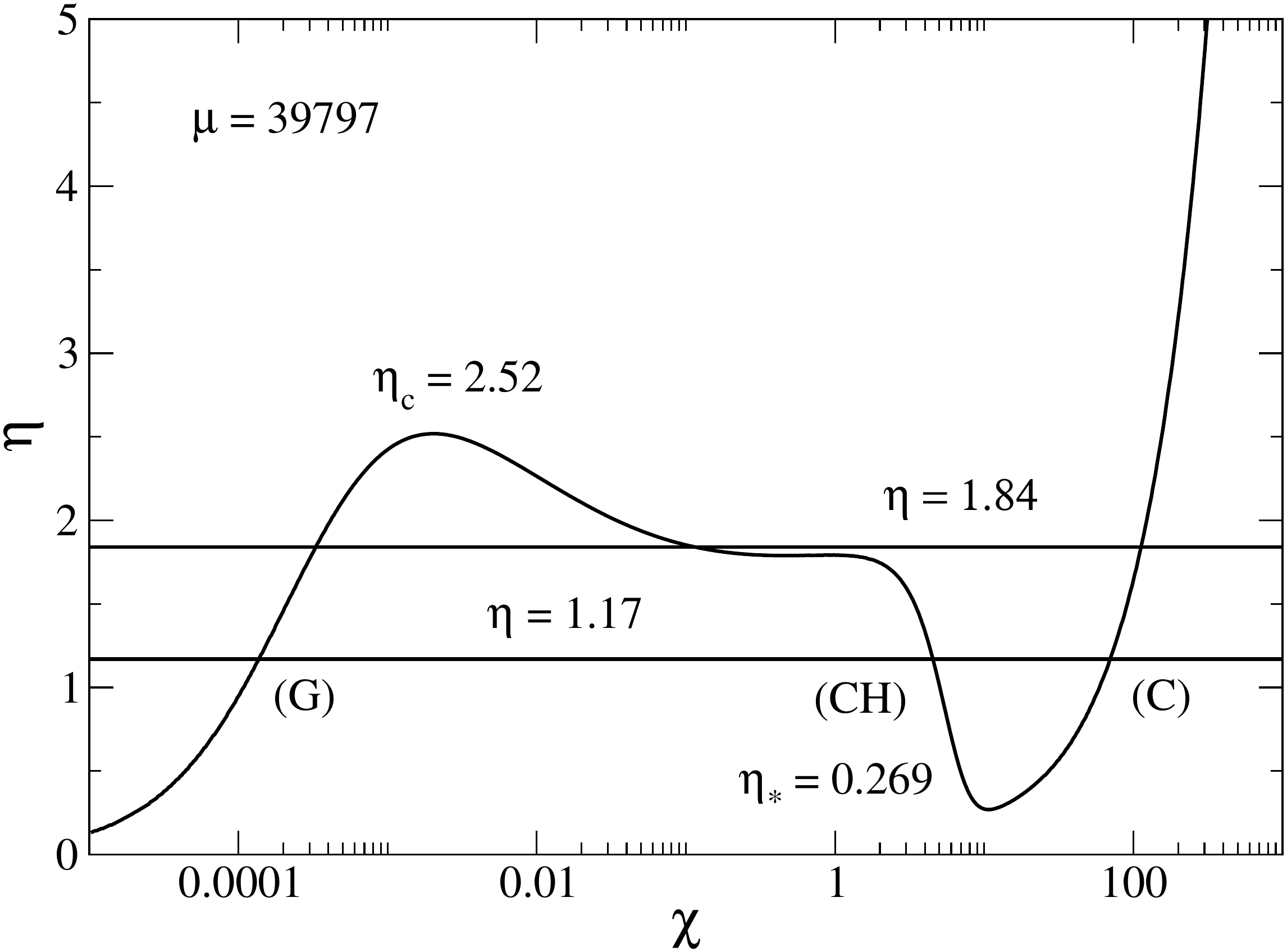}
\caption{Series of equilibria $\eta(\chi)$ for $\mu>\mu_{\rm CCP}\simeq 130$ not
too
large (here $\mu=39797$). For $\eta=1.17$ and $\eta=1.84$
there are three 
equilibrium states: (G) is the stable gaseous phase, (C) is the stable
condensed phase and
(CH) is the unstable core-halo phase.}
\label{chietaMU39797}
\end{center}
\end{figure}

\begin{figure}
\begin{center}
\includegraphics[clip,scale=0.3]{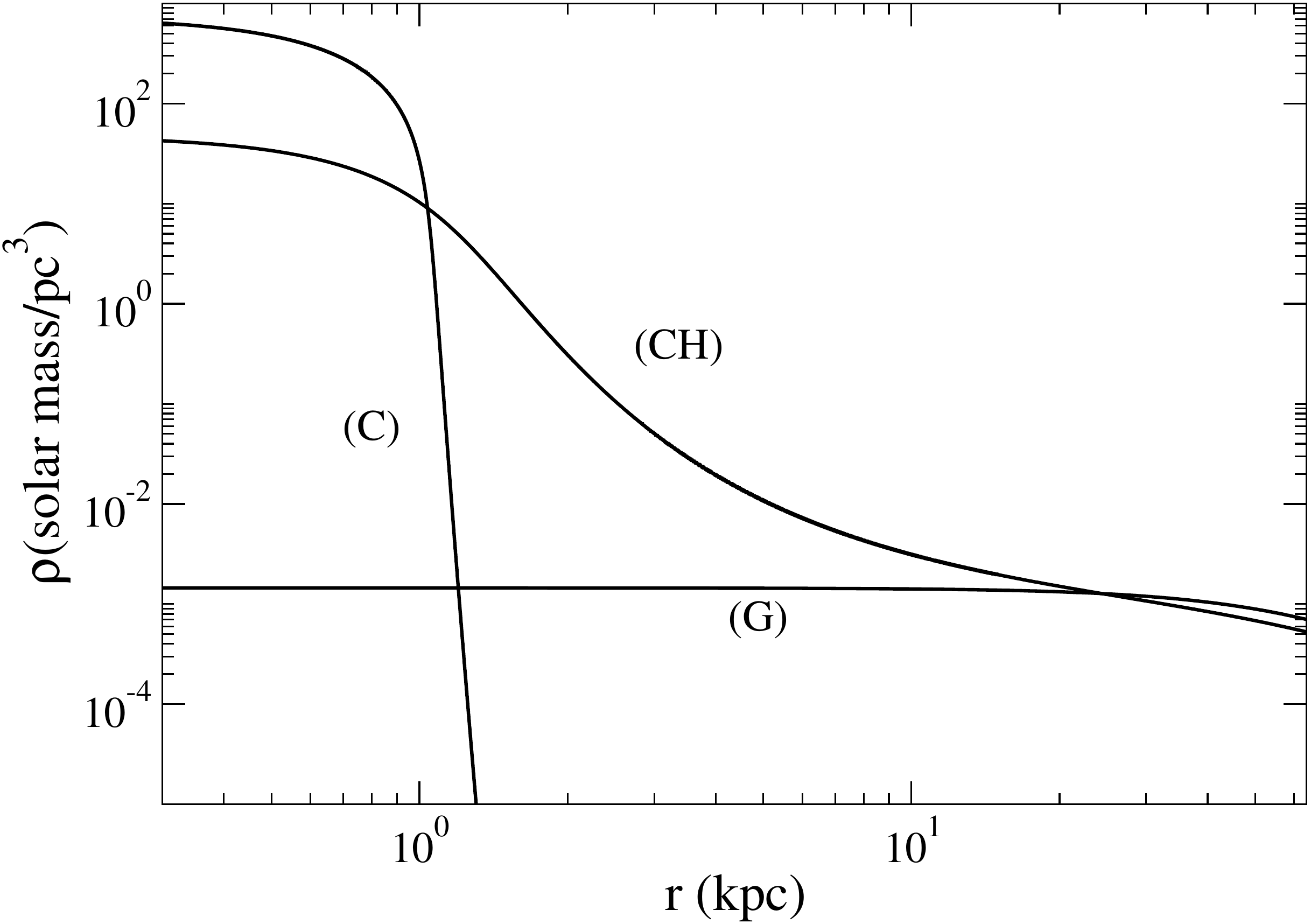}
\caption{Density profiles of the three solutions corresponding to the inverse
temperature $\eta=1.17$ for $\mu=39797$. They correspond to a DM halo of
mass $M=10^{12}\, M_{\odot}$ and size $R=63.5\, {\rm
kpc}$.
}
\label{prof}
\end{center}
\end{figure}

When $\mu>\mu_{\rm CCP}\simeq 130$ there is a canonical phase transition
associated with a
multiplicity of equilibrium states with the same temperature. Let us first
consider the case where $\mu$ is not too large. We take $\mu=39797$
(see Fig. \ref{chietaMU39797}). This value of $\mu$
corresponds to a DM halo of mass $M=10^{12}\, M_{\odot}$ and size $R=63.5\, {\rm
kpc}$ similar to the DM halo that surrounds the Milky Way (see Sec.
\ref{sec_ac} and Sec. \ref{sec_connap} below). For the
inverse temperature $\eta=1.17$ (the choice of this value is explained in Sec.
\ref{sec_connap} below)  there are
three solutions with different concentration parameters $\chi$. Their density
profiles are represented in Fig. \ref{prof}. For each of
these solutions, the core mass $M_c$ can be determined by Eq. (\ref{be8}).

The first solution (G) with $\chi^{\rm (G)}=1.37\times 10^{-4}$ and
$M_c^{\rm (G)}/M=1.84\times 10^{-6}\ll 1$ corresponds to an isothermal halo
having a negligible solitonic core. This is the gaseous
phase.

The
third solution (C)  with $\chi^{\rm (C)}=69.8$  and
$M_c^{\rm (C)}/M=0.939$  corresponds to a very
compact halo (pure solitonic core) having a negligible
atmosphere. This is the condensed
phase.

Finally, the second solution (CH)  with $\chi^{\rm (CH)}=4.54$  and
$M_c^{\rm (CH)}/M=0.0611$  has a
core-halo structure with a relatively massive solitonic core and a large
isothermal atmosphere. This is the core-halo
phase.

Using the Poincar\'e turning point criterion
\cite{poincare,katzpoincare},\footnote{See
\cite{ijmpb} for an application of the Poincar\'e turning point criterion in the
case of self-gravitating fermions.} we can deduce from Fig. \ref{chietaMU39797}
that
the solutions (G) and (C) are thermodynamically stable (minima of free
energy at fixed mass)  while
the solution (CH) is thermodynamically unstable (saddle point of free
energy at fixed mass) in the canonical ensemble.

We note that the multiplicity of the solutions depends on the temperature.
When $\eta<\eta*$ there is only one solution: the gaseous
phase (G). When $\eta>\eta_c$ there is only one solution: the
condensed phase (C). When $\eta_*<\eta<\eta_c$ there are three solutions: the
gaseous phase (G), the core-halo phase (CH) and the condensed phase (C).

There exists a transition
temperature $\eta_t$ (not represented) such that when $\eta<\eta_t$ the gaseous
state is fully
stable (global minimum of free energy) and the condensed phase is metastable
(local   minimum of free
energy). When  $\eta>\eta_t$ the situation is reversed (see \cite{ijmpb} for a
detailed discussion of phase transitions in the case of self-gravitating
fermions). However, for
systems with long-range interactions such as self-gravitating systems, the
metastable states have a very long lifetime scaling as $e^N$ where $N$ is
the number of particles. This lifetime is generally much larger than the age of
the
Universe so that metastable states can be as much, or even more, relevant than
fully stable
states \cite{lifetime}. 
Therefore, the first order phase transition at $\eta_t$ does not take place in
practice. As a result, we shall not distinguish between fully stable and
metastable states.  The selection of the gaseous or condensed phase depends on
the initial conditions and on a notion of basin of attraction \cite{ijmpb}.

\subsubsection{$\mu\gg \mu_{\rm CCP}$}
\label{sec_cc}

\begin{figure}
\begin{center}
\includegraphics[clip,scale=0.3]{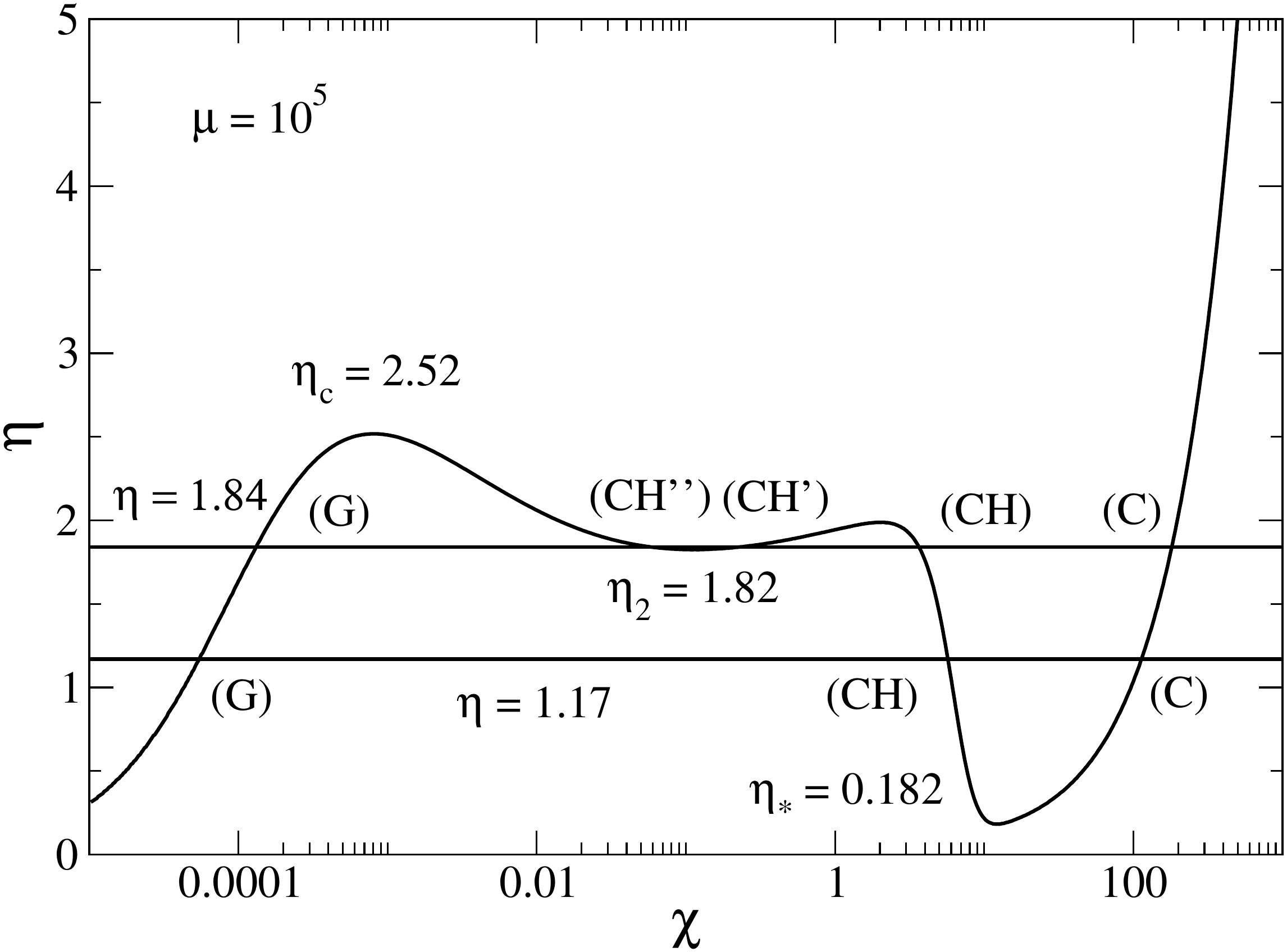}
\caption{Series of equilibria $\eta(\chi)$ for $\mu\gg \mu_{\rm CCP}$ (here
$\mu=10^5$). For $\eta=1.84$ there are five intersections:
(G) is the gaseous phase, (C) is the condensed phase and (CH),(CH') and (CH'')
are
unstable phases with a core-halo structure. (CH) is the less unstable of them.
For $\eta=1.17$ there are only three solutions as before. }
\label{chietaMU1e5}
\end{center}
\end{figure}

For very large values of $\mu$, new equilibrium states may appear in a certain
range
of temperatures. For example, when $\mu=10^5$, we have five solutions at
$\eta=1.84$
(see Fig. \ref{chietaMU1e5}). Using the 
Poincar\'e turning point criterion, we can show that (G) is stable, (CH'') has
one mode of instability, (CH')
has two modes of instability, (CH) has one mode of instability, and (C) is
stable.

As before, the multiplicity of the solutions depends on the value of $\eta$. 
When $\eta<\eta_{*}$, there is only one solution: the gaseous phase (G).  When
$\eta>\eta_{c}$ there is only one solution: the condensed phase (C). When
$\eta_*<\eta<\eta_2$ there are three solutions: the gaseous phase (G), the
core-halo phase (CH) and the condensed phase (C). When $\eta_2<\eta<\eta_c$
there
are three or more solutions.

\subsection{Caloric curves}
\label{sec_calc}

We can also visualize the multiplicity of the solutions by  plotting
the
caloric curves $\eta(\Lambda)$ giving the inverse temperature $\eta=\beta
GMm/R$ as a function of the opposite of the energy $\Lambda=-ER/GM^2$. The
caloric curves of a thermal self-gravitating BEC will be given in a forthcoming
paper but they are similar to those obtained in the case of
self-gravitating fermions \cite{ijmpb}.

When $\mu<\mu_{\rm CCP}$, the caloric curve is monotonic (see Fig. 14 of
\cite{ijmpb}) leading to the results of Sec. \ref{sec_aa}.

When $\mu>\mu_{\rm CCP}$ is not too large, the caloric curve has an $N$-shape
structure (see Fig.
31 of \cite{ijmpb}) leading to the results of Sec. \ref{sec_bb}.

When $\mu\gg \mu_{\rm CCP}$, the caloric curve has a more complicated structure,
corresponding to a thick spiral (see Fig.
22 of \cite{ijmpb}). We call it ``thick'' because it is made of
two branches that almost superimpose: a direct spiral and an inverse  spiral.
This leads 
to the results of Sec. \ref{sec_cc}. For $\eta\simeq 2$,
corresponding to the center of the spiral (see below), the number of solutions
increases (up to an infinity!) as $\mu$ increases. 

When $\mu\rightarrow +\infty$, we recover the classical spiral  (see Fig.
8 of \cite{ijmpb}).\footnote{The quantum caloric curves with a  finite value of
$\mu$ described previously correspond to the unwiding of the
classical spiral (see Fig. 14 of
\cite{ijmpb}).} In that case, the direct and inverse branches exactly
superimpose. We get $\eta_*\rightarrow 0$, $\eta_c\rightarrow 2.52$
(maximum of the classical isothermal spiral), $\eta_2\rightarrow 1.84$ (minimum
of the classical isothermal spiral), and $\eta_s\rightarrow 2$ (center of the
spiral).

\subsection{Application to real DM halos}
\label{sec_app}

In this section, we apply the box model to real DM halos. We
shall recover from the box model the bifurcation obtained in Sec.
\ref{sec_aar} and the
relation between the core mass $M_c$ and the halo mass $M_h$ obtained in Sec.
\ref{sec_ap}.

\subsubsection{Connection to astrophysical parameters}
\label{sec_connap}

In order to apply the box model to real DM halos, we identify 
the box radius $R$ to the halo radius $r_h$ and the mass  $M$ to the halo
mass $M_h$. For a given DM halo, we can compute the parameter $\mu$
given by Eq. (\ref{be5}).\footnote{We note that for the model of BECDM
considered in this paper $\mu$ depends  only on $R$.} We have previously seen
that large DM
halos appear
``from the outside'' as being essentially isothermal (the solitonic core - if
there is any - does not affect their external structure). As a result, the
halo mass is related to the halo radius by the relation 
\begin{eqnarray}
\label{app1}
M_h=1.76\, \Sigma_0 r_h^2.
\end{eqnarray} 
As in Sec. \ref{sec_ac}, we consider a halo of mass
$M=10^{12}\, M_\odot$ and radius $R=63.5\, {\rm kpc}$ similar to the DM halo
that surrounds the Milky Way. For such a halo we get $\mu=39797$. The
corresponding series of equilibria $\eta(\chi)$ is represented in Fig.
\ref{chietaMU39797}.

Using Eqs. (\ref{i11}) and (\ref{i13}), we find that  the normalized
inverse temperature of the halo is\footnote{Remarkably, this
value turns out to be very close to the value $\eta_2=1.84$ corresponding to the
minimum of the classical isothermal spiral (see Sec. \ref{sec_calc}).} 
\begin{eqnarray}
\label{app2}
\frac{\beta GM_hm}{r_h}=\xi_h\psi'_h=1.84.
\end{eqnarray} 
Therefore, if we want to make the connection between the box model and real DM
halos, we should consider a value of  $\eta$
equal to $1.84$. The intersection between the series of equilibria $\eta(\chi)$
and the line level $\eta=1.84$ determines the possible equilibrium states.
It is reassuring to note that $\eta=1.84$ is smaller than $\eta_c\simeq 2.52$
(corresponding to the maximum  of the classical isothermal spiral) implying
that there always exists a gaseous equilibrium state (G).
Actually, we should not give too much importance on the precise value of
$\eta$. It is sufficient to consider that $\eta$
is of the order of $1-2$. This is essentially a consequence of the virial
theorem. Considering Fig. \ref{chietaMU39797}, we see that the gaseous
solution (G) and the condensed solution (C) do not strongly depend on $\eta$ in
the interval $\eta\sim 1-2$ while the core-halo solution (CH) is more sensitive
to
its precise value. This is even more true for the case considered in Fig.
\ref{chietaMU1e5}, with a larger value of $\mu$, where several core-halo
solutions (CH), (CH'), (CH'') may exist at the same temperature. In
order to always clearly identify the less unstable
core-halo solution (corresponding to the last but one intersection), we find it
convenient to select a value of $\eta$ smaller than $\eta_2$. In this manner, we
are
guaranted to have at most three solutions: a gaseous solution (G), a core-halo
solution (CH), and a condensed solution (C). To be specific, and guided by the
results obtained in Sec. \ref{sec_ac}, we choose $\eta=1.17$. In that case, the
core-halo
solution for $\mu=39797$  has $\chi^{\rm (CH)}=4.54$ (see Fig.
\ref{chietaMU39797})
as in Sec. \ref{sec_ac}.

\subsubsection{The critical mass $(M_{h})_{\rm CCP}$ for the onset\\
 of the
bifurcation}

We have seen in Sec. \ref{sec_soe} that phase transitions (associated with the
multiplicity of the solutions for a given temperature) appear for $\mu>\mu_{\rm
CCP}\simeq 130$.  If we identify $R$ with $r_h$ in Eq. (\ref{be5}), we get
\begin{equation}
\label{app3}
\frac{r_h}{R_c}=\frac{\sqrt{\mu}}{\pi}.
\end{equation}
Combining Eqs. (\ref{app1}) and (\ref{app3}), we obtain
\begin{equation}
\label{app4}
M_h=1.76\, \Sigma_0 \frac{\mu}{\pi^2}R_c^2.
\end{equation}
Therefore, the halo mass corresponding to the canonical critical point is
\begin{equation}
\label{app5}
\frac{(M_h)_{\rm CCP}}{\Sigma_0 R_c^2} =1.76\,  \frac{\mu_{\rm
CCP}}{\pi^2}\simeq 23.2.
\end{equation}
Using Eq. (\ref{chp2}), we obtain 
\begin{equation}
\label{app6}
(M_h)_{\rm CCP}=3.27\times 10^9\, M_{\odot}.
\end{equation}
When $M_h<(M_h)_{\rm CCP}$ there is only one
equilibrium state corresponding to a
solitonic core surrounded by a tiny isothermal atmosphere. When 
$M_h>(M_h)_{\rm
CCP}$ there are generically three equilibrium states with the same temperature:
a
purely isothermal halo (gaseous phase G), an almost purely solitonic halo
(condensed
phase C), and an isothermal halo containing a small but relatively massive
solitonic core (core-halo phase CH). Eliminating the almost purely
solitonic solution which is not consistent with the observations of large
DM halos, it remains the gaseous solution (G) and the core-halo solution (CH).
Therefore, the box model predicts a bifurcation
similar to the one predicted in Sec. \ref{sec_aar} from different
arguments. We note that
the critical mass  $(M_h)_{\rm CCP}=3.27\times 10^9\, M_{\odot}$ obtained in
the present section is relatively close to the critical mass 
$(M_h)_{\rm b}\sim 10^9$ obtained in Sec. \ref{sec_aar} indicating that the two
approaches are consistent. The critical mass $(M_h)_{\rm CCP}=3.27\times 10^9\,
M_{\odot}$ is also extremely close to  $(M_h)_{*}=3.30\times 10^9\, M_{\odot}$.
Therefore, it is convenient to identify $(M_h)_*$, $(M_h)_b$ and  $(M_h)_{\rm
CCP}$. As a
result, the halo parameters at the bifurcation are given by Eqs.
(\ref{pdp4})-(\ref{pdp8}).

{\it Remark:} Using Eq. (\ref{app4}), we note that the halo mass
associated with $\mu_{\rm min}=\pi^2=9.87$ is
$M_h=1.76\Sigma_0R_c^2=2.48\times 10^8\, M_{\odot}$
which can be identified with the minimum halo mass $(M_h)_{\rm
min}=1.32\, \Sigma_0 R_c^2=1.86\times 10^8\, M_{\odot}$.

\subsubsection{The $M_c-M_h$ relation}

\begin{figure}
\begin{center}
\includegraphics[clip,scale=0.3]{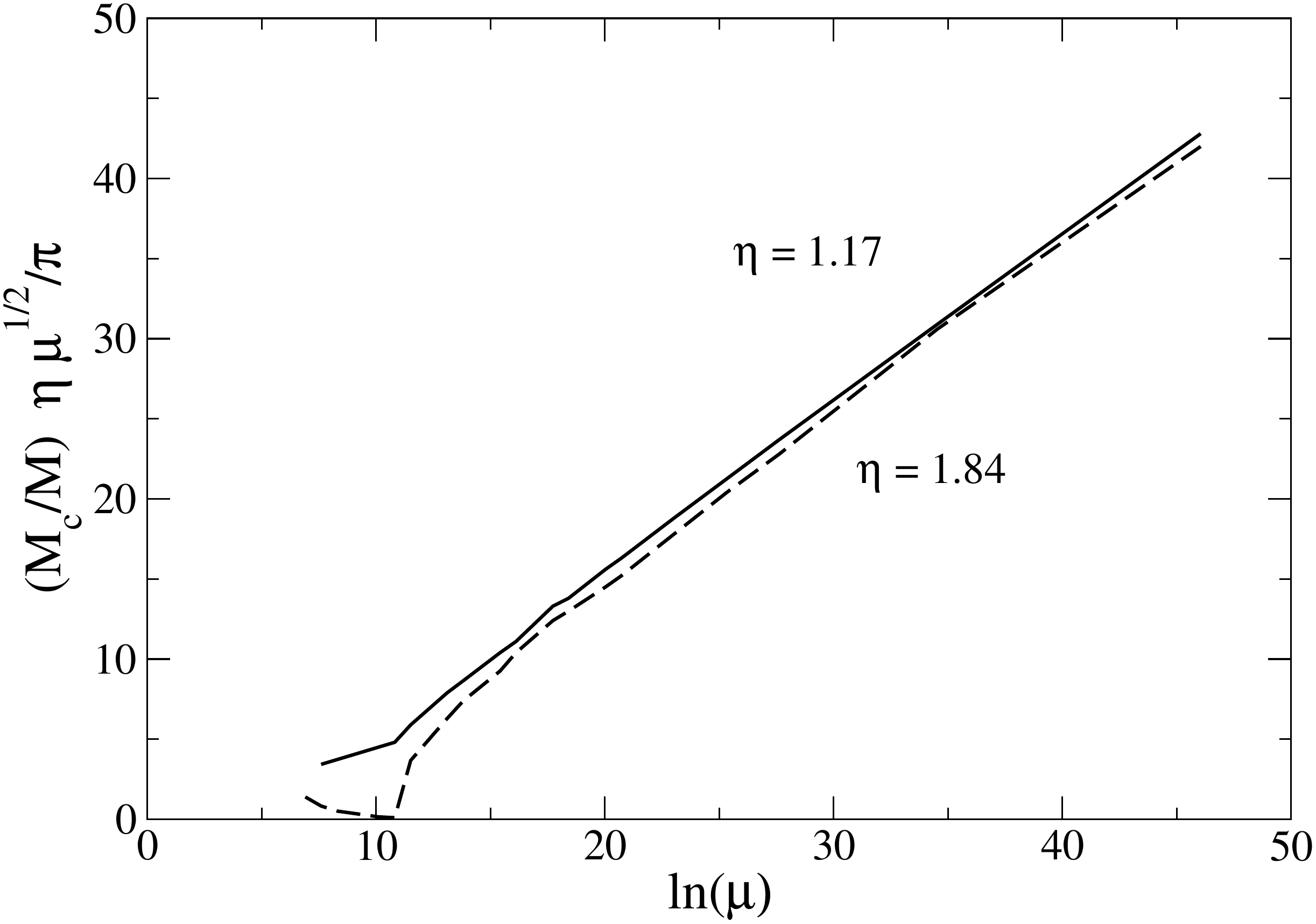}
\caption{Relation between the solitonic core mass normalized by the total mass,
$M_c/M$, and the size of the system measured by the dimensionless parameter
$\mu$
for two different values of the dimensionless temperature $\eta$ in the
framework of the box model. This curve shows that the concentration
parameter  $\chi^{\rm (CH)}$  of the core-halo phase behaves as $\ln\mu$. This
leads to the scaling given by Eq. (\ref{app7}).
}
\label{mumctotNEWlogcorrection}
\end{center}
\end{figure}

The $M_c-M_h$ relation can be obtained from the box model as follows.
For $\eta=1.17$ and for a given value of $\mu$, we can determine the
concentration parameter $\chi^{\rm (CH)}$ of the core-halo
solution by the intersections illustrated in Fig.
\ref{chietaMU39797}. Then, we can  obtain the solitonic core
mass normalized by the total mass, $M_c/M$, from Eq. (\ref{be8}). We can repeat
this procedure for different values of $\mu$ and obtain the curve $M_c/M$ as
a function of $\mu$. In Fig. \ref{mumctotNEWlogcorrection}, we plot
$(M_c/M)\eta\sqrt{\mu}/\pi=\chi^{\rm (CH)}(\mu)$ as a function of
$\ln\mu$. In the dominant approximation, we numerically find that\footnote{This
scaling law can be understood analytically as follows. We have found in Sec.
\ref{sec_another} that $\chi\sim
2\ln(r_h/R_c)$ (see footnote 29). From Eq. (\ref{be5}), we have
$\mu=\pi^2(r_h/R_c)^2$. Combining these two relations, we get $\chi\sim
\ln\mu$. Substituting this result into Eq. (\ref{be8}) yields Eq.
(\ref{app7}). }
\begin{equation}
\label{app7}
\frac{M_c}{M}\sim \frac{\pi}{\eta}\frac{\ln\mu}{\mu^{1/2}}.
\end{equation}
We have repeated the same procedure
with  $\eta=1.84$ being careful to define  $\chi^{\rm (CH)}$ as being the last
but
one intersection in Fig. \ref{chietaMU1e5}. Although the results
differ for small
values of $\mu$ (showing the sensibility of the core-halo solution with $\eta$
mentioned previously), we find essentially the same scaling for
large values  of $\mu$. From this scaling
law, identifying $M$ with the
halo mass and using Eqs. (\ref{be5}) and (\ref{app1}), we find that
\begin{equation}
\label{app8}
\frac{M_c}{\Sigma_0R_c^2}=
\frac{1.33}{\eta}\ln\left(\frac{M_h}{\Sigma_0R_c^2}\right )\left
(\frac{M_h}{\Sigma_0R_c^2}\right )^{1/2},
\end{equation}
with $\eta\sim 1-2$. This returns the scaling $M_c\propto M_h^{1/2}$ obtained
in Sec. \ref{sec_ap}
from a
different method.  The prefactor is also consistent with the one
obtained from Eq. (\ref{ap1}), displaying a logarithmic correction.  Since the
core mass increases with the halo mass, the solitonic core is persistant in the
core-halo phase of large DM halos.

Performing the same study with the concentration parameter  $\chi^{\rm (G)}$ of
the gaseous phase, we numerically find that
\begin{equation}
\label{app7g}
\frac{M_c}{M}\sim \frac{14.7}{\mu^{3/2}},
\end{equation}
leading to
\begin{equation}
\label{app8g}
\frac{M_c}{\Sigma_0R_c^2}\sim 1.11 \sqrt{\frac{\Sigma_0R_c^2}{M_h}}
\end{equation}
in qualitative agreement with Eq. (\ref{pcd2zb}) of Model I. Since the
core mass decreases with the halo mass, there is no solitonic core
in the gaseous phase of large DM halos.

Finally, performing the same study with the concentration parameter  $\chi^{\rm
(C)}$ of
the condensed phase, we numerically find that
\begin{equation}
\label{app7c}
\frac{M_c}{M}\simeq 1
\end{equation}
meaning that all the mass is in the solitonic core. As previously mentioned
this solution is not in agreement with the observed structure of large DM halos.

\subsubsection{Simple analytical model}
\label{sec_sim}

In Appendix \ref{sec_ana} we develop a simple analytical model of
self-gravitating BECs with an isothermal atmosphere in a box. In that model, the
mass of the
solitonic core $M_c$ is obtained by extremizing the free
energy $F(M_c)$ for a given value of $T$, $M$ and $R$. We find that, above a
canonical critical point $\mu_{\rm CCP}$, the free
energy $F(M_c)$ has generically three extrema in agreement with the results of
Sec. \ref{sec_soe}:

(i) A minimum at $M_c=0$ corresponding to an isothermal
halo without a solitonic core (gaseous phase G). This solution is
thermodynamically stable.

(ii) A minimum at some $M_c\simeq M$ corresponding to a soliton without
halo (condensed phase C). This solution is thermodynamically  stable.

(iii) A maximum at some $M_c$ satisfying
\begin{equation}
\label{app9}
\frac{M_c}{M}\propto \frac{1}{\mu^{1/2}},
\end{equation}
corresponding to a solitonic core surrounded by a large isothermal halo 
(core-halo
phase CH).  This solution is thermodynamically unstable in the canonical
ensemble. We note that the
scaling
law from Eq. (\ref{app9}) is consistent with the numerical result from Eq.
(\ref{app7}) up to logarithmic corrections offering therefore an alternative
derivation of this result. The
analytical study of Appendix
\ref{sec_ana} also confirms that the core-halo solution is thermodynamically
unstable in the canonical ensemble in
agreement with the result obtained in Sec. \ref{sec_soe} from the Poincar\'e
turning point criterion.

\subsubsection{Ensembles inequivalence}
\label{sec_ineq}

In the previous sections, we have worked in the canonical ensemble. This is the
statistical ensemble associated with the generalized (coarse-grained) GPP
equations (\ref{intro3}) and (\ref{intro4}) where the
temperature $T$ is fixed. However, the microcanonical ensemble, where the energy
$E$ is fixed, may also be relevant. Actually, it may even be more relevant
than the canonical ensemble since the total energy should be conserved as in
the original (fine-grained) GP equations (\ref{intro1}) and (\ref{intro2}). As
explained in Appendix I of \cite{bdo}, the GPP
equations  (\ref{intro3}) and (\ref{intro4}) could be modified in
order to conserve the energy. In that case, the 
statistical ensemble
associated to these equations would be the microcanonical one.

The equilibrium states in the microcanonical ensemble are the same as in the
canonical ensemble. However, their stability may be different in the canonical
and in the microcanonical ensembles. An equilibrium state that is canonically
stable is always microcanonically stable, but the converse is wrong. This
corresponds to the concept of ensembles inequivalence for systems with
long-range interaction \cite{paddy,ijmpb,campa}. In particular, the core-halo
states (CH)
that we have found previously are
always unstable in the canonical ensemble but they may be stable in the
microcanonical ensemble. This is the case in particular if there is no
turning point of energy in the caloric curve, or if the core-halo state stands
before
the first turning point of energy. In that case, the core-halo state  has a
negative specific
heat $C<0$. This is forbidden in the canonical ensemble but this is allowed in
the microcanonical ensemble \cite{paddy,ijmpb,campa}.

There exists a  microcanonical critical point 
\begin{equation}
\mu_{\rm MCP}\sim 10^5
\end{equation}
above which the caloric curve presents at least one turning point of
energy.\footnote{For bosonic DM, we have not
computed the microcanonical critical point $\mu_{\rm MCP}$ precisely but its
value should be close to the
situation where the caloric curve presents three turning points of temperature.
Therefore, according to Figs. \ref{chietaMU39797} and \ref{chietaMU1e5} we have 
$39797<\mu_{\rm MCP}<10^5$. To be specific, we shall take $\mu_{\rm MCP}\sim
10^5$.} Using Eq. (\ref{app4}) it corresponds to a halo mass
\begin{equation}
\label{crit1}
\frac{(M_h)_{\rm MCP}}{\Sigma_0 R_c^2} =1.76\,  \frac{\mu_{\rm
MCP}}{\pi^2}\sim 2\times 10^4.
\end{equation}
Using Eq. (\ref{chp2}), we obtain 
\begin{equation}
\label{crit2}
(M_h)_{\rm MCP}\sim 2\times 10^{12}\, M_{\odot}.
\end{equation}
When $\mu<\mu_{\rm MCP}$, the caloric curve $\eta(\Lambda)$ is
univalued. All the equilibrium states are stable in the
microcanonical ensemble, even the core-halo states that are unstable in the
canonical ensemble. This is the case in particular for the value $\mu=39797$
corresponding to a DM halo of mass $M=10^{12}\, M_{\odot}$ (see
Sec. \ref{sec_connap}). Therefore, the solitonic core of mass $M_c=6.39\times
10^{10}\, M_{\odot}$ (bulge) that this halo may harbor (see
Sec. \ref{sec_ac}) is part of a core-halo structure that is stable in
the microcanonical ensemble while being unstable in the canonical ensemble.
When $\mu>\mu_{\rm MCP}$, the caloric curve has
a $Z$-shape structure (see Fig. 21 of \cite{ijmpb}). Only
the equilibrum states before the first turning point of energy and after the
last turning point of energy are stable in the microcanonical ensemble. Using
these arguments, we can show that the core-halo configurations
with a large value of $\chi$ (high central density) constructed
in
Sec. 
\ref{sec_cheat}  are unstable
both in the canonical and in the microcanonical ensembles\footnote{For a halo of
mass $M_h=10^{11}\, M_{\odot}$ and radius $r_h=20.1\,
{\rm kpc}$, taking $R_c=6.98\times 10^{-2}\, {\rm pc}$ as in Sec.
\ref{sec_cheat} we get
$\mu=8.18\times 10^{11}\gg \mu_{\rm MCP}$. Taking $R_c=6\times 10^{-4}\, {\rm
pc}$ as in footnote 36, we get  $\mu=1.11\times 10^{16}\gg
\mu_{\rm MCP}$. Since $\mu$ is large in the two cases, the caloric curve  has a
complex structure
(see Fig. 44 of \cite{clm2}) and the core-halo configurations
with a
large $\chi$  considered in
Sec. 
\ref{sec_cheat} are located between the first and the last turning points of
energy. As a result, they are thermodynamically unstable.} (see the
similar discussion for fermions in Secs. VI-VIII of \cite{clm2}). Therefore, not
only the
solitonic core cannot mimic a black hole (see the arguments given in Sec.
\ref{sec_cheat}) but the core-halo configuration to which it belongs is unstable
in all thermodynamical
ensembles.

\section{Astrophysical applications}
\label{sec_astapp}

We now discuss several potential scenarios that are suggested by the
previous results.

\subsection{Small DM halos with $(M_h)_{\rm
min}<M_h<(M_h)_{\rm CCP}$: Solitonic core $+$ tenuous
isothermal atmosphere (quantum solution)}
\label{sec_small}

\begin{figure}
\begin{center}
\includegraphics[clip,scale=0.3]{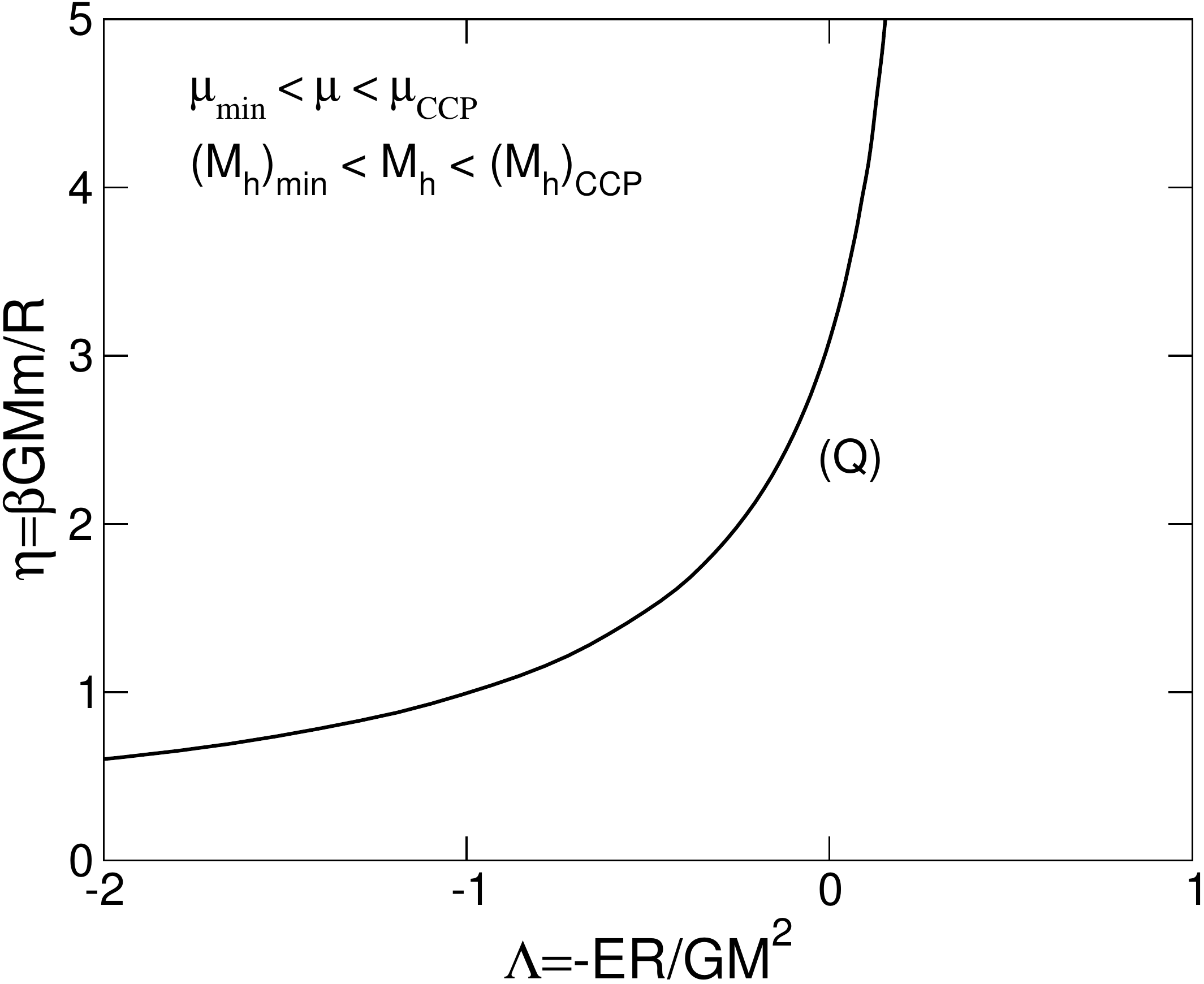}
\caption{For $(M_h)_{\rm min}<M_h<(M_h)_{\rm CCP}$, the
caloric curve is monotonic. The quantum solutions (Q) are stable.}
\label{multimu2}
\end{center}
\end{figure}

We consider a small DM halo with $(M_h)_{\rm
min}=1.86\times 10^8\, M_{\odot}<M_h<(M_h)_{\rm CCP}=3.27\times 10^9\,
M_{\odot}$.  Since $\mu_{\rm min}<\mu<\mu_{\rm CCP}$ the caloric curve
is monotonic (see Fig. \ref{multimu2}).\footnote{In this section, the series of
equilibria are indicative. They have been taken from our previous work
\cite{ijmpb} related to self-gravitating fermions but similar curves would be
obtained for self-gravitating bosons in the framework of the present model
\cite{forthcoming}.} The
DM halo is
made of a solitonic core surrounded by a tenuous isothermal atmosphere. This
quantum solution (Q) is thermodynamically stable.

\subsection{Large DM halos with $(M_h)_{\rm CCP}<M_h<(M_h)_{\rm MCP}$}
\label{sec_large}

\begin{figure}
\begin{center}
\includegraphics[clip,scale=0.3]{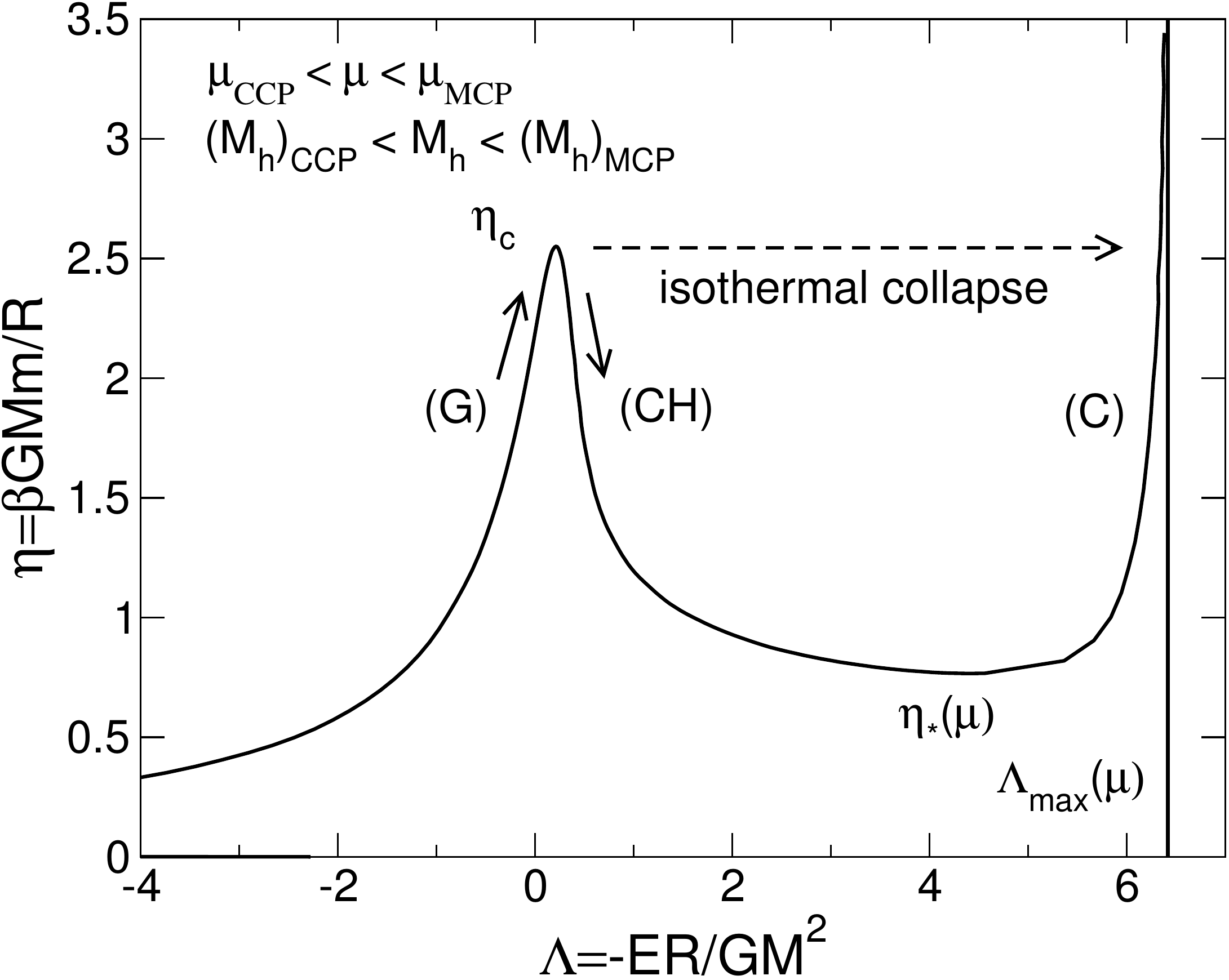}
\caption{For $(M_h)_{\rm CCP}<M_h<(M_h)_{\rm MCP}$, the caloric curve has a
$N$-shape structure. In the canonical ensemble, the gaseous phase (G) and the
condensed phase (C) are stable while the core-halo phase (CH) is unstable. The
system can evolve collisionally in the gaseous phase (G) up to
the turning point of temperature $T_c$ and collapse towards the condensed phase
(C). This corresponds to an isothermal collapse (dotted arrow).  As explained in
the text, this canonical scenario does not seem to be realistic for large DM
halos. In the
microcanonical ensemble, all the equilibrium states are stable. The system can
evolve collisionally (solid arrows) from the gaseous solution (G) to the
core-halo solution (CH).}
\label{fel}
\end{center}
\end{figure}

We consider a large DM halo with $(M_h)_{\rm CCP}=3.27\times
10^9\, M_{\odot}<M_h<(M_h)_{\rm MCP}\sim 2\times
10^{12}\, M_{\odot}$.  Specifically, we consider a
DM halo of mass $M=10^{12}\,
M_{\odot}$ and size $R=63.5\, {\rm kpc}$ similar to the one that surrounds our
Galaxy. Since $\mu_{\rm CCP}<\mu<\mu_{\rm MCP}$ the caloric curve has a
$N$-shape structure (see Fig. \ref{fel}).

\subsubsection{Isothermal halo $+$ solitonic bulge possibly
triggering the formation of a supermassive black hole (core-halo solution)}
\label{sec_chann}

We first assume that the DM halo is in the core-halo (CH) phase  (see Fig.
\ref{fel}).\footnote{The system may reach this core-halo phase
directly from a process of collisionless violent relaxation or, more slowly,
from a
``collisional'' evolution as discussed in Sec. \ref{sec_gch} and in the caption
of
Fig. \ref{fel}.} It is
therefore made
of a solitonic core of mass $M_c=6.39\times 10^{10}\, M_{\odot}$  and size
$R=1\,
{\rm kpc}$  surrounded by an isothermal halo (see Sec. \ref{sec_ac}). 
We have shown in Secs. \ref{sec_soe} and \ref{sec_sim} that this core-halo
solution is 
thermodynamically unstable in the canonical ensemble in the sense that it is a
saddle point of free
energy,
not a (local) minimum of free energy.  However, the timescale for the
development of the instability may be
very large (possibly larger than the age of the Universe) so that this core-halo
structure may be long-lived.\footnote{A saddle
point of free energy can persist for a long time as long as the fluctuations
(or the environment) have not generated
the dangerous perturbations that destabilize it (see an explicit illustration
of this result in Ref.
\cite{naso} in the context of two-dimensional
turbulence). Moreover, as discussed in Sec. \ref{sec_ineq}, this
core-halo solution (with a negative specific heat) is stable in the
microcanonical ensemble. Therefore, it is fully stable for 
perturbations that conserve the energy.}
 Actually, these
core-halo structures are
observed in the
numerical simulations of Schive {\it et al.} \cite{ch2,ch3} so they appear to be
robust and physical. In the case of real galaxies, the solitonic core may have
existed in
the past as a
temporary state,  or may still possibly exist. We have mentioned in Secs.
\ref{sec_ac} and \ref{sec_cheat}  that the solitonic core cannot mimic a
supermassive black hole. However, it can represent a bulge  providing a
favorable environment for triggering the formation of a supermassive black
hole.\footnote{In that context, the solitonic core would be a sort of
``critical droplet'' (in the canonical ensemble) allowing for the transition to
a more compact structure, e.g., a supermassive black hole.} 
The final outcome of this scenario would be an isothermal halo containing
either a solitonic bulge or a
supermassive black hole that would be the remnant of the original solitonic
bulge.

\subsubsection{Isothermal halo without solitonic core (gaseous solution)}
\label{sec_g}

We now assume that the DM halo  is in
the gaseous (G) phase (see Fig.
\ref{fel}). In that case, it has the form of a purely isothermal halo
without solitonic core. This gaseous solution is thermodynamically stable. In
this sense, this is the most probable state in the canonical ensemble. A first
possibility is that the DM
halo remains in this phase. This is not inconsistent with the observations since
we have shown in Sec. \ref{sec_comp} that, in many cases, an isothermal halo is
almost indistinguishable from the observational Burkert profile, especially if
we account for
tidal effects (see Appendix \ref{sec_diff} and
\cite{clm1,clm2}). However, this scenario
does not account for the presence of a compact object, such
as a
supermassive black hole, at the centers of the galaxies. Of course, we
can always add a primordial supermassive black hole at the
centre of our isothermal halo but this is almost assuming
the result. In order  to explain the presence of the black hole we consider
another possibility. Following our previous work \cite{clm1,clm2}, we assume
that the DM halo evolves dynamically due to collisions between DM particles.
These collisions are not
two-body gravitational encounters because the relaxation time would be too long,
but they can have another origin.\footnote{In the context of
BECDM this ``collisional'' evolution may be due to the formation of
``granules'' or ``quasiparticles'' (arising from the wave nature of the system
\cite{ch2,ch3}) which can lead to a ``collisonal'' relaxation as
suggested by Hui {\it et al.} \cite{hui}. This scenario has been developed very
recently by Bar-Or {\it et al.} \cite{bft} who showed that the DM halos behave
similarly to classical $N$-body systems like globular clusters. We note that
these results give further support
to our study in which we model the halo as an
isothermal gas following \cite{bdo,nottalechaos}. We argued that this isothermal
halo
arises from a process of
violent collisionless relaxation but it can also be due to (or maintained by)
``collisions'' of quasiparticles. The process of violent collisionless
relaxation
(or gravitational cooling) may explain the rapid formation of a core-halo (CH)
structure with a solitonic core (bulge) and an isothermal halo, or simply the
formation of an isothermal halo (G). The process
of collisional relaxation may justify why the halo evolves slowly along a series
of equilibria due to collisions among pseudoparticles and evaporation (tidal
effects), possibly leading to the formation of a solitonic core (see Sec.
\ref{sec_gch}). Finally, we note that the self-interaction of the bosons
($a_s>0$) may also be responsible for a collisional evolution of the system
and justify a (quasi) isothermal distribution (see footnote 53).} Because of
this
dynamical evolution, the
central density of the
halo increases until it
reaches a critical value at which the halo becomes
thermodynamically unstable and undergoes a gravitational collapse. Since the
statistical ensembles are inequivalent for
self-gravitating systems we have to consider two possibilities (canonical and
microcanonical) as detailed in the following sections.

\subsubsection{Canonical evolution: Isothermal collapse from the gaseous phase
to the condensed phase}
\label{sec_gc}

In the canonical ensemble, the temperature slowly decreases and the series of
equilibria becomes unstable at the
turning point of temperature $T_c$ (see Fig.
\ref{fel}). At that point,
the halo undergoes an isothermal
collapse \cite{aa}  which is eventually halted by quantum mechanics (in the
present model
by the repulsive self-interaction of the bosons). This takes the system from the
gaseous phase (G) to the condensed phase (C) in which almost all the mass of the
halo forms a compact soliton (see the analogous discussion for
fermions in \cite{ijmpb}). The
final outcome of this scenario is therefore a
pure soliton of radius $R_c=1\, {\rm kpc}$ and  mass  $M_c\sim 10^{12}\,
M_{\odot}$
without atmosphere. Such a structure is not observed (a pure soliton is
expected to have a much
smaller mass $(M_h)_{\rm min}=1.86\times 10^8\, M_{\odot}$ corresponding to the
ground state of the BECDM model) so this scenario should be rejected. A possible
reason for the failure of this scenario is that the
canonical ensemble is not relevant for our model (see Sec. \ref{sec_ineq}).
Therefore, the microcanonical evolution discussed in the next section may be
more relevant.

{\it Remark:} This scenario (isothermal collapse) could be valid
in a different
context in order to explain the formation of a supermassive boson star from the
gravitational collapse of a dilute gaseous cloud of bosons (see
Appendices \ref{sec_bhell} and \ref{sec_prob}). The boson star could mimic a
supermassive BH (without
DM halo) of mass $\sim 10^9\
M_{\odot}$ at the center of an elliptical
galaxy.   In that
case, we have to change the values of the model parameters (i.e. the
characteristics of the DM particle) and take general relativity into account.
The possibility of this scenario presupposes that DM may be made of
different types of particles which is not impossible.

\subsubsection{Microcanonical evolution: from the gaseous solution to the
core-halo solution}
\label{sec_gch}

In the microcanonical ensemble, the energy slowly decreases while  the
temperature increases as
the system enters in the region of negative specific heats. The whole series
of equilibria represented in
Fig.
\ref{fel} is stable. Therefore, if the system evolves microcanonically under
the effect of collisions, the DM halo can go smoothly from the gaseous phase
(G) to the core-halo phase (CH).  This may be a
mechanism
which explains how the
system reaches the core-halo phase (CH). The core-halo
phase contains a
solitonic core which may represent a bulge. This bulge is stable and
persistent (in the microcanonical ensemble) and can be present at the centers of
galaxies (see the Remark at the end of Sec. \ref{sec_ac}).

\subsection{Very large DM halos with $M_h>(M_h)_{\rm MCP}$}
\label{sec_vlh}

\begin{figure}
\begin{center}
\includegraphics[clip,scale=0.3]{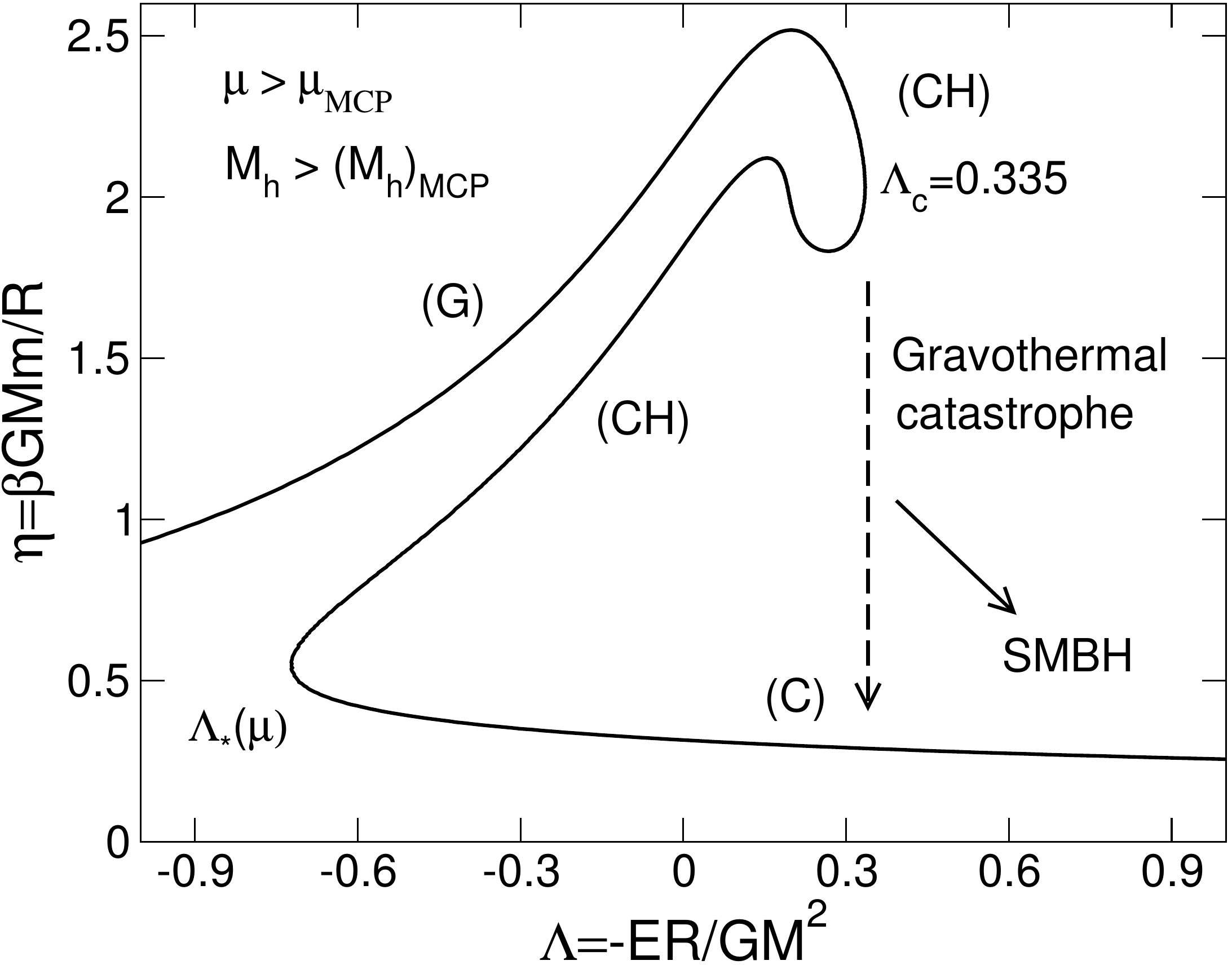}
\caption{For
$M_h>(M_h)_{\rm MCP}$, the caloric curve has a
$Z$-shape structure. In the microcanonical ensemble, the gaseous phase (G) and
the
condensed phase (C) are stable. The core-halo phase (CH) before the first
turning point of energy is also stable while the core-halo phase (CH) between
the first
and last turning points of energy is unstable. The
system can evolve collisionally in the gaseous phase (G) up to
the turning point of energy $E_c$ and collapse towards
the condensed phase
(C). This corresponds to the gravothermal
catastrophe (dotted arrow) \cite{lbw}. As explained in the text, this scenario
does not
seem realistic. Another
possibility is that the gravothermal catastophe triggers a dynamical
instability of general relativistic origin leading to the formation of a
SMBH \cite{balberg}.}
\label{le5}
\end{center}
\end{figure}

We consider a very large DM halo with $M_h>(M_h)_{\rm MCP}\sim 2\times
10^{12}\, M_{\odot}$.  Since $\mu>\mu_{\rm MCP}$ the caloric curve
has a $Z$-shape structure (see Fig. \ref{le5}). We restrict ourselves to the
microcanonical ensemble since the discussion in the canonical ensemble is the
same as before. Again, we assume that the system evolves collisionally along
the series of equilibria.

\subsubsection{Gravothermal catastrophe and expulsion
of an envelope}
\label{sec_halt}

In the microcanonical ensemble, the system first evolves slowly from the
gaseous phase (G) to the core-halo phase (CH) as before. Then, the series of
equilibria becomes unstable at the
turning point of energy $E_c$ (see Fig. \ref{le5}). At that point,
the DM halo undergoes a gravothermal 
catastrophe \cite{lbw} which is eventually halted by quantum mechanics (here,
the
repulsive self-interaction of the bosons). This takes the system
from the gaseous phase (G) to the condensed phase (C) in which only a fraction
($\sim 1/4$) of the mass of the
DM halo forms a compact solitonic core while the rest of the mass forms a hot
halo (see the analogous discussion for fermions in \cite{ijmpb,ac}).
In the box model, the halo is held by the walls of the box. In more realistic
models where the box is absent (see \cite{clm2} in the case of fermionic DM),
the halo is expelled at very
large distances and forms a very extended atmosphere (see Fig 41 of
\cite{clm2}). The final outcome of this scenario is therefore a pure soliton of
radius $R_c=1\, {\rm kpc}$ and mass  $M_c\lesssim 10^{12}\, M_{\odot}$ with the
ejection of a hot atmosphere of mass $M-M_c$. This  core-halo structure is
reminiscent of red-giant structure and supernovae in the context of compact
stars (white dwarfs and neutron stars).
However, this extreme core-halo structure is not observed in the case of DM
halos (see the discussion in \cite{clm2}) so this scenario
should be rejected. A possible
reason for the failure of this scenario is that the microcanonical evolution
(gravothermal catastrophe) leads to another possibility as detailed below.

{\it Remark:} We can make the same comment as at the end of
Sec. \ref{sec_gc}. This scenario (gravothermal catastrophe) is discussed in more
detail in Appendices \ref{sec_bhell}  and \ref{sec_prob}, and in Ref.
\cite{ijmpb,ac} in the case of
fermions.

\subsubsection{Gravothermal catastrophe and black hole formation leaving the
isothermal envelope undisturbed}
\label{sec_balberg}

As in the previous section we assume that the halo undergoes a
gravothermal catastrophe at $E_c$ but we consider another evolution in which the
system is not affected by quantum mechanics (the validity of this hypothesis is
considered in the following section). This scenario (already advocated in
\cite{clm1,clm2}) is based on
the self-interacting DM model of Balberg {\it et al.} \cite{balberg} who
developed 
the idea of ``avalanche-type contraction'' towards a SMBH initially suggested by
Zeldovich and
Podurets \cite{zp}, improved by Fackerell {\it et al.} \cite{fit}, and
confirmed numerically  by Shapiro and Teukolsky
\cite{st2,st3,st4}. The
initial stage of the gravothermal catastrophe is well-known. The core collapses
and reaches high densities and high
temperatures while the halo is not sensibly affected by the collapse of the
core and maintains its initial structure.\footnote{The
gravothermal catastrophe has been studied in detail in the case of
globular clusters evolving via two-body gravitational encounters.
The dynamical evolution of the system is due to
the gradient of temperature (velocity dispersion) between the core and the halo
and the fact that the core  has a negative specific heat.
The core loses heat to the profit of the halo, becomes hotter, and
contracts. If the temperature increases more rapidly in the
core than in the halo there is no possible equilibrium
and we get a runaway: this is the gravothermal catastrophe \cite{lbw}. The
collapse of the core is self-similar and leads to a
singularity in which the central density and the temperature become infinite in
a finite time (core collapse) \cite{lbe,cohn}. However, the mass contained
in the core tends to zero at the collapse time. The evolution may continue
in a
postcollapse regime
with the formation of a binary star \cite{inagaki}. The energy released by the
binary can stop the collapse and
induce a reexpansion of the system. Then a series of gravothermal oscillations
should follow \cite{oscillations,hr}. It has to be noted that, for
globular clusters, this process is very long,
taking place on a collisional relaxation timescale (of the order of the age of
the Universe) since it is due to two-body gravitational encounters. In the model
of Balberg {\it et al.} \cite{balberg},
the dynamical evolution of the system is due to the self-interaction of the
DM particles. In that case, a typical halo has sufficient time
to thermalize and acquire a gravothermal profile consisting of a flat core
surrounded by an extended halo. The same idea may
apply to our bosonic model (and also to the fermionic model of
Refs. \cite{clm1,clm2}) where the self-interaction of the bosons $a_s>0$ could
be responsible for the collisional evolution of the system
(considering a cross section per unit of mass $\sigma/m=
1.25\, {\rm cm^2/g}$ which corresponds to the constraint set by the Bullet
Cluster \cite{bullet} and is consistent with the estimate of Dav\'e {\it
et al.} \cite{dave} used by Balberg {\it et
al.} \cite{balberg}, and using Eq. (\ref{pv7}), we obtain a boson mass
$m=1.10\times 10^{-3}\, {\rm eV/c^2}$ with a scattering length $a_s= 4.41\times
10^{-6}\, {\rm fm}$). Balberg {\it et
al.} \cite{balberg} show that,
during the gravothermal catastrophe, the core of the self-interacting DM
halos passes from a long mean free path (LMFP) limit to a  short mean free
path (SMFP) limit. In the LMFP limit, the system displays a 
self-similar collapse similar to that of globular clusters in which the
core mass decreases rapidly. In the SMFP limit, the core
mass decreases more slowly (and almost saturates) so that a relatively large
mass can 
ultimately collapse into a SMBH (see below).} Now,
Balberg {\it et al.} \cite{balberg} argue that during the gravothermal
catastrophe, when the central density and the temperature increase above a
critical value, the system
undergoes a dynamical instability of general relativistic
origin leading to the
formation of a SMBH
on a
dynamical time scale. Only the central region of the DM halo (not its outer
part) is affected by this process so the final outcome of this scenario is 
an isothermal halo (possibly with a critical King profile \cite{clm1,clm2}, see
Appendix \ref{sec_diff}) containing a central SMBH.

{\it Remark:} For large DM halos with $M_h>(M_h)_{\rm
MCP}\sim 2\times
10^{12}\, M_{\odot}$ the core-halo solutions with a large value of
concentration $\chi$
(similar to those considered in Sec. \ref{sec_cheat}) are
thermodynamically unstable so they cannot be reached by the system during a
natural evolution (they lie well
after
the critical point of energy in the series of equilibria). On the other hand,
the core-halo
solutions with a small value of concentration $\chi$
(similar to those considered in Sec. \ref{sec_ac}) lying before the
critical point of energy in the
series of equilibria are thermodynamically stable. Therefore, they can be
reached by the system on a short timescale. However, on a long timescale, the
system may evolve
collisionally towards the critical point of energy $E_c$ and collapse towards a
SMBH. In
conclusion, large DM
halos with $M_h>(M_h)_{\rm MCP}\sim 2\times
10^{12}\, M_{\odot}$ should not contain a solitonic core (or only temporarily).
They should rather
contain a SMBH resulting from the process of Balberg {\it et
al.} \cite{balberg} described
previously.

\subsection{Criterion for the possible existence of a  black hole at the
centers of the galaxies}

The scenario discussed in Sec. \ref{sec_balberg} can lead to a black hole at
the centers of the galaxies only if the gravothermal catastrophe can take place
and only if it is sufficiently efficient to allow the core of the system to
develop high values of the density and of the temperature required to trigger
the relativistic dynamical instability. However, 
quantum mechanics can
prevent gravitational collapse and stop the gravothermal catastrophe. Therefore,
the previous scenario can lead to a black hole only if the parameter $\mu$ is
sufficiently larger than the microcanonical critical point  $\mu_{\rm MCP}\sim
10^5$ at which the gravothermal catastrophe appears. Using Eqs. (\ref{crit1})
and (\ref{crit2}), we conclude that only sufficiently 
large galaxies with $M_h>(M_h)_{\rm MCP}\sim 2\times 10^{12}\, M_{\odot}$ can
contain a supermassive black hole. Smaller halos may not contain black holes
because they do not experience the
gravothermal catastrophe. Indeed, the
gravothermal catastrophe is inhibited by
quantum mechanics. In that case, the halos can be either in the gaseous phase
(G) or in the
core-halo phase (CH) that are both thermodynamically stable in the
microcanonical ensemble. This result - the fact that black holes can form only
in
sufficiently large galaxies -  it consistent with the conclusion reached by
Ferrarese \cite{ferrarese} on the basis of observations. Furthermore, the order
of magnitude of the critical mass that we find in Eq. (\ref{crit2}) is
consistent with her estimate of $\sim 5\times 10^{11}\, M_{\odot}$. This
qualitative agreement is encouraging in view of the crudeness of our
theoretical model.

{\it Remark:} in the case of fermionic DM, the equivalent criterion
$\mu>\mu_{\rm MCP}$ for the
possible existence of a  black hole at the centers of the galaxies (see
Appendix H of \cite{clm2}) is
\begin{equation}
\label{app6b}
(M_h)_{\rm MCP}^{\rm F}=0.0106\left (\frac{\mu_{\rm
MCP}^4h^{12}\Sigma_0^3}{m^{16}G^6}\right )^{1/5}.
\end{equation}
If we take a fermion mass $m=1.23\, {\rm keV/c^2}$ as in \cite{clm2}, we get
$(M_h)_{\rm MCP}^{\rm F}=1.74\times 10^7\, M_{\odot}$. If we take the more
relevant value $m=170\, {\rm eV/c^2}$ obtained in  Appendix D
of \cite{suarezchavanis3}, we obtain $(M_h)_{\rm
MCP}^{\rm F}=9.78\times 10^9\, M_{\odot}$ in qualitative agreement with the
estimate of Ferrarese \cite{ferrarese}.

\subsection{Summary}

\begin{figure}
\begin{center}
\includegraphics[clip,scale=0.3]{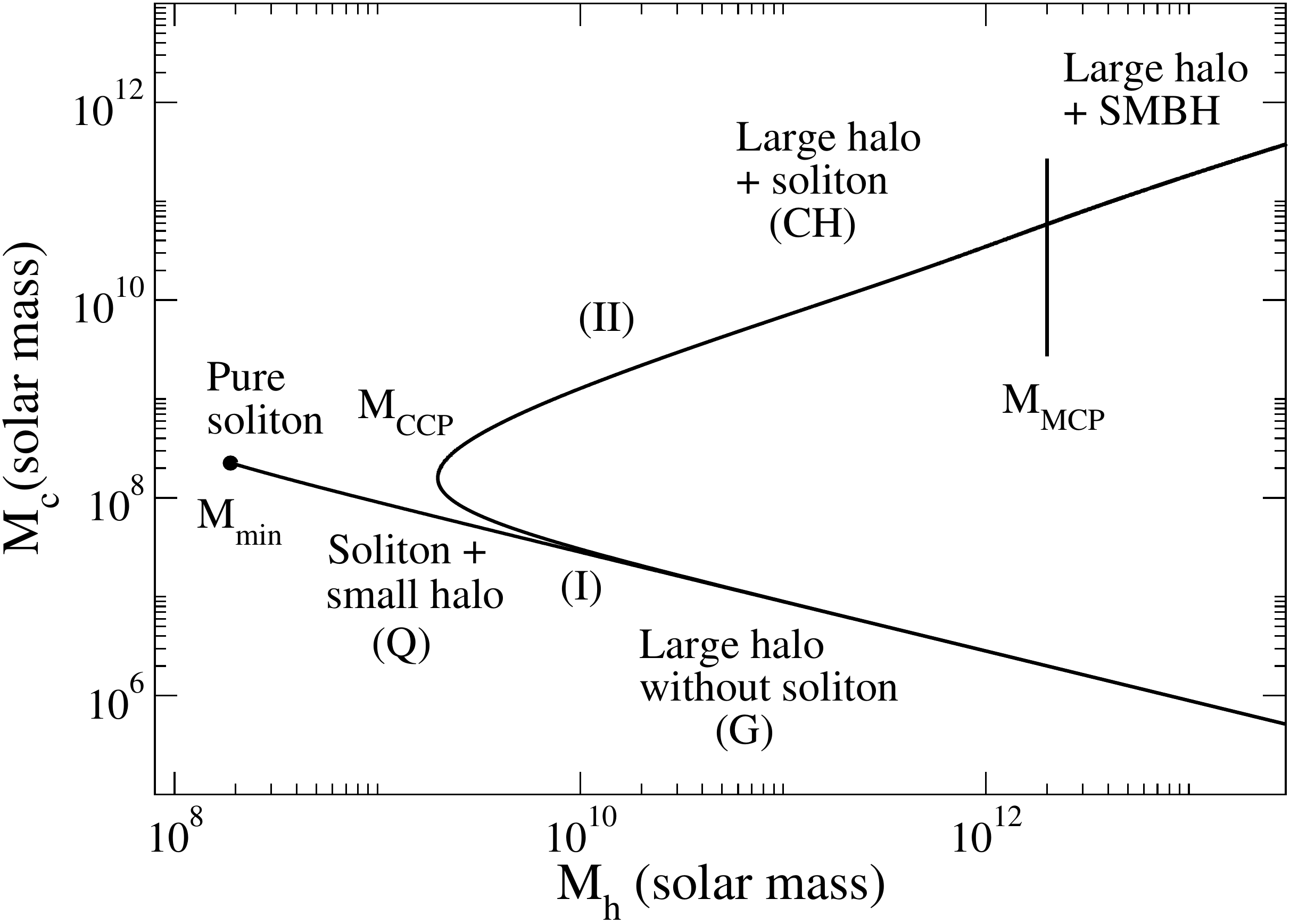}
\caption{Phase
diagram summarizing our main results. It displays the ground
state $(M_h)_{\rm min}=1.86\times
10^8\,
M_{\odot}$  where the DM halo is a pure
soliton without isothermal atmosphere. It also displays the canonical critical
point $(M_h)_{\rm CCP}=3.27\times
10^9\,
M_{\odot}$ at which there is a bifurcation between the gaseous branch (G) where
the
DM halos are
purely
isothermal without a central soliton and the core-halo branch (CH) where the DM
halos are made of a solitonic core (bulge) surrounded by a large isothermal
halo.
Finally, it displays the microcanonical
critical point $(M_h)_{\rm
MCP}\sim 2\times 10^{12}\,
M_{\odot}$  above which the DM
halos may undergo a gravothermal catastrophe leading to the formation of a
central SMBH.}
\label{conclusion}
\end{center}
\end{figure}

Our main results are summarized in the phase diagram of Fig. \ref{conclusion}.

For $M_h=(M_h)_{\rm min}=1.86\times 10^8\, M_{\odot}$ (ground state), the DM
halo
is a pure soliton without atmosphere. This is a purely quantum
object. Quantum mechanics (here, the repulsive self-interaction of the
bosons) fixes the minimum mass and the minimum radius of BECDM halos.
This situation  may describe ultracompact dSphs like Fornax.

For $(M_h)_{\rm min}=1.86\times 10^8\,
M_{\odot}<M_h<(M_h)_{\rm CCP}=3.27\times 10^9\, M_{\odot}$, the DM halo has
a solitonic core surrounded by a tenuous isothermal atmosphere.\footnote{We note
that the presence of an isothermal atmosphere, even tenuous, allows us to
satisfy
the observed mass-radius relation of DM halos  corresponding to a
constant surface density $\Sigma_0=\rho_0r_h=141\, M_{\odot}/{\rm pc}^2$ 
\cite{kormendy,spano,donato}. This important point is developed in Appendix
\ref{sec_paradox}.} This is
essentially a quantum (Q) object.  The caloric curve is monotonic ($\mu<\mu_{\rm
CCP}$; see Fig. \ref{multimu2}). There is only
one solution for any value of the temperature and of the
energy. This equilibrium state is stable in the microcanonical and canonical
ensembles. Even if the system evolves because of collisions, there is
no collapse, hence no black hole formation. This situation may describe dSphs.
Therefore, small halos like dSphs should not contain a SMBH.

For $(M_h)_{\rm CCP}=3.27\times 10^9\, M_{\odot}<M_h<(M_h)_{\rm
MCP}\sim 2\times 10^{12}\, M_{\odot}$, there are two principal solutions: a
gaseous
solution (G) corresponding to a purely isothermal halo without solitonic core
and a core-halo (CH) solution with a solitonic core surrounded by a massive
atmosphere.  The soliton may mimic a bulge, not a black hole (see Sec.
\ref{sec_ac}). The caloric
curve has a $N$-shape structure ($\mu_{\rm CCP}<\mu<\mu_{\rm MCP}$; see Fig.
\ref{fel}). The gaseous
solution is stable in both ensembles. The core-halo solution is unstable in the
canonical
ensemble but it is stable in the microcanonical ensemble. It has a 
negative specific heat. If the system evolves microcanonically because of
collisions, it can pass from the gaseous phase to the core-halo phase without
collapsing. There is no gravothermal catastrophe, hence no black hole
formation.
This situation may describe small spiral galaxies. In this sense, small
spiral galaxies should not contain a SMBH (at least according to the
scenario of Sec. \ref{sec_balberg}). Small spiral
galaxies in the core-halo phase should rather contain a solitonic bulge. We
note, however, this this bulge may itself induce the formation of a SMBH
(see Sec. \ref{sec_ac}).

For $M_h>(M_h)_{\rm MCP}\sim 2\times 10^{12}\, M_{\odot}$, there are
two principal solutions as before. However, the 
caloric curve now has a $Z$-shape structure ($\mu>\mu_{\rm MCP}$; see Fig.
\ref{le5}). If the
system evolves microcanonically because of collisions, it can trigger a 
gravothermal catastrophe leading to the formation of a SMBH by the
mechanism described in Sec. \ref{sec_balberg}.  This situation may apply to
large spiral galaxies and elliptical galaxies. Therefore, large spiral galaxies
and elliptical galaxies are expected to contain a SMBH, not a solitonic core.

The canonical critical point  $(M_h)_{\rm CCP}=3.27\times 10^9\,
M_{\odot}$ determines the bifurcation between
gaseous (G) solutions (without soliton) having a positive specific heat and
core-halo (CH) solutions (possessing a soliton) having a negative specific heat.
The microcanonical critical point $(M_h)_{\rm MCP}\sim 2\times 10^{12}\,
M_{\odot}$ determines the transition between DM halos possessing a solitonic
bulge and DM halos harboring a SMBH resulting from a gravothermal catastrophe
followed by a general relativistic dynamical instability.

{\it Remark:} We note that quantum mechanics is very important for small halos
on the  (Q) branch ($\mu<\mu_{\rm CCP}$; dSphs). In particular, it determines
the minimum mass $(M_h)_{\rm min}$ and the minimum radius $(r_h)_{\rm min}$
of BECDM halos ($\mu=\mu_{\rm min}$; ground state). Quantum mechanics is also
important in the core of
large halos on the (CH) branch (soliton). By contrast,
quantum mechanics is negligible for
large halos on the (G) branch (no soliton).  However, if these halos evolve
collisionally,
quantum mechanics determines whether they pass smoothly from the gaseous (G)
branch
without soliton to the core-halo (CH) branch with a soliton ($\mu_{\rm
CCP}<\mu<\mu_{\rm MCP}$; small spiral galaxies; quantum mechanics
important) or if they undergo a
gravothermal catastrophe and form a SMBH
($\mu>\mu_{\rm MCP}$; large spiral galaxies
and elliptical galaxies;  quantum mechanics noninmportant).

\section{Conclusion}

In this paper, we have developed the model of BECDM halos with a solitonic core
and an isothermal atmosphere proposed in \cite{bdo}. Following previous works,
we have assumed that the
thermodynamical temperature $T_{\rm th}$ is equal to zero, or is much smaller
than the condensation temperature $T_c$, so that the bosons form a pure BEC.
Therefore, the system is basically described by the GPP equations
(\ref{intro1}) and (\ref{intro2}). These
equations develop a complicated process of gravitational cooling
\cite{seidel94} and violent relaxation \cite{lb}  leading to a quasiequilibrium
state with a
core-halo structure \cite{ch2,ch3}. The core is a soliton, corresponding to a
stationary solution of GPP equations (ground state), and the halo arises from
quantum
interferences of excited states. Numerical simulations \cite{lin,moczchavanis}
show that the halo
is
relatively close to an isothermal halo (or a more refined fermionic King model
\cite{clm1,clm2}) which
is predicted from the theory of violent relaxation for
collisionless self-gravitating systems \cite{lb,mnras}. In any case, an
isothermal halo is a good working hypothesis to start with.

We have proposed to parametrize the complicated processes of
gravitational cooling and violent relaxation on the coarse-grained scale by the
generalized GPP equations (\ref{intro3}) and (\ref{intro4}). 
Through the Madelung transformation, these equations are
equivalent to the fluid equations (\ref{mad2})-(\ref{mad7}). They
generalize the hydrodynamic equations of the CDM model by accounting for a
quantum force due to the Heisenberg uncertainty principle, a pressure force due
to the self-interaction of the
bosons (scattering),  a temperature, and a friction. These
terms are due to quantum mechanics ($\hbar$ and $a_s$) and violent relaxation
($T$
and $\xi$). The friction term
accounts for the relaxation of the system towards an equilibrium state
in which
the gravitational attraction is balanced by the quantum pressure and by the
thermal pressure.\footnote{The GPP equations are able to
account for the damped oscillations of a system experiencing gravitational
cooling \cite{seidel94,gul0,gul}. In particular, the damping term
heuristically explains {\it how} a system of
self-gravitating bosons rapidly reaches an equilibrium state by
dissipating some free energy. This relaxation towards an equilibrium state is
encapsulated in the $H$-theorem of Eq. (\ref{ht1}).} This leads to the
formation of virialized DM halos at small cosmological scales (i.e. at galactic
scales). At large
cosmological scales, quantum mechanics and violent relaxation are negligible
(and coarse-graining is not
necessary) so we recover the hydrodynamic equations of the CDM model that
prove to be very relevant to explain the large-scale structure of the Universe.
This amounts to taking $\hbar=a_s=T=\xi=0$ in the generalized
hydrodynamic equations (\ref{mad2})-(\ref{mad7}). Therefore, quantum
mechanics is potentially able to solve the problems of the CDM model at small
scales without affecting the virtues of this model at large scales.

If we neglect the quantum pressure (TF approximation), as we have done in
this paper, the DM halos are described by an equation of state of the form
\begin{equation}
P=\frac{2\pi a_s\hbar^2\rho^2}{m^3}+\rho\frac{k_B T}{m}.
\end{equation}
This equation of state, which is at
the basis of our study, is interesting in its own right and could have been
introduced at the start without reference to the generalized GPP
equations (\ref{mad2})-(\ref{mad7}). It leads to DM halos presenting  a
core-halo structure with
a solitonic core and an isothermal halo. The polytropic equation of state
$P=2\pi
a_s\hbar^2\rho^{2}/m^3$ dominates in the core where the density is high and the
isothermal equation of state $P=\rho k_B T/m$ dominates in the halo where the
density is low (the transition occurs at $\rho_i\sim {k_B Tm^2}/({2\pi
a_s\hbar^2})$). As a result, the equilibrium state is made of a compact core
(BEC/soliton) with an equation of state $P=2\pi a_s\hbar^2\rho^{2}/m^3$, which
is a stable stationary solution of the GPP equations (\ref{intro1}) and
(\ref{intro2}) at $T=0$ (ground state), surrounded by
an isothermal atmosphere with an equation of state $P=\rho k_B T/m$
mimicking a halo of scalar radiation (quantum interferences) at an effective
temperature $T$. The solitonic core is stabilized against gravitational
collapse by quantum mechanics (here the repulsive self-interaction of the
bosons)
and has a smooth density profile replacing the $r^{-1}$ cusp of
CDM. On the other hand, the temperature term accounts
for the almost isothermal atmosphere of DM halos, where the density
approximately decreases as $r^{-2}$, leading to flat
rotation curves. Therefore, the solitonic core
solves the cusp-core problem and the isothermal halo leads to
flat rotation curves.

We have constrained our model by imposing the universal
value $\Sigma_0=141\, M_{\odot}/{\rm
pc}^2$ of the surface density of DM halos. On the other hand, we
have determined the ratio $a_s/m^3=3.28\times 10^3 \, {\rm
fm}/({\rm eV/c^2})^3$ of the DM particle by identifying the ground state of
the GPP equations with the most compact halo that has been observed (we took 
the dSph Fornax as a reference but this choice could be improved if necessary).
As a result, there is no free (undetermined) parameter in our
model.

We have first studied a model (Model I) which is particularly well adapted
to small DM halos. This model predicts three types of DM halos
depending on their mass:

(i) Dwarf DM halos with a mass $(M_h)_{\rm
min}=1.86\times 10^8\,
M_{\odot}$ are
ultracompact objects that are completely condensed without an atmosphere.
They represent
the ground state of the GPP equations (\ref{intro1}) and (\ref{intro2}) where
the halo is a pure soliton. Therefore, their size $(r_h)_{\rm
min}=788\, {\rm
pc}$  is equal to the size of the BEC/soliton.

(ii) Larger, but still small, DM halos  with a mass $(M_h)_{\rm
min}=1.86\times 10^8\,
M_{\odot}<M_h<(M_h)_*=3.30\times
10^9\,
M_{\odot}$ are extended
objects with a core-halo structure.  They have a condensed core
(BEC/soliton)
surrounded by a tenuous atmosphere made of scalar radiation
(quantum interferences) with an approximately isothermal
density profile decaying as $r^{-2}$ at large distances (or, more realistically,
with a NFW or Burkert profile decaying as $r^{-3}$). It is the
atmosphere that fixes their proper size while the soliton creates a central core
that solves the cusp problem. The atmosphere can be much larger than the size of
the soliton. The presence of the halo of scalar radiation explains why the size
of the DM halos increases with their mass contrary to what is predicted from the
ground state of the GPP
equations  according to which the size of DM halos has a constant value
$R_c=1\, {\rm kpc}$ in the TF limit (see Appendix \ref{sec_paradox}).

(iii) Large DM halos with a mass $M_h>(M_h)_*=3.30\times
10^9\,
M_{\odot}$ are
purely isothermal without a solitonic core. In
that case, the central core is due to effective thermal effects, not to quantum
mechanics. The size of the halos increases with their mass
according to the law $M_h=1.76\, \Sigma_0 r_h^2$.

In conclusion,  Model I predicts that DM halos are essentially classical
isothermal spheres except close to the ground state where quantum effects become
important. In other words,  quantum mechanics is essential to provide a
ground state corresponding to a minimum
halo mass $(M_h)_{\rm
min}=1.86\times 10^8\,
M_{\odot}$  and a minimum halo radius $(r_h)_{\rm
min}=788\, {\rm
pc}$. But as soon as $M_h>(M_h)_{\rm
min}=1.86\times 10^8\,
M_{\odot}$ quantum
mechanics  becomes negligible (the solitonic core disappears) and the halo is
purely isothermal.  This leads to the
mass-radius relation reported in Fig. \ref{massradius}.

We have then studied another model (Model II) which is particularly well adapted
to large DM halos. By redefining the notion of ``central density'' we
have found a new branch of solutions. A bifurcation from the branch
of Model I appears at a critical mass $(M_h)_b\sim
(M_h)_*=3.30\times
10^9\,
M_{\odot}$.
Above that
mass, the system may be purely isothermal without a solitonic core (as in Model
I) or have a well-developed core-halo structure with a solitonic core and an
isothermal envelope. The core mass scales with the halo as $M_c\propto
M_h^{1/2}$ [see Eq.
(\ref{ap1})]. The density profile presents a plateau between 
the core and the halo while the rotation curve presents a dip.
This core-halo solution is similar to the one found by numerous authors
\cite{wares,margrave,ht,bvriper,em,edwards,stella,cls,ir,gmr,merafina,
imrs,csmnras,bmtv,btv,rar,krut,clm2} in the case of fermionic DM. However, we
have found that
the solitonic core cannot mimic a supermassive BH at the center
of galaxies because it is to big. It may rather represent a bulge that may be
present now (see the Remark at the end of Sec. \ref{sec_ac}) or that, in the
past,  may have triggered the collapse of the surrounding gas, leading to a
supermassive black hole and a quasar.

Finally, we have been able to recover the bifurcation at 
$(M_h)_b$ from a box model
of
self-gravitating bosons, establishing an interesting connection between 
the statistical mechanics of self-gravitating bosons in a  box and real DM
halos. In this connection, the bifurcation point $(M_h)_b\sim
(M_h)_*=3.30\times
10^9\,
M_{\odot}$
corresponds to a canonical critical point $(M_h)_{\rm
CCP}=3.27\times 10^9\, M_{\odot}$ where the caloric curve takes an $N$-shape
structure leading to a region of negative specific heats associated with a
canonical phase transition and ensemble inequivalence. We have shown that
the core-halo solution is unstable in the canonical ensemble while it is
stable in the microcanonical ensemble.  In that last case (microcanonical
ensemble), if the DM halos evolve
collisionally, they can slowly pass from the gaseous  phase (without soliton)
to the
core-halo  phase (with a soliton). The core-halo phase may also be directly
formed by a process of violent collisionless relaxation. We have identified
another
critical mass  $(M_h)_{\rm MCP}=2\times 10^{12}\, M_{\odot}$ corresponding to a
microcanonical critical
point where the caloric curve takes a $Z$-shape structure leading to a
microcanonical phase transition. In that case, if the DM halos evolve
collisionally, they can undergo a gravothermal catastrophe ultimately leading to
the formation of a supermassive black hole on a dynamical timescale
\cite{balberg}. Our model therefore
predicts that black holes can form (by this process) only in sufficiently large
halos with
a mass $M_h>(M_h)_{\rm MCP}=2\times 10^{12}\, M_{\odot}$. Interestingly, this
typical mass is qualitatively
consistent with the results of Ferrarese \cite{ferrarese} obtained from
observations and leading to a critical mass  $\sim 5\times 10^{11}\,
M_{\odot}$.

In our model, the atmosphere is assumed to be isothermal in
agreement with very general thermodynamical arguments. This is the ``most
probable'' or ``most natural'' profile. However, the isothermal density profile
decreases as $r^{-2}$. Therefore, a purely isothermal atmosphere is clearly an
idealization since it has an infinite mass. Furthermore, the isothermal profile
($\rho\propto r^{-2}$) is apparently different from the observational Burkert
profile ($\rho\propto r^{-3}$). We have shown that for large halos
$M_h>(M_h)_c=6.86\times 10^{10}\, M_\odot$, the two profiles are
indistinguishable on the scale of observations ($r_h<100\, {\rm kpc}$). For
smaller halos, $(M_h)_*=3.30\times 10^{9}\, M_\odot<M_h<(M_h)_c=6.86\times
10^{10}\, M_\odot$, the two profiles show differences in slope. We have
suggested, following our previous works \cite{clm1,clm2}, that
the deviation from the (most probable) isothermal law may be explained by
incomplete violent relaxation, tidal effects, or stochastic forcing (see
Appendix \ref{sec_diff}). More precisely, we have argued that large halos,
instead of being described by the isothermal
profile, should be described by the King profile at the point of marginal
microcanonical stability. In that case, it almost coincides with the modified
Hubble profile which decreases as $\rho\propto r^{-3}$ like the Burkert profile.
For $r_h\sim 100\, {\rm kpc}$ the modified Hubble profile is much closer to the
Burkert profile than the 
isothermal profile. This may  explain the confinement of DM halos and the
observed logarithmic slope $-3$ of their density profile instead of the
ideal slope $-2$.

In forthcoming papers \cite{forthcoming}, we shall adapt our model to the case
of bosons without
self-interactions, to the case of bosons with attractive self-interactions, and
to fermions. Preliminary results, which are in good agreement with the
results of Schive {\it et al.} \cite{ch2,ch3} for noninteracting bosons
and to the results of Ruffini
{\it et al.} \cite{rar} for fermions, are presented at the end of Appendix
\ref{sec_ana}.

\appendix

\section{Effective thermal effects versus quantum mechanics}
\label{sec_eff}

In model I of our paper (see Sec. \ref{sec_mod1}), there is an
important distinction to make between small DM halos and large DM halos

(i) Small
DM halos have a core-halo structure with a solitonic core and an envelope. The
core is due to quantum mechanics. The envelope is expected to be identical to
that of a classical (nonquantum) collisionless self-gravitating system
described by the Vlasov equation.\footnote{In the context of the GPP equations
the envelope arises from quantum interferences of interaction-free excited
states. It is expected to match
the classical envelope arising from a process of collisionless violent
relaxation based on the Vlasov equation \cite{lin,moczchavanis}.} It may be
described by an isothermal or (fermionic) King profile. Such profiles are 
consistent with the Burkert and NFW profiles at large distances (see Sec.
\ref{sec_comp} and Appendix \ref{sec_diff}). 

(ii) Large DM
halos have no solitonic core. There are not quantum objects. Still they have a
core with a finite density (instead of a cusp) that is due to effective thermal
effects. They are well-described by an isothermal or King profile. Such profiles
are consistent with the Burkert profile at all distances (including
the core) or with the NFW profile at large distances (the cusp being regularized
by thermal effects). 

Therefore, the small DM halos of model I are similar to
those found by Schive {\it et al.} \cite{ch2,ch3} but the large DM halos,
being purely classical without a solitonic core, are different. In the case of
small DM halos, the core
is due to quantum mechanics, not to thermal effects. In the case of large DM
halos, the core is due to effective thermal pressure, not to quantum mechanics.
Quantum mechanics is negligible at large scales while it
provides a ground state at small scales (see Sec. \ref{sec_gs}).

In model II of our paper (see Sec. \ref{sec_another}), both small and large DM
halos have a core-halo structure with a  solitonic core due to quantum mechanics
and an essentially classical isothermal atmosphere. In that case, quantum
mechanics
(leading to the soliton) is important in the core of all types of DM halos
(small and large). The DM halos of model II are similar to those found by Schive
{\it et al.} 
\cite{ch2,ch3}.

\section{Some reasons of the difference between the isothermal profile and
the observational Burkert profile}
\label{sec_diff}

We have seen in Sec. \ref{sec_comp} that the isothermal DM halos of our model
with
a mass $(M_h)_*=3.30\times 10^{9} M_{\odot}<M_h<(M_h)_c=6.86\times 10^{10}
M_{\odot}$ exhibit a pronounced
difference with the observational Burkert profile in the
sense that their density profiles decrease at large distances as $r^{-2}$
(isothermal) instead of $r^{-3}$ (Burkert). Our point of view is that the
isothermal profile is the ``ideal'' profile that a self-gravitating BEC is
expected to reach through violent relaxation, gravitational cooling, or through
successive mergings
with other halos (in a process of hierarchical clustering). Indeed, the
isothermal distribution is predicted from general thermodynamical
considerations, whatever the origin of the relaxation (collisional,
collisionless, stochastic...) \cite{paddy,ijmpb}. It corresponds to the ``most
probable
state'', i.e., to the maximum entropy state. In this sense, our isothermal model
is ``ideal''. However, in practice, there are many ``nonideal'' effects that
prevent the system from reaching the isothermal distribution.\footnote{This is
actually obvious for self-gravitating systems since the isothermal
profile, decreasing at large distances as $r^{-2}$, has an infinite mass
\cite{bt}. In other words, there is no maximum entropy state for 
self-gravitating systems in an unbounded domain \cite{paddy,ijmpb}.
In reality, the density of the halos is steeper than what is predicted by
statistical mechanics. We note in this respect that
the exponent $\alpha=-3$  (NFW/Burkert) of the density profile $\rho\sim
r^{-\alpha}$ of observed DM halos is the closest exponent to the ``ideal''
exponent $\alpha=2$ (isothermal) that yields a halo with a (marginal) finite
mass. This rough argument may explain why the exponent $\alpha=3$ is selected.}
Let us briefly discuss some of these effects.

\subsection{Incomplete violent relaxation}

In the Introduction, we have developed a parallel between the process of
gravitational cooling \cite{seidel94} for self-gravitating
bosons and the process of violent
relaxation \cite{lb} for collisionless stellar systems (or collisionless DM
halos). Indeed, it is reasonable
to consider that the formation of the atmosphere that results from 
gravitational cooling or hierarchical clustering is similar to the process
of violent relaxation in stellar dynamics. Far from the core, quantum mechanics
effects are negligible and the system behaves as a classical
collisionless gas. Complete violent relaxation leads
to the Lynden-Bell distribution that is similar to the Fermi-Dirac distribution
(for a reason different from quantum mechanics).
It leads to configutrations with  a core-halo structure made of a fermionic
core (fermion ball) surrounded by an isothermal halo. At large distances, the
density should decrease as $r^{-2}$ \cite{lb}. However, in practice, this
isothermal
profile (that would have an infinite mass) is not reached because of incomplete
relaxation. Direct numerical simulations
of collisionless stellar systems \cite{henonVR,albada,roy,joyce} and theoretical
models of incomplete relaxation \cite{bertin1,bertin2,hjorth}
lead to a density profile that decreases as $r^{-4}$ at large distances. These
configurations are relatively close to
H\'enon's isochrone profile \cite{isochrone}. This $r^{-4}$
profile is steeper than the Burkert profile. Therefore, other reasons must
be advocated to explain the observed  $r^{-3}$ profile.

\subsection{Stochastic forcing}

In practice, a DM halo is never completely isolated from the surrounding but is
permanently subjected to perturbations caused by its environment (infall,
accretion, merger, bars, tidal
fields, resonances...). These
perturbations can be modeled by  a stochastic forcing that can alter the density
profiles of the halos. We suggest that the observational Burkert profile may be
(partly)
justified by this stochastic forcing resulting from the interaction of the
system with its environment.

\subsection{Tidal effects: King and Hubble profiles}

\begin{figure}[!h]
\begin{center}
\includegraphics[clip,scale=0.3]{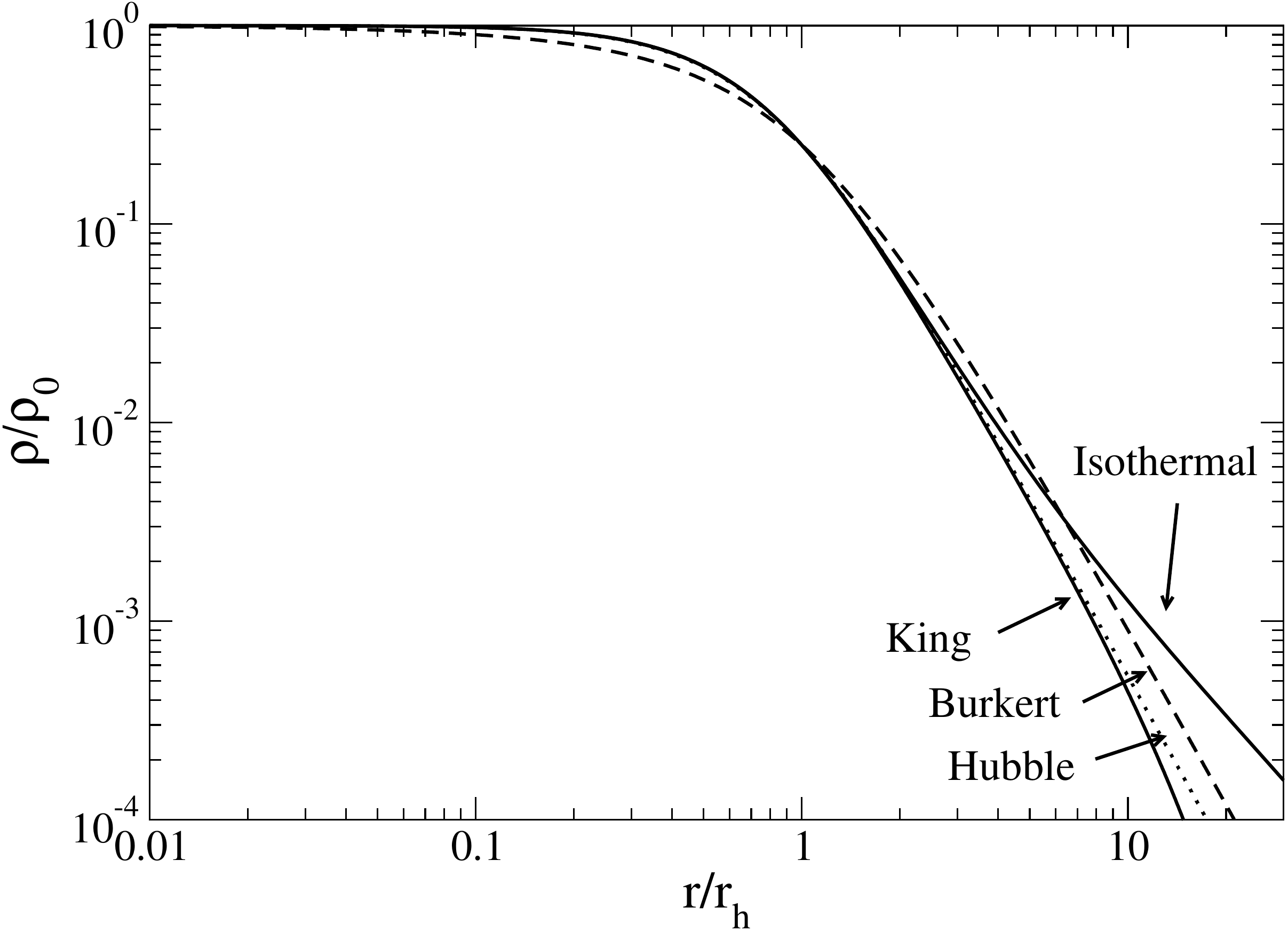}
\caption{Normalized density profiles up to $30 r_h$. We have plotted the
isothermal profile (upper solid line), the critical King profile (lower solid
line), the modified Hubble profile (dotted line), and the Burkert profile
(dashed line). At large distances, the critical King, modified Hubble and
Burkert profiles
decrease as $r^{-3}$ while the isothermal profile decreases as $r^{-2}$.
Among these profiles, only the King profile is physical and relies on a
rigorous theoretical modeling taking tidal effects into account (it improves
upon the ideal isothermal profile
that has an infinite mass). The modified Hubble and Burkert profiles are
empirical profiles that provide a good fit of DM halos but do not have
a physical justification.}
\label{kingburkert}
\end{center}
\end{figure}

DM halos may experience tidal interactions from other halos and galaxies. Tidal
effects have been extensively studied in astrophysics in the context of globular
clusters \cite{bt}. It was shown that, because of tidal interactions, the
isothermal
distribution is replaced by the King distribution \cite{king}. The same ideas
can be
exported to
the case of DM halos.\footnote{Globular clusters evolve through collisional
relaxation driven by two-body gravitational encounters. By contrast, DM halos
are essentially collisionless for what concerns gravitational encounters (the
Chandrasekhar time exceeds the age of the Universe by several orders of
magnitudes). However, there can be other sources of evolution (e.g. strong
collisions due to the self-interaction of the particles in the core of the
system \cite{balberg} or collisions between quasiparticles \cite{hui,bft}) that
can justify a King - or close to King - distribution for DM halos
\cite{clm1,clm2}.} In Refs.
\cite{clm1,clm2}, we have given
arguments according to which large DM halos should
be described by the King profile at the point of
 marginal stability in the microcanonical ensemble.\footnote{In the present
model they correspond to DM halos of mass $M_h>(M_h)_{\rm MCP}\sim 2\times
10^{12}\, M_{\odot}$. The point of  marginal microcanonical stability in the
King model is analogous to the point $\Lambda_c$ in Fig. \ref{le5}.} We call it
below the
``critical King profile''. We have shown that the critical King profile
is well fitted by the  modified Hubble profile (see Appendix \ref{sec_hubble})
on the range
$0\le r\le 30r_h$. The modified
Hubble profile decreases at large distances as $r^{-3}$ like the Burkert
profile. Therefore, tidal interactions can produce a $r^{-3}$ density profile.
The isothermal, critical King, modified Hubble
and Burkert profiles are plotted in
Fig. \ref{kingburkert}. For $r/r_h\le 6$, they are
very close to each other. By contrast, in the range $6\le r/r_h\le 30$, the
critical King
and modified Hubble profiles are closer to the Burkert profile than the
isothermal
profile because they display a slope $-3$ instead of a slope $-2$. We argue
that, physically, large DM halos are described by the  King profile at the point
of marginal microcanonical stability (critical King profile) which turns out to
be relatively close to the empirical modified Hubble and Burket profiles.
Therefore,
tidal effects may explain why the DM halos are more confined than the
isothermal profile and, consequently, why they are well-fitted by the Burkert
profile. As argued in \cite{clm1,clm2}, the critical King profile may provide a
justification of the observed logarithmic slope $\alpha=3$ of the density
profile of DM halos from
physical, instead of empirical (fit), arguments. This is confirmed by the 
recent paper of Arg\"uelles {\it et al.} \cite{krut} who consider the
fermionic King model \cite{clm1,clm2} and show that tidal effects are important
to match observational data.

\section{Complex hydrodynamic representation of the generalized Schr\"odinger
equation}
\label{sec_nott}

In this Appendix, we show that the generalized Schr\"odinger equation
\begin{eqnarray}
\label{nott1}
i\hbar \frac{\partial\psi}{\partial
t}=&-&\frac{\hbar^2}{2m}\Delta\psi+m\Phi\psi+2k_B
T\ln|\psi|\psi\nonumber\\
&-&i\frac{\hbar}{2}\xi\left\lbrack \ln\left (\frac{\psi}{\psi^*}\right
)-\left\langle \ln\left (\frac{\psi}{\psi^*}\right
)\right\rangle\right\rbrack\psi
\end{eqnarray}
can be
written as
a hydrodynamic equation involving a complex velocity field and an imaginary
viscosity. We then briefly mention the connection between this equation and the
theory of scale relativity developed by Nottale \cite{nottale} and with the
stochastic interpretation of quantum mechanics developed by Nelson
\cite{nelson}. A more detailed discussion is given in separate paper
\cite{chavnot}
where we adopt the opposite presentation, i.e., we derive
the generalized  Schr\"odinger equation from a complex
hydrodynamic equation.

\subsection{Complex Burgers equation}

It is easy to check that the generalized  Schr\"odinger equation
(\ref{nott1}) can be rewritten as
\begin{equation}
\label{nott2}
i\hbar\frac{\partial\psi}{\partial
t}=-\frac{\hbar^2}{2m}\Delta\psi+m\Phi\psi+V\psi+\hbar \, {\rm
Im}(\gamma\ln\psi)\psi,
\end{equation}
where
\begin{equation}
\label{nott3}
V(t)=i\frac{\hbar}{2}{\rm Re}(\gamma)\left\langle \ln\left
(\frac{\psi}{\psi^*}\right
)\right\rangle
\end{equation}
is a real function of time and
\begin{equation}
\label{nott4}
\gamma=\xi+i\frac{2k_B T}{\hbar}
\end{equation}
is a complex friction coefficient. If we make the WKB transformation 
\begin{equation}
\label{nott5}
\psi=e^{i{\cal S}/\hbar},
\end{equation}
where ${\cal S}$ is a complex action, we 
obtain the complex Hamilton-Jacobi equation
\begin{equation}
\label{nott6}
\frac{\partial {\cal S}}{\partial t}+\frac{1}{2m}(\nabla {\cal
S})^2-i\frac{\hbar}{2m}\Delta{\cal S}+m\Phi+V(t)+{\rm Re} (\gamma {\cal S})=0.
\end{equation}
When $\hbar=\gamma=0$ we recover the classical Hamilton-Jacobi equation (in that
case
${\cal S}$ is real). If we introduce the complex velocity
\begin{equation}
\label{nott7}
{\bf U}=\frac{\nabla {\cal S}}{m},
\end{equation}
and take the gradient of Eq. (\ref{nott6}), we obtain
\begin{equation}
\label{nott8}
\frac{\partial {\bf U}}{\partial t}+({\bf U}\cdot \nabla){\bf U}=i
\frac{\hbar}{2m}\Delta{\bf U}-\nabla\Phi-{\rm Re} (\gamma {\bf U}).
\end{equation}
This equation can be interpreted as a damped viscous Burgers equation
(no pressure term) involving a complex velocity field and an imaginary
viscosity
\begin{equation}
\label{nott9}
\nu=\frac{i\hbar}{2m}
\end{equation}
proportional to the Planck constant and inversely proportional to the mass of
the particle.

\subsection{Relation to the work of Nottale}

The complex hydrodynamic equation (\ref{nott8}) can be written as
\begin{equation}
\label{nott10}
\frac{D{\bf U}}{Dt}=-\nabla\Phi-{\rm Re} (\gamma {\bf U}),
\end{equation}
where
\begin{equation}
\label{nott11}
\frac{D{\bf U}}{Dt}=\frac{\partial {\bf U}}{\partial t}+({\bf U}\cdot
\nabla){\bf U}-i {\cal D}\Delta{\bf U}
\end{equation}
is the scale covariant derivative and
\begin{equation}
\label{nott12}
{\cal D}=\frac{\hbar}{2m},
\end{equation}
is the quantum diffusion coefficient. When $\gamma=0$, Eq.
(\ref{nott10})
can be interpreted as a scale covariant equation of dynamics (Newton's law). 
Nottale
\cite{nottale}
has shown that a particle that has a nondifferentiable trajectory is described
by an equation of this form.  He considered the conservative
case $\gamma=0$ where Eq. (\ref{nott10}) leads to the ordinary Schr\"odinger
equation. If we take into account dissipative effects ($\gamma\neq 0$) in Eq.
(\ref{nott10}), we
obtain the generalized Schr\"odinger equation (\ref{nott1}) involving an
effective temperature term ($T$) and a friction term ($\xi$).\footnote{The fact
that $T$ and $\xi$ in Eq. (\ref{nott1}) can be interpreted as a temperature and
a friction coefficient is explained in Sec. \ref{sec_prop} using the Madelung
transformation (see also \cite{bdo,chavnot}).} In this
formulation, the
temperature and the dissipation are two manifestation of the same phenomenon,
i.e., they represent the real and the imaginary parts of the complex friction
coefficient $\gamma$ (see Eq. (\ref{nott4})).

{\it Remark:} We note that, in Nottale's theory, ${\cal D}$ may be different
from $\hbar/2m$. In other words, the (generalized)  Schr\"odinger Eq.
(\ref{nott1}) may be valid in a more general context than quantum mechanics.
Indeed, it may apply to particles that have nondifferentiable trajectories due
to
their chaotic motion or due to the fractal structure of spacetime itself at
large (cosmological) scales. This opens new perspectives for the interpretation
of the Schr\"odinger-Poisson and GPP equations for DM as discussed in
\cite{nottalechaos}.

\subsection{Relation to the work of Nelson}

If we write ${\bf U}={\bf u}-i{\bf u}_Q$ where
\begin{equation}
\label{nott13}
{\bf u}=\frac{\nabla S}{m}\qquad {\rm and}\qquad {\bf
u}_Q=\frac{\hbar}{2m}{\nabla\ln\rho}
\end{equation}
are  the classical and quantum  velocities, and
take the real and imaginary parts of the generalized
complex viscous Burgers equation (\ref{nott8}), we obtain the two real coupled
equations
\begin{eqnarray}
\label{nott14}
\frac{\partial {\bf u}}{\partial t}+({\bf u}\cdot \nabla){\bf u}-({\bf u}_Q\cdot
\nabla){\bf u}_Q=\frac{\hbar}{2m}\Delta{\bf u}_Q-\nabla\Phi\nonumber\\
-{\rm
Re}(\gamma)
{\bf
u}-{\rm Im}(\gamma) {\bf u}_Q,
\end{eqnarray}
\begin{equation}
\label{nott15}
\frac{\partial {\bf u}_Q}{\partial t}+({\bf u}_Q\cdot \nabla){\bf u}+({\bf
u}\cdot \nabla){\bf u}_Q= -\frac{\hbar}{2m}\Delta{\bf u}.
\end{equation}
When $\gamma=0$ these equations coincide with those introduced by Nelson
\cite{nelson} in his stochastic interpretation of quantum
mechanics.\footnote{We note that Eqs. (\ref{nott14}) and (\ref{nott15}) are
equivalent to the Madelung hydrodynamic equations (\ref{mad2}) and
(\ref{mad3}); see \cite{chavnot} for more details.} In that
case, ${\bf u}_Q$ is called the osmotic velocity (see footnote 41 in
\cite{chavnot}).
Nelson derived these equations  from Newton's law and showed their equivalence
with the ordinary Schr\"odinger equation. Equations (\ref{nott14}) and 
(\ref{nott15})  can
therefore be viewed as a generalization of Nelson's equations taking dissipative
effects into account.

\subsection{Generalized Einstein relation}

It is interesting to note that the complex nature of the
friction coefficient $\gamma=\gamma_{\rm R}+\gamma_{\rm I}$ [see Eq.
(\ref{nott4})] leads to a generalized Schr\"odinger equation (\ref{nott1})
exhibiting {\it simultaneously} a
friction term 
and an effective temperature term. They correspond to the
real and imaginary parts of $\gamma$. This may be viewed as a new form of
fluctuation-dissipation theorem. In this respect, we note that
the relation
\begin{equation}
{\cal D}=\frac{k_B T}{m\gamma_I}
\end{equation}
resulting from Eqs. (\ref{nott4}) and (\ref{nott12}) is similar to the Einstein
relation of
Brownian motion \cite{einstein}.

On the other hand, if we assume that $\gamma_{\rm R}=\gamma_{\rm I}$ (see
the argument given in Appendix D of \cite{nottalechaos}), we obtain the
relation
\begin{eqnarray}
\label{einstein7}
{\cal D}=\frac{k_B T}{m\xi}.
\end{eqnarray}
Explicitly,
\begin{eqnarray}
\label{einstein7b}
\frac{\hbar}{2m}=\frac{k_B T}{m\xi}\qquad {\rm or}\qquad
\frac{\hbar}{2}=\frac{k_B T}{\xi}.
\end{eqnarray}
Again, this can be viewed as a sort of generalized Einstein relation expressing
a form
of  fluctuation-dissipation theorem. Here, the diffusion
coefficient has a quantum origin.

\section{Particular profiles of self-gravitating systems}
\label{sec_famous}

In this Appendix, we consider particular profiles  of self-gravitating
systems that are useful in our study to interpret
the structure of DM halos.

\subsection{Basic equations and definitions}
\label{sec_hr}

The condition of hydrostatic equilibrium of a
self-gravitating system described by a barotropic equation of state
$P(\rho)$ is 
\begin{eqnarray}
\label{hr6}
\nabla P+\rho\nabla\Phi={\bf 0}.
\end{eqnarray}
Dividing this equation by $\rho$, taking its divergence and using the Poisson
equation (\ref{mad4}), we obtain the fundamental  differential equation 
\cite{chandra}:
\begin{eqnarray}
\label{hr7}
\nabla\cdot \left (\frac{1}{\rho}\nabla P\right )=-4\pi G\rho.
\end{eqnarray}
Depending on the equation of state this equation can be solved analytically or
numerically to obtain the density profile $\rho(r)$. 

The halo radius $r_h$ is defined as the distance at which the
central density  $\rho_0$  is divided by $4$,
\begin{eqnarray}
\label{hr1}
\frac{\rho(r_h)}{\rho_0}=\frac{1}{4}.
\end{eqnarray} 
The mass $M(r)$ contained within a sphere of radius $r$ is given by
\begin{eqnarray}
\label{hr2}
M(r)=\int_0^r\rho(r')4\pi{r'}^2\, dr'.
\end{eqnarray} 
The halo mass is 
\begin{eqnarray}
\label{hr3}
M_h=M(r_h).
\end{eqnarray} 
The circular velocity is defined by
\begin{eqnarray}
\label{hr4}
v^2(r)=\frac{GM(r)}{r}.
\end{eqnarray} 
The circular velocity at the halo radius is
\begin{eqnarray}
\label{hr5}
v_h^2=v^2(r_h)=\frac{GM_h}{r_h}.
\end{eqnarray} 
We note the identity
\begin{eqnarray}
\label{hr8}
\frac{v_h^2}{G\rho_0r_h^2}=\frac{M_h}{\rho_0 r_h^3}.
\end{eqnarray}

\subsection{Isothermal profile}
\label{sec_i}

We consider the isothermal equation
of state \cite{chandra}:
\begin{eqnarray}
\label{i1}
P=\rho\frac{k_B T}{m},
\end{eqnarray}
where $T$ is the temperature. The
fundamental equation of
hydrostatic equilibrium (\ref{hr7}) can be rewritten as
\begin{eqnarray}
\label{i2}
\frac{k_B T}{m}\Delta\ln\rho=-4\pi G\rho.
\end{eqnarray}
Writing
\begin{equation}
\label{i3}
\rho=\rho_0
e^{-\psi},
\end{equation}
where $\rho_0$ is the central density, introducing the normalized radial
distance
\begin{equation}
\label{i3b}
\xi=r/r_0, \qquad r_0=\left (\frac{k_B T}{4\pi G\rho_0
m}\right )^{1/2}
\end{equation}
where $r_0$ is the thermal core radius, and assuming
spherical
symmetry, we obtain the Emden equation \cite{chandra}:
\begin{equation}
\label{i5}
\frac{1}{\xi^2}\frac{d}{d\xi}\left
(\xi^2\frac{d\psi}{d\xi}\right
)=e^{-\psi},
\end{equation}
\begin{equation}
\label{i5b}
\psi(0)=\psi'(0)=0.
\end{equation}
Using Eqs. (\ref{hr2}), (\ref{i3}), (\ref{i3b})
and (\ref{i5}), the mass
contained within the
sphere of radius $r$ is given by
\begin{eqnarray}
\label{i7}
M(r)=4\pi \rho_0 r_0^3 {\xi}^2\psi'(\xi).
\end{eqnarray} 
According to Eqs. (\ref{hr4}), (\ref{i3b}) and (\ref{i7}), the circular velocity
is
\begin{equation}
\label{i7q}
v^2(r)=4\pi G\rho_0r_0^2\xi \psi'(\xi).
\end{equation}
Using Eq. (\ref{i3b}), we find that the temperature satisfies the
relation 
\begin{eqnarray}
\frac{k_B T}{m}=4\pi G\rho_0 r_0^2.
\end{eqnarray} 
Therefore, we can rewrite  Eq. (\ref{i7q}) as
\begin{equation}
\frac{m v^2(r)}{k_B T}=\xi \psi'(\xi).
\end{equation}
The halo radius defined by Eq. (\ref{hr1}) is given by $r_h=\xi_h r_0$, where
$\xi_h$
is determined by the equation
\begin{eqnarray}
\label{i9}
e^{-\psi(\xi_h)}=\frac{1}{4}.
\end{eqnarray} 
Solving the Emden equation (\ref{i5}) numerically, we find
\begin{eqnarray}
\label{i10}
\xi_h=3.63,\qquad \psi'(\xi_h)=0.507.
\end{eqnarray}
The normalized halo mass is
\begin{eqnarray}
\label{i11}
\frac{M_h}{\rho_0 r_h^3}=4\pi\frac{\psi'(\xi_h)}{{\xi_h}}=1.76.
\end{eqnarray} 
The normalized circular velocity at the halo radius is 
\begin{eqnarray}
\label{i12}
\frac{v_h^2}{4\pi G\rho_0r_h^2}=\frac{\psi'(\xi_h)}{\xi_h}=0.140.
\end{eqnarray} 
The normalized temperature is 
\begin{eqnarray}
\label{i13}
\frac{k_B T}{Gm\rho_0 r_h^2}=\frac{4\pi}{\xi_h^2}=0.954.
\end{eqnarray} 

\subsection{Polytropic profiles}
\label{sec_p}

We consider the polytropic equation 
of state \cite{chandra}:
\begin{eqnarray}
\label{p1}
P=K\rho^{\gamma},\qquad \gamma=1+\frac{1}{n},
\end{eqnarray}
where $K$ is the polytropic constant and $\gamma$ (or $n$) is the polytropic
index. The fundamental equation of
hydrostatic equilibrium (\ref{hr7}) can be rewritten as
\begin{eqnarray}
\label{p2}
K(n+1)\Delta\rho^{1/n}=-4\pi G\rho.
\end{eqnarray}
Writing 
\begin{eqnarray}
\label{p3}
\rho=\rho_0 \theta^{n},
\end{eqnarray}
where $\rho_0$ is the central density, introducing the normalized radial 
distance
\begin{eqnarray}
\label{p4}
\xi=r/r_0,\qquad r_0=\left \lbrack \frac{K(n+1)}{4\pi G\rho_0^{1-1/n}}\right
\rbrack^{1/2},
\end{eqnarray}
where $r_0$ is the polytropic core radius, and assuming spherical
symmetry, we obtain the
Lane-Emden equation
\cite{chandra}:
\begin{eqnarray}
\label{p5}
\frac{1}{\xi^2}\frac{d}{d\xi}\left (\xi^2\frac{d\theta}{d\xi}\right
)=-\theta^{n},
\end{eqnarray}
\begin{eqnarray}
\label{p6}
\theta(0)=1,\qquad \theta'(0)=0.
\end{eqnarray}
Using Eqs. (\ref{hr2}), (\ref{p3}), (\ref{p4}) and (\ref{p5}), the mass
contained within the
sphere of radius $r$ is given by
\begin{eqnarray}
\label{p8}
M(r)=-4\pi \rho_0 r_0^3 {\xi}^2\theta'(\xi).
\end{eqnarray}
According to Eqs. (\ref{hr4}), (\ref{p4})  and (\ref{p8}), the circular
velocity is
\begin{eqnarray}
\label{p9}
v^2(r)=-4\pi G\rho_0 r_0^2\xi\theta'(\xi).
\end{eqnarray} 
The halo radius  defined by Eq. (\ref{hr1}) is given by $r_h=\xi_h r_0$, where
$\xi_h$ is determined by the equation
\begin{eqnarray}
\label{p10}
\theta(\xi_h)^{n}=\frac{1}{4}.
\end{eqnarray} 
The value of $\xi_h$ can be obtained by solving the Lane-Emden equation
(\ref{p5}) for a given value of $n$. The normalized  halo mass is
\begin{eqnarray}
\label{p11}
\frac{M_h}{\rho_0 r_h^3}=-4\pi\frac{\theta'(\xi_h)}{{\xi_h}}.
\end{eqnarray} 
The normalized circular velocity at the halo radius is
\begin{eqnarray}
\label{p12}
\frac{v_h^2}{4\pi G\rho_0 r_h^2}=-\frac{\theta'(\xi_h)}{\xi_h}.
\end{eqnarray} 

When $n<5$, the polytropes are self-confined (their density has a compact
support). We denote by $\xi_1$ the normalized radius at which the density
vanishes: $\theta_1=0$. Their total mass $M$ and their radius $R$ are
given by
\begin{eqnarray}
M=-4\pi \rho_0 r_0^3 \xi^2_1\theta'_1,\qquad R=\xi_1 r_0.
\end{eqnarray}
Eliminating the central density between these two equations, we obtain the
mass-radius relation \cite{chandra}:
\begin{eqnarray}
M^{(n-1)/n}R^{(3-n)/n}=\frac{K(n+1)}{G(4\pi)^{1/n}}\omega_n^{(n-1)/n},
\end{eqnarray}
where $\omega_n=-\xi_1^{(n+1)/(n-1)}\theta'_1$.

For the polytrope $n=1$ the Lane-Emden equation (\ref{p5}) can be solved
analytically. The solution is \cite{chandra}:
\begin{equation}
\frac{\rho(r)}{\rho_0}=\theta=\frac{\sin(\xi)}{\xi}.
\label{p13}
\end{equation}
The normalized radial distance is $\xi=r/r_0$ where $r_0=(K/2\pi G)^{1/2}$ is
independent of the central density. The density vanishes at
$\xi_1=\pi$. This corresponds to a radius 
\begin{equation}
R=\pi\left (\frac{K}{2\pi G}\right )^{1/2}.
\label{p14}
\end{equation}
We can then write $\xi=\pi r/R$. The central density is related to the total
mass by
\begin{equation}
\rho_0=\frac{\pi M}{4R^3}=\frac{M}{4\pi^2}\left (\frac{2\pi G}{K}\right
)^{3/2}.
\end{equation}
The halo radius is $r_h=\xi_hR/\pi$ where
$\xi_h$ is the smallest root of
$\sin(\xi_h)/\xi_h=1/4$. We find
\begin{equation}
\xi_h=2.4746,\qquad \theta'(\xi_h)=-0.41853.
\label{p15}
\end{equation}
The mass profile and the circular velocity profile can be written as
\begin{equation}
M(r)=\frac{4\pi\rho_0 r_h^3}{\xi_h^3}\left\lbrack \sin(\xi)-\xi\cos(\xi)\right
\rbrack,
\label{p16}
\end{equation}
\begin{equation}
v^2(r)=\frac{4\pi G \rho_0 r_h^2}{\xi_h^2}\left\lbrack
\frac{\sin(\xi)}{\xi}-\cos(\xi)\right \rbrack.
\label{p18}
\end{equation}
The normalized halo mass and the
normalized circular velocity at the halo
radius are
\begin{equation}
\label{p18b}
\frac{M_h}{\rho_0 r_h^3}=2.12534,\qquad \frac{v_h^2}{4\pi G \rho_0
r_h^2}=0.169129.
\end{equation}

\subsection{Burkert profile}
\label{sec_burkert}

The Burkert profile \cite{observations} is given by the empirical law 
\begin{equation}
\frac{\rho(r)}{\rho_0}=\frac{1}{(1+x)(1+x^2)},\qquad x=\frac{r}{r_h}.
\label{bu1}
\end{equation}
The corresponding rotation curve is
\begin{equation}
v^2(r)=2\pi G\frac{\rho_0 r_h^2}{x}\left\lbrack \ln(1+x)-\arctan
x+\frac{1}{2}\ln(1+x^2)\right \rbrack.
\label{bu2}
\end{equation}
The normalized halo mass and the normalized circular velocity at the halo
radius are
\begin{equation}
\frac{M_h}{\rho_0 r_h^3}=1.60,\qquad \frac{v_h^2}{4\pi G \rho_0
r_h^2}=0.127.
\label{bu4}
\end{equation}

\subsection{Pseudo-isothermal profile}
\label{sec_pseudo}

The pseudo-isothermal profile is given
by 
\begin{equation}
\frac{\rho(r)}{\rho_0}=\frac{1}{1+3x^2},\qquad x=\frac{r}{r_h}.
\label{ps1}
\end{equation}
The corresponding rotation curve is
\begin{equation}
v^2(r)=\frac{4\pi G\rho_0 r_h^2}{3}\left\lbrack
1-\frac{\arctan(\sqrt{3}x)}{\sqrt{3}x}\right\rbrack.
\label{ps4}
\end{equation}
The normalized halo mass and the normalized circular velocity at the halo
radius are
\begin{equation}
\frac{M_h}{\rho_0 r_h^3}=1.66,\qquad \frac{v_h^2}{4\pi G \rho_0
r_h^2}=0.132.
\label{ps5}
\end{equation}

\subsection{Modified Hubble profile}
\label{sec_hubble}

The modified Hubble model \cite{bt}  is given by
\begin{equation}
\frac{\rho(r)}{\rho_0}=\frac{1}{(1+a x^2)^{3/2}},\qquad x=\frac{r}{r_h},
\label{mod1}
\end{equation}
where $a=4^{2/3}-1=1.52$.  The corresponding rotation curve
is
\begin{equation}
v^2(r)=4\pi G\frac{\rho_0 r_h^2}{x}\left\lbrack
\frac{\sinh^{-1}(\sqrt{a}x)}{a^{3/2}}-\frac{x}{a\sqrt{1+a x^2}}\right \rbrack.
\label{mod2}
\end{equation}
The normalized halo mass and the normalized circular velocity at the halo
radius are
\begin{equation}
\frac{M_h}{\rho_0 r_h^3}=1.75,\qquad \frac{v_h^2}{4\pi G \rho_0
r_h^2}=0.139.
\label{mod3}
\end{equation}

\subsection{Natarajan and Lynden-Bell profile}
\label{sec_nlb}

The Natarajan and Lynden-Bell profile \cite{nlb} is given by
\begin{equation}
\frac{\rho}{\rho_0}=\frac{A}{a^2+\xi^2}-\frac{B}{b^2+\xi^2},
\label{nlb1}
\end{equation}
where $\xi$ is defined by Eq. (\ref{i3b}). The corresponding rotation curve is
\begin{eqnarray}
v^2(r)=\frac{4\pi G\rho_0r_0^2}{\xi}\Biggl\lbrace A
a\left\lbrack\frac{\xi}{a}-\tanh^{-1}\left (\frac{\xi}{a}\right
)\right\rbrack\nonumber\\
-B b\left\lbrack\frac{\xi}{b}-\tanh^{-1}\left
(\frac{\xi}{b}\right )\right\rbrack\Biggr\rbrace.
\label{nlb2}
\end{eqnarray}
The halo radius  defined by Eq. (\ref{hr1}) is given by $r_h=\xi_h r_0$, where
$\xi_h$ is determined by
\begin{equation}
\xi_h^4+(a^2+b^2+4B-4A)\xi_h^2+a^2b^2-4Ab^2+4Ba^2=0.
\label{nlb3}
\end{equation}
A good approximation of the isothermal profile is obtained by taking $A=50$,
$a^2=10$, $B=48$, and $b^2=12$ \cite{nlb}. This gives $\xi_h= 3.64$. The
normalized halo mass and the normalized circular velocity at the halo
radius are then given by
\begin{equation}
\frac{M_h}{\rho_0 r_h^3}=1.75,\qquad \frac{v_h^2}{4\pi G \rho_0
r_h^2}=0.139,
\label{nlb4}
\end{equation}
in very good agreement with the exact results from Appendix \ref{sec_i}.

\section{Fundamental differential equation of our model}
\label{sec_gde}

In our model \cite{bdo,nottalechaos}, the density of the DM halos is
determined by the fundamental 
differential equation
\begin{eqnarray}
\label{gde0}
\frac{\hbar^2}{
2m^2}\Delta
\left (\frac{\Delta\sqrt{\rho}}{\sqrt{\rho}}\right
)-\frac{K\gamma}{\gamma-1}\Delta\rho^{\gamma-1}-\frac{k_B
T}{m}\Delta\ln\rho\nonumber\\
=4\pi
G\rho+3\omega_0^2.
\end{eqnarray}
If we define
\begin{equation}
\label{gde2}
\rho=\rho_0
e^{-\psi},\qquad  \xi=\left (\frac{4\pi G\rho_0
m}{k_B T}\right )^{1/2}r, 
\end{equation}
\begin{equation}
\label{gde3}
 \chi=\frac{K\gamma m\rho_0^{\gamma-1}}{k_B T},\qquad \epsilon=\frac{2\pi
G\rho_0\hbar^2}{(k_B T)^2},\qquad
\Omega^2=\frac{3\omega_0^2}{4\pi G\rho_0},
\end{equation}
we find that Eq. (\ref{gde0}) takes the form of a
generalized Emden equation
\begin{equation}
\label{gde4}
\epsilon \Delta\left (e^{\psi/2}\Delta
e^{-\psi/2}\right )+\Delta\psi+\chi\nabla\cdot\left \lbrack
e^{-(\gamma-1)\psi}\nabla\psi\right
\rbrack=e^{-\psi}+\Omega^2.
\end{equation}
The ordinary Emden equation (\ref{i5}) is recovered for
$\epsilon=\chi=\Omega=0$. Alternatively, if we define
\begin{equation}
\label{gde5}
\rho=\rho_0\theta^n,\qquad \xi=\left\lbrack \frac{4\pi
G}{K(n+1)\rho_{0}^{1/n-1}}\right\rbrack^{1/2}r,
\end{equation}
we find that Eq. (\ref{gde0}) takes
the form of a
generalized Lane-Emden equation
\begin{equation}
\label{gde7}
-\frac{\epsilon}{n^2\chi^2}\Delta\left
(\frac{\Delta\theta^{n/2}}{\theta^{n/2}}\right )+\frac{1}{\chi}
\Delta\ln\theta+\Delta\theta=-\theta^n-\Omega^2.
\end{equation}
The ordinary Lane-Emden equation (\ref{p5}) is recovered for
$\epsilon=1/\chi=\Omega=0$.

\section{Comparison with the model of Slepian and Goodman (2012)}
\label{sec_sg}

Our model of BECDM halos shows some analogies with the  model of Slepian and
Goodman \cite{sg} but it is fundamentally different, thereby escaping the
problems mentioned by these authors to construct BECDM halos consistent with the
observations.

Slepian and Goodman \cite{sg}  consider a self-gravitating boson gas at finite
temperature,
corresponding to a true statistical equilibrium state of bosons resulting from
a collisional relaxation. They take into account the repulsive self-interaction
of the bosons and the possibility that the bosons form a BEC above a
critical density $\rho_{\rm c}$ and derive the equation of state of this
system. It behaves as $P\sim \rho k_B T_{\rm th}/m$
(isothermal)
at low densities and as $P\sim 2\pi a_s\hbar^2\rho^2/m^3$ (condensate) at high
densities. The normalized equation of state depends on a dimensionless parameter
$\theta\sim a_s/\Lambda_{\rm dB}$ where $\Lambda_{\rm dB}=h/\sqrt{2\pi mk_B
T_{\rm th}}$
is the thermal de Broglie wavelength. Importantly, it shows a plateau after
$\rho_{\rm c}$ when $\theta\ll 1$. Slepian and Goodman \cite{sg} numerically
determine the density and circular velocity
profiles corresponding to this equation of state. They find that
the circular velocity profile presents a dip which increases as $\theta$
decreases and argue that, in order to match the observations (which do not
exhibit a strong dip), we must have $\theta\ge
10^{-4}$.\footnote{We note that Slepian and Goodman \cite{sg}
impose $\theta \hat{\nu}(0)=1$ which corresponds to $\chi\sim 1$ in our
notations. According to Eq. (\ref{chp4}) this is equivalent to the equality
between the soliton radius and
the thermal core radius: $r_0\sim R_c$.} This implies that
$m\ge 10\, {\rm eV/c^2}$ (assuming $v_{\infty}=100 \, {\rm km/s}$ and
$R_c=1\, {\rm kpc}$). However, this constraint is not consistent with the
constraint $m<10^{-3}\, {\rm eV/c^2}$ implied by the  Bullet Cluster (see
Appendix D of \cite{suarezchavanis3}). They conclude therefore that the thermal
BECDM model is ruled out.

Their model is physically different from ours because it describes the true
statistical
equilibrium state of self-gravitating bosons at finite temperature while our
model is a heuristic parametrization of the GPP equations at $T_{\rm th}=0$ (or
$T_{\rm th}\ll T_c$) taking into
account violent relaxation and gravitational cooling. In their model, $T_{\rm
th}$
represents the true thermodynamic temperature while in our model $T$
is an
effective
out-of-equilibrium temperature (like in Lynden-Bell's theory of violent
relaxation). In the same way, their equation of state aims at representing the
true equation of
state of a self-interacting boson gas at statistical equilibrium while our
equation of state (\ref{mad6bq}) is a heuristic equation of state of an
out-of-equilibrium
self-interacting boson gas (again like in Lynden-Bell's theory of violent
relaxation). Therefore, their equation of state is different, and more complex,
than ours (although they both have the same asymptotic behaviors). In
particular, it presents a plateau between the condensed phase and the
uncondensed phase which is responsible for the problems that they encounter
to constuct a DM halo satisfying all the observational constraints. In
our out-of-equilibrium equation of state there is no such plateau so there is no
problem to obtain solutions satisfying the observational constraints.

Slepian and Goodman \cite{sg} we careful to mention that their conclusions only
apply to self-gravitating bosons at statistical equilibrium. Since we
consider out-of-equilibrium (but still virialized) self-gravitating bosons
described by a different equation of state their critics do
not apply to our model.

\section{Condensation temperature}
\label{sec_cond}

We have seen that large DM halos have an isothermal, or almost isothermal, 
atmosphere which is responsible for the flat, or almost flat, rotation
curves of the galaxies. The temperature $T$ is related to the circular velocity
at
infinity $v_{\infty}$ by the relation
\begin{equation}
\frac{k_B T}{m}=\frac{v_{\infty}^2}{2}.
\label{cond1}
\end{equation}
For the Medium Spiral,
$v_{\infty}=153\, {\rm km/s}$ (see Sec. \ref{sec_lh}). For bosons with a
repulsive self-interaction, the boson mass must
be in the range  $2.92\times 10^{-22}\, {\rm eV}/c^2<m<1.10\times 10^{-3}\, {\rm
eV}/c^2$ in order to account
for the mass and size of ultracompact dwarf halos at $T=0$ such as Fornax as
well as the constraint set by the Bullet Cluster
(see Appendix D of Ref. \cite{suarezchavanis3}). In that case, we find
from Eq. (\ref{cond1}) that the temperature of large halos such as the Medium
Spiral is $4.41\times 10^{-25}\, {\rm K}<T<1.66\times
10^{-6}\, {\rm K}$. Such a small temperature may not be
physical.
This strongly
suggests that $T$ is not the true thermodynamic temperature. It may
rather
represent an effective temperature as we have suggested in the present paper.
In that case, $T$ has not a real physical meaning. Only $k_B T/m$ has a physical
meaning.

The condensation temperature of a boson gas is given by 
\begin{equation}
T_c=\frac{2\pi \hbar^2\rho^{2/3}}{m^{5/3}k_B\zeta(3/2)^{2/3}},
\label{cond2}
\end{equation}
where $\zeta(3/2)=2.6124...$. The bosons are uncondensed for $T_{\rm th}>T_c$
while they form a BEC for $T_{\rm th}<T_c$. 
Evaluated at the center of large DM halos such as the Medium
Spiral where $\rho_0=7.02\times 10^{-3}M_{\odot}/{\rm
pc}^3$ (see Sec. \ref{sec_lh}),  we get $5.29\times 10^{5}\, {\rm
K}<T_c<4.82\times 10^{36}\, {\rm
K}$. This value of the condensation temperature is considerably larger
than the thermodynamic temperature of radiation $T_{\rm th}\sim 3\, {\rm K}$,
than the
effective temperature of the DM halos $4.41\times 10^{-25}\, {\rm
K}<T<1.66\times
10^{-6}\, {\rm K}$ and than any
reasonable temperature. This indicates that the bosons are
completely condensed and that we can consider that $T_{\rm th}=0$. This
justifies our starting hypothesis.

Of course, in a given halo, the condensation temperature decreases as the
density decreases. For a given temperature, we can define a critical
density 
\begin{equation}
\rho_c=\frac{\zeta(3/2)}{(2\pi)^{3/2}}\frac{(k_B T_{\rm
th})^{3/2}m^{5/2}}{\hbar^3}
\label{cond3}
\end{equation}
above which the bosons form a BEC. Taking $T_{\rm th}\sim 3\, {\rm K}$
(thermodynamic
temperature of radiation), we get $3.44\times 10^{-57}\, M_{\odot}/{\rm pc}^3 < 
  \rho_c< 9.48\times 10^{-11}\, M_{\odot}/{\rm pc}^3$. This is much smaller than
the typical densities represented in Fig. \ref{pd} indicating that the bosons
are
always completely condensed.  Therefore, in all relevant cases, we
can assume that $T_{\rm
thermo}=0$.

{\it Remark:} If we assume that DM halos are made of
fermions, like sterile neutrinos, then the fermion mass must be  $m\sim 170\,
{\rm eV}/c^2$ in order to account
for the mass and size of ultracompact dwarf halos at $T=0$ such as Fornax (see
Appendix D of \cite{suarezchavanis3}). In that case, we find from Eq.
(\ref{cond1})
that the temperature of large halos such as the Medium Spiral
is $T\sim 0.257\, {\rm K}$. This temperature is more physical
suggesting that,
if DM is made of fermions, $T$ may represent the true thermodynamic
temperature. There remains, however, the timescale problem to reach a
statistical equilibrium state, as discussed in the Introduction.

\section{Proof that the solitonic core in our model is always nonrelativistic}
\label{sec_mimic}

The soliton of mass $M_c$ and radius $R_c$ studied in Sec. \ref{sec_cheat} would
be strongly
relativistic, and could mimic a supermassive black hole, if its radius were of
the order of the Schwarzschild radius: 
\begin{equation}
\label{mimic1}
R_c\sim R_S=\frac{2GM_c}{c^2}.
\end{equation}
Using Eq. (\ref{ap1}), we find that
\begin{equation}
\label{mimic2}
\frac{R_c}{R_S}=\frac{R_c
c^2}{2GM_c}=0.695\frac{c^2}{G \sqrt{\Sigma_0 M_h}}\frac{1}{ \ln \left
(\frac{M_h}{\Sigma_0 R_c^2}\right )}.
\end{equation}
Interestingly, in our model, the compactness $R_c/R_S$ of the soliton is
independent of the properties of
the DM particle ($a_s$ and $m$), except for a logarithmic correction (the
logarithmic factor depends on $R_c$, hence on $a_s/m^3$). For a halo of mass
$M_h=10^{11}\, M_{\odot}$, similar to the one that surrounds our Galaxy, we get
$R_c/R_S=5.89\times 10^5 \gg 1$. Therefore, the soliton is not a black hole,
not even a relativistic object. We find that $R_c/R_S$ becomes of order unity
for a halo mass
\begin{equation}
\label{mimic3}
(M_h)_{\rm crit}=0.121\, \frac{c^4}{\Sigma_0 G^2}\frac{1}{\ln^2\left
(\frac{c^2}{G\Sigma_0R_c}\right )}.
\end{equation}
When $M_h\ll (M_h)_{\rm crit}$ the soliton is nonrelativistic. When
$M_h$ approaches  $(M_h)_{\rm crit}$ the soliton becomes strongly relativistic
and may mimic
a black hole (in that case, a general relativistic treatment becomes
mandatory). Using Eqs. (\ref{observation}) and (\ref{pv4}),
we
obtain 
\begin{equation}
\label{mimic4}
(M_h)_{\rm crit}\sim 10^{21}\, M_{\odot}.
\end{equation} 
This value is independent  of the properties of
the DM particle.
This critical mass is much larger than any relevant mass of DM halos in the
Universe. We therefore conclude that the solitonic core in our model is always
nonrelativistic and
cannot mimic a black hole. This justifies a posteriori our Newtonian approach.

Another, sensibly equivalent, argument can be given as follows. When general
relativity is
taken into account,
we know that a self-interacting boson star in the TF regime is stable only
below a maximum mass \cite{colpi,tkachev,chavharko}: 
\begin{equation}
\label{mimic5}
(M_c)_{\rm max}=0.307\frac{\hbar c^2\sqrt{a_s}}{(Gm)^{3/2}}=9.78\times 10^{-2}
\frac{c^2 R_c}{G}.
\end{equation}
Using Eq. (\ref{pv4}), we get 
\begin{equation}
\label{mimic6}
(M_c)_{\rm max}=2.04\times  10^{15}\, M_{\odot}.
\end{equation}
When $M_c\sim (M_c)_{\rm max}$, the boson star is strongly relativistic and
when  $M_c>(M_c)_{\rm max}$ it collapses into a black hole. Inversely, when 
 $M_c\ll (M_c)_{\rm max}$, the boson star is nonrelativistic. For a halo of mass
$M_h=10^{11}\, M_{\odot}$, the mass of the soliton is $M_c= 1.77\times 10^{10}\,
M_{\odot}$ (see Sec. \ref{sec_cheat}). Since $M_c\ll (M_c)_{\rm max}$, the
soliton is nonrelativistic. The soliton would collapse into a black hole if
its mass satisfied $M_c>(M_c)_{\rm
max}$.
From Eqs. (\ref{ap1}) and (\ref{mimic5}) we find that
\begin{equation}
\label{mimic7}
\frac{M_c}{(M_c)_{\rm max}}=7.35\frac{G \sqrt{\Sigma_0
M_h}}{c^2} \ln \left
(\frac{M_h}{\Sigma_0 R_c^2}\right ).
\end{equation}
Therefore, $M_c$ would be larger that $(M_c)_{\rm
max}$ in a halo of mass $M_h>(M_h)'_{\rm crit}$
where $(M_h)'_{\rm crit}$ (obtained from Eq. (\ref{mimic7})) is essentially the
same mass as in Eqs. (\ref{mimic3})
and (\ref{mimic4}). Therefore, we conclude that
the soliton is always nonrelativistic (for all the halos in the Universe) and
that it cannot collapse into a black hole. This does not prevent, however,
the possibility that the solitonic bulge attracts  the gas around it and creates
a situation favorable to the formation of a supermassive black hole and a
quasar as discussed in Sec. \ref{sec_ac}.

{\it Remark:} We can similarly compute the maximum soliton mass in the case of
noninteracting bosons. Using $(M_c)_{\rm max}=0.633\hbar c/Gm$
\cite{kaup,rb} and $m=2.92\times 10^{-22}\, {\rm eV/c^2}$ (see
Sec. \ref{sec_gs}), we obtain $(M_c)_{\rm max}=2.90\times 10^{11}\, M_{\odot}$.
On the other hand, the maximum mass of the fermion ball $(M_c)_{\rm
max}=0.384(\hbar c/G)^{3/2}/m^2$ \cite{ov} in the case of fermions of mass
$m=170\, {\rm eV/c^2}$  (see
Sec. \ref{sec_gs}) is  $(M_c)_{\rm
max}=2.16\times 10^{13}\, M_{\odot}$. These maximum masses are much larger than
the core masses  of DM halos ($M_c\ll (M_c)_{\rm max}$) so the cores of DM halos
are generally nonrelativistic. We note that Ruffini {\it et al.} \cite{rar}
reach a different
conclusion because they take a much larger mass of the fermionic particle,
$m=48\, {\rm keV/c^2}$, which is not consistent with the arguments given in 
Sec. \ref{sec_gs}.

\section{Analytical model of a self-gravitating
BEC with an isothermal atmosphere in a box}
\label{sec_ana}

In this Appendix, we develop an analytical model of a self-gravitating
BEC with an isothermal atmosphere enclosed within a box of radius
$R$.\footnote{This model is directly
inspired by the analytical model developed in
Refs. \cite{csmnras,pt,ac} for
self-gravitating fermions.} This
model returns the gaseous (G), core-halo (CH) and condensed (C)
phases obtained in Sec. \ref{sec_ptb}. It allows us to analytically obtain the
relation $M_c(M_h)$ between the core mass and the halo mass by extremizing the
free energy $F(M_c)$ with respect to $M_c$. Furthermore, it
shows that the gaseous and condensed solutions are thermodynamically stable
(minima of free energy) while the core-halo solution is thermodynamically
unstable in the canonical ensemble (maximum of free energy).\footnote{It is
possible to generalize this model in the microcanonical ensemble. In that case,
it can be shown that the core-halo phase may be microcanonically stable in
agreement with the discussion
of Sec. \ref{sec_ineq}.}

We modelize the core by a pure soliton of mass $M_c$ and radius $R_c$ as in
Sec. \ref{sec_tfw}. For a self-interacting BEC in the TF approximation, we
recall that the
soliton radius  $R_c$ has a unique value given by
Eq. (\ref{rc}). On the other hand, the internal energy and the gravitational
energy of
the soliton are given by \cite{prd1}: 
\begin{equation}
\label{ana1}
U_c=\frac{GM_c^2}{4R_c},\qquad W_c=-\frac{3GM_c^2}{4R_c}.
\end{equation}

We modelize the halo by an isothermal atmosphere of uniform
density and mass $M-M_c$ contained between the spheres of radius
$R_c$ and $R$. The
internal energy of the atmosphere is given by 
\begin{equation}
\label{ana2}
U_h=\frac{k_B T}{m}(M-M_c)\left\lbrack \ln(M-M_c)-\ln V-1\right\rbrack,
\end{equation}
and its  gravitational energy (in the presence of the solitonic core) by
\begin{equation}
\label{ana3}
W_h=-\frac{3GM_c(M-M_c)}{2R}-\frac{3G(M-M_c)^2}{5R}.
\end{equation}
To obtain these results, we have assumed that $R_c\ll R$
\cite{pt}.

The free energy of the system  is therefore
\begin{eqnarray}
\label{ana4}
F&=&-\frac{GM_c^2}{2R_c}+\frac{k_B T}{m}(M-M_c)\left\lbrack
\ln(M-M_c)-\ln
V-1\right\rbrack\nonumber\\
&-&\frac{3GM_c(M-M_c)}{2R}-\frac{3G(M-M_c)^2}{5R}.
\end{eqnarray}
This is a function $F(M_c)$ of the core mass for a given value of
$M$, $R$ and $T$. The extrema of this function determine the possible 
equilibrium states of the system. More precisely, they determine the
possible equilibrium
core masses, $M_c^{(i)}$, as a function of $M$, $R$ and $T$. This is
valid both in the
canonical and microcanonical ensembles \cite{pt}. In the canonical
ensemble,
a minimum of
$F(M_c)$ corresponds to a stable equilibrium state (most probable state) while
a maximum of $F(M_c)$
corresponds to an unstable equilibrium state (less probable state).

It is convenient to introduce the dimensionless
quantities 
\begin{eqnarray}
\label{ana5}
x=\frac{M_c}{M},\qquad \eta=\frac{\beta GMm}{R},
\end{eqnarray}
\begin{eqnarray}
\label{ana6}
\mu=\pi^2\left
(\frac{R}{R_c}\right )^2, \qquad f(x)=\frac{F(M_c)R}{GM^2},
\end{eqnarray}
so that Eq. (\ref{ana4}) can be rewritten as 
\begin{eqnarray}
\label{ana7}
f(x)&=&-\frac{\sqrt{\mu}}{2\pi}x^2+\frac{1}{\eta}(1-x)\left\lbrack \ln \left
(\frac{M}{V}\right
)+\ln(1-x)-1\right\rbrack\nonumber\\
&-&\frac{3}{2}x(1-x)-\frac{3}{5}(1-x)^2,
\end{eqnarray}
with $0\le x\le 1$. 

\begin{figure}[!h]
\begin{center}
\includegraphics[clip,scale=0.3]{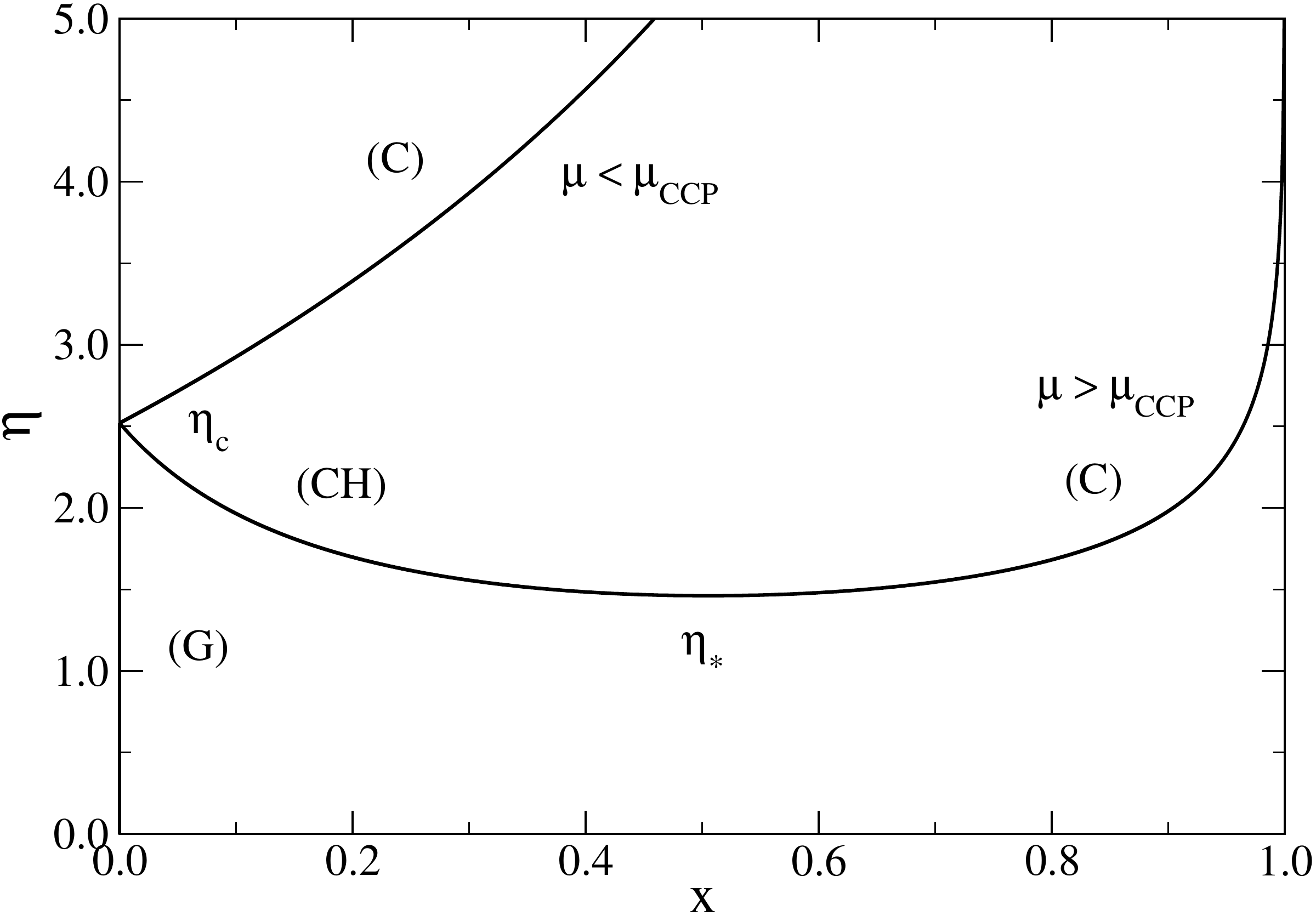}
\caption{The function $\eta(x)$ for $\mu<\mu_{\rm CCP}^{\rm app}=47.6$
(specifically $\mu=30$)
and for $\mu>\mu_{\rm CCP}^{\rm app}$ (specifically $\mu=100$). We have
represented the
gaseous phase (G), the condensed phase (C) and the core-halo phase (CH).}
\label{xeta}
\end{center}
\end{figure}

The equilibrium states, corresponding to $f'(x)=0$, are the solutions of the
equation
\begin{equation}
\label{ana8}
\ln(1-x)+\left (\frac{\sqrt{\mu}}{\pi}-\frac{9}{5}\right
)x\eta+\frac{3}{10}\eta+\ln\left (\frac{M}{V}\right )=0.
\end{equation}
This equation determines the core mass $x=M_c/M$ as a function of $\eta$, $\mu$
and $M/V$. For $x=0$ (purely gaseous phase) we find $\eta(0)= -({10}/{3})\ln
({M}/{V})$ and, for
reasons that will become clear below, we shall identify this value with
$\eta_c=2.52$, the minimum temperature of a
classical self-gravitating isothermal gas. Therefore, we set
\begin{equation}
\label{ana9b}
\ln \left (\frac{M}{V}\right)=-\frac{3}{10}\eta_c.
\end{equation}
We can then rewrite Eq.
(\ref{ana8}) as
\begin{equation}
\label{ana9}
\ln(1-x)+\left (\frac{\sqrt{\mu}}{\pi}-\frac{9}{5}\right
)x\eta+\frac{3}{10}(\eta-\eta_c)=0.
\end{equation}
The solutions of this equation can be easily found by studying the
inverse function
\begin{equation}
\label{ana10}
\eta(x)=\frac{\eta_c-\frac{10}{3}\ln(1-x)}{1+\frac{10}{3}\left
(\frac{\sqrt{\mu}}{\pi}-\frac{9}{5}\right
)x}
\end{equation}
for a given value of $\mu$ (see Fig \ref{xeta}).  For $x\rightarrow 0$, we
get
\begin{equation}
\label{ana11}
\eta(x)=\eta_c+\frac{10}{3}\left\lbrack 1-\left
(\frac{\sqrt{\mu}}{\pi}-\frac{9}{5}\right
)\eta_c\right\rbrack x+...
\end{equation}
Close to $x=0$, the curve $\eta(x)$ is increasing when $\mu<\mu_{\rm CCP}^{\rm
app.}$ and decreasing when $\mu>\mu_{\rm CCP}^{\rm
app.}$, where
\begin{eqnarray}
\label{ana15}
\mu_{\rm CCP}^{\rm app.}=\pi^2\left (\frac{1}{\eta_c}+\frac{9}{5}\right
)^2=47.6.
\end{eqnarray}
This value can be identified with the canonical
critical point. For $x\rightarrow 1$, we get
\begin{eqnarray}
\label{ana12}
\eta\sim \frac{-\ln(1-x)}{\frac{\sqrt{\mu}}{\pi}-\frac{3}{2}}\rightarrow
+\infty,
\end{eqnarray}
where we have assumed $\mu>(3\pi/2)^2=22.2$ to avoid unphysical results.

\begin{figure}[!h]
\begin{center}
\includegraphics[clip,scale=0.3]{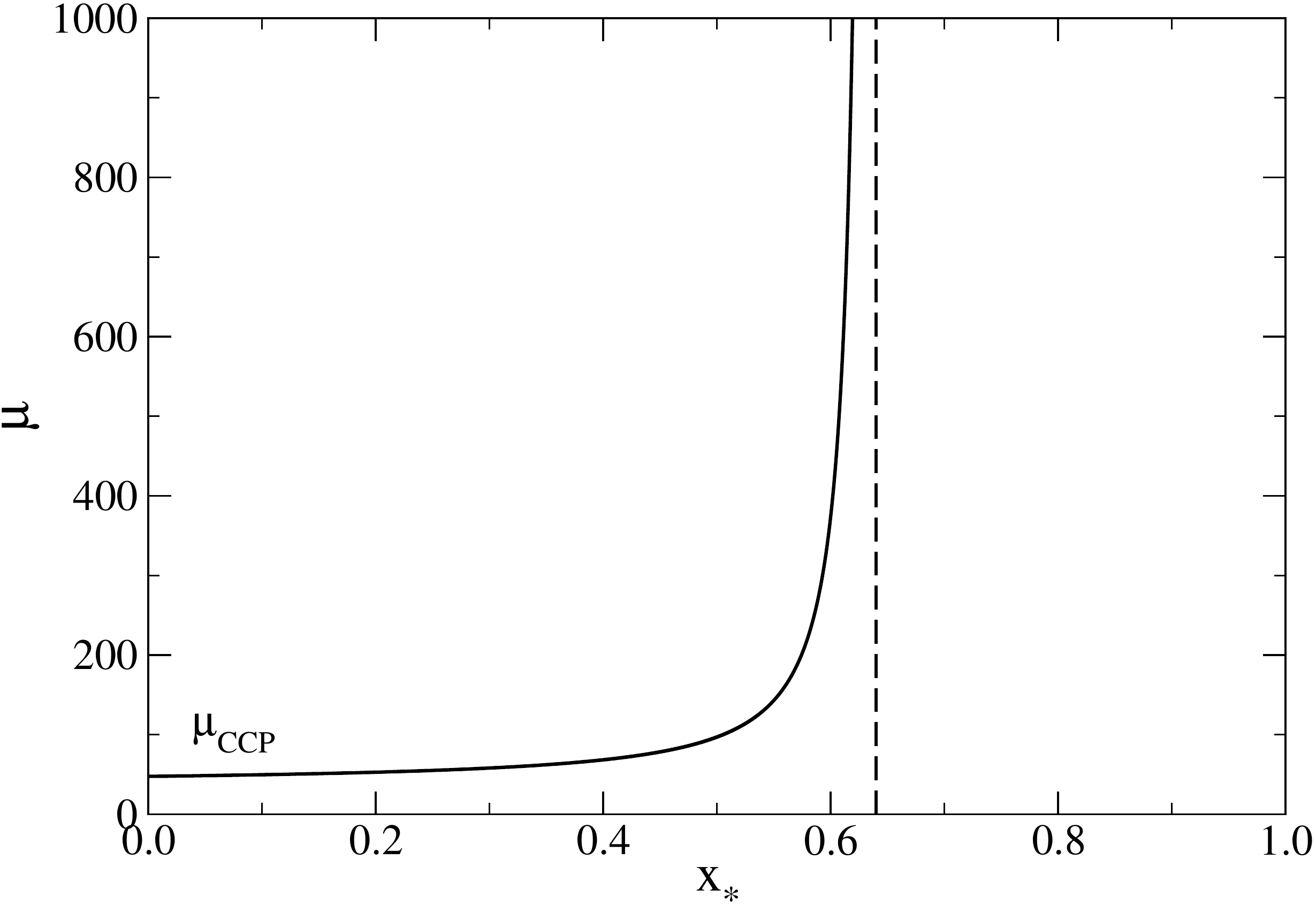}
\caption{The function $\mu(x_*)$. By inversion, it gives the value of $x_*$ as a
function of $\mu$.}
\label{xmu}
\end{center}
\end{figure}

\begin{figure}[!h]
\begin{center}
\includegraphics[clip,scale=0.3]{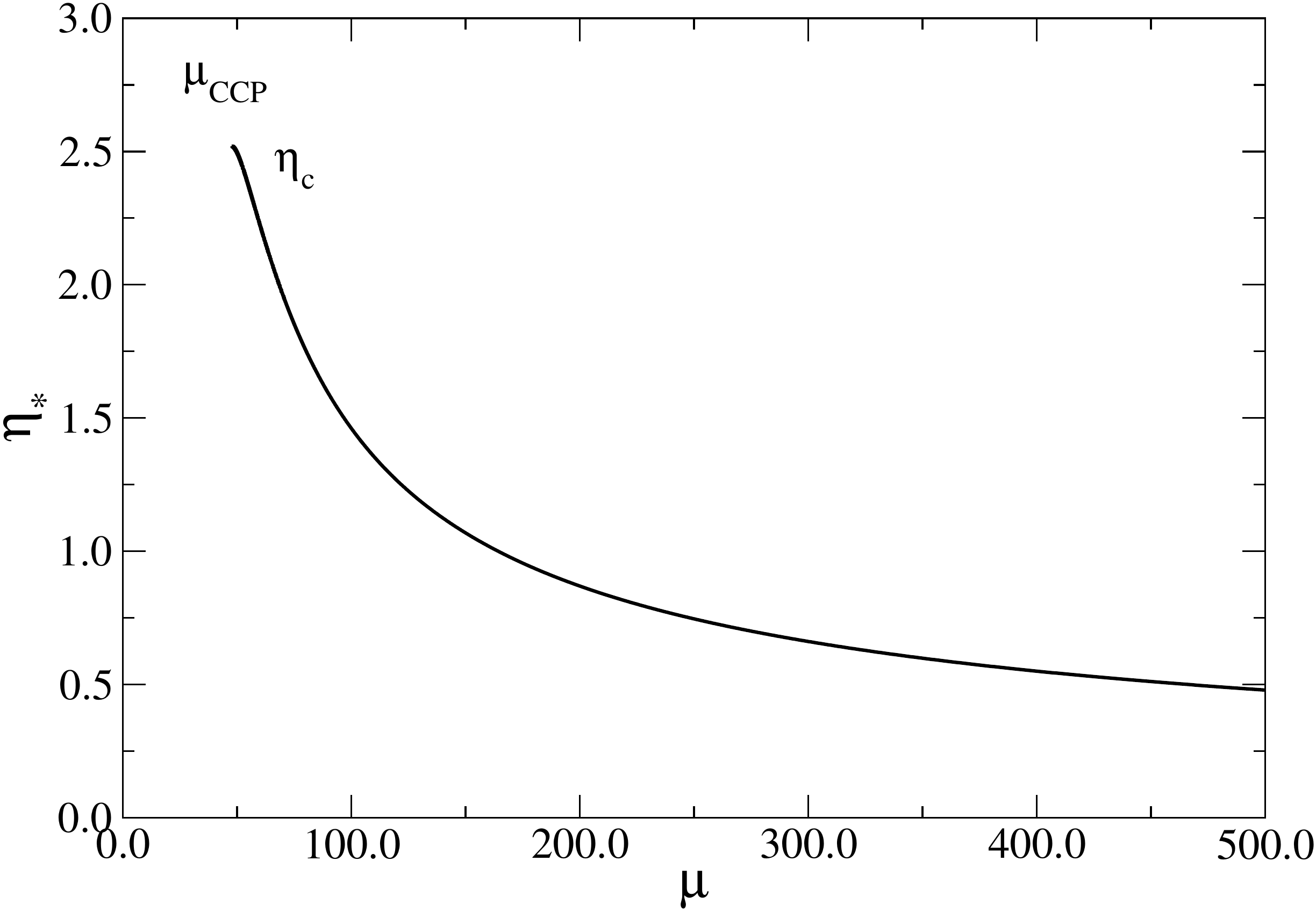}
\caption{The value of $\eta_*$ as a function of $\mu$.}
\label{muetastar}
\end{center}
\end{figure}

For $\mu>\mu_{\rm CCP}^{\rm app.}$, the minimum of the curve $\eta(x)$, denoted
$(x_*,\eta_*)$, is determined by the equations
\begin{eqnarray}
\label{ana13}
&&\ln(1-x_*)+\frac{x_*}{1-x_*}\nonumber\\
&+&\frac{3}{10\left
(\frac{\sqrt{\mu}}{\pi} -\frac{9}{5}\right
)(1-x_*)}-\frac{3}{10}\eta_c=0
\end{eqnarray}
and
\begin{eqnarray}
\label{ana14}
\eta_*=\frac{1}{\left
(\frac{\sqrt{\mu}}{\pi} -\frac{9}{5}\right
)(1-x_*)}.
\end{eqnarray}
Instead of solving Eq. (\ref{ana13}) for $x_*$ as a function of $\mu$, it is
simpler to study the inverse function
\begin{equation}
\label{ana16}
\mu(x_*)=\left \lbrack\frac{9\pi}{5}+\frac{\frac{3\pi}{10(1-x_*)}}{\frac
{ 3 }{10}\eta_c-\ln(1-x_*)-\frac{x_*}{1-x_*}}\right\rbrack^2.
\end{equation}
The values of $x_*$ and $\eta_*$
characterizing the minimum of the curve $\eta(x)$ as a function of
$\mu$ are plotted in Figs. \ref{xmu} and \ref{muetastar}. For $\mu\rightarrow
\mu_{\rm CCP}^{\rm app.}=47.6$, we find that $x_*\rightarrow 0$ and
$\eta_*\rightarrow \eta_c$. For $\mu\rightarrow +\infty$, we find that
$x_*\rightarrow x_*^c$, where $x_*^c$ is the solution of the equation
\begin{eqnarray}
\label{ana17}
\ln(1-x_*)+\frac{x_*}{1-x_*}-\frac{3}{10}\eta_c=0.
\end{eqnarray}
We numerically obtain $x_*^c\simeq 0.640$. We then find that 
\begin{eqnarray}
\label{ana18}
\eta_*\sim \frac{\pi}{\sqrt{\mu}(1-x_*)}\sim \frac{8.73}{\sqrt{\mu}}\rightarrow
0.
\end{eqnarray}

\begin{figure}[!h]
\begin{center}
\includegraphics[clip,scale=0.3]{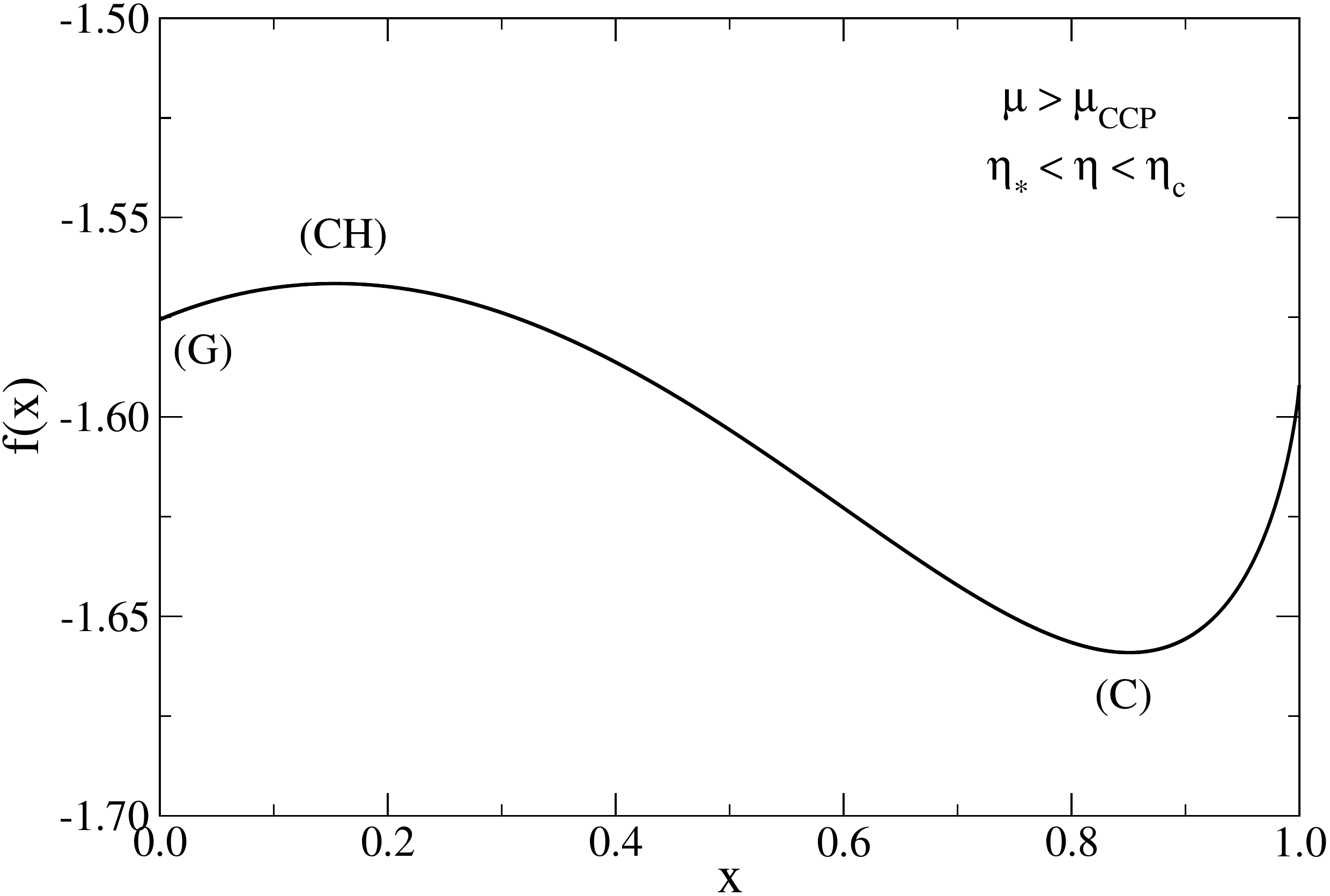}
\caption{The function $f(x)$ for $\mu>\mu_{\rm CCP}^{\rm app}$ and
$\eta_*<\eta<\eta_c$ (specifically $\mu=100$ and $\eta=1.8$) where three
equilibrium states exist.}
\label{xf}
\end{center}
\end{figure}

Let us now consider more specifically the function $f(x)$ giving the free
energy as a function of the core mass $x$ for a given value of $\mu$ and
$\eta$ (see Fig. \ref{xf}). Using Eq. (\ref{ana9b}), we can rewrite Eq.
(\ref{ana7}) as
\begin{eqnarray}
\label{ana19}
f(x)&=&-\frac{\sqrt{\mu}}{2\pi}x^2+\frac{1}{\eta}
(1-x)\left\lbrack -\frac{3}{10}\eta_c+\ln(1-x)-1\right\rbrack\nonumber\\
&-&\frac{3}{2}x(1-x)-\frac{3}{5}(1-x)^2.
\end{eqnarray}
Its first derivative is 
\begin{equation}
\label{ana20b}
f'(x)=-\frac{\sqrt{\mu}}{\pi}x+\frac{3}{10}\frac{\eta_c}{\eta}-\frac{1}{\eta}
\ln(1-x)-\frac{3}{10}+\frac{9}{5}x.
\end{equation}
The condition $f'(x)=0$ determines the equilibrium states as we have just
seen. The stability of these equilibrium states in the canonical ensemble is
then
determined by the sign of the second derivative of the free energy 
\begin{eqnarray}
\label{ana20}
f''(x)=-\frac{\sqrt{\mu}}{\pi}+\frac{1}{\eta(1-x)}+\frac{9}{5}.
\end{eqnarray}
An equilibrium state is stable when $f''(x)>0$, corresponding to a  minimum of
free energy, and  unstable when $f''(x)<0$, corresponding to a maximum of free
energy. Coming back to the function $f(x)$, its values  at
$x=0$ and $x=1$ are
\begin{eqnarray}
\label{ana21}
f(0)=-\frac{1}{\eta}\left (\frac{3}{10}\eta_c+1\right
)-\frac{3}{5}
\end{eqnarray}
and
\begin{eqnarray}
\label{ana22}
f(1)= -\frac{\sqrt{\mu}}{2\pi}.
\end{eqnarray}
For $x\rightarrow 0$, we find that 
\begin{equation}
\label{ana23}
f(x)=f(0)+\frac{3}{10}\left (
\frac{\eta_c}{\eta}-1\right ) x+...
\end{equation}
The term in parenthesis is positive when $\eta<\eta_c$ and negative  when
$\eta>\eta_c$. Since the function $f(x)$ is
defined for $x\ge 0$, the solution $x=0$ (purely gaseous phase) is a local
minimum of $f(x)$ for
$\eta<\eta_c$ even though $f'(0)\neq 0$. We shall therefore consider that this
solution is a stable equilibrium state.

We are now ready to perform the complete analysis of the equilibrium states of
our simple analytical model. We note that the function $\eta(x)$ defined by
Eq. (\ref{ana10}) is the counterpart of the function $\eta(\chi)$ defined in
Sec. \ref{sec_ptb}.

For $\mu<\mu_{\rm CCP}^{\rm app}$, the curve $\eta(x)$ is made of a vertical
branch at $x=0$ up to $\eta=\eta_c$, then it increases monotonically up to
infinity (see Fig. \ref{xeta}). This is similar to Figs.
\ref{chietaMU100} and \ref{multimu2}. For $\eta<\eta_c$ there is a unique
equilibrium
state with $x=0$ corresponding to the gaseous phase (G). For
$\eta>\eta_c$ there is a unique equilibrium state with $x>0$ corresponding to
the condensed phase (C). They are both stable (minima of free energy).
There is no phase transition in the present situation. Here, $\eta_c$ just
separates the gaseous configurations from the condensed configurations.

For $\mu>\mu_{\rm CCP}^{\rm app}$,  the curve $\eta(x)$ is made of a vertical 
branch at $x=0$ up to $\eta=\eta_c$, then it decreases up to $\eta_*$ and
finally it increases up to infinity (see Fig. \ref{xeta}). This is similar to
Figs. \ref{chietaMU39797} and \ref{fel}. When 
$\eta<\eta_*$, there is a unique equilibrium state with $x=0$. It corresponds to
the gaseous phase (G). 
When 
$\eta>\eta_c$, there is a unique equilibrium state with $x\simeq 1$. It
corresponds to the condensed phase (C). They are both stable (minima of free
energy).
When $\eta_*<\eta<\eta_c$ there are three equilibrium states (see Fig.
\ref{xf}): (i) a gaseous
phase (G) with $x=0$; (ii) a core-halo phase (CH) with $x\ll 1$;
(iii) a condensed phase (C) with
$x\simeq 1$. Let us analyze these solutions in more detail  in the limit
$\mu\rightarrow
+\infty$:

(i) The gaseous solution (G) corresponds to a purely isothermal  halo
without soliton. The core mass is equal to zero: $x_1=0$. This solution is
stable, being a minimum of free energy, although the derivative of $f(x)$ is not
defined at $x=0$ (as explained above).

(ii) The core-halo solution (CH) corresponds to an isothermal  halo
harboring a central soliton with a small mass. From Eq. (\ref{ana9}), we find
that the core
mass scales as
\begin{eqnarray}
\label{ana24}
x_2\propto \frac{1}{\sqrt{\mu}},
\end{eqnarray}
leading to the results of Sec. \ref{sec_sim}. 
Substituting Eq. (\ref{ana24}) into Eq. (\ref{ana20}) we find that
$f''(x_2)=-\sqrt{\mu}/\pi<0$. Therefore, the core-halo solution is
unstable in the canonical ensemble. It may, however, be stable in the
microcanonical ensemble (see Sec. \ref{sec_ineq}).

(iii) The condensed solution (C) corresponds to a solitonic core surrounded
by a tenuous atmosphere. From Eq. (\ref{ana9}), we find that the core mass
scales as 
\begin{eqnarray}
\label{ana30}
1-x_3\propto e^{-\sqrt{\mu}\eta/\pi},
\end{eqnarray}
showing that the soliton contains almost all the mass. Substituting Eq.
(\ref{ana30}) into Eq. (\ref{ana20}) we find that $f''(x_3)\sim
(1/\eta)e^{\eta\sqrt{\mu}/\pi}\rightarrow +\infty$. Therefore, the condensed
solution is stable.

We now show that the result (\ref{ana24}) can be obtained from the ``velocity
dispersion tracing'' relation
\begin{eqnarray}
\label{ana31b}
v_c^2\sim v_h^2
\end{eqnarray}
stating that the
velocity dispersion in the core $v_c^2\sim GM_c/R_c$ is of the same order as the
velocity dispersion in the halo $v_h^2\sim GM_h/r_h$. This condition gives
\begin{eqnarray}
\label{ana31}
M_c\propto \frac{R_c}{r_h}M_h.
\end{eqnarray}
Using $\mu\propto ({r_h}/{R_c})^2$ it can be rewritten as Eq. (\ref{ana24}).
Therefore, Eq.  (\ref{ana24}) is fully consistent with the ``velocity
dispersion tracing'' relation (\ref{ana31b}) which, in the noninteracting case
($a_s=0$), leads to the core mass - halo mass relation found by Schive {\it et
al.} \cite{ch3} (see the
discussion in  Mocz {\it et al.} \cite{mocz}). In the present case
(self-interacting bosons), since $R_c$ is independent of $M_c$ and since
$M_h\propto \Sigma_0 r_h^2$, we get
\begin{eqnarray}
\label{ana32}
M_c\propto R_c\sqrt{\Sigma_0 M_h}\propto \left (\frac{a_s\hbar^2\Sigma_0
M_h}{Gm^3}\right )^{1/2}\propto M_h^{1/2}
\end{eqnarray}
in agreement with Eqs. (\ref{ap1b}) and (\ref{app8}) obtained by two
different methods (in total, we have presented four independent arguments
leading to this relation).

We conclude this Appendix by presenting preliminary results obtained for
noninteracting bosons and fermions (they will be developed in a specific paper
\cite{forthcoming}). 

The relation $M_c(M_h)$ can be obtained either by minimizing the free energy
with respect to $M_c$ \cite{forthcoming} or, more directly, by assuming
the  ``velocity dispersion tracing'' relation (\ref{ana31b}) or
(\ref{ana31}).\footnote{We note that this relation is not obvious {\it a priori}
and
that other relations are possible such as the ``energy tracing'' relation
\cite{mocz}. The fact that relation (\ref{ana31b}) can be justified from a free
energy minimization principle as shown in Ref. \cite{forthcoming} provides a
physical basis for it.} In the case of
noninteracting bosons, using the mass-radius relation $M_cR_c=9.95\,
\hbar^2/Gm^2$ \cite{membrado,prd1,prd2}, we
obtain 
\begin{eqnarray}
\label{ana33}
M_c\propto \left (\frac{\hbar^2M_h}{Gm^2r_h}\right )^{1/2}\propto \left
(\frac{\hbar^4\Sigma_0M_h}{G^2m^4}\right )^{1/4}\propto M_h^{1/4}.
\end{eqnarray}
In the case of fermions, using the
mass-radius relation  $M_cR_c^3=1.49\times
10^{-3} h^6/G^3m^8$ \cite{chandra}, we get 
\begin{equation}
\label{ana34}
M_c\propto \frac{\hbar^{3/2}}{m^2}\left (\frac{M_h}{Gr_h}\right
)^{3/4}\propto \left (\frac{\hbar^3}{m^4}\right )^{1/2}\left (\frac{M_h
\Sigma_0}{G^2}\right )^{3/8}\propto M_h^{3/8}.
\end{equation}
 We now note that the
mass-radius relation $M_h\propto r_h^2$ used in the
present paper (based on the observation that the surface
density of DM halos is constant \cite{kormendy,spano,donato}) is
different from the  mass-radius relation $M_v\propto r_v^3$  used by Schive {\it
et al.}  \cite{ch3}. This suggests that the halo
mass $M_v$
considered by these authors is different from the halo mass $M_h$ considered
here. Using the relation $GM_h/r_h\sim GM_v/r_v$ (consistent with the velocity
dispersion tracing relation), we get $M_h\propto M_v^{4/3}$.
Using this
relation\footnote{From
$M_h\propto r_h^2$ and $v_h^2=GM_h/r_h \propto r_h$, we get $v_h\propto
M_h^{1/4}$ which is the Tully-Fisher relation \cite{tf,mcgaugh}. Using 
$M_h\propto M_v^{4/3}$ we get
$v_h\propto M_v^{1/3}$ which is consistent with the scaling reported in
\cite{ferrarese,rbb}. This gives some confidence to the relation $M_h\propto
M_v^{4/3}$.} together with the $M_c(M_h)$ relations obtained previously, we
obtain for
self-interacting bosons:
\begin{eqnarray}
\label{ana35}
M_c\propto M_v^{2/3},
\end{eqnarray}
for noninteracting bosons:
\begin{eqnarray}
\label{ana36}
M_c\propto M_v^{1/3},
\end{eqnarray}
and for fermions:
\begin{equation}
\label{ana37}
M_c\propto M_v^{1/2}.
\end{equation}
The scaling of Eq.
(\ref{ana36}) is consistent with the numerical results of Schive {\it et al.}
\cite{ch3}. The scaling of Eq. (\ref{ana37}), previously
given in the form of Eq. (\ref{ana34}) in Appendix H of \cite{clm2}, is
consistent with the scaling of Ruffini {\it et al.} \cite{rar} who find $0.52$
instead of $1/2$. We have also shown that our procedure of minimizing the free
energy
is consistent with the velocity dispersion tracing relation 
(\ref{ana31b}), leading to the relation of  Schive {\it et al.} \cite{ch3} for
noninteracting bosons, as
explained in \cite{mocz}. Therefore, our approach provides a
 justification of the results of  Schive {\it et al.} \cite{ch3} and 
 Ruffini {\it et al.} \cite{rar} from thermodynamical arguments.
The prefactor in Eqs.
(\ref{ana35})-(\ref{ana37}) can
be obtained from our approach [like Eq. (\ref{ap1b}) for self-interacting
bosons] but this requires additional calculations that will be presented
in a future work \cite{forthcoming}.

{\it Remark:} It is interesting to study how the mass $M_c$, the radius $R_c$,
the velocity dispersion $GM_c/R_c$ and the energy $GM_c^2/R_c$  in the core
behave in the classical limit $\hbar\rightarrow 0$. For self-interacting
bosons, using Eq. (\ref{ana32}), we find
$M_c\sim R_c\sim GM_c^2/R_c\sim
\hbar\rightarrow 0$ and $GM_c/R_c\sim 1$. For noninteracting bosons, using Eq.
(\ref{ana33}), we find $M_c\sim R_c\sim GM_c^2/R_c\sim
\hbar\rightarrow 0$ and $GM_c/R_c\sim 1$. For fermions, using Eq.
(\ref{ana34}), we find $M_c\sim
R_c\sim GM_c^2/R_c\sim
\hbar^{3/2}\rightarrow 0$ and $GM_c/R_c\sim 1$.

\section{Phase transitions leading to a boson or fermion ball mimicking a
supermassive black hole at the centers of elliptical galaxies}
\label{sec_bhell}

In this Appendix, we consider the possibility that the supermassive black holes
of mass  $M\sim 10^9\ M_{\odot}$  that reside at the centers of elliptical
galaxies are actually boson or fermion balls
corresponding to a purely condensed phase (C) without halo.

Let us consider a dilute gas of bosons or fermions with a mass $M \sim 10^9\
M_{\odot}$ and a sufficiently large radius $R$ so that a canonical phase
transition
can take place (this requires that $\mu>\mu_{\rm CCP}$ so that the caloric
curve has the shape of Figs. \ref{fel} and \ref{le5}). In that case, below the
critical temperature $T_c$ (corresponding to $\eta_c\simeq 2.52$), the gas
undergoes a gravitational collapse (isothermal collapse) and forms a
compact object (completely condensed boson or fermion star) of about the same
mass $M\sim 10^9\
M_{\odot}$ but with a much smaller radius $R_*\ll R$. This corresponds to a
zeroth
order phase transition from a gaseous phase (G) to a condensed phase (C). The
condensed solution (C) represents a pure
boson or fermion star without DM halo.

The boson or fermion star (compact object) may mimic a supermassive BH  of mass
$M \sim 10^9\ M_{\odot}$  at the center of an
elliptical galaxy if its maximum mass $M_{\rm
max}$ is close to $M \sim 10^9\ M_{\odot}$. In that case, the boson or fermion
star is very relativistic and general relativity must be taken into account.

Using the results of Appendix F of \cite{bectcoll} we can estimate the
characteristics  of the corresponding DM particle.\footnote{Similar numerical
applications have been made in  Appendix F
of \cite{bectcoll} to model the compact object Sgr $A^*$ of mass
$M=4.2\times 10^{6}\, M_{\odot}$ at the center of our Galaxy (purported to be a
supermassive black hole) by a general relativistic boson or fermion star.}  For
noninteracting bosons,
using $M_{\rm
max}=0.633\, \hbar c/Gm$ and $R_*=9.53\,
GM_{\rm max}/c^2$ \cite{kaup,rb} we find that  $M_{\rm
max}=10^9\ M_{\odot}$ (with $R_*=4.56\times 10^{-4}\, {\rm pc}$) provided
that $m=8.46\times 10^{-20}\,
{\rm eV/c^2}$.  For self-interacting
bosons in the TF limit, using  $M_{\rm
max}=0.307\, \hbar c^2\sqrt{a_s}/(Gm)^{3/2}$ and $R_*=6.25\, GM_{\rm max}/c^2$
\cite{colpi,chavharko}
we find that $M_{\rm
max}= 10^9\ M_{\odot}$ (with $R=2.99\times
10^{-4}\, {\rm pc}$) provided that $a_s/m^3=7.86\times
10^{-10}\,
{\rm fm/(eV/c^2)^3}$. For fermions, using 
$M_{\rm
max}=0.384\, (\hbar c/G)^{3/2}/m^2$ and $R_{*}=8.73\, GM_{\rm max}/c^2$
\cite{ov}, we
find that $M_{\rm
max}= 10^9\ M_{\odot}$ (with
$R_*=4.18\times 10^{-4}\, {\rm pc}$) provided that $m=25.0\, {\rm keV/c^2}$.

The results are essentially the same in the microcanonical ensemble but the
existence of a microcanonical phase transition requires an initially  larger
system size ($\mu>\mu_{\rm MCP}$) so that the caloric curve has the shape of
Fig. \ref{le5}. On the other hand,
the compact object resulting from the gravitational collapse (gravothermal
catastrophe) at $E_c$ (corresponding to $\Lambda_c\simeq 0.335$) contains a
fraction ($\sim 1/4$) of the initial mass \cite{ijmpb,ac}. The
formation of the compact object (implosion) is accompanied by the expulsion of a
hot envelope (explosion). This core-halo structure is reminiscent of the red
giant structure and supernova phenomenon  in the context of compact
stars (white dwarfs and neutron stars).

To study these phase transitions in detail we have to use general relativity.
This has been done in the case of fermions in Refs. \cite{bvr,ac}. Numerical
applications have been made for fermionic particles of mass $m=17.2\,
{\rm keV/c^2}$. It is shown that they can 
form fermion balls of mass  $M\sim 10^9\ M_{\odot}$ similar to the mass of the
presumed black holes that reside at the centers of elliptical galaxies.

{\it Remark:} it is important to note that, in the present Appendix, we are
considering the purely condensed solution (C), not the core-halo solution (CH).
Since we are not trying to construct a self-consistent ``core $+$
halo'' solution we do not face the difficulties encountered in Sec.
\ref{sec_cheat}.
Furthermore, the solution
(C) is canonically stable while the solution (CH) is canonically unstable.
Therefore, the results of this Appendix suggest that large galaxies may contain
a dark matter compact object (boson or fermion star) mimicking a supermassive
black hole but that this object is {\it not} surrounded by a dark matter halo.
If this scenario is correct, large galaxies should not contain a dark matter
halo,  just a DM
compact object.  Note that other scenarios are possible such as those
considered
in Sec. \ref{sec_astapp} in which large galaxies contain a bulge (soliton), or
a central black hole, surrounded by a DM halo.

\section{Problems
with the boson or fermion ball scenario
to mimic a supermassive black hole}
\label{sec_prob}

In this Appendix, we show the impossibility for a noninteracting boson or
fermion ball to simultaneously mimic a supermassive BH of
mass $M \sim 10^9\, M_{\odot}$ at the centers of elliptical galaxies (see
Appendix \ref{sec_bhell})
and  a compact object (Sgr $A^*$) of mass $M=4.2\times 10^{6}\,
M_{\odot}$ and sufficiently small radius $R<R_{\rm P}=6\times 10^{-4}\, {\rm
pc}$ at the center of our Galaxy (see Sec.
\ref{sec_cheat}).\footnote{If a boson or fermion
ball can mimic a BH of mass $M \sim 10^9\,
M_{\odot}$ this means that $M_{\rm max}\sim 10^9\,
M_{\odot}$. In that case, it cannot mimic a BH of smaller mass, $M\sim
10^6-10^9$, because it would be nonrelativistic ($M\ll M_{\rm max}$) and it
cannot mimic a BH of larger mass  because it would be
unstable ($M>M_{\rm max}$). However, the compact object of mass $M=4.2\times
10^{6}\,
M_{\odot}\ll M_{\rm max}\sim 10^9\, M_{\odot}$ that resides at the center of our
Galaxy is not necessarily a black hole, not even a relativistic object. We just
require that it has a radius $R<R_{\rm P}=6\times 10^{-4}\, {\rm
pc}=1492\, R_S$ in order to be consistent with the observations. Therefore, we
can  use the nonrelativistic mass-radius relation of a boson or fermion
ball to describe this object.} Our discussion confirms and extends the
arguments given in Ref.
\cite{genzel}.

In Appendix \ref{sec_bhell}, we have determined the characteristics that the DM
particle must have so that the maximum mass of the associated boson or fermion
ball is
$M_{\rm max}= 10^9\, M_{\odot}$. Below, we show that the associated
boson or fermion ball of mass $M=4.2\times 10^{6}\, M_{\odot}$ has a radius
$R>R_{\rm P}=6\times 10^{-4}\, {\rm pc}$ so that it cannot account for the
compact object (Sgr $A^*$) at the center of our
Galaxy.

For noninteracting bosons, using the mass-radius relation $MR=9.95\,
\hbar^2/Gm^2$ \cite{membrado,prd1,prd2} and taking $m=8.46\times
10^{-20}\, {\rm eV/c^2}$ (see
Appendix \ref{sec_bhell}) we find  $R=0.283\, {\rm
pc}>R_{\rm P}$. The
constraint $R<R_{\rm P}$ implies 
$m>1.84\times 10^{-18}\, {\rm eV/c^2}$. However, in this case, the boson star
cannot mimic a supermassive BH of
mass $M_{\rm max}\sim 10^9\, M_{\odot}$ at the centers of elliptical galaxies
since this requires $m=8.46\times 10^{-20}\, {\rm eV/c^2}$. Indeed, if we take
$m>1.84\times 10^{-18}\, {\rm eV/c^2}$ we find $M_{\rm max}=4.60\times 10^7\,
M_{\odot}<10^9\,
M_{\odot}$.

For fermions, using the mass-radius relation $MR^3=1.49\times 10^{-3}
h^6/G^3m^8$ \cite{chandra} and taking $m=25.0\, {\rm keV/c^2}$ (see Appendix
\ref{sec_bhell}) we find $R=4.81\times 10^{-3}\, {\rm
pc}>R_{\rm P}$. The constraint
$R<R_{\rm P}$ implies $m>54.5\, {\rm keV/c^2}$.
However, in this case, the fermion star cannot mimic a supermassive BH of
mass $M_{\rm max}\sim 10^9\, M_{\odot}$ at the centers of elliptical galaxies
since this requires $m=25.0\, {\rm keV/c^2}$. Indeed, if we take $m>54.5\,
{\rm keV/c^2}$ we find $M_{\rm max}=2.11\times 10^8\, M_{\odot}<10^9\,
M_{\odot}$.\footnote{In relation to this fundamental incompatibility,
we note that Arg\"uelles and Ruffini
\cite{ar} consider a fermion of mass $10\, {\rm keV/c^2}$ to mimic a SMBH of
mass $10^9 M_{\odot}$ at the centers of elliptical galaxies while Arg\"uelles
{\it et al.} \cite{krut} consider another fermion of larger mass, $48\, {\rm
keV/c^2}$, to mimic the compact object at the center of our Galaxy.}

Interestingly, it turns out that self-interacting boson stars can
simultaneously mimic a supermassive BH of
mass $M \sim 10^9\, M_{\odot}$ and a
compact object like Sgr $A^*$. To our knowledge, this result has not been
pointed out previously. For self-interacting bosons in the TF limit, using the
fact that their radius
is
$R=\pi(a_s\hbar^2/Gm^3)^{1/2}$ \cite{tkachev,maps,leekoh,goodman,arbey,bohmer,
prd1} and taking $a_s/m^3=7.86\times
10^{-10}\,
{\rm fm/(eV/c^2)^3}$ (see Appendix
\ref{sec_bhell}) we find $R=4.90\times 10^{-4}\, {\rm
pc}<R_{\rm P}$. The
constraint
$R<R_{\rm P}$ implies
$a_s/m^3<1.18\times 10^{-9}\, {\rm fm/(eV/c^2)^3}$. In that case, the
bosonic
particle can mimic a supermassive BH of
mass $M_{\rm max}\sim 10^9\, M_{\odot}$ at the centers of elliptical galaxies
since this requires $a_s/m^3=7.86\times 10^{-10}\,
{\rm fm/(eV/c^2)^3}$. Indeed, if we take $a_s/m^3\lesssim 1.18\times 10^{-9}\,
{\rm
fm/(eV/c^2)^3}$
 we find $M_{\rm max}=1.22\times 10^9\,
M_{\odot}\gtrsim 10^9\,
M_{\odot}$.

However, there are important problems with the boson or fermion ball model that
also concern the case of self-interacting bosons. In particular, the
characteristics
of the DM particle that we find in Appendix 
\ref{sec_bhell} and in this Appendix are not consistent with the 
characteristics of the DM particle obtained from the  minimum halo
model of  Sec. \ref{sec_gs}. Indeed, we have argued that the most
compact halos (dSphs like Fornax) should correspond to the ground state of the
self-gravitating boson or fermion gas. This immediately fixes the
characteristics of the DM particle. Comparing the results found in Sec.
\ref{sec_gs} with the results found above, we see that they are not consistent.
As a result, if we determine the characteristics of
the DM particle from the minimum halo model (see Sec.
\ref{sec_gs}), then the boson or fermion ball corresponding to the
self-consistent core-halo (CH) solution that we obtain for a DM halo similar to
the Milky Way has the form of a large bulge (see Sec. \ref{sec_ac}), not the
form of a small compact object like  a BH (see Sec. \ref{sec_cheat}). In
addition, even if we relax the constraint from the minimum halo model we cannot
get a self-consistent core-halo solution mimicking a BH as shown in Sec.
\ref{sec_cheat}.

{\it Remark:} It is not excluded (actually it is even very
likely) that DM is made of different types of particles (bosons and
fermions) with
different characteristics. Some of these particles (like bosons) could form
a solitonic bulge and other particles (like fermions) could form a fermion ball
mimicking SMBHs. However, accounting for several species
obviously introduces arbitrariness in the models and limits therefore their
predictive power. This is why, in this paper, we have just considered one
type of
DM particle.

\section{Solution of an apparent paradox related to the mass-radius relation of
dark matter halos}
\label{sec_paradox}

We have seen that the ground state of the GPP equations (\ref{intro1}) and
(\ref{intro2}) corresponds to a soliton.  For noninteracting bosons, the
mass-radius relation of the soliton is given by 
\cite{membrado,prd1,prd2}
\begin{eqnarray}
\label{paradox1}
R=9.95\frac{\hbar^2}{GMm^2},
\end{eqnarray}
implying that the radius decreases as the mass increases. For self-interacting
bosons, in the TF approximation, the soliton has
a unique radius 
\begin{eqnarray}
\label{paradox2}
R=\pi \left (\frac{a_s\hbar^2}{Gm^3}\right )^{1/2}
\end{eqnarray}
which is independent of its
mass \cite{tkachev,maps,leekoh,goodman,arbey,bohmer,
prd1}. Clearly, these results are in contradiction with the
universality
of the surface density of DM halos
[see Eq. (\ref{observation})] implying that  the radius increases with
the mass as $r_h\propto M_h^{1/2}$.

This apparent paradox was pointed out in Appendix F of Ref. \cite{clm2} and in
the
Introduction of Ref. \cite{bdo}. This difficulty was rediscussed later by Deng
{\it et al.}
\cite{deng} who conclude that ultralight dark matter may not
be
able to solve
the core-cusp problem. Alternatively, we had suggested in our previous works
\cite{clm2,bdo} that the above-mentioned apparent paradox could be solved by
accounting for the
presence of an (isothermal) atmosphere surrounding the solitonic core of large
DM halos.

More precisely, we argued that a pure soliton describes only the ground state
of the GPP equations (\ref{intro1}) and (\ref{intro2}) corresponding to
ultracompact halos such as Fornax (see Sec. \ref{sec_gs}). 
The mass-radius relations (\ref{paradox1}) and (\ref{paradox2})  apply only to
these ultracompact halos. Larger halos contain a
solitonic core plus an atmosphere resulting from quantum interferences 
related to the complicated processes of gravitational cooling
\cite{seidel94,gul0,gul} and violent
relaxation \cite{lb}. The atmosphere is approximately isothermal. It is the
atmosphere
that fixes the size of large dark matter halos. As a result, we cannot apply the
mass-radius
relations (\ref{paradox1}) and (\ref{paradox2}) to large DM halos, but just
to their solitonic core. In this
sense, there is no paradox anymore and  ultralight dark matter may be able
to solve the core-cusp problem. 

The ideas sketched in Refs. \cite{clm2,bdo} have been confirmed in the present
paper. As
soon as
there is an (isothermal) atmosphere surrounding the solitonic core it is
possible to satisfy the constraint from Eq. (\ref{observation}). The BECDM
halos that we have constructed in this paper all satisfy this contraint. We thus
find that the mass-radius relation of large DM halos is given by Eq. (\ref{pv8})
in agreement with the observations \cite{kormendy,spano,donato}.

In the present paper, the universality of $\Sigma_0$ is imposed to our model as
an observational constraint (see Sec. \ref{sec_cst}). This implies that the
temperature of the
atmosphere of our DM halos must change precisely in order to satisfy this
constraint (see Fig. \ref{rayontemp}). This leads to the relations of Sec.
\ref{sec_pet}.

However, in Ref. \cite{logotrope} we have shown that it is possible
to predict the universal value of $\Sigma_0$ from a cosmological model based on
a logotropic equation of state. This model can be derived from
a generalized GPP equation similar to Eq. (\ref{intro3}) but involving a
nonlinearity of the form $-Am/|\psi|^2$ instead of $2k_B T\ln|\psi|$ [see Eq.
(C.56) of \cite{logotrope}]. In that case, we can
theoretically predict that 
\begin{equation}
\Sigma_0^{\rm th}=\left (\frac{B}{32}\right
)^{1/2}\frac{\xi_h}{\pi}\frac{\sqrt{\Lambda}c}{G}\simeq 133 \,
M_{\odot}/{\rm
pc}^2,
\label{lt6}
\end{equation}
where $\Lambda=1.00\times 10^{-35}\, {\rm s}^{-2}$ is the
effective cosmological constant of the model while
 $\xi_h=5.8458...$ and $B=3.53\times 10^{-3}$ are coefficients derived from the
theory (the consequences of this
relation are further discussed in \cite{forthcoming}). This suggests replacing
the
isothermal atmosphere of the present paper by a logotropic atmosphere. In that
case, our model will
be characterized by a universal constant $\Lambda$  instead of being
characterized by a temperature $T$ changing from halo to halo. This
logotropic model will be considered in a future contribution \cite{forthcoming}.
For the present, we think that it is better to develop our model with a more
conventional isothermal atmosphere as presented in this paper.

{\it Remark:} The same arguments apply to the fermionic model with a fermion
ball (core) and an isothermal atmosphere (halo). Again, the isothermal
atmosphere could be replaced by a logotropic atmosphere.


\begin{thebibliography}{99}

\bibitem{susy}  {\small  G. Jungman, M. Kamionkowski, K. Griest, Phys. Rep. {\bf
267}, 195 (1996)}
\bibitem{peeblesbook}{\small P.J.E. Peebles, {\it The Large-Scale
Structure of the Universe} (Princeton University Press, 1980)}
\bibitem{planck2013}{\small Planck Collaboration, Astron.
Astrophys. {\bf 571}, 66 (2014)}
\bibitem{planck2016}{\small Planck Collaboration, Astron.
Astrophys. {\bf 594}, A13 (2016)}
\bibitem{nfw}{\small J.F. Navarro, C.S. Frenk, S.D.M. White, Astrophys. J.
{\bf 462}, 563 (1996)}
\bibitem{observations}{\small A. Burkert, Astrophys. J. {\bf
447}, L25 (1995)}
\bibitem{satellites}{\small G. Kauffmann, S.D.M. White, B.
Guiderdoni, Mon. Not. R. astr. Soc. {\bf 264}, 201 (1993); A. Klypin, A.V.
Kravtsov, O. Valenzuela, Astrophys. J. {\bf 522}, 82 (1999); M. Kamionkowski,
A.R. Liddle, Phys. Rev. Lett. {\bf 84}, 4525 (2000)}
\bibitem{spergelsteinhardt}{\small D.N. Spergel, P.J. Steinhardt, Phys. Rev.
Lett.  {\bf 84}, 3760 (2000)}
\bibitem{wdm}{\small P. Bode, J.P. Ostriker, N. Turok, Astrophys. J. {\bf 556},
93 (2001)}
\bibitem{romano} {\small E. Romano-D\'iaz, I. Shlosman, Y. Hoffman, C. Heller,
Astrophys. J. {\bf 685}, L105 (2008); A. Pontzen, F. Governato, Nature {\bf
506}, 171 (2014); J. O\~norbe {\it et al.}  Mon. Not. R. astr. Soc. {\bf 454},
2092 (2015)}
\bibitem{markov}{\small M.A. Markov, Phys. Lett. {\bf  10}, 122
(1964)}
\bibitem{cmc1}{\small R. Cowsik and J. McClelland, Phys. Rev. Lett. {\bf  29},
669
(1972)}
\bibitem{cmc2}{\small R. Cowsik and J. McClelland, Astrophys. J. {\bf  180}, 7
(1973)}
\bibitem{gao}{\small J.G. Gao, R. Ruffini, Phys. Lett. {\bf  97B}, 388
(1980)}
\bibitem{stella}{\small R. Ruffini, L. Stella, Astron. Astrophys. {\bf 119}, 35
(1983)}
\bibitem{zcls}{\small J.L. Zhang, W.Y. Chau, K. Lake, J. Stone, Astrophys.
Space Sci. {\bf 96}, 417
(1983)}
\bibitem{cls}{\small W.Y. Chau, K. Lake, J. Stone, Astrophys. J. {\bf 
281}, 560 (1984)}
\bibitem{cl}{\small W.Y. Chau, K. Lake, Phys. Lett. {\bf 
134B}, 409 (1984)}
\bibitem{ir}{\small G. Ingrosso, R. Ruffini, Nuovo Cimento {\bf
101}, 369 (1988)}
\bibitem{gmr}{\small J.G. Gao, M. Merafina, R. Ruffini, Astron. Astrophys. {\bf
235}, 1  (1990)}
\bibitem{merafina}{\small M. Merafina, Nuovo Cimento {\bf
105}, 985 (1990)}
\bibitem{imrs}  {\small G. Ingrosso, M. Merafina, R. Ruffini and F.
Strafella, Astron. Astrophys. {\bf 258}, 223 (1992)}
\bibitem{vtt}{\small R.D. Viollier, D. Trautmann, G.B. Tupper,  Phys. 
Lett. B {\bf 306}, 79 (1993)}
\bibitem{bvn}{\small N. Bilic, R.D. Viollier, Phys. Lett. B {\bf  408}, 75
(1997)}
\bibitem{bmv}{\small N. Bilic, F. Munyaneza, R.D. Viollier,  Phys. Rev. D {\bf
59}, 024003 (1998)}
\bibitem{csmnras}{\small P.H. Chavanis, J. Sommeria, Mon. Not. R. Astron. Soc.
{\bf 296}, 569 (1998)}
\bibitem{bvr}{\small N. Bilic, R.D. Viollier, Eur. Phys. J. C {\bf  11}, 173
(1999)}
\bibitem{pt}  {\small P.H. Chavanis, Phys. Rev. E {\bf 65}, 056123 (2002)}
\bibitem{dark}  {\small P.H. Chavanis, {\it The self-gravitating Fermi gas}, in
Dark Matter in Astro- and Particle Physics, edited by H.V. Klapdor-Kleingrothaus
and R.D. Viollier (Springer, 2002)}
\bibitem{bmtv}  {\small N. Bilic, F. Munyaneza, G.B. Tupper, R.D. Viollier,
Prog. Part. Nucl. Phys.  {\bf 48}, 291 (2002)}
\bibitem{btv}{\small N. Bilic, G.B. Tupper, R.D. Viollier, Lect. Notes Phys.
{\bf  616}, 24 (2003)}
\bibitem{rieutord}  {\small P.H. Chavanis, M. Rieutord, Astron. Astrophys. {\bf
412}, 1 (2003)}
\bibitem{ijmpb}  {\small P.H. Chavanis, Int. J. Mod. Phys. B {\bf 20}, 3113
(2006) }
\bibitem{dvs1}  {\small C. Destri, H.J. de Vega, N.G. Sanchez, New Astronomy
{\bf  22}, 39 (2013)}
\bibitem{dvs2}  {\small C. Destri, H.J. de Vega, N.G. Sanchez, Astroparticle
Physics
{\bf  46}, 14 (2013)}
\bibitem{vss}  {\small  H.J. de Vega, P. Salucci, N.G. Sanchez, Mon. Not. R.
Astron. Soc. {\bf 442}, 2717 (2014) }
\bibitem{urbano}  {\small V. Domcke, A. Urbano, JCAP {\bf 01},
002 (2015)}
\bibitem{rar}{\small R. Ruffini, C.R. Arg\"uelles, J.A. Rueda, Mon. Not. R.
Astron. Soc. {\bf 451}, 622 (2015)}
\bibitem{clm1}{\small P.H. Chavanis, M. Lemou, F. M\'ehats, Phys. Rev. D {\bf
91}, 063531 (2015)}
\bibitem{clm2}{\small P.H. Chavanis, M. Lemou, F. M\'ehats,  Phys. Rev. D {\bf
92}, 123527 (2015)}
\bibitem{vs1}  {\small  H.J. de Vega, N.G. Sanchez, Int. J. Mod. Phys. A {\bf
31}, 1650073 (2016)}
\bibitem{krut}  {\small C.R. Arg\"uelles, A. Krut, J.A. Rueda,
R. Ruffini, arXiv:1606.07040}
\bibitem{vs2}  {\small  H.J. de Vega, N.G. Sanchez, Eur. Phys. J. C {\bf 77},
81 (2017)}
\bibitem{rsu}  {\small  L. Randall, J. Scholtz, J. Unwin, Mon. Not. R.
Astron. Soc. {\bf 467}, 1515 (2017) }
\bibitem{flat}{\small  V. C. Rubin, W.K. Ford, Astrophys. J. {\bf 159}, 379
(1970); V.C. Rubin, W.K. Ford, N. Thonnard, Astrophys. J. {\bf
238}, 471 (1980); A. Bosma, Astron. J. {\bf 86}, 1791 (1981); M. Persic, P.
Salucci, F. Stel, Mon. Not. R. astr. Soc.
 {\bf 281}, 27 (1996)}
\bibitem{marsh}{\small D. Marsh, Phys. Rep. {\bf 643}, 1 (2016)}
\bibitem{baldeschi}{\small M.R. Baldeschi, G.B. Gelmini, R. Ruffini, Phys. Lett.
B {\bf  122}, 221 (1983)}
\bibitem{khlopov}{\small M.Yu. Khlopov, B.A. Malomed, Ya.B. Zeldovich, Mon. Not.
R. astr. Soc. {\bf  215}, 575 (1985) }
\bibitem{membrado}{\small M. Membrado, A.F. Pacheco, J. Sanudo, Phys. Rev. A
{\bf  39}, 4207 (1989)}
\bibitem{sin}{\small S.J. Sin, Phys. Rev. D {\bf  50}, 3650 (1994)}
\bibitem{jisin}{\small S.U. Ji, S.J. Sin, Phys. Rev. D {\bf  50}, 3655 (1994)}
\bibitem{leekoh}{\small J.W. Lee, I. Koh, Phys. Rev. D {\bf  53}, 2236 (1996)}
\bibitem{schunckpreprint}{\small F.E. Schunck, [astro-ph/9802258]}
\bibitem{matosguzman}{\small T. Matos,
F.S. Guzm\'an, F. Astron. Nachr. {\bf 320}, 97 (1999)}
\bibitem{sahni}{\small V. Sahni, L. Wang
Phys. Rev. D {\bf 62}, 103517 (2000)}
\bibitem{guzmanmatos}{\small F.S. Guzm\'an,
T. Matos, Class. Quantum Grav.  {\bf 17}, L9 (2000)}
\bibitem{hu}{\small W. Hu, R. Barkana, A. Gruzinov, Phys. Rev. Lett. {\bf  85},
1158 (2000)}
\bibitem{peebles}{\small P.J.E. Peebles, Astrophys. J. {\bf 534}, L127 (2000)}
\bibitem{goodman}{\small J. Goodman, New Astronomy {\bf 5}, 103 (2000)}
\bibitem{mu}{\small T. Matos, L.A. Ure\~na-L\'opez,
Phys. Rev. D {\bf 63}, 063506 (2001)}
\bibitem{arbey1}{\small A. Arbey, J. Lesgourgues, P. Salati, Phys. Rev. D {\bf
64}, 123528 (2001)}
\bibitem{silverman1}{\small M.P. Silverman, R.L. Mallett, Class. Quantum Grav.
{\bf  18}, L103 (2001)}
\bibitem{matosall}{\small M. Alcubierre, F.S. Guzm\'an, T. Matos, D. N\'u\~nez,
L.A. Ure\~na-L\'opez, P.  Wiederhold, Class. Quantum. Grav. {\bf 19}, 5017
(2002)}
\bibitem{silverman}{\small M.P. Silverman, R.L. Mallett, Gen. Rel. Grav. {\bf
34}, 633 (2002)}
\bibitem{lesgourgues}{\small J. Lesgourgues,  A. Arbey, P. Salati, New Astron.
Rev. {\bf 46}, 791 (2002)}
\bibitem{arbey}{\small A. Arbey, J. Lesgourgues, P. Salati, Phys. Rev. D {\bf
68}, 023511 (2003)}
\bibitem{fm1}{\small T. Fukuyama, M. Morikawa, Prog. Theor. Phys.
{\bf 115}, 1047 (2006)}
\bibitem{bohmer}{\small C.G. B\"ohmer, T. Harko, J. Cosmol. Astropart. Phys.
{\bf 06}, 025 (2007)}
\bibitem{fm2}{\small T. Fukuyama, M. Morikawa, T. Tatekawa, J. Cosmol.
Astropart. Phys. {\bf 06}, 033 (2008)}
\bibitem{bmn}{\small A. Bernal, T. Matos, D. N\'u\~nez, Rev. Mex. Astron.
Astrofis.  {\bf 44}, 149 (2008)}
\bibitem{fm3}{\small T. Fukuyama, M. Morikawa, Phys. Rev. D {\bf 80}, 063520
(2009)}
\bibitem{sikivie}{\small P. Sikivie, Q. Yang, Phys. Rev. Lett. {\bf  103},
111301 (2009)}
\bibitem{mvm}{\small T. Matos, A. V\'azquez-Gonz\'alez,
J. Maga\~na, Mon. Not. R. Astron. Soc. {\bf 393}, 1359 (2009)}
\bibitem{lee09}{\small J.W. Lee, Phys. Lett. B {\bf 681}, 118 (2009)}
\bibitem{ch1}{\small T.P. Woo, T. Chiueh, Astrophys. J. {\bf 697}, 850 (2009)}
\bibitem{lee}{\small J.W. Lee, S. Lim, J. Cosmol. Astropart. Phys.  {\bf 01},
007 (2010)}
\bibitem{prd1}{\small P.H. Chavanis, Phys. Rev. D {\bf 84}, 043531 (2011)}
\bibitem{prd2}{\small P.H. Chavanis, L. Delfini, Phys. Rev. D {\bf 84}, 043532
(2011)}
\bibitem{prd3}{\small P.H. Chavanis, Phys. Rev. D {\bf 84}, 063518 (2011)}
\bibitem{briscese}{\small F. Briscese, Phys. Lett. B
{\bf 696}, 315 (2011)}
\bibitem{harkocosmo}{\small T. Harko, Mon. Not. R. Astron. Soc. {\bf 413}, 3095
(2011)}
\bibitem{harko}{\small T. Harko, J. Cosmol. Astropart. Phys. {\bf 05}, 022
(2011)}
\bibitem{abrilMNRAS}{\small A. Su\'arez, T. Matos, Mon. Not. R. Astron. Soc.
{\bf 416}, 87 (2011)}
\bibitem{aacosmo}{\small P.H. Chavanis, Astron. Astrophys. {\bf 537}, A127
(2012)}
\bibitem{velten}{\small H. Velten, E. Wamba, Phys. Lett. B {\bf 709}, 1
(2012)}
\bibitem{pires}{\small M.O.C. Pires, J.C.C. de Souza, J. Cosmol. Astropart.
Phys. {\bf 11} (2012) 024}
\bibitem{park}{\small C.-G. Park, J.-C. Hwang, H. Noh, Phys. Rev. D {\bf 86},
083535 (2012)}
\bibitem{rmbec}{\small V.H. Robles, T. Matos, Monthly Not. Roy. Astron. Soc.
{\bf 422},
282 (2012)}
\bibitem{rindler}{\small T. Rindler-Daller, P. R. Shapiro, Monthly Not. Roy.
Astron.
Soc. {\bf 422}, 135 (2012)}
\bibitem{lora}{\small V. Lora, J. Maga\~na, A. Bernal, F.J. S\'anchez-Salcedo,
E.K.
Grebel, J. Cosmol. Astropart. Phys.  {\bf  02}, 011 (2012)}
\bibitem{abrilJCAP}{\small J. Maga\~na, T. Matos, A. Su\'arez,
F. J. S\'anchez-Salcedo, JCAP {\bf 10}, 003 (2012)}
\bibitem{mhh}{\small G. Manfredi, P.A. Hervieux, F. Haas, Class. Quantum Grav.
{\bf 30}, 075006 (2013)}
\bibitem{lensing}{\small A.X. Gonz\'alez-Morales, A. Diez-Tejedor, L.A.
Ure\~na-L\'opez, O. Valenzuela, Phys. Rev. D {\bf 87}, 021301(R) (2013)}
\bibitem{glgr1}{\small F.S. Guzm\'an, F.D. Lora-Clavijo, J.J.
Gonz\'alez-Avil\'es, F.J. Rivera-Paleo, J. Cosmol. Astropart. Phys. {\bf 09}
(2013) 034}
\bibitem{ch2}{\small H.Y. Schive, T. Chiueh, T. Broadhurst, Nature Physics {\bf
10}, 496 (2014)}
\bibitem{ch3}{\small H.Y. Schive {\it et al.}, Phys. Rev. Lett. {\bf 113},
261302 (2014)}
\bibitem{shapiro}{\small B. Li, T. Rindler-Daller, P.R. Shapiro, Phys. Rev. D
{\bf 89}, 083536 (2014)}
\bibitem{bettoni}{\small D. Bettoni, M. Colombo, S. Liberati, JCAP
{\bf 02}, 004 (2014)}
\bibitem{lora2}{\small V. Lora, J. Maga\~na, JCAP
{\bf 09}, 011 (2014)}
\bibitem{mlbec}{\small P.H. Chavanis,  Eur. Phys. J. Plus {\bf 130}, 180 (2015)}
\bibitem{madarassy}{\small E.J.M. Madarassy, V.T. Toth,  Phys. Rev. D {\bf 91},
044041 (2015)}
\bibitem{abrilph}{\small A. Su\'arez, P.H. Chavanis,  Phys. Rev. D {\bf 92},
023510 (2015)}
\bibitem{playa}{\small A. Su\'arez, P.H. Chavanis, J. Phys.: Conf. Series {\bf
654}, 012088 (2015)}
\bibitem{stiff}{\small P.H. Chavanis,  Phys. Rev. D {\bf 92},
103004 (2015)}
\bibitem{guth}{\small A.H. Guth, M.P. Hertzberg, C. Prescod-Weinstein,  Phys.
Rev. D {\bf 92},
103513 (2015)}
\bibitem{souza}{\small J.C.C. de Souza, M. Ujevic, Gen. Relat. Grav. {\bf 47},
100 (2015)}
\bibitem{freitas}{\small R.C. de Freitas, H. Velten, Eur. Phys. J. C {\bf 75},
597 (2015)}
\bibitem{alexandre}{\small J. Alexandre,  Phys. Rev. D {\bf 92},
123524 (2015)}
\bibitem{schroven}{\small K. Schroven, M. List, C. L\"ammerzahl,  Phys. Rev. D
{\bf 92}, 124008 (2015)}
\bibitem{pop}{\small D. Marsh, A.R. Pop, Monthly Not. Roy. Astron. {\bf 451},
2479 (2015)}
\bibitem{eby}{\small J. Eby, C. Kouvaris, N.G. Nielsen, L.C.R. Wijewardhana, 
JHEP {\bf 02}, 028 (2016)}
\bibitem{cembranos}{\small J.A.R. Cembranos, A.L. Maroto, S.J. N\'u\~nez
Jare\~no, JHEP {\bf 03}, 013 (2016)}
\bibitem{braaten}{\small E. Braaten, A. Mohapatra, H. Zhang, Phy. Rev. Lett.
{\bf 117}, 121801 (2016)}
\bibitem{davidson}{\small S. Davidson, T. Schwetz, Phys. Rev. D {\bf 93},
123509 (2016)}
\bibitem{schwabe}{\small B. Schwabe, J. Niemeyer, J. Engels, Phys. Rev. D {\bf
94}, 043513 (2016)}
\bibitem{fan}{\small J. Fan, Phys. Dark Univ. {\bf 14}, 84 (2016)}
\bibitem{calabrese}{\small E. Calabrese, D.N. Spergel, Monthly Not. Roy. Astron.
Soc. {\bf 460}, 4397 (2016)}
\bibitem{bectcoll}{\small P.H. Chavanis,  Phys. Rev. D {\bf 94},
083007 (2016)}
\bibitem{cotner}{\small E. Cotner, Phys. Rev. D {\bf 94}, 063503 (2016)}
\bibitem{chavmatos}{\small P.H. Chavanis, T. Matos, Eur. Phys. J. Plus {\bf
132}, 30 (2017)}
\bibitem{hui}{\small L. Hui, J. Ostriker, S. Tremaine, E. Witten, Phys. Rev. D
{\bf 95}, 043541 (2017)}
\bibitem{zhang}{\small J. Zhang, Y.S. Tsai, K. Cheung, M. Chu,
arXiv:1611.00892}
\bibitem{tkachevprl}{\small D.G. Levkov, A.G.  Panin, I.I.
Tkachev, Phys. Rev. Lett. {\bf 118}, 011301 (2017)}
\bibitem{suarezchavanis3}  {\small A. Su\'arez, P.H. Chavanis,
Phys. Rev. D {\bf 95}, 063515 (2017)}
\bibitem{shapironew}{\small B. Li, T. Rindler-Daller, P.R. Shapiro,
arXiv:1611.07961}
\bibitem{phi6}{\small P.H. Chavanis, arXiv:1710.06268}
\bibitem{abriljeans}{\small A. Su\'arez, P.H. Chavanis, arXiv:1710.10486}
\bibitem{desjacques}  {\small V. Desjacques, A. Kehagias, A. Riotto,
Phys. Rev. D {\bf 97}, 023529 (2018)}
\bibitem{kc}{\small J.E. Kim, G. Carosi, Rev. Mod. Phys. {\bf 82}, 557
(2010)}
\bibitem{pq}{\small R.D. Peccei, H.R. Quinn, Phys. Rev. Lett. {\bf 38}, 1440
(1977)}
\bibitem{seidel94}{\small E. Seidel, W.M. Suen, Phys. Rev. Lett.
{\bf  72}, 2516 (1994)}
\bibitem{gul0}{\small  F.S. Guzm\'an, L.A. Ure\~na-L\'opez, Phys. Rev. D {\bf
69}, 124033  (2004)}
\bibitem{gul}{\small  F.S. Guzm\'an, L.A. Ure\~na-L\'opez, Astrophys. J. {\bf
645}, 814  (2006)}
\bibitem{bt}{\small J. Binney, S. Tremaine, Galactic Dynamics (Princeton, NJ:
Princeton University Press, 1987)}
\bibitem{lb}{\small D. Lynden-Bell, Mon. Not. R. Astron. Soc. {\bf 136}, 101
(1967)}
\bibitem{csr} {\small P.H. Chavanis, J. Sommeria, R. Robert,
Astrophys. J. {\bf 471}, 385 (1996)}
\bibitem{mnras}  {\small P.H. Chavanis,  Mon. Not. R. Astron. Soc. {\bf 300},
981 (1998)}
\bibitem{dubrovnik}  {\small P.H. Chavanis, {\it Statistical mechanics of
violent relaxation in stellar systems}, in Multiscale Problems in Science and
Technology, edited by N. Antoni\'c, C.J. van Duijn, W. J\"ager, and A. Mikeli\'c
(Springer, 2002)}
\bibitem{bdo}{\small P.H. Chavanis, Eur. Phys. J. Plus {\bf 132}, 248
(2017)}
\bibitem{nottalechaos}{\small P.H. Chavanis, arXiv:1706.05900}
\bibitem{forthcoming}{\small P.H. Chavanis, in preparation}
\bibitem{gross1}{\small E.P. Gross, Ann. of Phys. {\bf 4}, 57 (1958)}
\bibitem{gross2}{\small E.P. Gross, Nuovo
Cimento {\bf 20}, 454 (1961)}
\bibitem{gross3}{\small E.P. Gross, J.
Math. Phys. {\bf 4}, 195 (1963)}
\bibitem{pitaevskii2}{\small L.P. Pitaevskii, Sov. Phys. JETP {\bf
13}, 451 (1961)}
\bibitem{nottale}  {\small  L. Nottale, {\it Scale Relativity and Fractal
Space-Time}, Imperial College Press (2011) }
\bibitem{madelung}{\small E. Madelung, Zeit. F. Phys. {\bf 40}, 322 (1927)}
\bibitem{pre11}{\small P.H. Chavanis, Phys. Rev. E {\bf 84}, 031101
(2011)}
\bibitem{nfp}{\small P.H. Chavanis, Eur. Phys. J. B {\bf 62}, 179 (2008)}
\bibitem{entropy}{\small P.H. Chavanis, Entropy {\bf 17}, 3205 (2015)}
\bibitem{sp}{\small P.H. Chavanis, C. Rosier, C. Sire, Phys. Rev. E {\bf
66}, 036105 (2002); C. Sire, P.H. Chavanis, Phys. Rev. E {\bf
66}, 046133 (2002); P.H. Chavanis, C. Sire,  Phys. Rev. E {\bf 69}, 016116
(2004); C. Sire, P.H. Chavanis, Phys. Rev. E {\bf
69}, 066109 (2004)}
\bibitem{paddy}  {\small T. Padmanabhan,  Phys. Rep.  {\bf 188}, 285 (1990)}
\bibitem{ledoux}{\small P. Ledoux, C.L. Pekeris, Astrophys. J. {\bf 94}, 124
(1941)}
\bibitem{rb}{\small R. Ruffini, S. Bonazzola, Phys. Rev. {\bf  187}, 1767
(1969)}
\bibitem{chandra}{\small S. Chandrasekhar, An Introduction to the Study of
Stellar Structure (Dover, 1958)}
\bibitem{tkachev}{\small I.I. Tkachev, Sov. Astron. Lett. {\bf  12}, 305
(1986)}
\bibitem{maps}{\small M. Membrado, J. Abad, A.F. Pacheco, J. Sa\~nudo, Phys.
Rev. D {\bf 40}, 2736 (1989)}
\bibitem{nlb}{\small P. Natarajan, D. Lynden-Bell, Mon. Not. R. Astron. Soc.
{\bf 286}, 268 (1997)}
\bibitem{wares}{\small G.W. Wares, Astrophys. J. {\bf 100}, 158 (1944)}
\bibitem{margrave}{\small T. Margrave, Astrophys. Space Sci. {\bf 2}, 504
(1968)}
\bibitem{ht}  {\small  P. Hertel and W. Thirring,  in: Quanten und Felder,
edited by H.P. D\"urr (Vieweg, Braunschweig, 1971)}
\bibitem{bvriper}{\small S.A. Bludman, K.A. Van Riper, Astrophys. J. {\bf 212},
859
(1977)}
\bibitem{em}{\small T.W. Edwards, M.P. Merilan, Astrophys. J. {\bf 244}, 600
(1981)}
\bibitem{edwards}{\small T.W. Edwards, Astrophys. J. {\bf 288}, 630
(1985)}
\bibitem{sg}{\small Z. Slepian, J. Goodman, Mon. Not. R. Astron. Soc.
{\bf 427}, 839 (2012)}
\bibitem{lin}{\small S.C. Lin, H.Y. Schive, S.K. Wong, T. Chiueh, Phys. Rev. D
{\bf 97}, 103523 (2018)}
\bibitem{moczchavanis}{\small P. Mocz, L. Lancaster, A. Fialkov, F. Becerra,
P.H. Chavanis, Phys. Rev. D {\bf 97}, 083519 (2018)}
\bibitem{kormendy}{\small J. Kormendy, K.C. Freeman, in S.D. Ryder, D.J. Pisano,
M.A. Walker, K.C. Freeman, eds.,  Proc. IAU Symp. 220, Dark Matter in Galaxies.
Astron. Soc. Pac., San Francisco, p. 377 (2004)}
\bibitem{spano}{\small M. Spano, M. Marcelin, P. Amram, C. Carignan, B. Epinat,
O. Hernandez, Mon. Not. R. Astron. Soc. {\bf 383}, 297 (2008)}
\bibitem{donato}{\small F. Donato {\it et al.}, Mon. Not. R. Astron. Soc. {\bf
397}, 1169 (2009)}
\bibitem{strigari}{\small L.E. Strigari {\it et al.}, Nature {\bf 454}, 1096
(2008)}
\bibitem{martino}{\small I. De Martino {\it et al.}, arXiv:1807.08153}
\bibitem{zoccali}{\small M. Zoccali {\it et al.}, Astron. Astrophys. {\bf 562},
A66 (2014)}
\bibitem{portail}{\small M. Portail {\it et al.},  Mon. Not. R. Astron. Soc.
{\bf 465}, 1621 (2017)}
\bibitem{gillessen}{\small S. Gillessen {\it et al.}, Astrophys.
J. {\bf 707}, L114 (2009)}
\bibitem{nature} {\small R. Sch\"odel {\it et al.}, Nature {\bf 419}, 694
(2002)}
\bibitem{reid} {\small M.J. Reid, Int. J. Mod. Phys. D {\bf 18}, 889 (2009)}
\bibitem{genzel}{\small R. Genzel, F. Eisenhauer, S. Gillessen, Rev. Mod. Phys.
{\bf 82}, 3121 (2010)}
\bibitem{torres2000}{\small  D.F. Torres, S. Capozziello, G. Lambiase, Phys.
Rev. D {\bf 62}, 104012  (2000)}
\bibitem{guzmanbh}{\small  F.S. Guzm\'an, Phys. Rev. D {\bf 73}, 021501  (2006)}
\bibitem{ac}{\small G. Alberti, P.H. Chavanis, arXiv:1808.01007}
\bibitem{poincare}  {\small H. Poincar\'e, Acta Math. {\bf 7}, 259 (1885)}
\bibitem{katzpoincare}  {\small J. Katz,  Mon. Not. R. Astron. Soc. {\bf 183},
765 (1978)}
\bibitem{lifetime}  {\small P.H. Chavanis, Astron. Astrophys. {\bf 432}, 117
(2005)}
\bibitem{campa}{\small A. Campa, T. Dauxois, D. Fanelli, S. Ruffo, {\it
Physics of Long-Range Interacting Systems} (Oxford University Press, 2014)}
\bibitem{naso}  {\small A. Naso, P.H. Chavanis, B. Dubrulle, Eur. Phys. J. B 
{\bf 77}, 187 (2010)}
\bibitem{bft}  {\small B. Bar-Or, J.B. Fouvry, S. Tremaine, arXiv:1809.07673}
\bibitem{aa}  {\small P.H. Chavanis, Astron. Astrophys. {\bf 381}, 340 (2002)}
\bibitem{lbw}  {\small  D. Lynden-Bell, R. Wood, Mon. Not. R. Astron. Soc.
{\bf 138}, 495 (1968)}
\bibitem{balberg} {\small S. Balberg, S.L. Shapiro, S. Inagaki, Astrophys. J.
{\bf 568}, 475 (2002)}
\bibitem{zp} {\small Ya. B. Zel'dovich, M.A. Podurets, Sov. Astron.
{\bf 9}, 742 (1966)}
\bibitem{fit}  {\small D. Fackerell, J. Ipser, K. Thorne, Comments Ap. and Space
Phys. {\bf 1}, 134 (1969)}
\bibitem{st2} {\small S.L. Shapiro, S.A. Teukolsky,  Astrophys. J.
{\bf 298}, 58 (1985)}
\bibitem{st3} {\small S.L. Shapiro, S.A. Teukolsky,  Astrophys. J. 
{\bf 292}, L41 (1985)}
\bibitem{st4} {\small S.L. Shapiro, S.A. Teukolsky,  Astrophys. J. 
{\bf 307}, 575 (1986)}
\bibitem{lbe}  {\small D. Lynden-Bell, P.P. Eggleton, Mon. Not. R. Astron. Soc.
{\bf 191}, 483 (1980)}
\bibitem{cohn} {\small H. Cohn, Astrophys. J. {\bf 242}, 765 (1980)}
\bibitem{inagaki}  {\small S. Inagaki, D. Lynden-Bell, Mon. Not. R. Astron. Soc.
{\bf 205}, 913 (1983)}
\bibitem{oscillations} {\small D. Sugimoto, E. Bettwieser,   Mon. Not. R.
Astron. Soc. {\bf 204}, 19 (1983)}
\bibitem{hr} {\small D. Heggie, N. Ramamani, Mon. Not. R. Astron. Soc. {\bf
237}, 757 (1989)}
\bibitem{bullet}{\small S.W. Randall, M. Markevitch, D. Clowe, A.H. Gonzalez, M.
Bradac, Astrophys. J. {\bf 679}, 1173 (2008)}
\bibitem{dave}{\small R. Dav\'e, D.N. Spergel, P.J. Steinhardt, B.J. Wandelt,
Astrophys. J. {\bf 547}, 574 (2001)}
\bibitem{ferrarese}  {\small L. Ferrarese, Astrophys. J. {\bf 578}, 90
(2002)}
\bibitem{henonVR}  {\small M. H\'enon, Ann. Astrophys. {\bf 27}, 83 (1964)}
\bibitem{albada} {\small T.S. van Albada, Mon. Not. R. Astron. Soc. {\bf 201},
939 (1982)}
\bibitem{roy} {\small F. Roy, J. Perez, Mon. Not. R. Astron. Soc. {\bf 348},
62 (2004)}
\bibitem{joyce} {\small M. Joyce, B. Marcos, F. Sylos Labini, Mon. Not. R.
Astron. Soc. {\bf 397},
775 (2009)}
\bibitem{bertin1} {\small G. Bertin, M. Stiavelli, Astron. Astrophys. {\bf 137},
26 (1984)}
\bibitem{bertin2} {\small M. Stiavelli, G. Bertin, Mon. Not. R. Astron. Soc.  
{\bf 229},
61 (1987)}
\bibitem{hjorth} {\small J. Hjorth, J. Madsen, Mon. Not. R. Astron. Soc.   {\bf
253},
703 (1991)}
\bibitem{isochrone}  {\small M. H\'enon, Ann. Astrophys. {\bf 22}, 126 (1959)}
\bibitem{king} {\small I.R. King, Astron. J. {\bf 70}, 376 (1965)}
\bibitem{nelson} {\small E. Nelson, Phys. Rev.  {\bf 150}, 1079 (1966)}
\bibitem{chavnot}{\small P.H. Chavanis,  Eur. Phys. J. Plus 
{\bf
132}, 286 (2017)}
\bibitem{einstein}  {\small A. Einstein, Ann. Physik  {\bf 17}, 549 (1905)}
\bibitem{colpi}{\small M. Colpi, S.L. Shapiro, I. Wasserman, Phys. Rev. Lett.
{\bf  57}, 2485 (1986)}
\bibitem{chavharko}{\small P.H. Chavanis, T. Harko, Phys. Rev. D {\bf 86},
064011 (2012)}
\bibitem{kaup}{\small D.J. Kaup, Phys. Rev. {\bf  172}, 1331 (1968)}
\bibitem{ov}{\small J.R. Oppenheimer, G.M. Volkoff, Phys. Rev. {\bf 55}, 374
(1939)}
\bibitem{mocz}{\small P. Mocz {\it et al.}, Mon. Not. R. Astron. Soc. 
{\bf 471}, 4559 (2017)}
\bibitem{tf}{\small R.B. Tully, J.R. Fisher,  Astron. Astrophys.
{\bf 54}, 661 (1977)}
\bibitem{mcgaugh}{\small S.S. McGaugh, Astron. J.
{\bf 143}, 40 (2012)}
\bibitem{rbb}{\small V.H. Robles, J.S. Bullock, M. Boylan-Kolchin,
arXiv:1807.06018}
\bibitem{ar}{\small C.R. Arg\"uelles, R. Ruffini, Internat. J. Modern Phys. D 
{\bf 23}, 42020 (2014)}
\bibitem{deng}{\small H. Deng {\it et al.}, Phys. Rev. D {\bf 98}, 023513
(2018)}
\bibitem{logotrope}{\small P.H. Chavanis, Eur. Phys. J. Plus {\bf 130}, 130
(2015)}














\end{thebibliography}
\end{document}